\documentclass[letterpaper,twoside,12pt]{report}

\usepackage[margin=3.1cm]{geometry}
\usepackage{fancyhdr}

\usepackage{hyperref}
\usepackage{appendix}
\usepackage[sort]{natbib}
\usepackage{ifthen}
\usepackage{multirow}
\usepackage{color}

\usepackage{times}
\usepackage{graphicx,epic,eepic,epsfig,amsmath,latexsym,amssymb,amstext,verbatim,revsymb}

\renewcommand{\comment}[1]{}

\usepackage{amsmath,amssymb,amsfonts}
\usepackage{amsthm}

\newtheorem{theorem}{Theorem}[section]

\newtheorem{corollary}[theorem]{Corollary}

\newtheorem{definition}[theorem]{Definition}
\newtheorem{lemma}[theorem]{Lemma}

\newtheorem{proposition}[theorem]{Proposition}

\theoremstyle{remark}
\newtheorem*{claim}{Claim}
\newtheorem*{myremark}{Remark}

\theoremstyle{theorem}

\def\squareforqed{\hbox{\rlap{$\sqcap$}$\sqcup$}}
\def\qed{\ifmmode\squareforqed\else{\unskip\nobreak\hfil
\penalty50\hskip1em\null\nobreak\hfil\squareforqed
\parfillskip=0pt\finalhyphendemerits=0\endgraf}\fi}
\def\endenv{\ifmmode\;\else{\unskip\nobreak\hfil
\penalty50\hskip1em\null\nobreak\hfil\;
\parfillskip=0pt\finalhyphendemerits=0\endgraf}\fi}

\renewenvironment{proof}{\noindent \textbf{{Proof~} }}{\qed\medskip}
\newenvironment{proof+}[1]{\noindent \textbf{{Proof #1~} }}{\qed\medskip}
\newenvironment{remark}{\noindent \textit{{Remark.~}}}{\qed}

\mathchardef\ordinarycolon\mathcode`\:
\mathcode`\:=\string"8000
\def\vcentcolon{\mathrel{\mathop\ordinarycolon}}
\begingroup \catcode`\:=\active
  \lowercase{\endgroup
  \let :\vcentcolon
  }

\newcounter{protoCount}
\newcounter{protoList}
\newsavebox{\tmpbox}
\newlength{\protobox}

\newenvironment{protocol}[3]{
\bigskip
\addtocounter{protoCount}{1}
\noindent \begin{lrbox}{\tmpbox}
\setlength{\protobox}{\textwidth}
\addtolength{\protobox}{-0.4cm}
\begin{minipage}[c]{\protobox}
\begin{bfseries}Protocol #1: #2\end{bfseries}
\ifthenelse{\equal{#3}{\empty}}{}{\\ #3}
\begin{list}{\begin{bfseries}\arabic{protoList}:\end{bfseries}}
{\usecounter{protoList}}
}{
\end{list}
\end{minipage}\end{lrbox}
\fbox{\usebox{\tmpbox}}
\bigskip
}

\newcommand{\norm}[1]{\|{ #1 }\|_2}

\newcommand{\nc}{\newcommand}
\nc{\rnc}{\renewcommand}
\nc{\beq}{\begin{equation}}
\nc{\eeq}{{\end{equation}}}
\nc{\beqa}{\begin{eqnarray}}
\nc{\eeqa}{\end{eqnarray}}
\nc{\lbar}[1]{\overline{#1}}
\nc{\bra}[1]{\langle#1|}
\nc{\ket}[1]{|#1\rangle}
\nc{\ketbra}[2]{|#1\rangle\!\langle#2|}
\nc{\braket}[2]{\langle#1|#2\rangle}
\nc{\inp}[2]{\langle#1|#2\rangle}
\nc{\proj}[1]{| #1\rangle\!\langle #1 |}
\nc{\avg}[1]{\langle#1\rangle}
\nc{\smfrac}[2]{\mbox{$\frac{#1}{#2}$}}
\nc{\tr}{\operatorname{tr}}
\nc{\tracedist}[1]{\Delta_{}\!\left( #1 \right)}
\nc{\fid}[1]{F\!\left( #1 \right)}
\nc{\pd}{P}
\nc{\gtd}{\mathrm{D}}
\nc{\gfid}{\bar{\mathrm{F}}}

\def\01{\{0,1\}}

\newcommand{\states}{\mathcal{S}}
\newcommand{\substates}{\mathcal{S}_{\leq}}
\newcommand{\hmin}{\entHmin}

\newcommand{\meas}{{\mathcal{M}}}

\newcommand{\hil}{\mathcal{H}}

\nc{\ox}{\otimes}
\nc{\dg}{\dagger}
\nc{\dn}{\downarrow}
\nc{\cA}{{\cal A}}
\nc{\cB}{{\cal B}}
\nc{\cC}{{\cal C}}
\nc{\cD}{{\cal D}}
\nc{\cE}{{\mathcal E}}
\nc{\cF}{{\cal F}}
\nc{\cG}{{\cal G}}
\nc{\cH}{{\cal H}}
\nc{\cI}{{\cal I}}
\nc{\cJ}{{\cal J}}
\nc{\cK}{{\cal K}}
\nc{\cL}{{\cal L}}
\nc{\cM}{{\cal M}}
\nc{\cN}{{\cal N}}
\nc{\cO}{{\cal O}}
\nc{\cP}{{\cal P}}
\nc{\cQ}{{\cal Q}}
\nc{\cR}{{\cal R}}
\nc{\cS}{{\cal S}}
\nc{\cT}{{\cal T}}
\nc{\cU}{{\cal U}}
\nc{\cV}{{\cal V}}

\nc{\cX}{{\cal X}}
\nc{\cY}{{\cal Y}}
\nc{\cZ}{{\cal Z}}

\nc{\sA}{{\sf A}}
\nc{\sB}{{\sf B}}
\nc{\sU}{{\sf U}}
\nc{\sD}{{\sf D}}
\nc{\sH}{{\sf H}}
\nc{\sL}{{\sf L}}
\nc{\sR}{{\sf R}}
\nc{\sV}{{\sf V}}
\nc{\sM}{{\sf M}}

\nc{\entI}{{\bf I}}
\nc{\entIarrow}{{\bf I}^{\leftarrow}}
\nc{\entH}{{\bf H}}
\nc{\entS}{{\bf S}}
\nc{\entHmin}{\mathbf{H}_{\min}}
\nc{\entHtwo}{\mathbf{H}_{2}}
\nc{\entHmax}{\mathbf{H}_{\max}}
\nc{\pguess}{P_{\rm guess}}

\nc{\binent}{h_2}

\nc{\divz}{\mathbf{D}_{0}}
\nc{\macdiv}{\mathbf{M}_{0}}

\newcommand{\hin}{\mathcal{H}_{\rm in}}
\newcommand{\hout}{\mathcal{H}_{\rm out}}
\newcommand{\msg}{M}

\newcommand{\sym}{ { \rm sym } }

\nc{\entF}{{\bf E}_f}

\nc{\isom}{\simeq}

\nc{\re}{\mathrm{Re}}
\nc{\im}{\mathrm{Im}}

\nc{\rank}{\operatorname{rank}}
\nc{\rar}{\rightarrow}
\nc{\lrar}{\longrightarrow}
\nc{\polylog}{\operatorname{polylog}}
\nc{\poly}{\operatorname{poly}}
\nc{\1}{\id}

\nc{\weight}{\textbf{w}}
\nc{\hamdist}{d_{H}}

\nc{\eps}{\epsilon}

\def\e{\epsilon}

\def\ph{\varphi}

\nc{\Sp}{{{\mathbb S}}}
\nc{\RR}{{{\mathbb R}}}
\def\Real{\mathbb{R}}

\nc{\CC}{{{\mathbb C}}}
\nc{\FF}{{{\mathbb F}}}
\nc{\NN}{{{\mathbb N}}}
\nc{\ZZ}{{{\mathbb Z}}}
\nc{\PP}{{{\mathbb P}}}
\nc{\QQ}{{{\mathbb Q}}}
\nc{\UU}{{{\mathbb U}}}
\nc{\OO}{{{\mathbb O}}}
\nc{\EE}{{{\mathbb E}}}
\nc{\id}{{\operatorname{id}}}

\nc{\qubitchannel}{\id_2}
\nc{\bitchannel}{\overline{\id}_2}
\nc{\clchannel}{\overline{\id}}

\newcommand{\epsball}{\mathcal{B}^{\eps}}

\nc{\be}{\begin{equation}}
\nc{\ee}{{\end{equation}}}
\nc{\bea}{\begin{eqnarray}}
\nc{\eea}{\end{eqnarray}}
\nc{\<}{\langle}
\rnc{\>}{\rangle}
\nc{\Hom}[2]{\mbox{Hom}(\CC^{#1},\CC^{#2})}
\nc{\rU}{\mbox{U}}

\nc{\ob}[1]{#1}

\nc{\Ext}{Ext}

\newcommand{\eqdef}	{\stackrel{\textrm{def}}{=}}

\newcommand{\ex}[1]	{\mathbf{E}\left\{ #1 \right\}}
\newcommand{\exc}[2]	{\mathbf{E}_{#1}\left\{ #2 \right\}}

\newcommand{\prob}[1]	{\mathbf{Pr}\left\{ #1 \right\}}
\newcommand{\pr}[1]	{\prob{#1}}

\newcommand{\probc}[2]	{\mathbf{Pr}_{#1}\left\{ #2 \right\}}

\newcommand{\event}[1]	{\left[ #1 \right]}

\renewcommand{\exp}[1]	{\operatorname{exp}\left( #1 \right)}

\newcommand{\floor}[1]	{\left\lfloor #1 \right\rfloor}
\newcommand{\ceil}[1]	{\left\lceil #1 \right\rceil}

\newcommand{\comp}{\circ}
\newcommand{\concat}{\cdot}
\nc{\unif}{\textrm{unif}}

\begin{document}


\begin{titlepage}
\begin{center}
\huge Uncertainty relations for multiple measurements with applications \\
\vfill
\large Omar Fawzi \\
\vfill
\textit{School of Computer Science\\
McGill University, Montr\'eal\\}
August, 2012\\
\vfill
A thesis submitted to McGill University in partial fulfillment of the requirements of the degree of PhD.\\
\vfill
Copyright \copyright Omar Fawzi 2012.
\end{center}
\vspace*{0.5 in}
\end{titlepage}

\pagenumbering{roman}
\setcounter{page}{1}


\tableofcontents

\thispagestyle{plain}
\chapter*{Abstract}
\addcontentsline{toc}{section}{Abstract}
\vspace{-0.7cm}
Uncertainty relations express the fundamental incompatibility of certain observables in quantum mechanics. Far from just being puzzling constraints on our ability to know the state of a quantum system, uncertainty relations are at the heart of why some classically impossible cryptographic primitives become possible when quantum communication is allowed. This thesis is concerned with strong notions of uncertainty relations and their applications in quantum information theory.

One operational manifestation of such uncertainty relations is a purely quantum effect referred to as \emph{information locking}. A locking scheme can be viewed as a cryptographic protocol in which a uniformly random $n$-bit message is encoded in a quantum system using a classical key of size much smaller than $n$. Without the key, no measurement of this quantum state can extract more than a negligible amount of information about the message, in which case the message is said to be  ``locked''.  Furthermore, knowing the key, it is possible to recover, that is ``unlock'', the message.
We give new efficient constructions of bases satisfying strong uncertainty relations leading to the first explicit construction of an information locking scheme. We also give several other applications of our uncertainty relations both to cryptographic and communication tasks.

In addition, we define objects called QC-extractors, that can be seen as strong uncertainty relations that hold against quantum adversaries. We provide several constructions of QC-extractors, and use them to prove the security of cryptographic protocols for two-party computations based on the sole assumption that the parties' storage device is limited in transmitting quantum information. In doing so, we resolve a central question in the so-called noisy-storage model by relating security to the quantum capacity of storage devices.

\chapter*{\vspace{-2cm}R\'{e}sum\'{e} \vspace{-1cm}}
\addcontentsline{toc}{section}{R\'{e}sum\'{e}}

Les relations d'incertitude expriment l'incompatibilit\'{e} de certaines observables en m\'{e}canique quantique. Les relations d'incertitude sont utiles pour comprendre pourquoi certaines primitives cryptographiques impossibles dans le monde classique deviennent possibles avec de la communication quantique. Cette th\`{e}se \'{e}tudie des notions fortes de
relations d'incertitude et leurs applications \`{a} la th\'{e}orie de l'information quantique.

Une manifestation op\'{e}rationnelle de telles relations d'incertitude est un effet purement quantique appel\'{e} verrouillage d'information. Un syst\`{e}me de verrouillage peut \^{e}tre consid\'{e}r\'{e} comme un protocole cryptographique dans lequel un message al\'{e}atoire compos\'{e} de $n$ bits est encod\'{e} dans un syst\`{e}me quantique en utilisant une cl\'{e} classique de taille beaucoup plus petite que $n$. Sans la cl\'{e}, aucune mesure sur cet \'{e}tat quantique ne peut extraire plus qu'une quantit\'{e} n\'{e}gligeable d'information sur le message, auquel cas le message est ``verrouill\'{e}''. Par ailleurs, connaissant la cl\'{e}, il est possible de r\'{e}cup\'{e}rer ou ``d\'{e}verrouiller'' le message. Nous proposons de nouvelles constructions efficaces de bases v\'{e}rifiant de fortes relations d'incertitude conduisant \`{a} la premi\`{e}re construction explicite d'un syst\`{e}me de verrouillage. Nous exposons \'{e}galement plusieurs autres applications de nos relations d'incertitude \`{a} des t\^{a}ches cryptographiques et des t\^{a}ches de communication.

Nous d\'{e}finissons \'{e}galement des objets appel\'{e}s QC-extracteurs, qui peuvent \^{e}tre consid\'{e}r\'{e}s comme de fortes relations d'incertitude qui tiennent contre des adversaires quantiques. Nous fournissons plusieurs constructions de QC-extracteurs, que nous utilisons pour prouver la s\'{e}curit\'{e} de protocoles cryptographiques pour le calcul s\'{e}curis\'{e} \`{a} deux joueurs en supposant uniquement que la m\'{e}moire des joueurs soit limit\'{e}e en ce qui concerne la transmission d'information quantique. Ce faisant, nous r\'{e}solvons une question centrale dans le mod\`{e}le de m\'{e}moire bruit\'{e}e en mettant en relation la s\'{e}curit\'{e} et la capacit\'{e} quantique de la m\'{e}moire.

\chapter*{Contents of the thesis}
\addcontentsline{toc}{section}{Contents of the thesis}

This thesis is mainly based on two papers. The first one is joint work with Patrick Hayden and Pranab Sen \citep{FHS11} and is presented in Chapters \ref{chap:uncertainty-relations} and \ref{chap:applications-ur}. The second paper is presented in Chapter \ref{chap:qc-extractor} and is joint work with Mario Berta and Stephanie Wehner \citep{BFW11}.

\chapter*{\vspace{-2.2cm} Acknowledgements \vspace{-1cm}}
\addcontentsline{toc}{section}{Acknowledgements}

I was very fortunate to be supervised by Luc Devroye and Patrick Hayden during the last four years. I wish to thank them for their generosity with their time and for their guidance. I learned almost all of what I know about probability from Luc and almost all what I know about classical and quantum information theory from Patrick. I also thank them for their financial support. I also consider Pranab Sen as my supervisor. I thank him for all his extremely clear explanations.  I should also thank Claude Cr\'{e}peau, Aram Harrow and Louis Salvail for accepting to be on my committee.

I had the chance of collaborating with many other great researchers. I would like to thank my other co-authors for all what they taught me: Anil Ada, Mario Berta, Nicolas Broutin, Arkadev Chattopadhyay, Nicolas Fraiman, Phuong Nguyen, Ivan Savov, Stephanie Wehner, Mark Wilde. Special thanks to Anil and Ivan for countless hours of  entertaining discussions and Mark for carefully reading this thesis. Part of the work presented in this thesis was done while I was visiting the Center for Quantum Technologies in Singapore. I wish to thank Mikl\'{o}s S\'{a}ntha for inviting me. I also benefited from various useful discussions with many colleagues including Louigi Addario-Berry, Abdulrahman Al-lahham, Kamil Bradler, Fr\'{e}d\'{e}ric Dupuis, Nicolas Dutil, Laszlo Egri, Hamza Fawzi, Hussein Fawzi, Jan Florjanczyk, Hamed Hatami, Kassem Kalach, Ross Kang, Marc Kaplan, Jamie King, Sean Kennedy, Pascal Koiran, Christian Konrad, Nima Lashkari, Debbie Leung, Zhentao Li, Fr\'{e}d\'{e}ric Magniez, Joe Renes, Renato Renner, Mikl\'{o}s S\'{a}ntha, Thomas Vidick, Andreas Winter, J\"{u}rg Wullschleger. 

I am also grateful to all the staff of the School of Computer Science for providing an excellent environment. I certainly have to thank many friends who made life in Montr\'{e}al so enjoyable. At the risk of forgetting many, I'll try to mention some of them: Ahmad, Alaa, Kosai, Mazen, Mohammed's, Mustafa, Nour, Omar, Qusai, Rayhan, Saeed, Tamer, Tarek. And it is a pleasure to thank my parents and brothers for everything.


\chapter*{Notation}
\addcontentsline{toc}{section}{Notation}
\noindent
\begin{tabular}{p{3.7cm}p{10.5cm}}
\hline
\textbf{Common} &  \\
\hline
$\log$ & Binary logarithm. \\
$\ln$ & Natural logarithm. \\
$\RR$ & Real numbers. \\
$\CC$ & Complex numbers. \\
$M^{\dagger}$ & Conjugate transpose of the matrix $M$. \\
$[n]$ & Set $\{1, \dots, n\}$. \\
$\hamdist$ & Hamming distance $\hamdist(x,y) = \{ i : x_i \neq y_i\}$. \\
$\weight$ & Hamming weight $\weight(x) = \{ i : x_i \neq 0\}$. \\
$p_X$ & The distribution of a random variable $X$. \\
$\pr{E}$ & Probability of the event $E$. \\
$\ex{X}$ & Expectation of a random variable $X$. \\
$\exc{y}{f(x,y)}$ & Expectation over $y$ and fixed $x$. \\
$f \circ g$ & Composition of the functions $f$ and $g$. \\
\end{tabular} \\
\noindent
\begin{tabular}{p{3.7cm}p{10.5cm}}
\hline
\multicolumn{2}{l}{\textbf{Spaces}} \\
\hline
$A, B, C, \ldots$ & Hilbert spaces associated with the systems $A, B, C, \ldots$ \\
$A \isom A'$ & $A'$ is a copy of $A$. \\ 
$d_A$ & Dimension of the space $A$. \\
$AB$ &  Tensor product $A \otimes B$ or composite system $AB$.\\
$\cL(A,B)$ & Space of linear operators from $A$ to $B$. \\
$\cL(A)$ & $\cL(A,A)$. \\
\end{tabular} \\
\noindent
\begin{tabular}{p{3.7cm}p{10.5cm}}
\hline
\multicolumn{2}{l}{\textbf{Vectors}} \\
\hline
$\ket{\psi}^A, \ket{\phi}^A, \ldots$ & Vectors belonging to $A$.\\
$\bra{\psi}^A$ & Dual vectors in $\cL(A, \CC)$. \\
$\braket{\psi}{\phi}$ & Inner product of the vectors $\ket{\psi}$ and $\ket{\phi}$. \\
\end{tabular}\\
\noindent
\begin{tabular}{p{3.7cm}p{10.5cm}}
\hline
\multicolumn{2}{l}{\textbf{Operators}} \\
\hline
$\cS(A)$ & Set of density operators on $A$. \\
${\cal S}_{\leq}(A)$ & Set of sub-normalized density operators on $A$. \\
$\rho^A = \rho_A, \psi^A, \ldots$ & Density operators on $A$. \\
$\mathrm{id}_A = \id^A$ & Identity map on $A$ or $\cL(A)$.\\
$\|X\|_1$ & Trace norm of the operator $X$. \\
$\|X\|_2$ & Hilbert-Schmidt norm of the operator $X$. \\
\end{tabular}\\
\noindent
\begin{tabular}{p{3.7cm}p{10.5cm}}
\hline
\multicolumn{2}{l}{\textbf{Distance measures for operators}} \\
\hline
$\tracedist{\rho, \sigma}$ & Trace distance between $\rho$ and $\sigma$. \\
$F(\rho,\sigma)$ & Fidelity between $\rho$ and $\sigma$. \\
$\bar{F}(\rho,\sigma)$ & Generalized fidelity between $\rho$ and $\sigma$. \\
$\pd(\rho,\sigma)$ & Purified distance between $\rho$ and $\sigma$.
\end{tabular}
\\
\noindent
\begin{tabular}{p{3.7cm}p{10.5cm}}
\hline
\multicolumn{2}{l}{\textbf{Measures of information}} \\
\hline
$\entH(A)_{\rho}$ & von Neumann entropy of the density operator $\rho^A$. \\
$\entH(A|B)_{\rho}$ & Conditional von Neumann entropy of $\rho^{AB}$. \\
$\entI(A;B)_{\rho}$ & Mutual information of the density operator $\rho^{AB}$. \\
$\entHmin(A|B)_{\rho|\sigma}$ & Min-entropy of $\rho^{AB}$ relative to $\sigma^{B}$. \\
$\entHmin(A|B)_{\rho}$ & Conditional min-entropy of $\rho^{AB}$ given $B$. \\
$\entHmax(A|B)_{\rho}$ & Conditional max-entropy of $\rho^{AB}$ given $B$. \\
$\entHmin^{\e}(A|B)_{\rho}$ & Smooth min-entropy of $\rho^{AB}$ given $B$. \\
$\entHmax^{\e}(A|B)_{\rho}$ & Smooth max-entropy of $\rho^{AB}$ given $B$. \\
$\entHtwo(A|B)_{\rho|\sigma}$ & Collision entropy of $\rho^{AB}$ relative to $\sigma^{B}$. \\ 
$\binent(\e)$	& Binary entropy $\binent(\e) = -\e \log \e - (1-\e) \log(1-\e)$. \\
\end{tabular}

\chapter{Introduction}

\pagenumbering{arabic}
\pagestyle{fancy}                       
\fancyfoot{}                            
\renewcommand{\chaptermark}[1]{         
\markboth{\MakeUppercase{\chaptername\ } \thechapter.\ #1}{}}
\renewcommand{\sectionmark}[1]{        
\markright{\thesection.\ #1}}
\fancyhead[LO,RE]{\thepage}    
\fancyhead[LE]{\rightmark}      
\fancyhead[RO]{\leftmark}     


\section{Quantum information science}
Even though Turing machines are abstract mathematical constructions, they are widely believed to capture a universal notion of computation in our physical world. This is reflected by the Church-Turing thesis, which states that any computation performed on a physical device can also be performed by a Turing machine. The main reason the Church-Turing thesis is believed is that all known models for (reasonable) physical computation mechanisms were shown to be simulatable by a Turing machine. In fact, the strong Church-Turing thesis states that any computation performed efficiently by some physical device can be computed efficiently by a Turing machine.

Consider now the problem of information transmission through a physical channel. How to model such a channel? A natural answer is to associate for each possible input a probability distribution on the possible outputs of the channel. The randomness is used to model our ignorance or lack of control of some phenomena happening in the transmission. There is a feeling that a better understanding of the physical process can always be incorporated in the model by adjusting the probabilities assigned to each outcome. As for Turing machines, there is a belief that the most general way to model a physical information channel is using probability distributions.

When taking into account quantum theory, these assumptions should be re-examined. According to quantum physics, the state of a physical system, e.g., a potential computing device, need not be represented by some string of characters written on a tape, but can potentially be a superposition of many strings. Like for waves, different parts of the system could interfere with each other. Could such a model define a different notion of computation? There is by now significant evidence that this might be the case. \cite{Sho97} showed that one could make use of wave-like properties in a quantum system to factor integers efficiently, a problem that is believed to be hard for classical computers. In addition, many other computational tasks seem to be much more natural and efficiently implementable for a quantum computer, in particular concerning the simulation of quantum mechanical systems~\citep{Fey82}. For information transmission, imagine a channel that carries information using photon polarization, a property which is known to be best described by a quantum state. In this case, modelling the channel as a distribution over the outputs for each possible input is incomplete. In fact, it turns out that one can use quantum mechanical properties not only to increase the rate at which information is transmitted but also to perform tasks that are simply impossible using only ``classical'' communication.

\section{Conjugate coding}

An example of a task that becomes possible when quantum properties are used is \emph{key distribution}. Suppose Alice and Bob are far apart and they want to exchange a secret over email. To achieve unconditional security,\footnote{Unconditional security means that security doesn't rest on unproven computational assumptions.
} it is well-known that they have to share a large private key about which the adversary does not have any information. How can they obtain such a key by communicating over a public channel? This task is impossible to achieve with unconditional security using only classical communication.

In groundbreaking work, \cite{BB84} based on an idea of \cite{Wie83} devised a simple protocol for key distribution using quantum communication.\footnote{Note that this protocol can be and is implemented with today's technology. In fact, encryptors based on quantum key distribution can actually be bought from a handful of companies.} One of the key ideas of the protocol is to use ``conjugate coding'' \citep{Wie83}. Even though it cannot store (reliably) more than one bit of information, there are several ways of encoding one bit in the polarization of a photon. We can encode in the ``rectilinear'' basis, e.g., $0 \mapsto \sH$ (horizontal) and $1 \mapsto \sV$ (vertical), or in the ``diagonal'' basis, e.g., $0 \mapsto \sM$ (main diagonal) and $1 \mapsto \sA$ (anti-diagonal). This is a valid encoding because in both cases, the states corresponding to $0$ and $1$ are perfectly distinguishable. However, an observer that does not know which one of the two encodings was used cannot recover the encoded bit perfectly. In fact, if he performs a measurement in the rectilinear basis and the actual state was $\sM$ (which belongs to the diagonal basis and encodes $0$), then the result will be $\sH$ (corresponding to $0$) with probability $1/2$ and $\sV$ (corresponding to $1$) with probability $1/2$. 

We stress that this type of encoding in the polarization of a single photon does not have a classical analogue. Assume we have two perfectly distinguishable classical states $\sA$ and $\sB$. One can define two possible encodings for bits: $0 \mapsto \sA$ and $1 \mapsto \sB$, or $0 \mapsto \sB$ and $1 \mapsto \sA$. An adversary who ignores which encoding was used cannot obtain any information about the encoded bit by seeing $\sA$ or $\sB$. But given that we see $\sA$, we know that the encoded bit is $0$ if the first encoding was chosen and it is $1$ if the second encoding was chosen. For the quantum encoding described above, for all possible states $\sH, \sV, \sM$ or $\sA$, it is not possible to have a definite encoded value for both the rectilinear and diagonal bases. This is a form of the uncertainty principle: either the ``rectilinear value'' or the ``diagonal value'' of a state has to be undetermined. This idea of encoding in conjugate bases is at the heart of the whole field of quantum cryptography that takes advantage of the uncertainty principle and related ideas to guarantee privacy properties; see \citep{GRTZ02, Sca+09} for surveys. The results in this thesis can be seen as stronger versions of conjugate coding that use multiple (more than two) encodings.

The following more technical sections describe the context and the main results in this thesis.  

\section{Uncertainty relations for quantum measurements}

\subsection{Context}
The uncertainty principle was first formulated by \cite{Hei27} and it states that the position and momentum of a particle cannot both have definite values. It was then generalized by \cite{Rob29} to arbitrary observables that do not commute. 
Here, we consider modern formulations of the uncertainty principle for which the measure of uncertainty is an entropic quantity. Entropic uncertainty relations were introduced in \cite{Hir57, BM75,Deu83} and have found many applications in quantum information theory. For example, such relations are the main ingredients in the proofs of security of protocols for two-party computations in the bounded and noisy quantum storage models \citep{DFSS05,DFRSS07,KWW09}. A simple example of an entropic uncertainty relation was given by \cite{MU88}. Let $\cB_{+}$ denote a ``rectilinear'' or computational basis of $\CC^2$ and $\cB_{\times}$ be a ``diagonal'' or Hadamard basis and let $\cB_{+^n}$ and $\cB_{\times^n}$ be the corresponding bases obtained on the tensor product space $(\CC^2)^{\otimes n}$. All vectors in the rectilinear basis $\cB_{+^n}$ have an inner product with all vectors in the diagonal basis $\cB_{\times^n}$ upper bounded by $2^{-n/2}$ in absolute value. The uncertainty relation of \cite{MU88} states  that for \emph{any} quantum state on $n$ qubits described by a unit vector $\ket{\psi} \in (\CC^2)^{\otimes n}$, the average measurement entropy satisfies
\begin{equation}
\label{eq:entropic-ur-n-intro-thesis}
\frac{1}{2} \left( \entH(p_{\cB_{+^n} , \ket{\psi}}) + \entH(p_{\cB_{\times^n}, \ket{\psi}}) \right) \geq \frac{n}{2},
\end{equation}
where $p_{\cB , \ket{\psi}}$ denotes the outcome probability distribution when $\ket{\psi}$ is measured in basis $\cB$ and $\entH$ denotes the Shannon entropy. Equation \eqref{eq:entropic-ur-n-intro-thesis} expresses the fact that the outcome of at least one of two measurements cannot be well predicted, even after knowing which measurement was performed.

A surprising application of entropic uncertainty relations is the effect known as \emph{information locking} \citep{DHLST04} (see also \cite{Leu09}). Suppose Alice holds a uniformly distributed random $n$-bit string $X$. She chooses a random basis $K \in_u \{+^n,\times^n\}$ and encodes $X$ in the basis $\cB_K$. This random quantum state $\cE(X,K)$ is then given to Bob. How much information about $X$ can Bob, who does not know $K$, extract from this quantum system via a measurement? To better appreciate the quantum case, observe that if $X$ were encoded in a classical state $\cE_c(X,K)$, then $\cE_c(X,K)$ would ``hide'' at most one bit about $X$; more precisely, the mutual information between $X$ and $\cE_c(X,K)$ is at least $n-1$. For the quantum encoding $\cE$, one can show that for \emph{any measurement} that Bob applies on $\cE(X,K)$ whose outcome is denoted $I$, the mutual information between $X$ and $I$ is at most $n/2$ \citep{DHLST04}. 
The $n/2$ missing bits of information about $X$ are said to be \emph{locked} in the quantum state $\cE(X,K)$. If Bob had access to $K$, then $X$ can be easily obtained from $\cE(X,K)$: The one-bit key $K$ can be used to \emph{unlock} $n/2$ bits about $X$.

A natural question is whether it is possible to lock more than $n/2$ bits in this way. In order to achieve this, the key $K$ has to be chosen from a larger set. In terms of uncertainty relations, this means that we need to consider $t > 2$ bases to achieve an average measurement entropy larger than $n/2$ (equation \eqref{eq:entropic-ur-n-intro-thesis}). In this case, the natural candidate is a set of $t$ \emph{mutually unbiased bases}, the defining property of which is a small inner product between any pair of vectors in different bases. Surprisingly, it was shown by \cite{BW07} and \cite{Amb09} that there are up to $t = 2^{n/2}$ mutually unbiased bases $\{ \cB_1, \cB_2, \dots, \cB_{t}\}$ that only satisfy an average measurement entropy of $n/2$, which is only as good as what can be achieved with two measurements \eqref{eq:entropic-ur-n-intro-thesis}. In other words, looking at the pairwise inner product between vectors in different bases is not enough to obtain uncertainty relations stronger than \eqref{eq:entropic-ur-n-intro-thesis}.

To achieve an average measurement entropy of $(1-\e) n$ for small $\e$ while keeping the number of bases subexponential in $n$, the only known constructions are probabilistic and computationally inefficient \citep{HLSW04}.  

\subsection{Summary of the contributions}

\paragraph{Chapter \ref{chap:uncertainty-relations}}
We introduce the notion of a \emph{metric uncertainty relation} and connect it to low-distortion embeddings of $\ell_2$ into $\ell_1$. A metric uncertainty relation also implies an entropic uncertainty relation. We prove that random bases satisfy uncertainty relations with a stronger definition and better parameters than previously known. Our proof is also considerably simpler than earlier proofs. We give \emph{efficient} constructions of bases satisfying metric uncertainty relations. The bases are computable by quantum circuits of almost linear size. These constructions are obtained by adapting an explicit norm embedding due to \cite{Ind07} and an extractor construction of \cite{GUV09}.

\paragraph{Chapter \ref{chap:applications-ur}}
We prove that any metric uncertainty relation leads to a locking scheme. Applying the results of Chapter \ref{chap:uncertainty-relations}, we show the existence of locking schemes with key size independent of the message length. Moreover, using the efficient constructions, we give the first explicit strong information locking scheme. Moreover, we present a locking scheme that can in principle be implemented with current technology. We use our locking schemes to construct hiding fingerprints as defined by \cite{GI10}.

We also apply our metric uncertainty relations to exhibit communication protocols that perform equality testing of $n$-qubit states. We prove that this task can be performed by a single message protocol using $O(\log(1/\e))$ qubits and $n$ bits of communication, where $\e$ is an error parameter. We also give a single message protocol that uses $O(\log^2 n)$ qubits, where the computation of the sender is efficient.

\section{Uncertainty relations in the presence of quantum side information}

\subsection{Context}

Suppose that we are now looking for a stronger notion of uncertainty. We want the outcome to be unpredictable even if the adversary, who is trying to predict the outcome of the measurement, holds a system that is \emph{entangled} with the system being measured. Let Alice hold a system $A$ and Eve hold $E$, and the two systems are maximally entangled. How well can Eve predict the outcome of measurements in bases $\{\cB_1, \dots, \cB_t\}$? It turns out that because Alice and Eve are maximally entangled, Eve can \emph{perfectly} predict the outcome that Alice obtains. In this case, there is no uncertainty at all from the point of view of Eve. The interesting question is then: Can we obtain some uncertainty if Eve holds some quantum side information about the system $A$ but is not maximally entangled with it? The amount of uncertainty in the measurement outcomes should then be a function of some quantum correlation measure between Alice and Eve. We should note here that unlike classical side information which can usually be handled easily, quantum side information can behave in unexpected ways; see for example the work on randomness extractors against quantum adversaries \citep{KMR05, RK05, GKKRW07}. In beautiful recent work, \cite{RB09} and \cite{BCCRR10} showed that in fact one can extend the uncertainty relation in equation \eqref{eq:entropic-ur-n-intro-thesis} to allow for quantum side information. Related uncertainty relations that hold in the presence of quantum memory have proven to be a very useful tool in security proofs for quantum key distribution \citep{TR11, TLGR12, FFBSTW12}. 

But as in the previous section, just two measurements are in many cases not sufficient to obtain the desired amount of uncertainty. Before this work, uncertainty relations that hold when the adversary has a quantum memory were known only for two measurements.





\subsection{Summary of contributions}
\paragraph{Chapter \ref{chap:qc-extractor}} We introduce QC-extractors by analogy to classical randomness extractors, which are objects that found many applications in theoretical computer science, and relate them to uncertainty relations with quantum side information. Using techniques similar to the ones used for proving decoupling results, we give several constructions of QC-extractors based on unitary two-designs, complete sets of mutually unbiased bases and single-qudit unitaries. These naturally lead to uncertainty relations in terms of the min-entropy and in terms of the von Neumann entropy. This gives the first uncertainty relations in the presence of quantum side information for more than two measurements. Moreover, we use the uncertainty relation for single-qubit measurements to finally link the security of two-party secure function evaluation to the ability of the parties' storage device to store \emph{quantum} information \citep{WST08}. Previously, the security could only be shown when the classical capacity \citep{KWW09} or entanglement cost \citep{entCost} of the storage device was limited.

\chapter{Preliminaries}
\label{chap:prelim}

The objective of this chapter is to introduce some notations and results that will be used throughout this thesis. We start with a very brief section about classical information theory before moving to the description of quantum systems.

\section{Classical information theory}
Random variables are usually denoted by capital letters $X, K, \dots$, while $p_X$ denotes the distribution of $X$, i.e., $\pr{X = x} = p_X(x)$. The notation $X \sim p$ means that $X$ has distribution $p$. $\unif(S)$ is the uniform distribution on the set $S$. To measure the distance between probability distributions on a finite set $\cX$, we use the total variation distance or trace distance $\tracedist{p,q} = \frac{1}{2} \sum_{x \in \cX} |p(x) - q(x)|$. We also have
 $\tracedist{p,q} = \max_{A \subseteq \cX} (\sum_{x \in A} p(x) - \sum_{x \in A} q(x))$. 
 
 We will also write $\tracedist{X,Y}$ for $\tracedist{p_X, p_Y}$. When $\tracedist{X,Y} \leq \e$, we say that $X$ is $\e$-close to $Y$. A useful characterization of the trace distance is $\tracedist{p,q} = \max_{X \sim p, Y \sim q} \pr{X = Y}$ (this equality is sometimes attributed to \cite{Doe38}). Another useful measure of closeness between distributions is the fidelity $\fid{p,q} = \sum_{x \in \cX} \sqrt{p(x)q(x)}$ also known as the Bhattacharyya distance and related to the Hellinger distance. We have the following relation between the fidelity and the trace distance:
\begin{equation}
\label{eq:ineq-tracedist-fid}
1 - \fid{p,q} \leq \tracedist{p,q} \leq \sqrt{1-\fid{p,q}^2}.
\end{equation}
The Shannon entropy of a distribution $p$ on $\cX$ is defined as $\entH(p) = -\sum_{x \in \cX} p(x) \log p(x)$ where the $\log$ is taken here and throughout the thesis to be base two. We will also write $\entH(X)$ for $\entH(p_X)$. The conditional entropy is defined by $\entH(X|Y) = \entH(XY) - \entH(Y)$. It also has the property that $\entH(X|Y) = \exc{y}{\entH(X|Y=y)}$. The mutual information between two random variables $X$ and $Y$ is defined as $\entI(X; Y) = \entH(X) + \entH(Y) - \entH(X,Y)$. The min-entropy of a distribution $p$ is defined as $\entHmin(p) = -\log \max_x p(x)$. We say that a random variable $X$ is a $k$-source if $\entHmin(X) \geq k$.


\section{Representation of physical systems}

We briefly describe the notation and the basic facts about quantum theory that will be used in this thesis. We refer the reader to \cite{NC00, Wil11} for more details.

\subsection{Quantum states}
The state of a (pure) quantum system is represented by a unit vector in a Hilbert space. For the purpose of this thesis, a Hilbert space is a finite-dimensional complex inner product space. Quantum systems are denoted $A, B, C\dots$ and are identified with their corresponding Hilbert spaces. The dimension of $A$ is denoted $d_A$. It is important to note that all unit vectors represent valid physical states and for any two different vectors,\footnote{Technically, quantum states are actually rays rather than unit vectors in the Hilbert space, so two vectors that only differ by a global phase represent the same state.} one can perform an experiment for which the two states have a different observable behaviour.
Vectors in $A$ are denoted by ``kets'' $\ket{\psi}^A \in A$ and dual vectors (i.e., linear functions from $A$ to $\CC$) are denoted by ``bras'' $\bra{\phi}$, so that $\bra{\phi} \left( \ket{\psi} \right) = \inp{\phi}{\psi}$ is simply the inner product between the vectors $\ket{\phi}$ and $\ket{\psi}$. Performing the product in the other direction  $\ket{\psi} \bra{\phi}$, we obtain a linear transformation mapping $A$ to itself. In particular, $\proj{\psi}$ is the orthogonal projector onto the span of $\ket{\psi}$. If we fix a basis of the Hilbert space, then we can represent $\ket{\psi}$ as a column vector $\vec{v}$ and the dual vector $\bra{\psi}$ can be represented by $\vec{v}^{\dagger}$, where $M^{\dagger}$ represents the conjugate transpose of the matrix $M$. In this thesis, every Hilbert space $A$ comes with a preferred orthonormal basis $\{\ket{a}\}_{a \in [d_A]}$ that we call the computational basis. The elements of this basis are labeled by integers in $[d_A] \eqdef \{1, \dots, d_A\}$. Often, the Hilbert spaces we consider are composed of $n$ qubits, i.e., have the form $(\CC^{2})^{\otimes n}$. In this case, the computational basis will also be labeled by strings in $\{0,1\}^n$. 




In order to model our ignorance of the description of a quantum system, we can consider distributions $\{p_1, \dots, p_r\}$ over quantum states $\{\ket{\psi_1}, \dots, \ket{\psi_r}\}$. It is well known that such a distribution over states is best described by a \emph{density operator} $\rho = \sum_{i=1}^r p_i \proj{\psi_i}$ acting on $A$. We denote by $\cL(A,B)$ the set of linear transformations from $A$ to $B$ and we write $\cL(A)$ for $\cL(A,A)$. Observe that a density operator is a Hermitian positive semidefinite operator with unit trace. Conversely, any unit trace Hermitian operator $\rho$ with non-negative eigenvalues is a valid density operator. If we write $\rho^A = \sum_i p_i \proj{\psi_i}$ where $\{\ket{\psi_i}\}_i$ form an orthonormal eigenbasis for $\rho$, we can interpret the state of $A$ as being $\ket{\psi_i}$ with probability $p_i$. In particular, the density operator associated with a pure state $\ket{\psi}$ is $\proj{\psi}$ and it will be abbreviated by omitting the ket and bra: $\psi \eqdef \proj{\psi}$. We use $\cS(A)$ to denote the set of density operators acting on $A$. The Hilbert space on which a density operator $\rho \in \cS(A)$ acts is sometimes denoted by a superscript or subscript, as in $\rho^A$ or $\rho_A$. This notation is also used for pure states $\ket{\psi}^A \in A$.

In order to describe the joint state of a system $AB$, the associated state space is the tensor product Hilbert space $A \otimes B$, which is sometimes simply denoted $AB$.  If $\rho^{AB}$ describes the joint state on $AB$, the state on the system $A$ is described by the \emph{partial trace} $\rho^A \eqdef \tr_B \rho^{AB}$. The partial trace $\tr_B : \cL(A \otimes B) \to \cL(A)$ is defined as $\tr_B[\rho^{AB}] = \sum_{b} \left(\id_{A} \otimes \bra{b} \right) \rho^{AB} \left(\id_A \otimes \ket{b} \right)$, where $\{ \ket{b} \}$ is an orthonormal basis of $B$.

A classical system can easily be described using this formalism. A distribution $\{p_i\}$ over $[d]$ is represented by $\rho = \sum_{i \in [d]} p_i \proj{i}$. A state on $XB$ is said to be classical on $X$ if there exists a basis $\{\ket{x}\}$ of $X$ and a set of (non-normalized) operators $\rho_x$ on $B$ such that 
\begin{equation}
\label{eq:cq-state}
\rho^{XB} = \sum_{x} \proj{x}^{X} \ox \rho^{B}_x.
\end{equation}


\subsection{Evolution of quantum systems}

The operations that change the state of a closed quantum system $A$ are unitary transformations on $A$. Recall that $U \in \cL(A)$ is unitary if $U U^{\dagger} = U^{\dagger} U = \1$. After applying such a transformation, the state of system $A$ evolves from $\rho$ to $U \rho U^{\dagger}$. We can also consider a system $AB$ and act by a unitary on $A$ to obtain the state $U_A \rho_{AB} U_A^\dagger = (U\otimes \id_B) \rho_{AB} (U \otimes \id_B)^\dagger$.

Another important class of quantum operations are measurements. The most general way to obtain classical information from a quantum state is by performing a measurement. A measurement is described by a positive operator-valued measure (POVM), which is a set $\{P_1, \dots, P_s\}$ of positive semidefinite operators that sum to the identity.  If the state of the quantum system is represented by the density operator $\rho$, the probability of observing the outcome labeled $i$ is $\tr[P_i \rho]$ for all $i \in \{1, \dots, s\}$. Whenever $\{P_i\}$ are orthogonal projectors, we say that $\{P_i\}$ is a projective measurement. A simple class of measurements that will be extensively used in this thesis are measurements in a basis $\cB$. The measurement in the basis $\cB = \{\ket{e_i}\}_{i \in [d_A]}$ is defined by the POVM described by the operators $\{ \proj{e_i} \}_{i \in [d_A]}$ so that we obtain outcome $i$ with probability $\tr[ \proj{e_i} \rho ] = \bra{e_i} \rho \ket{e_i}$ whenever the state of the system is $\rho$. In particular, if the measurement is in the computational basis, we use the special notation $p_{\rho}(a) = \tr[ \proj{a} \rho]$, and $p_{\ket{\psi}}(a) = |\braket{a}{\psi}|^2$ whenever the state $\rho = \proj{\psi}$ is pure.

More generally, we can represent the evolution of any quantum system by a \emph{completely positive trace preserving} (CPTP) map $\cE_{A \to C}$. A map is called positive if for any positive operator $\rho$, $\cE(\rho)$ is also positive. It is called completely positive if for any quantum system $B$, the map $\cE \otimes \id_B : \cL(A \otimes B) \to \cL(C \otimes B)$ is positive. Because this is the most general kind of quantum operation, a CPTP map is also called a \emph{quantum channel}.


%
%

We can view a measurement as a quantum channel that maps a quantum system to a classical one. In particular, the map that performs a measurement in the computational basis can be written as:
\begin{align}
\label{eq:meas-computational-basis}
\cM(.)_{A \to X} = \sum_{a} \bra{a} (.) \ket{a} \proj{a} \ ,
\end{align}
where $\{\ket{a}\}$ is the computational basis of $A$.
Note that we renamed the system $X$ to emphasize that it is a classical system.
We will also use extensively in Chapter \ref{chap:qc-extractor} the map 
\begin{align}
\label{eq:meas-map-intro}
\cT(.)_{A\rightarrow A_1}= \sum_{a_{1}a_{2}} \bra{a_{1}a_{2}}(.)\ket{a_{1}a_{2}} \proj{a_1}\ ,
\end{align}
where $\{\ket{a_{1}}\},\{\ket{a_{2}}\}$ are the computational bases of $A_{1},A_{2}$ respectively.  A small calculation readily reveals that this map can be understood as tracing out $A_2$, and then measuring the remaining system $A_{1}$ in the basis $\{\ket{a_{1}}\}$. Note that the outcome of the measurement map is classical in the basis $\{\ket{a_{1}}\}$ on $A_{1}$.

 

\subsection{Distance measures}

We will employ two well known distance measures between quantum states. The first is the distance induced by the $\ell_{1}$-norm defined by $\| M \|_1 = \tr\left[\sqrt{M^\dagger M}\right]$. For $\rho, \sigma \in \cS(A)$, $\| \rho - \sigma \|_1$ is the sum of the absolute values of the eigenvalues of $\rho - \sigma$. As in the classical case, one half of the $\ell_1$-norm of a difference of two density operators, also known as the \emph{trace distance} $\tracedist{\rho, \sigma} = \frac{1}{2} \cdot \| \rho - \sigma\|_1$, is related to the success probability of distinguishing two states $\rho$ and $\sigma$ given with a priori equal probability~\citep{Hel67}:
\begin{equation}
\label{eq:tracedist-max}
\tracedist{\rho, \sigma} = \max_{0 \leq \Lambda \leq \1} \tr[\Lambda (\rho - \sigma)].
\end{equation}
The second distance measure we use is the \emph{purified distance}. To define it, we first define the \emph{fidelity} between two states $\rho, \sigma \in \cS(A)$ by $F(\rho,\sigma)=\|\sqrt{\rho}\sqrt{\sigma}\|_1$. Note that if $\rho = \proj{\psi}$ is pure, then $F(\rho, \sigma) = \sqrt{ \bra{\psi} \sigma \ket{\psi}}$. 
Another useful characterization of the fidelity is with Uhlmann's theorem. Before stating the theorem, we need to define the important notion of a \emph{purification}. A purification of a density operator $\rho \in \cS(A)$ is a pure state $\ket{\rho} \in AR$ such that $\tr_R[\rho^{AR}] = \rho^A$. Such a purification always exists, for example one can choose $R$ to be a copy of $A$ and $\ket{\rho} = \sum_i \sqrt{p_i} \ket{\psi_i}^A \ket{\psi_i}^R$, where $\{ \ket{\psi_i} \}_i$ is an eigenbasis for $\rho^A$.
\begin{theorem}[Uhlmann's theorem \citep{Uhl76}]
\label{thm:uhlmann}
Let $\rho, \sigma \in \cS(A)$ and let $\ket{\rho}^{AR}$ and $\ket{\sigma}^{AR}$ be purifications of $\rho$ and $\sigma$. Then we have
\[
\fid{\rho, \sigma} = \max_{U} | \bra{\rho} U^R \otimes \1^A \ket{\sigma} |.
\]
\end{theorem}
See e.g., \cite[Theorem 9.2.1]{Wil11} for a proof. We will also need the concept of \emph{generalized fidelity} between two possibly sub-normalized positive operators $\rho,\sigma$, which can be defined as~\citep{TCR10},
\begin{align*}
\bar{F}(\rho,\sigma)=F(\rho,\sigma)+\sqrt{\left(1-\tr[\rho]\right)\left(1-\tr[\sigma]\right)}.
\end{align*}
Note that if at least one of the states is normalized, then the generalized fidelity is the same as the fidelity, i.e.,~$\bar{F}(\rho,\sigma)=F(\rho,\sigma)$. The purified distance between two possibly subnormalized states $\rho,\sigma$ is then defined as:
\begin{align}
\pd(\rho,\sigma)=\sqrt{1-\bar{F}(\rho,\sigma)^2}\ ,
\end{align}
and is a metric on the set of sub-normalized states~\citep{TCR10, Tom12}. 

Observe that for pure states $\sqrt{1 - F(\proj{\rho},\proj{\sigma})^2} = \frac{1}{2} \| \proj{\rho} - \proj{\sigma}\|_1$. Hence, by Uhlmann's theorem, we can think of the purified distance between two normalized states as the minimal trace distance between any two purifications of the states $\rho$ and $\sigma$. The purified distance is indeed closely related to the trace distance, as for any two states $\rho,\sigma$ we have~\citep{FvdG99, TCR10}:
\begin{align}\label{eq:purifiedVStrace}
\frac{1}{2} \| \rho - \sigma \|_1 \leq \pd(\rho,\sigma) \leq \sqrt{2 \| \rho - \sigma \|_1}\ .
\end{align}
It is furthermore easy to see that for normalized states the factor $2$ on the right hand side can be improved to $1$.

For any distance measure, we can define an $\eps$-ball of states around $\rho$ as the states at a distance of at most $\eps$ from $\rho$. For the purified distance, we write
\begin{align*}
\cB^{\eps}(\rho_{A})=\{\sigma_{A}\in\substates(A)\mid \pd(\rho_{A},\sigma_{A})\leq\eps\}\ ,
\end{align*}
where $\cS_{\leq}(A)$ is the set of positive operators on $A$ with trace at most $1$.

All the distances we introduced have the property that they cannot increase by applying a completely positive trace preserving map $\cF : \cS(A) \to \cS(C)$. For any $\rho, \sigma \in \cS_{\leq}(A)$, we have
\begin{equation}
\label{eq:monotonicity-pd}
\pd(\rho, \sigma) \geq \pd(\cF(\rho), \cF(\sigma)),
\end{equation}
and
\begin{equation}
\label{eq:monotonicity-tracedist}
\|\rho - \sigma \|_1 \geq \|\cF(\rho) -  \cF(\sigma)\|_1.
\end{equation}


\subsection{Information measures}

The \textit{von Neumann entropy} of $\rho\in\cS(A)$ is defined as $\entH(A)_{\rho}=-\tr[\rho \log\rho]$. 
Note that for a classical state $\rho_X$ this is simply the Shannon entropy defined earlier.
The \emph{conditional von Neumann entropy of $A$ given $B$} for $\rho_{AB}\in\cS(AB)$ is defined as
\begin{align*}
\entH(A|B)_{\rho}=\entH(AB)_{\rho}-\entH(B)_{\rho}\ .
\end{align*}
There is an important difference with the classical case: $\entH(A|B)_{\rho}$ can be negative when the state $\rho$ is entangled between $A$ and $B$.
The \emph{conditional min-entropy} of a state $\rho_{AB}\in\states(AB)$ defined as\footnote{We write $\max$ instead of $\sup$ as we work with finite dimensional Hilbert spaces.}
\begin{align}
\label{eq:def-min-entropy}
\entHmin(A|B)_{\rho}=\max_{\sigma_B \in \states(B)}\entHmin(A|B)_{\rho|\sigma}\ ,
\end{align}
with
\begin{align*}		
\entHmin(A|B)_{\rho|\sigma}=\max\left\{\lambda \in \Real: \rho_{AB} \leq 2^{-\lambda}\cdot\id_A \otimes \sigma_B \right\}\ .
\end{align*}
For the special case where $B$ is trivial, we obtain $\hmin(A)_{\rho}=-\log\|\rho_{A}\|_{\infty}$, where $\| \rho \|_{\infty}$ denotes the largest singular value of $\rho$. For the case where we are conditioning on classical side information, we can write the conditional min-entropy as:
\begin{equation}
\label{eq:cond-min-entropy-classical}
\entHmin(X|QJ) = -\log \exc{j}{2^{- \entHmin(X|Q,J=j)}}.
\end{equation}
The min-entropy is known to have interesting operational interpretations~\citep{krs:operational}. If $A$ is classical, then the min-entropy can be expressed as 
\begin{equation}
\label{eq:pguess-hmin}
\entHmin(A|B)_{\rho}=-\log \pguess(A|B),
\end{equation} 
where $\pguess(A|B)$ is the average probability of guessing the classical symbol $A=a$ maximized over all possible measurements on $B$. If $A$ is quantum, then $\entHmin(A|B)_{\rho}$ is directly related to the maximal singlet fraction achievable by performing an operation on $B$:
\begin{equation}
\label{eq:singlet-fraction}
\entHmin(A|B)_{\rho} = - \log |A| \max_{\Lambda_{B \rightarrow A'}} F(\Phi_{AA'}, (\id_A \otimes \Lambda) (\rho_{AB}))\ ,
\end{equation}
where $\Phi_{AA'} = \frac{1}{|A|} \sum_{a, a' \in [|A|]} \ket{a a} \bra{a' a'}$ is a maximally entangled state. 

As the information theoretic tasks we wish to study usually allow for some error $\e \geq 0$, the relevant entropy measures are often \emph{smoothed entropies}. For the conditional min-entropy this takes the form
\begin{align}\label{eq:smoothMinDef}
\entHmin^{\eps}(A|B)_{\rho}=\max_{\tilde{\rho}_{AB} \in \epsball(\rho_{AB})} \entHmin(A|B)_{\tilde{\rho}}\ .
\end{align}

\subsubsection{More technical properties of entropic quantities}
\label{app:coll}
In this section, we state some additional entropic quantities that will be needed for some proofs. 

It will sometimes be more convenient to work with a version of the min-entropy in which instead of maximizing over all states $\sigma_B$ on $B$, we simply take $\sigma_B = \rho_B$. The reason the standard definition of the conditional min-entropy involves a maximization as in equation \eqref{eq:def-min-entropy} is to obtain the nice operational interpretation presented above. In particular, if the systems $A$ and $B$ are classical taking discrete values $\{a\}$ and $\{b\}$, then $\entHmin(A|B)_{\rho|\rho}~=~-\log \max_{a,b} \bra{ab} \rho \ket{ab}$, which is in general different from equation \eqref{eq:cond-min-entropy-classical}.
The smoothed version of this alternative definition becomes
\begin{align*}
\entHmin^{\eps}(A|B)_{\rho|\rho}=\max_{\tilde{\rho}_{AB} \in \epsball(\rho_{AB})} \entHmin(A|B)_{\tilde{\rho}|\tilde{\rho}}\ .
\end{align*}
\cite{TSSR10} showed that the smoothed versions of the two different definitions cannot be too far apart from each other.
\begin{lemma}[{\cite[Lemma 18]{TSSR10}}]
\label{lem:hmin-rhorho}
Let $\eps'\geq0$, $\eps'>0$, and $\rho_{AB} \in \cS(AB)$. Then
\begin{align*}
\entHmin^{\eps}(A|B)_{\rho} - \log\left(\frac{2}{\eps'^2} + \frac{1}{1-\eps}\right) \leq \entHmin^{\eps+\eps'}(A|B)_{\rho|\rho} \leq \entHmin^{\eps+\eps'}(A|B)_{\rho}\ .
\end{align*}
\end{lemma}

The max-entropy is defined by
\begin{align}\label{eq:maxentropy}
\entHmax(A|B)_{\rho}=\max_{\sigma_{B}\in\cS(B)}\log F(\rho_{AB},\id_{A}\otimes\sigma_{B})^{2}\ ,
\end{align}
and its smooth version
\begin{align}\label{eq:smoothmaxentropy}
\entHmax^{\eps}(A|B)_{\rho}=\min_{\tilde{\rho}_{AB}\in\cB^{\eps}(\rho_{AB})}\entHmax(A|B)_{\tilde{\rho}}\ .
\end{align}
The following lemma shows that the conditional min- and max-entropies are dual to one another.
\begin{lemma}[\cite{TCR10}]\label{lem:duality}
Let $\rho_{AB}\in\cS(AB)$, $\eps\geq0$, and $\rho_{ABC}$ be an arbitrary purification of $\rho_{AB}$. Then
\begin{align*}
\entHmax^{\eps}(A|B)_{\rho}=-\entHmin^{\eps}(A|C)_{\rho}\ .
\end{align*}
\end{lemma}

Finally, the quantum conditional collision entropy, which is closely related to the min-entropy, will be used in the proofs in Chapter \ref{chap:qc-extractor}. For a state $\rho_{AB} \in \states(AB)$ relative to a state $\sigma_B \in \states(B)$, it is defined as
\begin{align}\label{eq:coll}
\entHtwo(A|B)_{\rho|\sigma}=-\log\tr\left[(\id_A \ox \sigma_B^{-1/4}) \rho_{AB} (\id_A \ox \sigma_B^{-1/4})\right]^{2}\ ,
\end{align}
where the inverses are generalized inverses. For $M\in\cL(A)$, $M^{-1}$ is a generalized inverse of $M$ if $MM^{-1}=M^{-1}M= \Pi_S$, where $\Pi_S$ denotes the projector onto the support of $M$. In particular, if $M = \sum_i \alpha_i \proj{v_i}$ and the vectors $\ket{v_i}$ are orthogonal with unit norm, then $M^{-1} = \sum_{i : \alpha_i \neq 0 } \alpha_i^{-1} \proj{v_i}$. 

The following lemma relates the collision and the min-entropy.
\begin{lemma}\label{lem:hmin-h2-rhorho}
Let $\rho_{AB}\in\cS_{\leq}(AB)$ and $\sigma_{B}\in\cS(B)$ with $\mathrm{supp}(\rho_{AB})\subseteq\id_{A}\otimes\mathrm{supp}(\sigma_{B})$, where $\mathrm{supp}(.)$ denotes the support. Then
\begin{align*}
\hmin(A|B)_{\rho|\sigma}\leq \entHtwo(A|B)_{\rho|\sigma}\ .
\end{align*}
\label{collision}
\end{lemma}
\begin{proof}
We have $\mathrm{supp}(\rho_{AB})\subseteq\id_{A}\otimes\mathrm{supp}(\rho_{B})$ and hence by~\cite[Lemma B.2]{BCR09}
\begin{align*}
\hmin(A|B)_{\rho|\sigma}=-\log\max_{\omega_{AB}\in\cS(AB)}\tr\left[\omega_{AB}\left(\id_{A}\otimes\sigma_{B}^{-1/2}\right)\rho_{AB}\left(\id_{A}\otimes\sigma_{B}^{-1/2}\right)\right]\ ,
\end{align*}
where the inverses are generalized inverses. But for $\hat{\rho}_{AB}=\frac{\rho_{AB}}{\tr\left[\rho_{AB}\right]}\in\cS(AB)$ we have,
\begin{align*}
\entHtwo(A|B)_{\rho|\sigma}&=-\log\tr\left[\rho_{AB}\left(\id_{A}\otimes\sigma_{B}^{-1/2}\right)\rho_{AB}\left(\id_{A}\otimes\sigma_{B}^{-1/2}\right)\right]\\
&=-\log\tr\left[\rho_{AB}\right]-\log\tr\left[\hat{\rho}_{AB}\left(\id_{A}\otimes\sigma_{B}^{-1/2}\right)\rho_{AB}\left(\id_{A}\otimes\sigma_{B}^{-1/2}\right)\right]\\
&\geq-\log\max_{\omega_{AB}\in\cS(AB)}\tr\left[\omega_{AB}\left(\id_{A}\otimes\sigma_{B}^{-1/2}\right)\rho_{AB}\left(\id_{A}\otimes\sigma_{B}^{-1/2}\right)\right] \\
&=\hmin(A|B)_{\rho|\sigma}\ .
\end{align*}
\end{proof}

We finish with three diverse lemmas that will be used several times. First the Alicki-Fannes inequality states that two states that are close in trace distance have von Neumann entropies that are close. 
\begin{lemma}[\cite{AF03}]
\label{lem:alicki-fannes}
For any states $\rho^{AB}$ and $\sigma^{AB}$ such that $\| \rho^{AB} - \sigma^{AB} \|_1 \leq \e$ with $\e \leq 1/2$, we have
\[
|\entH(A|B)_{\rho} - \entH(A|B)_{\sigma}| \leq 4 \e \log d_A + 2 \binent(\e) \ ,
\]
where $\binent(\e) = -\e \log \e - (1-\e) \log(1 - \e)$ is the binary entropy function.
\end{lemma}
For a reference, see \cite[Theorem 11.9.4]{Wil11}. Note that such a statement is not true of the min- and max-entropies, and it is for this reason that it is useful to define smoothed versions.

The next lemma says that if you discard a classical system, the min-entropy can only decrease.
\begin{lemma}[{\cite[Lemma C.5]{BFS11}}]\label{lem:monotonicity-minentropy-classical}
Let $\rho_{AXB}\in\cS(AXB)$, $\eps\geq0$, with $X$ classical. Then
\begin{align*}
\entHmin^{\eps}(AX|B)_{\rho} \geq \entHmin^{\eps}(A|B)_{\rho}\ .
\end{align*}
\end{lemma}

The last lemma we present here states that for states of the form $\rho^{\otimes n}$, the smooth min-entropy converges to the von Neumann entropy when the number of copies $n$ grows. This is called the asymptotic equipartition property (AEP) for the smooth conditional min-entropy.
\begin{lemma}[{\cite[Remark 10]{TCR09}}]
\label{lem:aep}
Let $\rho_{AB}\in\cS(AB)$, $\eps>0$, and $n\geq2\left(1-\eps^{2}\right)$. Then,
\begin{align*}
\frac{1}{n}\hmin^{\eps}(A|B)_{\rho^{\otimes n}|\rho^{\otimes n}}\geq H(A|B)_{\rho}-\frac{4\sqrt{1-2\log\eps}\left(2+\frac{\log|A|}{2}\right)}{\sqrt{n}}\ .
\end{align*}
\end{lemma}

For a more detailed discussion of smooth entropies we refer to~\cite{Ren05,Tom12}.


\section{Quantum computation}
The most widely used model for quantum computation is the quantum circuit model. Let $U$ be a unitary acting on an $n$-qubit space. The objective is to implement $U$ with a small number of fixed gates. The main measure of efficiency is then the size of the circuit, which is the number of elementary gates that are used to perform the unitary. We say that a circuit is efficient if the size of the circuit is polynomial in $n$. 

There are many standard choices of sets of one and two-qubit gates that allow the approximation of all unitary transformations on $n$ qubits. This choice is not important here. The properties of quantum circuits we use here are the following. The Hadamard single-qubit gate defined by
\[
H = \frac{1}{\sqrt{2}} \left( \begin{array}{cc}
1 & 1 \\
1 & -1
\end{array} \right)
\] 
is part of our elementary gates. And any reversible classical circuit on $n$ bits can be directly extended to a quantum circuit with the same size that acts on the computational basis elements in the same way as the classical circuit.



\chapter[Uncertainty relations: Definitions and constructions]{Uncertainty relations for quantum measurements: Definition and constructions}
\label{chap:uncertainty-relations}


\paragraph{Outline of the chapter}
In this chapter, we start by introducing uncertainty relations and setting up some notation (Section \ref{sec:background-ur}). Then, we define metric uncertainty relations in Section \ref{sec:metric-ur}. In Section \ref{sec:existence-ur}, we prove the existence of strong metric uncertainty relations. Explicit constructions are given in Section \ref{sec:explicit-ur}.

\section{Background}
\label{sec:background-ur}
In quantum mechanics, an uncertainty relation is a statement about the relationship between measurements (or observables).\footnote{In physics language, it probably makes more sense to use the word \emph{observable} rather than measurement, but as we have not given a mathematical definition of an observable, we mostly use the word measurement.} 
Heisenberg's uncertainty principle \citep{Hei27} is one of the cornerstones of quantum mechanics. It states that the position and the momentum of a quantum particle cannot both have definite values. The uncertainty principle is a feature of quantum theory that makes it different from classical physics: having both a localized position and momentum is \emph{not} a valid state according to quantum theory.

Heisenberg's uncertainty relation was generalized in several ways. The most common way of presenting the uncertainty principle today is due to \citet{Rob29}. It gives a lower bound on the product of the variances of two observables as a function of their commutator, which quantifies how compatible the two observables are. Later, \cite{Hir57} and \cite{BM75} gave a formulation of an uncertainty relation in terms of the \emph{entropy} of the measurement outcomes. \cite{Deu83} pointed out that using an entropy instead of the variance is a more desirable way of expressing uncertainty. He proved that for any state $\ket{\psi}$, we have $\entH(p_{\cB_1, \ket{\psi}}) + \entH(p_{\cB_2, \ket{\psi}}) \geq -2 \log \left( \frac{1+c(\cB_1, \cB_2)}{2} \right)$ where $c(\cB_1, \cB_2) = \max_{\ket{b_1} \in \cB_1, \ket{b_2} \in \cB_2} |\braket{b_1}{b_2}|$ and $\cB_1$ and $\cB_2$ are bases of the ambient Hilbert space. $p_{\cB, \ket{\psi}}$ denotes the outcome distribution when performing a measurement in $\cB$ on the state $\ket{\psi}$ and $\entH$ denotes the Shannon entropy. This uncertainty relation was later improved by \cite{MU88} who showed that for all $\ket{\psi}$,
\begin{equation}
\label{eq:entropic-ur-intro}
\frac{1}{2}\left(\entH(p_{\cB_1, \ket{\psi}}) + \entH(p_{\cB_2, \ket{\psi}})\right) \geq - \log c(\cB_1, \cB_2).
\end{equation}
Observe that by using the properties of the Shannon entropy, we can rewrite equation \eqref{eq:entropic-ur-intro} as $\entH(X|K) \geq -\log c(\cB_1, \cB_2)$, where $K$ is uniformly distributed on $\{1, 2\}$ and $X$ is the outcome of a measurement in the computational basis for the state $U_K \ket{\psi}$. This says that even given the measurement $K$ that was performed, there is some uncertainty about the outcome.
If $\cB_1$ and $\cB_2$ are mutually unbiased, i.e., $c(\cB_1, \cB_2) \leq 2^{-n/2}$ where $2^n$ is the dimension of the ambient Hilbert space, we obtain a lower bound of $\frac{n}{2}$ on the average measurement entropy. It is easy to see that such a lower bound cannot be improved: For any bases $\cB_1, \cB_2$, one can always choose a state $\ket{\psi_1}$ that is aligned with one of the vectors of $\cB_1$ so that $\entH(p_{\cB_1, \ket{\psi_1}}) = 0$, in which case $\frac{1}{2} \left( \entH(p_{\cB_1 , \ket{\psi_2}}) + \entH(p_{\cB_2 , \ket{\psi_1}}) \right) \leq \frac{n}{2} $. More generally when considering $t$ basis, the best lower bound on the average measurement entropy one can hope for is $(1-1/t) n$.

For many applications, an average measurement entropy of $\frac{n}{2}$ is not good enough. In this chapter, we want to find bases for which the average measurement entropy is larger than $\frac{n}{2}$ and close to the maximal value of $n$. As mentioned earlier, in order to achieve this, one has to consider a larger set of measurements.
In this case, the natural candidate is a set of $t$ mutually unbiased bases, the defining property of which is a small inner product between any pair of vectors in different bases, more precisely $c(\cB_i, \cB_j) \leq 2^{-n/2}$ for all $i \neq j$. For $2^n+1$ measurements, \cite{Lar90, Iva92, San93} showed for $t = 2^n+1$ mutually unbiased bases, the average entropy is at least $\log(2^n+1) - 1$, which is close to the best possible. In fact, their result is stronger: it even holds for the collision entropy (R\'{e}nyi entropy of order $2$), which is in general smaller than the Shannon entropy. For $2 < t < 2^n+1$, the behaviour of mutually unbiased bases is not well understood. The best general bound for an incomplete set of mutually unbiased bases was proved by \cite{DPS04} and \cite{Aza04}:
\begin{equation}
\label{eq:entropic-ur-incomplete}
\frac{1}{t} \sum_{k=1}^{t} \entH(p_{\cB_1, \ket{\psi}}) \geq n + \log \left(\frac{ t }{ 2^n + t - 1} \right).
\end{equation}
Observe that this bound is not useful for $t \leq 2^{n/2}$, because in this case the term $\log(t/(2^n+t - 1)) \leq -n/2$, which makes \eqref{eq:entropic-ur-incomplete} at best as good as the uncertainty relation for two measurements in equation \eqref{eq:entropic-ur-intro}. Equation \eqref{eq:entropic-ur-incomplete} is known to also known to hold for the collision entropy. A similar bound for the min-entropy was also proved in \citep[Corollary 4.19]{Sch07}.
Surprisingly, it was shown by \cite{BW07} and \cite{Amb09} that there are up to $t = \sqrt{2^n}$ mutually unbiased bases $\{ \cB_1, \cB_2, \dots, \cB_{t}\}$ that only satisfy an average measurement entropy of $\frac{n}{2}$, which is only as good as what can be achieved with two measurements \eqref{eq:entropic-ur-intro}. In other words, looking at the pairwise inner product between vectors in different bases is not enough to obtain uncertainty relations stronger than \eqref{eq:entropic-ur-intro}.
To achieve an average measurement entropy of $(1-\e) n$ for small $\e$ while keeping the number of bases subexponential in $n$, the only known constructions are probabilistic and computationally inefficient. 
 \cite{HLSW04} prove that random bases satisfy entropic uncertainty relations of the form \eqref{eq:entropic-ur-intro} with $n^4$ measurements with an average measurement entropy of $n-3$.

\textbf{Brief word on applications of uncertainty relations}
Other than being one of the defining features of quantum mechanics, uncertainty relations have many applications particularly to proving the security of quantum cryptographic protocols. As an example, probably the simplest and most elegant proof of security for quantum key distribution known to date is based on a recently discovered uncertainty relation \citep{TR11}. Moreover, the proofs of the security of bit commitment and oblivious transfer in the bounded storage model are based on an uncertainty relation \citep{DFSS05, DFRSS07, KWW09}. We will describe several applications of uncertainty relations in Chapter~\ref{chap:applications-ur} and Section~\ref{sec:noisy}. For more details on entropic uncertainty relations and their applications, see the survey~\citep{WW10}.

\textbf{Notation} 
Instead of talking about uncertainty relations for a set of bases, it is more convenient here to talk about uncertainty relations for a set of unitary transformations. Let $\{\ket{x}^C\}_x$ be the computational basis of $C$. We associate to the unitary transformation $U$ the basis $\{U^{\dg} \ket{x}\}_x$. On a state $\ket{\psi}$, the outcome distribution is described by
\[
p_{U \ket{\psi}}(x) = |\bra{x} U \ket{\psi}|^2.
\]
As can be seen from this equation, we can equivalently talk about measuring the state $U \ket{\psi}$ in the computational basis. An entropic uncertainty relation for $U_1, \dots, U_{t}$ can be written as
\begin{equation}
\label{eq:entropic-ur}
\frac{1}{t} \sum_{k=1}^t \entH( p_{U_k \ket{\psi}} ) \geq h.
\end{equation}



\section{Metric uncertainty relations}
\label{sec:metric-ur}
Even though entropy is a good measure of randomness, it is usually easier to work with the distance to the uniform distribution when the distance is small. This will be our approach here: our measure of uncertainty will be the closeness in total variation distance to the uniform distribution. In other words, we are interested in sets of unitary transformations $U_1, \dots, U_{t}$ that for all $\ket{\psi} \in C$ satisfy
\[
\frac{1}{t} \sum_{k=1}^t \tracedist{p_{U_k \ket{\psi}}, \unif([d_C])} \leq \e
\]
for some $\e \in (0,1)$. $\Delta(p,q)$ refers to the total variation distance between distributions $p$ and $q$.
This condition is very strong, in fact too strong for our purposes, and we will see that a weaker definition is sufficient to imply entropic uncertainty relations. Let $C = A \otimes B$. (For example, if $C$ consists of $n$ qubits, $A$ might represent the first $n - \log n$ qubits and $B$ the last $\log n$ qubits.) Moreover, let the computational basis for $C$ be of the form $\{\ket{a}^A \otimes \ket{b}^B\}_{a,b}$ where $\{\ket{a}\}$ and $\{\ket{b}\}$ are the computational bases of $A$ and $B$. Instead of asking for the outcome of the measurement on the computational basis of the whole space to be uniform, we only require that the outcome of a measurement of the $A$ system in its computational basis $\{\ket{a}\}$ be close to uniform. More precisely, we define for $a \in [d_A]$,
\[
p^A_{U_k \ket{\psi}}(a) = \sum_{b=1}^{d_B} |\bra{a}^A \bra{b}^B U_k \ket{\psi}|^2.
\]
We can then define a metric uncertainty relation. Naturally, the larger the $A$ system, the stronger the uncertainty relation for a fixed $B$ system.
\begin{definition}[Metric uncertainty relation]
\label{def:metric-ur}
Let $A$ and $B$ be Hilbert spaces. We say that a set $\{U_1, \dots, U_{t}\}$ of unitary transformations on $AB$ satisfies an $\e$-metric uncertainty relation on $A$ if for all states $\ket{\psi} \in AB$,
\begin{equation}
\label{eq:marginalA}
\frac{1}{t} \sum_{k=1}^t \tracedist{p^A_{U_k \ket{\psi}},\unif([d_A]) } \leq \e.
\end{equation}
\end{definition}
\begin{myremark}[]
Observe that this implies that \eqref{eq:marginalA} also holds for mixed states: for any $\psi \in \cS(A \otimes B)$,
$
\frac{1}{t} \sum_{k=1}^t \tracedist{p^A_{U_k \psi U_k^{\dagger}},\unif([d_A]) } \leq \e.
$

Note that there is a reason we are looking at the average over the different values of $k$ rather that some other quantity. In fact we can rewrite the condition \eqref{eq:marginalA} as 
\begin{equation}
\label{eq:why-average}
\tracedist{q_{\ket{\psi}}, \unif([d_A]) \times \unif([t]) } \leq \e,
\end{equation}
where $q_{\ket{\psi}}$ is the distribution on $[d_A] \times [t]$ of the random variable $(X, K)$, where $X$ refers to the outcome of the computational basis measurement when it is performed on state $U_K \ket{\psi}$. This means that even given the measurement $K$ that was performed, the outcome of the measurement is still $\e$-close to uniform.
\end{myremark}

\paragraph{Metric uncertainty relations imply entropic uncertainty relations}
In the next proposition, we show that a metric uncertainty relation implies an entropic uncertainty relation. 
\begin{proposition}
\label{prop:metric-to-entropic}
Let $\e \in (0, 1/2)$ and $\{U_1, \dots, U_{t}\}$ be a set of unitaries on $AB$ satisfying an $\e$-metric uncertainty relation on $A$:
\[
\frac{1}{t} \sum_{k=1}^t \tracedist{p^A_{U_k \ket{\psi}}, \unif([d_A])}  \leq \e.
\]
Then
\[
\frac{1}{t} \sum_{k=1}^t \entH(p_{U_k \ket{\psi}}) \geq (1-8\e) \log d_A - 2\binent(2\e).
\]
where $\binent(\e) = -\e \log \e - (1-\e) \log(1-\e)$ is the binary entropy function.
\end{proposition}
\begin{proof}
Recall that the distribution $p^A_{U_k \ket{\psi}}$ (see equation \eqref{eq:marginalA} for a definition) on $[d_A]$ is a marginal of the distribution $p_{U_k\ket{\psi}}$. Thus $\entH(p_{U_k\ket{\psi}}) \geq \entH(p^A_{U_k, \ket{\psi}})$. Using Fannes' inequality (a special case of the Alicki-Fannes inequality \ref{lem:alicki-fannes}), we have for all $k$
\begin{align*}
\entH(p^A_{U_k, \ket{\psi}}) &\geq \log d_A - 8\tracedist{p^A_{U_k \ket{\psi}},\unif([d_A])} \log d_A - 2\binent\left(2\tracedist{p^A_{U_k \ket{\psi}},\unif([d_A])}\right).
\end{align*}
By averaging over $k$, and using the concavity of $\binent$, we obtain the desired result.
\end{proof}

\paragraph{Explicit link to low-distortion embeddings}
Even though we do not explicitly use the link to low-distortion embeddings, we describe the connection as it might have other applications. In the definition of metric uncertainty relations, the distance between distributions was computed using the trace distance. The connection to low-distortion metric embeddings is clearer when we measure closeness of distributions using the fidelity. We have
\begin{align*}
\fid{p^A_{U_k \ket{\psi}},\unif([d_A])} &= \frac{1}{\sqrt{d_A}} \sum_{a=1}^{d_A} \sqrt{p^A_{U_k \ket{\psi}}(a)} \\
											&= \frac{1}{\sqrt{d_A}}  \sum_{a=1}^{d_A} \sqrt{\sum_{b=1}^{d_B} |\bra{a}^A \bra{b}^B U_k \ket{\psi}|^2} \\
											&= \frac{1}{\sqrt{d_A}}  \| U_k \ket{\psi} \|_{\ell_1^A(\ell_2^B)}		
\end{align*}
where the norm $\ell_1^A(\ell_2^B)$ is defined by
\begin{definition}[$\ell_1(\ell_2)$ norm]
\label{def:l1l2}
For a state $\ket{\psi} = \sum_{a,b} \alpha_{a,b} \ket{a}^A \ket{b}^B$,
\[
\big \| \ket{\psi} \big \|_{\ell_1^{A}(\ell_2^B)} = \sum_{a} \big \| \{\alpha_{a,b}\}_b \big \|_2 = \sum_a \sqrt{\sum_b |\alpha_{a,b}|^2}.
\]
We use $\| \cdot \|_{12} \eqdef \| \cdot \|_{\ell_1^{A}(\ell_2^B)}$ when the systems $A$ and $B$ are clear from the context.
\end{definition}
Observe that this definition of norm depends on the choice of the computational basis. The $\ell^A_1(\ell^B_2)$ norm will always be taken with respect to the computational bases.

For $\{U_1, \dots, U_{t}\}$ to satisfy an uncertainty relation, we want
\[
\frac{1}{t}\sum_{k} \frac{1}{\sqrt{d_A}} \| U_k \ket{\psi} \|_{\ell^{A}_1(\ell_2^B)} \geq 1-\e.
\]
This expression can be rewritten by introducing a new register $K$ that holds the index $k$. We get for all $\ket{\psi}$
\begin{equation}
\label{eq:low-distortion1}
\left\| \frac{1}{\sqrt{t}} \sum_{k} U_k \ket{\psi}^C \ket{k}^K \right\|_{\ell^{AK}_1(\ell_2^B)} \geq (1-\e) \sqrt{t \cdot d_A}.
\end{equation}
Using the Cauchy-Schwarz inequality, we have that for all $\ket{\psi}$,
\begin{equation}
\label{eq:low-distortion2}
\left\| \frac{1}{\sqrt{t}} \sum_{k} U_k \ket{\psi}^C \ket{k}^K \right\|_{\ell^{AK}_1(\ell_2^B)} \leq \sqrt{t \cdot d_A} \left\| \frac{1}{\sqrt{t}} \sum_{k} U_k \ket{\psi}^C \ket{k}^K \right\|_2 = \sqrt{t \cdot d_A}.
\end{equation}
Rewriting \eqref{eq:low-distortion1} and \eqref{eq:low-distortion2} as
\[
(1-\e) \leq \frac{1}{ \sqrt{t \cdot d_A}} \cdot \frac{\left\| \frac{1}{\sqrt{t}} \sum_{k} U_k \ket{\psi}^C \ket{k}^K \right\|_{\ell^{AK}_1(\ell_2^B)}} { \left\| \frac{1}{\sqrt{t}} \sum_{k} U_k \ket{\psi}^C \ket{k}^K \right\|_2} \leq 1,
\]
we see that the image of $C$ by the linear map $\ket{\psi} \mapsto \frac{1}{\sqrt{t}} \sum_k U_k \ket{\psi} \otimes \ket{k}$ is an almost Euclidean subspace of $(A \otimes K \otimes B, \ell^{AK}_1(\ell_2^B))$. In other words, as the map $\ket{\psi} \mapsto \frac{1}{\sqrt{t}} \sum_k U_k \ket{\psi} \otimes \ket{k}$ is an isometry (in the $\ell_2$ sense), it is an embedding of $(C, \ell_2)$ into $(AKB, \ell^{AK}_1(\ell_2^B))$ with distortion $1/(1-\e)$.

Observe that a general low-distortion embedding of $(C, \ell_2)$ into $(AKB, \ell_1^{AK}(\ell_2^B))$ does not necessarily give a metric uncertainty relation as it need not be of the form $\ket{\psi} \mapsto \frac{1}{\sqrt{t}} \sum_k U_k \ket{\psi} \otimes \ket{k}$. When $t = 2$, a metric uncertainty relation is related to the notion of Kashin decomposition \citep{Kas77}; see also \citep{Pis89,Sza06}.

\paragraph{A remark on the composition of metric uncertainty relations}
There is a natural way of building an uncertainty relation for a Hilbert space from uncertainty relations on smaller Hilbert spaces. This composition property is also important for the cryptographic applications of metric uncertainty relations presented in Chapter \ref{chap:applications-ur}, in which setting it ensures the security of parallel composition of locking schemes.
\begin{proposition}
Consider Hilbert spaces $A_1$, $A_2$, $B_1$, $B_2$. For $i \in \{1,2\}$, let $\{U^{(i)}_{k_i}\}_{k_i \in [t_i]}$ be a set of unitary transformations of $A_i \otimes B_i$ satisfying an $\e$-metric uncertainty relation on $A_i$.
Then, $\{U^{(1)}_{k_1} \otimes U^{(2)}_{k_2}\}_{k_1, k_2 \in [t_1] \times [t_2]}$ satifies a $2\e$-metric uncertainty relation on $A_1 \otimes A_2$. 
\end{proposition}
\begin{proof}
Let $\ket{\psi} \in (A_1 \otimes B_1) \otimes (A_2 \otimes B_2)$ and let $p_{k_1, k_2}$ denote the distribution obtained by measuring $U^{(1)}_{k_1} \otimes U^{(2)}_{k_2} \ket{\psi}$ in the computational basis of $A_1 \otimes A_2$.
Our objective is to show that
\begin{equation}
\label{eq:parallel-compose}
\frac{1}{t_1 t_2} \sum_{k_1 \in [t_1], k_2 \in [t_2]} \tracedist{p_{k_1, k_2}, \unif([d_{A_1}] \times [d_{A_2}])} \leq 2\e.
\end{equation}
We have
\begin{align}
& \tracedist{p_{k_1, k_2}, \unif([d_{A_1}] \times [d_{A_2}])} \\
 &= \frac{1}{2} \sum_{a_1, a_2} \left|p_{k_1, k_2}(a_1, a_2) - \frac{1}{d_{A_1} d_{A_2}}\right| \notag \\
											&\leq \frac{1}{2} \sum_{a_1, a_2} \left|p_{k_1, k_2}(a_1, a_2) - \frac{ p^{A_1}_{k_1, k_2}(a_1) }{d_{A_2}} \right| +  \frac{1}{2} \sum_{a_1, a_2} \left| \frac{ p^{A_1}_{k_1, k_2}(a_1) }{d_{A_2}} - \frac{1}{d_{A_1} d_{A_2}} \right| \notag \\
											&= \frac{1}{2} \sum_{a_1} p^{A_1}_{k_1, k_2}(a_1) \sum_{a_2} \left|\frac{p_{k_1, k_2}(a_1, a_2)}{p^{A_1}_{k_1, k_2}(a_1) } - \frac{1}{d_{A_2}} \right| +  \frac{1}{2} \sum_{a_1} \left| p_{k_1, k_2}^{A_1}(a_1)  - \frac{1}{d_{A_1}} \right| \label{eq:parallel-compose2}
\end{align}
where $p^{A_1}_{k_1, k_2}(a_1) \eqdef \sum_{a_2} p_{k_1, k_2}(a_1, a_2)$ is the outcome distribution of measuring the  $A_1$ system of $U^{(1)}_{k_1} \otimes U^{(2)}_{k_2} \ket{\psi}$. The distribution $p_{k_1, k_2}$ can also be seen as the outcome of measuring the mixed state 
\[
U^{(1)}_{k_1} \psi^{A_1B_1} {U^{(1)}_{k_1}}^{\dagger}
\]
in the computational basis $\{ \ket{a_1} \}$. Thus, we have for any $k_2 \in [t_2]$,
\[
\frac{1}{t_1} \sum_{k_1} \tracedist{p^{A_1}_{k_1, k_2}, \unif([d_{A_1}])} \leq \e.
\]
Moreover, for $a_1 \in [d_{A_1}]$, the distribution on $[d_{A_2}]$ defined by $\frac{p_{k_1, k_2}(a_1, a_2)}{p^{A_1}_{k_1, k_2}(a_1) }$ is the outcome distribution of measuring in the computational basis of $A_2$ the state
\[
U^{(2)}_{k_2} \psi^{A_2B_2}_{k_1, a_1} {U^{(2)}_{k_2}}^{\dagger}
\]
where $\psi^{A_2B_2}_{k_1, a_1}$ is the density operator describing the state of the system $A_2B_2$ given that the outcome of the measurement of the $A_1$ system is $a_1$. We can now use the fact that $\{U^{(2)}_{k_2} \}$ satisfies a metric uncertainty relation. Taking the average over $k_1$ and $k_2$ in equation \eqref{eq:parallel-compose2}, we get
\[
\frac{1}{t_1 t_2} \sum_{k_1, k_2} \tracedist{p_{k_1, k_2}, \unif([d_{A_1}] \times [d_{A_2}])} \leq 2\e.
\]
\end{proof}
This observation is in the same spirit as \cite[Proposition 1]{IS10}, and can in fact be used to build large almost Euclidean subspaces of $\ell^A_1(\ell^B_2)$.



\section{Metric uncertainty relations: existence}
\label{sec:existence-ur}
In this section, we prove the existence of families of unitary transformations satisfying strong uncertainty relations. The proof proceeds by showing that choosing random unitaries according to the Haar measure defines a metric uncertainty relation with positive probability. The techniques used are quite standard and date back to Milman's proof of Dvoretzky's theorem \citep{Mil71,FLM77}. A version of Dvoretzky's theorem states that for any norm $\| \cdot \|$ over $\CC^d$, there exists a ``large'' subspace $E \subseteq \mathbb{C}^d$ which is almost Euclidean, i.e., for all $x \in E, ~ (1-\e) \| x \|_2 \leq s \| x \| \leq (1+\e) \| x \|_2$ for some constant $\e > 0$ and scaling factor $s$. Using the connection between uncertainty relations and embeddings of $\ell_2$ into $\ell_1(\ell_2)$ presented in the previous section, Theorem \ref{thm:existence-ur} can be viewed as a strengthening of Dvoretzky's theorem for the $\ell_1(\ell_2)$  norm \citep{MS86}. 

General techniques from asymptotic geometric analysis have recently found many applications in quantum information theory. For example, \cite{ASW10} show that the existence of large subspaces of highly entangled states follows from Dvoretzky's theorem for the Schatten $p$-norm\footnote{The Schatten $p$-norm of a matrix $M$ is defined as the $\ell_p$ norm of a vector of singular values of $M$.} for $p > 2$. This in turns shows the existence of channels that violate additivity of minimum output $p$-R\'enyi entropy as was previously demonstrated by \cite{HW08}. Using a more delicate argument, \cite{ASW10b} were also able to recover Hastings' counterexample to the additivity conjecture \citep{Has09}. The general strategy that is used to prove such results is to define a distribution over the set of objects one is looking for and use concentration of measure tools to prove that the desired properties can be satisfied with positive probability.


For Theorem \ref{thm:existence-ur}, we need to introduce the Haar measure over the unitary group $\cU(d)$. 
A natural way of defining a uniform measure over a group is to ask the measure of a subset to be invariant under multiplication by elements of the group. In particular, for the unitary group, consider measures $\mu$ on the unitary transformations of $\CC^d$ that satisfy $\mu(S) = \mu(\{ U \cdot M : M \in S\} )$ for all measurable sets $S \subseteq \cU(d)$ and unitaries $U \in \cU(d)$. It follows from Haar's theorem that there is a unique probability measure that satisfies this condition.


\begin{definition}[Haar measure]
The Haar measure $\mu_d$ on the set of unitary transformations on $\CC^d$ is the unique probability measure that is invariant under multiplication by a unitary operation.

We can then define a rotation invariant probability measure on pure states of $\CC^d$ by considering the distribution of $U \ket{0}$ where $U \sim \mu_d$ and $\ket{0}$ is any unit vector in $\CC^d$. We say that $U \ket{0}$ is a random pure state.
\end{definition}

We need another definition before stating the theorem. For some applications,\footnote{Quantum hiding fingerprints studied in Section \ref{sec:hiding-fingerprint}} we require an additional property for $\{U_1, \dots, U_{t}\}$. A set of unitary transformations $\{U_1, \dots, U_{t}\}$ of $\CC^d$ is said to define $\gamma$-approximately mutually unbiased bases ($\gamma$-MUBs) if for all elements $\ket{x}$ and $\ket{y}$ of the computational basis and all $k \neq k'$, we have
\begin{equation}
\label{eq:def-approx-mub}
|\bra{x} U_k^{\dagger} U_{k'} \ket{y}| \leq \frac{1}{d^{\gamma/2}}.
\end{equation}
$1$-MUBs correspond to the usual notion of mutually unbiased bases.

%

\begin{theorem}[Existence of metric uncertainty relations]
\label{thm:existence-ur}
Let $c = 9 \pi^2$ and $\e \in (0,1)$. Let $A$ and $B$ be Hilbert spaces with $\dim B \geq 9/\e^2$ and $d \eqdef \dim A \ox B \geq \frac{9c \cdot 16^2 \pi}{\e^2}$. Then, for all $t > \frac{4 \cdot 18 c \cdot \ln(9/\e)}{\e^2}$, there exists a set $\{U_{1}, \dots, U_{t}\}$ of unitary transformations of $AB$ satisfying an $\e$-metric uncertainty relation on $A$: for all states $\ket{\psi} \in AB$,
\[
\frac{1}{t} \sum_{k=1}^t \tracedist{p^A_{U_k \ket{\psi}}, \unif([d_A])} \leq \e.
\]
Moreover, for $\gamma \in (0,1)$ and $d$ such that $4t^2d^2 \exp{-d^{1-\gamma}} < 1/2$, the unitaries $\{U_1, \dots, U_{t}\}$ can be chosen to also form $\gamma$-MUBs.
\end{theorem}
\begin{myremark}[]
The proof proceeds by choosing a set of unitary transformations at random. See \eqref{eq:prur} and \eqref{eq:pr-approx-mub} for a precise bound on the probability that such a set does not form a metric uncertainty relation or a $\gamma$-MUB.
\end{myremark}
\begin{proof}
The first step is to evaluate the expected value of $\tracedist{p^A_{U\ket{\psi}}, \unif([d_A])}$ for a fixed state $\ket{\psi}$ when $U$ is a random unitary chosen according to the Haar measure. Then, we use a concentration of measure argument to show that with high probability, this distance is close to its expected value. After this step, we show that the additional averaging $\frac{1}{t} \sum_{k=1}^t \tracedist{p^A_{U_k\ket{\psi}}, \unif([d_A])}$ of $t$ independent copies results in additional concentration at a rate that depends on $t$. We conclude by showing the existence of a family of unitaries that makes this expression small for all states $\ket{\psi}$ using a union bound over a $\delta$-net. The four main ingredients of the proof are precisely stated here but only proved in Appendix \ref{sec:app-existence-ur}. 

We start by computing the expected value of the fidelity $\ex{\fid{p^A_{U\ket{\psi}}, \unif([d_A])}}$, which can be seen as an $\ell_1(\ell_2)$ norm. 

\begin{lemma}[Expected value of $\ell^A_1(\ell^B_2)$ over the sphere]
\label{lem:expl1l2}
Let $\ket{\ph}^{AB}$ be a random pure state on $AB$. Then,
\[
\ex{\fid{p^A_{\ket{\ph}}, \unif([d_A])}} 
\geq \sqrt{1 - \frac{1}{d_B}}.
\]
\end{lemma}

We then use the inequality $\tracedist{\rho, \sigma} \leq \sqrt{1 - \fid{\rho, \sigma}^2}$ to get 
\[
\ex{ \tracedist{ p^A_{\ket{\ph}}, \unif([d_A])} } \leq \ex{\sqrt{ 1 - \fid{p^A_{\ket{\ph}}, \unif([d_A])} ^2 } }.
\]
By the concavity of the function $x \mapsto \sqrt{1-x^2}$ on the interval $[0,1]$,
\begin{align*}
\ex{ \tracedist{ p^A_{\ket{\ph}}, \unif([d_A])} }
								&\leq \sqrt{ 1 - \ex{\fid{p^A_{\ket{\ph}}, \unif([d_A])}}^2 } \\
								&\leq \sqrt{ 1 - \left(1 - \frac{1}{d_B} \right) } \\
								&\leq \e/3. \\
\end{align*}
The last inequality comes from the hypothesis of the theorem that $d_B \geq 9/\e^2$.
In other words, for any fixed $\ket{\psi}$, the average over $U$ of the trace distance between $p^A_{U \ket{\psi}}$ and the uniform distribution is at most $\e/3$. The next step is to show that this trace distance is close to its expected value with high probability. For this, we use a version of L\'evy's lemma presented in \cite{MS86}.

\begin{lemma}[L\'evy's lemma]
\label{lem:levy}
Let $f : \CC^d \to \RR$ and $\eta > 0$ be such that for all pure states $\ket{\ph_1}, \ket{\ph_2}$ in $\CC^d$, 
\[
| f(\ket{\ph_1}) - f(\ket{\ph_2}) | \leq \eta \| \ket{\ph_1} - \ket{\ph_2} \|_2.
\]
Let $\ket{\ph}$ be a random pure state in dimension $d$. Then for all $0 \leq \delta \leq \eta$,
\[
\pr{| f(\ket{\ph}) - \ex{f(\ket{\ph})} | \geq \delta } \leq 4 \exp{- \frac{\delta^2 d}{c \eta^2} }
\]
where $c$ is a constant. We can take $c = 9 \pi^2$.
\end{lemma}

We apply this concentration result to $f : \ket{\ph}^{AB} \mapsto \tracedist{p^A_{\ket{\ph}}, \unif([d_A])}$. We start by finding an upper bound on the Lipshitz constant $\eta$. For any pure states $\ket{\ph_1}^{AB}$ and $\ket{\ph_2}^{AB}$, we have
\begin{align}
| f(\ket{\ph_1}) - f(\ket{\ph_2}) | 	&\leq \tracedist{p^A_{\ph_1}, p^A_{\ph_2}} \notag \\
			&\leq \frac{1}{2}\sum_{a,b} \left| |\bra{a}^A \bra{b}^B \ket{\ph_1}|^2 - \sum_{b}^{} |\bra{a}^A \bra{b}^B \ket{\ph_2}|^2  \right| \notag \\
			&= \tracedist{p_{\ket{\ph_1}}, p_{\ket{\ph_2}}} \notag \\
			&\leq \sqrt{1 - \fid{p_{\ket{\ph_1}}, p_{\ket{\ph_2}}}^2} \notag\\
			&\leq \sqrt{2 \left(1 - \fid{p_{\ket{\ph_1}}, p_{\ket{\ph_2}}}\right)} \notag \\
			&= \sqrt{2 - 2\sum_{a,b} |\bra{a} \bra{b} \ket{\ph_1}| \cdot |\bra{a} \bra{b} \ket{\ph_2} |} \notag \\
			&= \sqrt{ \sum_{a,b} \big| |\bra{a} \bra{b} \ket{\ph_1}|  - |\bra{a} \bra{b} \ket{\ph_2} | \big|^2 } \notag \\
			&\leq \| \ket{\ph_1} - \ket{\ph_2} \|_2. \label{eq:ineq-tracedist-l2}
\end{align}
The first two inequalities follow from the triangle inequality. The third inequality is an application of \eqref{eq:ineq-tracedist-fid}. The fourth inequality follows from the fact that $1-x^2 \leq 2(1-x)$ for all $x \in [0,1]$. The last inequality follows again from the triangle inequality. Thus, applying Lemma \ref{lem:levy}, we get for all $0 \leq \delta \leq 1$,
\begin{equation}
\label{eq:tail-bound-delta-less-1}
\pr{ \left|\tracedist{p^A_{\ket{\ph}}, \unif([d_A])}  - \mu \right|  \geq \delta} \leq 4\exp{- \frac{\delta^2 d}{c} }
\end{equation}
where $\mu = \ex{\tracedist{p^A_{\ket{\ph}}, \unif([d_A])}}$. The following lemma bounds the tails of the average of independent copies of a random variable.
\begin{lemma}[Concentration of the average]
\label{lem:avgconc}
Let $a, b \geq 1$, $\delta \in (0,1)$ and $t$ be a positive integer. Suppose $X$ is a random variable with $0$ mean satisfying the tail bounds
\[
\pr{X \geq \eta} \leq a e^{-b \eta ^2} \quad \text{ and } \quad \pr{X \leq -\eta} \leq ae^{-b \eta^2}.
\]
Let $X_1, \dots X_t$ be independent copies of $X$. Then if $\delta^2 b \geq 16 a^2 \pi $,
\[
\pr{\left| \frac{1}{t} \sum_{k=1}^t X_k \right| \geq \delta} \leq \exp{-\frac{\delta^2 bt }{2}}.
\]
\end{lemma}

We apply the above Lemma with $X_k = \tracedist{p^A_{U_k \ket{\psi}}, \unif([d_A])}  - \mu$ which satisfies the bound \eqref{eq:tail-bound-delta-less-1} in addition to being bounded in absolute value by $1$. Taking $\delta = \e/3$ and using Lemma \ref{lem:avgconc} (which we can apply because we have $(\e/3)^2 \cdot \frac{d}{c}  \geq 16 \cdot 4^2 \cdot \pi$), we get
\[
\pr{ \left| \frac{1}{t}\sum_{k=1}^t \tracedist{p^A_{U_k \ket{\psi}}, \unif([d_A])}  - \mu \right|  \geq \e/3 } \leq \exp{- \frac{1}{2} \frac{(\e/3)^2 t d}{c} }. 
\]
Using this together with Lemma \ref{lem:expl1l2}, we have
\begin{equation}
\label{eq:pravgfixedstate}
\pr{ \frac{1}{t}\sum_{k=1}^t \tracedist{p^A_{U_k\ket{\psi}}, \unif([d_A])} \geq 2\e/3 } \leq  \exp{- \frac{\e^2 t d}{18 c} }. 
\end{equation}

We would like to have the event described in \eqref{eq:pravgfixedstate} hold for all $\ket{\psi} \in AB$. For this, we construct a finite set $\cN$ of states (a $\delta$-net) for which we can ensure that $\frac{1}{t}\sum_{k=1}^t \tracedist{p^A_{U_k\ket{\psi}}, \unif([d_A])} < 2\e/3$ for all $\ket{\psi} \in \cN$ holds with high probability.
\begin{lemma}[$\delta$-net]
\label{lem:eps-net}
Let $\delta \in (0,1)$. There exists a set $\cN$ of pure states in $\CC^d$ with $|\cN| \leq (3/\delta)^{2 d}$ such that for every pure state $\ket{\psi} \in \CC^d$ (i.e., $\| \ket{\psi} \|_2 = 1$), there exists $\ket{\tilde{\psi}} \in \cN$ such that
\[
\| \ket{\psi} - \ket{\tilde{\psi}} \|_{2} \leq \delta.
\] 
\end{lemma}

Let $\cN$ be the $\e/3$-net obtained by applying this lemma to the space $AB$ with $\delta = \e/3$. We have
\begin{align*}
&\pr{ \exists \ket{\psi} \in \cN: \frac{1}{t}\sum_{k=1}^t \tracedist{p^A_{U_k\ket{\psi}}, \unif([d_A])} \geq 2\e/3 }  \\
&\leq |\cN| \cdot \exp{- \frac{\e^2 t d }{18c} } \\
																&\leq \exp{-d \left(\frac{\e^2 t}{18c}  - 2\ln(9/\e) \right)}.
\end{align*}
Now for an arbitrary state $\ket{\psi} \in AB$, we know that there exists $\ket{\tilde{\psi}} \in \cN$ such that $ \| \ket{\psi} - \ket{\tilde{\psi}} \|_{2} \leq \e/3$. As a consequence, for any unitary transformation $U$,
\begin{align*}
\tracedist{p^A_{U \ket{\psi}}, \unif([d_A])} &\leq \tracedist{ p^A_{U \ket{\tilde{\psi}}}, \unif([d_A])} +  \tracedist{p^A_{U \ket{\tilde{\psi}}}, p^A_{U \ket{\psi}}}   \\
							&\leq \tracedist{ p^A_{U \ket{\tilde{\psi}}}, \unif([d_A])} +  \| U \ket{\tilde{\psi}} - U \ket{\psi} \|_2 \\							 
							&\leq \tracedist{p^A_{U \ket{\tilde{\psi}}}, \unif([d_A])} + \e/3.
\end{align*}
In the first inequality, we used the triangle inequality and the second inequality can be derived as in \eqref{eq:ineq-tracedist-l2}.
Thus,
\begin{equation}
\label{eq:prur}
\pr{ \exists \ket{\psi} \in AB:  \frac{1}{t}\sum_{k=1}^t \tracedist{ p_{U_k \ket{\psi}}, \unif([d_A])} \geq \e } \leq \exp{-d \left(\frac{\e^2 t}{18 c}  - 2\ln(9/\e) \right)}.
\end{equation}
If $t > \frac{4 \cdot 18c \cdot \ln(9/\e)}{\e^2}$, this bound is strictly smaller than $1/2$ and the result follows.

To prove that we can suppose that $\{U_1, \dots, U_{t}\}$ define $\gamma$-MUBs, consider the function $f : \ket{\ph} \mapsto \braket{\psi}{\ph}$ for some fixed vector $\ket{\psi}$. Then, if $\ket{\ph}$ is a random pure state, we have $\ex{f(\ket{\ph})} = 0$. Moreover, using Levy's Lemma with $\delta = d^{-\gamma/2}$
\[
\pr{ |\braket{\psi}{\ph}| \geq d^{-\gamma/2}} \leq 4 \exp{-\frac{d^{1-\gamma}}{c}}.
\]
Thus,
\begin{equation}
\label{eq:pr-approx-mub}
\pr{ \exists k \neq k', x ,y \in [d], |\bra{x}  U^{\dagger}_k U_{k'} \ket{y}| \geq d^{-\gamma}} \leq 4t^2 d^2 \exp{-\frac{d^{1-\gamma}}{c}}
\end{equation}
which completes the proof.
\end{proof}

\begin{corollary}[Existence of entropic uncertainty relations]
\label{cor:existence-entropic-ur}
Let $C$ be a Hilbert space of dimension $d > 2$. There exists a constant $c' \geq 1$ such that for any integer $t > 2$ such that $\frac{9 \cdot 16^2 t}{5 \cdot 18 \log t} \leq d$, there exists a set $\{U_1, \dots, U_{t}\}$ of unitary transformations of $C$ satisfying the following entropic uncertainty relation: for any state $\ket{\psi}$,
\[
\frac{1}{t} \sum_{k=1}^t \entH(p_{U_k \ket{\psi}}) \geq \left(1-8\sqrt{\frac{c' \log t}{t}}\right) \log d - \log\left(\frac{18t}{c' \log t}\right) - 2 \binent\left(2\sqrt{\frac{c' \log t}{t}} \right) .
\]
In particular, in the limit $d \to \infty$, we obtain the existence of a sequence of sets of $t$ bases satisfying
\[
\lim_{d \to \infty} \frac{\frac{1}{t} \sum_{k=1}^t \entH(p_{U_k \ket{\psi}})}{\log d} \geq 1-\sqrt{\frac{c' \log t}{t}}.
\]
\end{corollary}
\begin{myremark}[]
Recall that the bases (or measurements) that constitute the uncertainty relation are defined as the images of the computational basis by $U^{\dagger}_k$.
Note that for any set of unitaries $\{U_1, \dots, U_{t}\}$, we have
\[
\frac{1}{t} \sum_{k=1}^t \entH(p_{U_k \ket{\psi}}) \leq \left(1- \frac{1}{t} \right) \log d.
\]
It is an open question whether there exists uncertainty relations matching this bound, even asymptotically as $d \to \infty$ \citep{WW10}. \cite{WW10} ask whether there even exists a growing function $f$ such that  
\[
\lim_{d \to \infty} \frac{1}{t} \frac{\sum_{k=1}^t \entH(p_{U_k \ket{\psi}})}{\log d} \geq 1- \frac{1}{f(t)}.
\]
The corollary answers this question in the affirmative with $f(t) = \sqrt{\frac{t}{c' \log t}}$. 
\end{myremark}
\begin{proof}
Define $c' = 5\cdot 18 c$ where $c$ comes from L\'{e}vy's Lemma \ref{lem:levy}, $\e = \sqrt{\frac{c' \log t}{t}}$ and decompose $C = A \otimes B$ with $d_B = \ceil{9/\e^2}$. As $d \geq \frac{9c \cdot 16^2 }{\e^2}$ and 
\[
\frac{4 \cdot 18 c \log(9/\e)}{\e^2} =  4\cdot18 c \log\left(\sqrt{\frac{t}{c' \log t}}\right) \cdot \frac{t}{5\cdot18 c \log t} \leq t,
\]
 we get a family $U_1, \dots, U_{t}$ of unitary transformations that satisfies
\[
\frac{1}{t} \sum_{k=1}^t \tracedist{p^A_{U_k \ket{\psi}}, \unif([d_A])} \leq \e.
\]
By Proposition \ref{prop:metric-to-entropic}, these unitary transformations also satisfy an entropic uncertainty relation:
\begin{align*}
\frac{1}{t} \sum_{k=1}^t \entH(p^A_{U_k \ket{\psi}}) &\geq (1-8\e) \log \left(\frac{d}{\ceil{9/\e^2}}\right) -2\binent(2\e) \\
											&\geq (1-8\e) \log d - \log(18/\e^2) - 2\binent(2\e).
\end{align*}
\end{proof}


\section{Metric uncertainty relations: explicit construction}
\label{sec:explicit-ur}

In this section, we are interested in obtaining families $\{U_1, \dots, U_{t}\}$ of unitaries satisfying metric uncertainty relations where $U_1, \dots, U_{t}$ are explicit and efficiently computable using a quantum computer. For this section, we consider for simplicity a Hilbert space composed of qubits, i.e., of dimension $d = 2^n$ for some integer $n$. This Hilbert space is of the form $A \otimes B$ where $A$ describes the states of the first $\log d_A$ qubits and $B$ the last $\log d_B$ qubits. Note that we assume that both $d_A$ and $d_B$ are powers of two.

We construct a set of unitaries by adapting an explicit low-distortion embedding of $(\RR^d, \ell_2)$ into $(\RR^{d'}, \ell_1)$ with $d' = d^{1+o(1)}$ by \cite{Ind07}. Indyk's construction has two main ingredients: a set of mutually unbiased bases and an extractor. Our construction uses the same paradigm while requiring additional properties of both the mutually unbiased bases and the extractor. 

In order to obtain a locking scheme that only needs simple quantum operations, we construct sets of \emph{approximately} mutually unbiased bases from a restricted set of unitaries that can be implemented with single-qubit Hadamard gates. Moreover, we impose three additional properties on the extractor: we need our extractor to be strong, to define a permutation and to be efficiently invertible. We want the extractor to be strong because we are constructing metric uncertainty relations as opposed to a norm embedding. The property of being a permutation extractor is needed to ensure that the induced transformation on $(\CC^{2})^{\otimes n}$ preserves the $\ell_2$ norm. We also require the efficient invertibility condition to be able to build an efficient quantum circuit for the permutation. See Definition \ref{def:perm-extractor} for a precise formulation.

The intuition behind Indyk's idea is as follows. Let $V_1, \dots, V_{r}$ be unitaries defining (approximately) mutually unbiased bases (see equation \eqref{eq:def-mub}) and let $\{P_y\}_{y \in S}$ be a permutation extractor (Definition \ref{def:perm-extractor}). The role of the mutually unbiased bases is to guarantee that for all states $\ket{\psi}$ and for most values of $j \in [r]$, most of the mass of the state $V_j\ket{\psi}$ is ``well spread'' in the computational basis. This spread is measured in terms of the min-entropy of the distribution $p_{V_j \ket{\psi}}$. Then, the extractor $\{P_y\}_y$ will ensure that on average over $y \in S$, the masses $\sum_{b} |\bra{a} \bra{b} P_y V_j \ket{\psi}|^2$ are almost equal for all $a \in [d_A]$. More precisely, the distribution $p^A_{P_y V_j \ket{\psi}}$ is close to uniform.

We start by recalling the definition of mutually unbiased bases. A set of unitary transformations $V_1, \dots, V_{r}$ is said to define $\gamma$-\emph{approximately mutually unbiased bases} (or $\gamma$-MUBs) if for $i \neq j$ and any elements $\ket{x}$ and $\ket{y}$ of the computational basis, we have
\begin{equation}
\label{eq:def-mub}
|\bra{x} V_i^{\dagger} V_j \ket{y}| \leq \frac{1}{d^{\gamma/2}}.
\end{equation}

As shown in the following lemma, there is a construction of mutually unbiased bases that can be efficiently implemented \citep{WF89}. 
%
%
\begin{lemma}[Quantum circuits for MUBs]
\label{lem:explicitmub}
Let $n$ be a positive integer and $d = 2^n$. For any integer $r \leq d+1$, there exists a family $V_1, \dots,  V_{r}$ of unitary transformations of $\CC^d$ that define mutually unbiased bases. Moreover, there is a randomized classical algorithm with runtime $O(n^2 \polylog n)$ that takes as input $j \in [r]$ and outputs a binary vector $\alpha_j \in \{0,1\}^{2n-1}$, and a quantum circuit of size $O(n \polylog n)$ that when given as input the vector $\alpha_j$ (classical input) and a quantum state $\ket{\psi} \in \CC^d$ outputs $V_j \ket{\psi}$.
\end{lemma}
\begin{myremark}[]
The randomization in the algorithm is used to find an irreducible polynomial of degree $n$ over $\FF_2[X]$. It could be replaced by a deterministic algorithm that runs in time $O(n^4 \polylog n)$. Observe that if $n$ is odd and $r \leq (d+1)/2$, it is possible to choose the unitary transformations to be real (see \cite{HSP06}).
\end{myremark}
\begin{proof} 
We define $V_1 = \1$, and the remaining unitaries are indexed by binary vectors $u \in \{0,1\}^n$, for example the binary representations of integers from $0$ to $r-2$. The construction is based on operations in the finite field $\FF_{2^n}$. The field $\FF_{2^n}$ can be seen as an $n$-dimensional vector space over $\FF_2$. Choose $\theta \in \FF_{2^n}$ such that $1, \theta, \dots, \theta^{n-1}$ form a basis of $\FF_{2^n}$. For any $x,y \in [n]$, $\theta^{x} \cdot \theta^{y} \in \FF_{2^n}$ can be decomposed in our chosen basis as
$\theta^{x} \cdot \theta^{y} = \sum_{\ell=0}^{n-1} m_{\ell}(x,y) \theta^{\ell}$ for some $m_{\ell}(x,y) \in \FF_2$. We can thus define the matrices $M_0, M_1, \dots, M_{n-1}$ from the multiplication table
\[
\left( \begin{array}{c}
1 \\
\theta  \\
\vdots \\
\theta^{n-1} 
\end{array} \right) \cdot
\left( \begin{array}{cccc}
1 & \theta & \hdots & \theta^{n-1}
\end{array} \right) = M_0  + M_1 \theta + \dots + M_{n-1} \theta^{n-1}. 
\]
where $M_{\ell} = (m_{\ell}(x,y))_{x,y \in [n]}$.
For a given $u \in \{0,1\}^n$, we define the matrix
\[
N_u = \sum_{\ell=0}^{n-1} u_{\ell} M_\ell.
\]
Notice that as $\theta^x \cdot \theta^y = \theta^{x+y}$, the entry $N_u(x,y)$ of $N_u$ only depends on $x+y$, i.e., $N_u(x,y) = N_u(x',y')$ if $x+y = x'+y'$. So we can represent this matrix by a vector $\alpha_u(x+y) = N_u(x,y)$ of length $2n-1$.
We then define a $\ZZ_4$-valued quadratic form by: for $v \in \{0,1\}^n$,
\[
T_u(v) = v^{T} N_u v \mod 4.
\]
Note that the operations $v^{T} N_u v$ are not performed in $\FF_2$ but rather in $\ZZ$. Using the vector $\alpha_u$, we can write
\[
T_u(v) = \sum_{x,y \in [n]} v_x N_u(x,y) v_y \mod 4 = \sum_{z=0}^{2n-2} \left(\sum_{x=0}^{z} v_x v_{z-x} \right) \alpha_u(z) \mod 4
\]
if we define $v_x = 0$ for $x \geq n$.
We then define the diagonal matrix $D_u = \textrm{diag}\left(i^{T_u(v)} \right)_{v \in \FF_2^n}$.
Finally, we define for $2 \leq j \leq r$,
\[
V_j = D_{\textrm{bin}(j-2)} H^{\otimes n}
\]
where $\textrm{bin}(j) \in \{0,1\}^n$ is the binary representation of length $n$ of the integer $j$.

The fact that these unitaries define mutually unbiased bases was proved in \cite{WF89}. We now analyse how fast these unitary transformations can be implemented. Note that we want a circuit that takes as input a state $\ket{\psi}$ together with  the index $j$ of the unitary transformation and outputs $V_j \ket{\psi}$.

Given the index $j$ as input, we show it is possible to compute $u = \textrm{bin}(j-2)$ and compute the vector $\alpha_j \eqdef \alpha_u$ in time $O(n^2 \polylog n)$. In fact, we start by computing a representation of the field $\FF_{2^n}$ by finding an irreducible polynomial $Q$ of degree $n$ in $\FF_2[X]$, so that $\FF_{2^n} = \FF_2[X]/Q$. This can be done in expected time $O(n^2 \polylog n)$ (Corollary 14.43 in the book \cite{GG99}). There also exists a deterministic algorithm for finding an irreducible polynomial in time $O(n^{4} \polylog n)$ \citep{Sho90}. We then take $\theta = X$. Computing the polynomial $X^x \cdot X^y = X^{x+y} \mod Q$ can be done in time $O(n \polylog n)$ using the fast Euclidean algorithm (see Corollary 11.8 in \cite{GG99}). As $x+y \in [0, 2n-2]$, we can explicitly represent all the polynomials $X^{z}$ for $0 \leq z \leq 2n-2$ in time $O(n^2 \polylog n)$. It is then simple to compute the vector $\alpha_u$ using the vector $u$ in time $O(n^2)$.


To build the quantum circuit, we first observe that applying a Hadamard transform only takes $n$ single-qubit Hadamard gates. 
Then, to design a circuit performing the unitary transformation $D_{\textrm{bin}(j-2)}$, we start by building a classical circuit that computes 
\[
T_u(v) = \sum_{z=0}^{2n-2} \left(\sum_{x=0}^{z} v_x v_{z-x} \right) \alpha_u(z) \mod 4
\]
on inputs $v$ and $\alpha_u$.
Observing that $\sum_{x=0}^{z} v_x v_{z-x}$ is the coefficient of $Y^z$ in the polynomial $\left(\sum_{x=0}^{n-1} v_x Y^x\right)^2$, we can use fast polynomial multiplication to compute $T_u(v)$ in time $O(n \polylog n)$ (Corollary 8.27 in \cite{GG99}). 
This circuit can be transformed into a reversible circuit with the same size (up to some multiplicative constant) that takes as input $(v, \alpha_j, g)$ where $v \in \{0,1\}^n$, $\alpha_j \in \{0,1\}^{2n-1}$ and $g \in \ZZ_4$, and outputs $(v, \alpha_j, g + T_u(v) \mod 4)$.

This reversible classical circuit can be readily transformed into a quantum circuit that computes the unitary transformation defined by $W: \ket{v} \ket{g} \mapsto \ket{v} \ket{g + T_u(v) \mod 4}$. Recall that we want to implement the transformation $D_u:\ket{v} \mapsto i^{T_u(v)} \ket{v}$ efficiently. This is simple to obtain using the quantum circuit for $W$. In fact, if we use a catalyst state $\ket{\phi} = \ket{0} - i \ket{1} - \ket{2} + i \ket{3}$, we have
\[
W \ket{v} \ket{\phi} = i^{T_u(v)} \ket{v} \ket{\phi} = D_{\textrm{bin}(j-2)} \ket{v} \ket{\phi}.
\]
Finally, $D_{\textrm{bin}(j-2)} H^{\otimes n}$ can be implemented by a quantum circuit of size $O(n \polylog n)$.\end{proof}

%
%
It is also possible to obtain approximately mutually unbiased bases that use smaller circuits. In fact, the following lemma shows that we can construct large sets of approximately mutually unbiased bases defined by unitaries in the restricted set
\[
\cH = \{ H^{v} \eqdef H^{v_1} \otimes \dots \otimes H^{v_n}, v \in \{0,1\}^n \},
\]
where $H$ is the Hadamard transform on $\CC^2$ defined by 
\[
H = \frac{1}{\sqrt{2}} \left( \begin{array}{cc}
1 & 1 \\
1 & -1 
\end{array} \right).
\]
In our construction of metric uncertainty relations (Theorem \ref{thm:explicit-ur1}), we could use the $1$-MUBs of Lemma \ref{lem:explicitmub} or the $(1/2-\delta)$-MUBs of Lemma \ref{lem:approx-mub}. As the construction of approximate MUBs is simpler and can be implemented with simpler circuits, we will mostly be using Lemma \ref{lem:approx-mub}.
%
%
\begin{lemma}[Approximate MUBs in $\cH$]
\label{lem:approx-mub}
Let $n'$ be a positive integer and $n = 2^{n'}$. 
\begin{enumerate}
\item For any integer $r \leq n$, there exists a family $V_1, \dots,  V_{r} \in \cH$ that define $1/2$-MUBs.
\item For any $\delta \in (0,1/2)$, there exists a constant $c > 0$ independent of $n$ such that for any $r \leq 2^{c n}$  there exists a family $V_1, \dots,  V_{r}$ of unitary transformations in $\cH$ that define $(1/2-\delta)$-MUBs.
\end{enumerate}
Moreover, in both cases, given an index $j \in [r]$, there is a polynomial time (classical) algorithm that computes the vector $v \in \{0,1\}^n$ that defines the unitary $V_j = H^{v}$.
\end{lemma}
\begin{proof}
Observe that for any $v \in \{0,1\}^n$ and any $y \in \{0,1\}^n$, we have
\[
H^v \left(\ket{y_1} \otimes \dots \otimes \ket{y_n} \right) = H^{v_1} \ket{y_1} \otimes \dots \otimes H^{v_n} \ket{y_n} =  \sum_{\substack{y'_i \in \{0,1\} \text{ for } v_i = 1 \\ y'_i = y_i \text{ for } v_i = 0 }} \frac{(-1)^{v \cdot y'}} {\sqrt{2}^{\weight(v)}} \ket{y'_1 \dots y'_n},
\]
where $\weight(v)$ is the number of non-zero components of $v$. Thus,
\begin{equation}
\label{eq:mub-had}
|\bra{x} H^{v} H^{v'} \ket{y}| = |\bra{x} H^{v+v'} \ket{y}| \leq \frac{1}{2^{\hamdist(v,v')/2}},
\end{equation}
where $\hamdist(v,v') = \{i : v_i \neq v'_i\}$ is the Hamming distance between the two vectors $v$ and $v'$.
Using this observation, we see that a binary code $C \subseteq \{0,1\}^n$ with minimum distance $\gamma n$ defines a set of $\gamma$-MUBs in $\cH$. It is now sufficient to find binary codes with minimum distance as large as possible. 
For the first construction, we use the Hadamard code that has minimum distance $n/2$. The Hadamard codewords are indexed by $x \in \{0,1\}^{n'}$; the codeword corresponding to $x$ is the vector $v \in \{0,1\}^n$ whose coordinates are $v_z = x \cdot z$ for all $z \in \{0,1\}^{n'}$. This code has the largest possible minimum distance for a non-trivial binary code but its shortcoming is that the number of codewords is only $n$. For our applications, it is sometimes desirable to have $r$ larger than $n$ (this is useful to allow the error parameter $\e$ of our metric uncertainty relation to be smaller than $n^{-1/2}$).

For the second construction, we use families of linear codes with minimum distance $(1/2 - \delta)n$ with a number of codewords that is exponential in $n$. For this, we can use Reed-Solomon codes concatenated with linear codes on $\{0,1\}^{\Theta(n')}$ that match the performance of random linear codes; see for example Appendix E in \cite{Gol08}. For a simpler construction, note that we can also get $2^{\Omega(\sqrt{n})}$ codewords by using a Reed-Solomon code concatenated with a Hadamard code.
\end{proof}

The next lemma shows that for any state $\ket{\psi}$, for most values of $j$, the distribution $p_{V_j \ket{\psi}}$ is close to a distribution with large min-entropy provided $\{V_j\}$ define $\gamma$-MUBs. This result might be of independent interest. In fact, \citet{DFRSS07} prove a lower bound close to $n/2$ on the min-entropy of a measurement in the computational basis of the state $U \ket{\psi}$ where $U$ is chosen uniformly from the full set of the $2^n$ unitaries of $\cH$. They leave as an open question the existence of small subsets of $\cH$ that satisfy the same uncertainty relation. When used with the $\gamma$-MUBs of Lemma \ref{lem:approx-mub}, the following lemma partially answers this question by exhibiting such sets of size polynomial in $n$ but with a min-entropy lower bound close to $n/4$ instead. This can be used to reduce the amount of randomness needed for many protocols in the bounded and noisy quantum storage models.
%
%
\begin{lemma}
\label{lem:minentropymub}
Let $n \geq 1, d = 2^n$ and $\e \in (0,1)$ and consider a set of $r = \ceil{\frac{2}{\e^2}}$ unitary transformations $V_1, \dots, V_{r}$ of $\CC^d$ defining $\gamma$-MUBs. For all $\ket{\psi} \in \CC^d$,
\[
\left| \left\{ j \in [r] : \exists  q_j, \tracedist{p_{V_j \ket{\psi}}, q_j} \leq \e \text{ and } \entHmin(q_j) \geq \frac{\gamma n}{2} - \log(8/\e^2) \right\} \right| \geq (1-\e) r.
\]
\end{lemma}
\begin{proof}
This proof proceeds along the lines of \cite[Lemma 4.2]{Ind07}. Similar results can also be found in the sparse approximation literature; see \cite[Proposition 4.3]{Tro04} and references therein. 

Consider the $rd \times d$ matrix $V$ obtained by concatenating the rows of the matrices $V_1, \dots, V_{r}$. For $S \subseteq [rd]$, $V_S$ denotes the submatrix of $V$ obtained by selecting the rows in $S$. The coordinates of the vector $V \ket{\psi} \in \CC^{rd}$ are indexed by $z \in [rd]$ and denoted by $(V \ket{\psi})_z$.
\begin{claim}[]
We have for any set $S \subseteq [rd]$ of size at most $d^{\gamma/2}$ and any unit vector $\ket{\psi}$,
\begin{equation}
\label{eq:upper-bound-vs}
\| (V \ket{\psi})_S \|_2^2 \leq 1 + \frac{ |S| }{d^{\gamma/2} }.
\end{equation}
\end{claim}
To prove the claim, we want an upper bound on the operator $2$-norm of the matrix $(V_S)$, which is the square root of the largest eigenvalue of $G = V^{\dagger}_S V_S$. As two distinct rows of $V$ have an inner product bounded by $\frac{1}{d^{\gamma/2}}$, the non-diagonal entries of $G$ are bounded by $\frac{1}{d^{\gamma/2}}$. Moreover, the diagonal entries of $G$ are all $1$. By the Gershgorin circle theorem, all the eigenvalues of $G$ lie in the disc centered at $1$ of radius $\frac{|S|-1}{d^{\gamma/2}}$. We conclude that \eqref{eq:upper-bound-vs} holds.

Now pick $S$ to be the set of indices of the $d^{\gamma/2}$ largest entries of the vector $\{|(V \ket{\psi})_z|^2\}_{z \in [rd]}$. Using the previous claim, we have
$\| (V \ket{\psi})_S \|_2^2 \leq 2.$
Moreover, since $S$ contains the $d^{\gamma/2}$ largest entries of $\{|(V \ket{\psi})_z|^2\}_z$, we have that for all $z \notin S$, $|(V\ket{\psi})_z|^2 d^{\gamma/2} \leq \| V \ket{\psi} \|_2^2 = \sum_{j=1}^{r} \|V_j \ket{\psi}\|_2^2 = r$. Thus, for all $z \notin S$, $|(V\ket{\psi})_z|^2 \leq \frac{r}{d^{\gamma/2}}$.

We now build the distributions $q_j$. For every $j \in [r]$, define 
\[
w_j = \sum_{z \in S \cap \{(j-1)d + 1, \dots, jd\}} |(V\ket{\psi})_z|^2,
\]
which is the total weight in $S$ of $V_j \ket{\psi}$. Defining $T_{\e} = \{j : w_j > \e\}$, we have $|T_{\e}| \e \leq \| (V \ket{\psi})_S \|^2_2 \leq 2$. Thus,
\[
|T_{\e}| \leq 2/\e \leq \e r.
\]
We define the distribution $q_j$ for $j \in [r]$ by
\[
q_j(x) = \left\{ \begin{array}{ll}
|\bra{x} V_j \ket{\psi}|^2 + \frac{w_j}{d} & \textrm{if $(j-1)d + x \notin S$}\\
\frac{w_j}{d} & \textrm{if $(j-1)d + x \in S$}.
\end{array} \right. 
\]
Since
\[
\sum_x q_j(x) = w_j + \sum_{x \in [d] : (j-1)d+x \notin S} |\bra{x} V_j \ket{\psi}|^2 = \sum_{x \in [d]} |\bra{x} V_j \ket{\psi}|^2 = 1,
\]
$q_j$ is a probability distribution. Moreover, we have that for $j \notin T_{\e}$
\[
\tracedist{p_{V_j \ket{\psi}}, q_j} \leq \frac{1}{2} \left( \sum_{x : (j-1)d+x \notin S} \frac{w_j}{d} + \sum_{x : (j-1)d+x \in S} \left( \frac{w_j}{d} + |\bra{x} V_j \ket{\psi}|^2 \right) \right) = w_j \leq \e.
\]
The distribution $q_j$ also has the property that 
for all $x \in [d]$, $q_j(x) \leq \frac{r}{d^{\gamma/2}} + \frac{1}{d} \leq \frac{2r}{d^{\gamma/2}}$. In other words, $\entHmin(q_j) \geq \frac{\gamma n}{2} - \log(8/\e^2)$.
\end{proof}

We now move to the second building block in Indyk's construction: randomness extractors.
Randomness extractors are functions that extract uniform random bits from weak sources of randomness. 

\begin{definition}[Strong permutation extractor]
\label{def:perm-extractor}
Let $n$ and $m \leq n$ be positive integers, $\ell \in [0,n]$ and $\e \in (0,1)$. A family of permutations $\{P_y\}_{y \in S}$ of $\{0,1\}^n$ where each permutation $P_y$ is described by two functions $P^E_y: \{0,1\}^n \to \{0,1\}^m$ (the first $m$ output bits of $P_y$) and $P^R_y: \{0,1\}^n \to \{0,1\}^{n-m}$ (the last $n-m$ output bits of $P_y$) is said to be an explicit $(n, \ell) \to_{\e} m$ strong permutation extractor if:
\begin{itemize}
\item For any random variable $X$ on $\{0,1\}^n$ such that $\entHmin(X) \geq \ell$, and an independent seed $U_S$ uniformly distributed over $S$, we have
\[
\tracedist{p_{\left(U_S, P_{U_S}^E(X)\right)}, \unif( S \times \{0,1\}^m)} \leq \e,
\]
which is equivalent to
\begin{equation}
\label{eq:strongext}
\frac{1}{|S|}\sum_{y \in S} \tracedist{p_{P_y^E(X)}, \unif(\{0,1\}^m)} \leq \e.
\end{equation}
\item For all $y \in S$, both the function $P_y$ and its inverse $P_y^{-1}$ are computable in time polynomial in $n$.
\end{itemize}
\end{definition}
\begin{myremark}[]
A similar definition of permutation extractors was used in \cite{RVW00} in order to avoid some entropy loss in an extractor construction. Here, the reason we use permutation extractors is different; it is because we want the induced transformation $P_y$ on $\CC^{2^n}$ to preserve the $\ell_2$ norm.
\end{myremark}

We can adapt an extractor construction of \cite{GUV09} to obtain a permutation extractor with the following parameters. The details of the construction are presented in Appendix \ref{sec:app-perm-extractor}.

\begin{theorem}[Explicit strong permutation extractors]
\label{thm:perm-extractor}
For all (constant) $\delta \in (0,1)$, all positive integers $n$, all $\ell \in [c \log(n/\e), n]$ ($c$ is a constant independent of $n$ and $\e$), and all $\e \in (0,1/2)$, there is an explicit $(n, \ell) \to_{\e} (1-\delta) \ell$ strong permutation extractor $\{P_y\}_{y \in S}$ with $\log |S| \leq O(\log(n/\e))$. Moreover, the functions $(x,y) \mapsto P_y(x)$ and $(x,y) \mapsto P^{-1}_y(x)$ can be computed by circuits of size $O(n \polylog(n/\e))$.
\end{theorem}

A permutation $P$ on $\{0,1\}^n$ defines a unitary transformation on $(\CC^2)^{\otimes n}$ that we also call $P$.  The permutation extractor $\{P_y\}$ will be seen as a family of unitary transformations over $n$ qubits. Moreover, just as we decomposed the space $\{0,1\}^n$ into the first $m$ bits and the last $n-m$ bits, we decompose the space $(\CC^2)^{\otimes n}$ into $A \otimes B$, where $A$ represents the first $m$ qubits and $B$ represents the last $n-m$ qubits. The properties of $\{P^E_y\}$ will then be reflected in the system $A$.
 
Combining Theorem \ref{thm:perm-extractor} and Lemma \ref{lem:minentropymub}, we obtain a set of unitaries satisfying a metric uncertainty relation.
%
%
\begin{theorem}[Explicit uncertainty relations: key optimized]
\label{thm:explicit-ur1}
Let $\delta > 0$ be a constant, $n$ be a positive integer, $\e \in (2^{-c'n},1)$ ($c'$ is a constant independent of $n$). Then, there exist $t \leq \left(\frac{n}{\e}\right)^{c}$ (for some constant $c$ independent of $n$ and $\e$) unitary transformations $U_1, \dots, U_{t}$ acting on $n$ qubits such that: if $A$ represents the first $(1-\delta)n/4 - O(\log(1/\e))$ qubits and $B$ represents the remaining qubits, then for all $\ket{\psi} \in AB$,
\[
\frac{1}{t}\sum_{k=1}^t \tracedist{p^A_{U_k \ket{\psi}}, \unif([d_A])} \leq \e.
\]
Moreover, the mapping that takes the index $k \in [t]$ and a state $\ket{\psi}$ as inputs and outputs the state $U_k \ket{\psi}$ can be performed by a classical computation with polynomial runtime and a quantum circuit that consists of single-qubit Hadamard gates on a subset of the qubits followed by a permutation in the computational basis. This permutation can be computed by (classical or quantum) circuits of size $O(n \polylog(n/\e))$.
\end{theorem}
\begin{myremark}[]
Observe that in terms of the dimension $d$ of the Hilbert space, the number of unitaries $t$ is polylogarithmic.
\end{myremark}
\begin{proof}
Let $\e' = \e/6$.
Lemma \ref{lem:approx-mub} gives $r = \ceil{2/\e'^2}$ unitary transformations $V_1, \dots, V_{r}$ that define $\gamma$-mutually unbiased bases with $\gamma = 1/2 - \delta/4$. Moreover, all theses unitaries can be performed by a quantum circuit that consists of single-qubit Hadamard gates on a subset of the qubits. Theorem \ref{thm:perm-extractor} with $\ell = (1-\delta/2)n/4 - \log(8/\e'^2)$ and error $\e'$ gives $|S| \leq 2^{c\log(n/\e')}$ permutations $\{P_y\}_{y \in S}$ of $\{0,1\}^n$ that define an $(n, \ell) \mapsto_{\e'} (1-\delta/2)\ell$ extractor and are computable by classical circuits of size $O(n \polylog(n/\e))$. We now argue that this classical circuit can be used to build a quantum circuit of size $O(n \polylog (n/\e))$ that computes the unitaries $P_y$. 

Given classical circuits that compute $P$ and $P^{-1}$, we can construct reversible circuits $C_P$ and $C_{P^{-1}}$ for $P$ and $P^{-1}$. The circuit $C_P$ when given input $(x,0)$ outputs the binary string $(x, P(x))$, so that it keeps the input $x$. Such a circuit can readily be transformed into a quantum circuit that acts on the computational basis states as the classical circuit. We also call these circuits $C_P$ and $C_{P^{-1}}$. Observe that we want to compute the unitary $P$, so we have to erase the input $x$. For this, we combine the circuits $C_P$ and $C_{P^{-1}}$ as described in Figure \ref{fig:qcircuit}. Note that the size of this quantum circuit is the same as the size of the original classical circuit up to some multiplicative constant. Thus, this quantum circuit has size $O(n \polylog (n/\e))$.
\begin{center}
\begin{figure}[h]
\begin{center}
\includegraphics[width=0.8\textwidth]{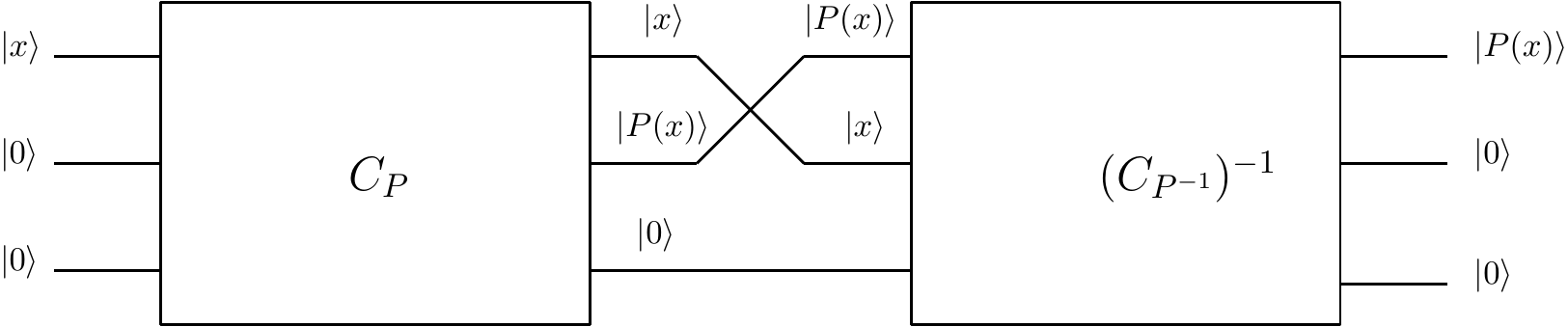}
\end{center}
\caption{Quantum circuit to compute the permutation $P$ using quantum circuits $C_P$ for $P$ and $C_{P^{-1}}$ for $P^{-1}$. $(C_{P^{-1}})^{-1}$ is simply the circuit $C_{P^{-1}}$ taken backwards. The bottom register is an ancilla register.}
\label{fig:qcircuit}
\end{figure}
\end{center}

The unitaries $\{U_1, \dots, U_{t}\}$ are obtained by taking all the possible products $P_y V_j$ for $j \in [r], y \in S$. Note that $t = r|S|$.
We now show that the set $\{U_1, \dots, U_{t}\}$ satifies the uncertainty relation property. Using Lemma \ref{lem:minentropymub}, for any state $\ket{\psi}$, the set 
\[
T_{\ket{\psi}} \eqdef \left\{ j : \exists q_j, \tracedist{p_{V_j \ket{\psi}}, q_j} \leq \e' \text{ and }  \entHmin(q_j) \geq (1-\delta/2)n/4 - \log(8/\e'^2)\right\}
\]
has size at least $(1-\e') r$. Moreover, for all $a \in [d_A]$, $p^A_{P_y V_i \ket{\psi}}(a) = \sum_{b} |\bra{a} \bra{b} P_y V_i \ket{\psi}|^2 = \pr{P^E_y(X) = a}$ where $X$ has distribution $p_{V_i \ket{\psi}}$. By definition, for $i \in T_{\ket{\psi}}$, we have $\tracedist{p_{V_i \ket{\psi}}, q_i} \leq \e'$ with $\entHmin(q_i) \geq (1-\delta/2)n/4 - \log(8/\e'^2)$. Using the fact that $\{P^E_y\}$ is a strong extractor (see \eqref{eq:strongext}) for min-entropy $(1-\delta/2)n/4 - \log(8/\e'^2)$, it follows that
\[
\frac{1}{|S|}\sum_{y \in S} \tracedist{p^A_{P_y V_i \ket{\psi}}, \unif([d_A])} \leq 2\e'
\]
%
%
for all $i \in T_{\ket{\psi}}$. As $|T_{\ket{\psi}}| \geq (1-\e')r$, we obtain
\[
\frac{1}{t}\sum_{k=1}^t \tracedist{p^A_{U_k \ket{\psi}}, \unif([d_A])} \leq 3\e' = \e/2.
\]
To conclude, we show that $t$ can be taken to be a power of two at the cost of multiplying the error by at most two. In fact, let $p$ be the smallest integer satisfying $t \leq 2^p$, so that $2^p \leq 2t$. By repeating $2^p - t$ unitaries, it is easily seen that we obtain an $\e$-metric uncertainty relation with $2^p$ unitaries from an $\e/2$-metric uncertainty relation with $t$ unitaries.
%
%
%
%
\end{proof}

Note that the $B$ system we obtain is quite large and to get strong uncertainty relations, we want the system $B$ to be as small as possible. For this, it is possible to repeat the construction of the previous theorem on the $B$ system. The next theorem gives a construction where the $A$ system is composed of $n-O(\log \log n) - O(\log(1/\e))$ qubits. Of course, this is at the expense of increasing the number of unitaries in the uncertainty relation.

\begin{theorem}[Explicit uncertainty relation: message length optimized]
\label{thm:explicit-ur2}
Let $n$ be a positive integer and $\e \in (2^{-c'n}, 1)$ where $c'$ is a constant independent of $n$. Then, there exist $t \leq \left(\frac{n}{\e}\right)^{c \log n}$ (for some constant $c$ independent of $n$ and $\e$) unitary transformations $U_1, \dots, U_{t}$ acting on $n$ qubits that are all computable by quantum circuits of size $O(n \polylog(n/\e))$ such that: if $A$ represents the first $n - O(\log \log n) - O( \log(1/\e))$ qubits and $B$ represents the remaining qubits, then for all $\ket{\psi} \in AB$,
\begin{equation}
\label{eq:cond-explicit-ur2}
\frac{1}{t} \sum_{k=1}^t \tracedist{p^A_{U_k \ket{\psi}}, \unif([d_A])} \leq \e.
\end{equation}
Moreover, the mapping that takes the index $k \in [t]$ and a state $\ket{\psi}$ as inputs and outputs the state $U_k \ket{\psi}$ can be performed by a classical precomputation with polynomial runtime and a quantum circuit of size $O(n \polylog(n/\e))$. The number of unitaries $t$ can be taken to be a power of two.
\end{theorem}

\begin{proof}
Using the construction of Theorem \ref{thm:explicit-ur1}, we obtain a system $A$ over which we have some uncertainty relation and a system $B$ that we do not control. In order to decrease the dimension of the system $B$, we can apply the same construction to that system. The system $B$ then gets decomposed into $A_2B_2$, and we know that the distribution of the measurement outcomes of system $A_2$ in the computational basis is close to uniform. As a result, we obtain an uncertainty relation on the system $AA_2$ (see Figure \ref{fig:composition-indyk}).

\begin{center}
\begin{figure}[htp]
\begin{center}
\includegraphics[width=0.6\textwidth]{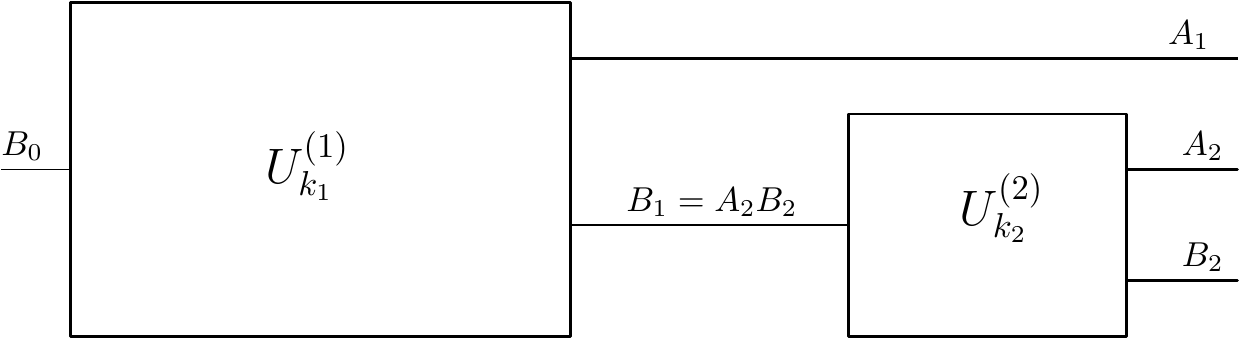}
\end{center}
\caption{Composition of the construction of Theorem \ref{thm:explicit-ur1}: In order to reduce the dimension of the $B$ system, we can re-apply the uncertainty relation to the $B$ system.}
\label{fig:composition-indyk}
\end{figure}
\end{center}

More precisely, we start by demonstrating a simple property about the composition of metric uncertainty relations. Note that this composition is different from the one described in \eqref{eq:parallel-compose}, but the proof is quite similar.
\begin{claim}[]
Suppose the set $\{U^{(1)}_{1}, \dots, U^{(1)}_{t_1}\}$ of unitaries on $A_1B_1$ satisfies a $(t_1, \e_1)$-metric uncertainty relation on system $A_1$ and the $\{U^{(2)}_{1}, \dots, U^{(2)}_{t_2}\}$ of unitaries on $B_1 = A_2B_2$ satisfies a $(t_2, \e_2)$-metric uncertainty relation on $A_2$. 
Then the set of unitaries $\left\{(\1^{A_1} \ox U^{(2)}_{k_2}) \cdot U^{(1)}_{k_1} \right\}_{k_1, k_2 \in [t_1] \times [t_2]}$ satisfies a $(t_1t_2, \e_1 + \e_2)$-metric uncertainty relation on $A_1 A_2$: for all $\ket{\psi} \in A_1A_2B_2$,
\[
\frac{1}{t_1t_2} \sum_{k_1, k_2 \in [t_1] \times [t_2]} \tracedist{ p_{U^{(2)}_{k_2} U^{(1)}_{k_1}\ket{\psi}}, \unif([d_{A_1}d_{A_2}]) } \leq \e_1 + \e_2.
\]
\end{claim}
For a fixed value of $k_1 \in [t_1]$ and $a_1 \in [d_{A_1}]$, we can apply the second uncertainty relation to the state $\frac{\bra{a_1}^{A_1} U_{k_1} \ket{\psi}}{\| \bra{a_1}^{A_1} U_{k_1} \ket{\psi} \|_2} = \frac{1}{\sqrt{p^{A_1}_{U_{k_1} \ket{\psi}}(a_1)}} \sum_{b_1} \left(\bra{a_1} \bra{b_1} U_{k_1} \ket{\psi} \right) \ket{b_1} \in B_1 = A_2B_2$. As $\{\ket{b_1}\}_{b_1} = \{\ket{a_2} \ket{b_2} \}_{a_2, b_2}$, we have
\begin{align*}
\frac{1}{t_2}\sum_{k_2} \sum_{a_2} \left| \frac{1}{p^{A_1}_{U_{k_1} \ket{\psi}}(a_1)}\sum_{b_2} | \bra{a_1}^{A_1} \bra{a_2}^{A_2} \bra{b_2}^{B_2} (\1^{A_1} \ox U_{k_2}) U_{k_1} \ket{\psi} |^2 - \frac{1}{d_{A_2}} \right| \leq \e_2. 
\end{align*}
We can then calculate, in the same vein as \eqref{eq:parallel-compose2}
\begin{align*}
&\frac{1}{t_1t_2}\sum_{k_1, k_2} \sum_{a_1, a_2} \left| \sum_{b_2} | \bra{a_1}^{A_1} \bra{a_2}^{A_2} \bra{b_2}^{B_2} (\1^{A_1} \ox U_{k_2}) U_{k_1} \ket{\psi} |^2 - \frac{1}{d_{A_1} d_{A_2}} \right| \\
	&\leq \frac{1}{t_1t_2} \sum_{k_1, k_2} \sum_{a_1} 
	\left| \sum_{b_2} | \bra{a_1}^{A_1} \bra{a_2}^{A_2} \bra{b_2}^{B_2} (\1^{A_1} \ox U_{k_2}) U_{k_1} \ket{\psi} |^2 - \frac{p^{A_1}_{U_{k_1} \ket{\psi}}(a_1)}{d_{A_2}} \right| \\
	&+ \frac{1}{t_1}\sum_{k_1} \sum_{a_1, a_2} \left| \frac{p^{A_1}_{U_{k_1} \ket{\psi}}(a_1)}{d_{A_2}} - \frac{1}{d_{A_1} d_{A_2}} \right|  \\
	& \leq \frac{1}{t_1} \sum_{k_1} \sum_{a_1} p^{A_1}_{U_{k_1} \ket{\psi}}(a_1) \e_2 + \e_1 \\
	&\leq \e_2+ \e_1.
\end{align*}
This completes the proof of the claim.

To obtain the claimed dimensions, we compose the construction of Theorem \ref{thm:explicit-ur1} $h$ times with an error parameter $\e' = \e/h$ and $\delta = 1/8$. Starting with a space of $n$ qubits, the dimension of the $B$ system (after one step) can be bounded by
\[
\frac{7}{8}n - O(\log(1/\e')) \leq \log d_B \leq \frac{7}{8}n.
\]
So after $h$ steps, we have
\[
\left(7/8\right)^h n - O(\log(1/\e')) \cdot 8(1-(7/8)^h) \leq \log d_{B_h} \leq \left(7/8\right)^h n.
\]
Thus,
\[
\left(7/8\right)^h n - O(\log(1/\e')) \leq \log d_{B_h} \leq \left(7/8\right)^h n.
\]
Note that $h$ cannot be arbitrarily large: in order to apply the construction of Theorem \ref{thm:explicit-ur1} on a system of $m$ qubits with error $\e'$, we should have $\e' \geq 2^{-c'm}$. In other words, if 
\begin{equation}
\label{eq:cond-repeat-cons}
\log d_{B_h} \geq \frac{1}{c'} \log(h/\e),
\end{equation}
then we can apply the construction $h$ times. 
%
Let $c''$ be a constant to be chosen later and $h = \floor{\frac{1}{\log(8/7)} \left(\log n - 
\log(c''\log \log n + c''\log(1/\e)) \right)}$.
This choice of $h$ satisfies \eqref{eq:cond-repeat-cons}. In fact,
\begin{align*}
\log d_{B_h} &\geq c'' \log \log n + c'' \log(1/\e) - O(\log(h/\e)) \\
			&\geq \frac{1}{c'} \log(h/\e)
\end{align*}
if $c''$ is chosen large enough. Moreover, we get 
\[
\log d_{B_h} = 2^{-\log n} \cdot 2^{ \log O\left(\log \log n + \log(1/\e)\right) } \cdot n = O( \log \log n + \log(1/\e))
\]
as stated in the theorem.

Each unitary of the obtained uncertainty relation is a product of $h$ unitaries each obtained from Theorem \ref{thm:explicit-ur1}. The overall number of unitaries is the product of the number of unitaries for each of the $h$ steps. As a result, we have $t \leq \left(\frac{n}{\e}\right)^{c \log n}$ for some constant $c$. $t$ can be taken to be a power of two as the number of unitaries at each step can be taken to be a power of two. As for the running time, every unitary transformation of the uncertainty relation is a product of $O(\log n)$ unitaries each computed by a quantum circuit of size $O( n \polylog (n/\e))$ and can thus be computed by a quantum circuit of size $O(n \polylog (n/\e))$.
\end{proof}

It is of course possible to obtain a trade-off between the key size and the dimension of the $B$ system by choosing the number of times the construction of Theorem \ref{thm:explicit-ur1} is applied. In the next corollary, we show how to obtain an explicit entropic uncertainty relation whose average entropy is $(1-\e) n$.

\begin{corollary}[Explicit entropic uncertainty relations]
\label{cor:explicit-entropic-ur}
Let $n \geq 100$ be an integer, and $\e \in ( 10n^{-1/2}, 1)$. Then, there exists $t \leq \left(\frac{n}{\e}\right)^{c \log (1/\e)}$ (for some constant $c$ independent of $n$ and $\e$) unitary transformations $U_1, \dots, U_{t}$ acting on $n$ qubits that are all computable by quantum circuits of size $O(n \polylog n)$ satisfying an entropic uncertainty relation: for all pure states $\ket{\psi} \in \left(\CC^2\right)^{ \otimes n} $,
\[
\frac{1}{t} \sum_{k=1}^t \entH(p_{U_k \ket{\psi}}) \geq (1 - 9\e) n - 2 \binent(2\e).
\]
Moreover, the mapping that takes the index $k \in [t]$ and a state $\ket{\psi}$ as inputs and outputs the state $U_k \ket{\psi}$ can be performed by a classical precomputation with polynomial runtime and a quantum circuit of size $O(n \polylog n)$. The number of unitaries $t$ can be taken to be a power of two.
\end{corollary}
\begin{proof}
The proof is basically the same as the proof of Theorem \ref{thm:explicit-ur2}, except that we repeat the construction $h = \ceil{\log(1/\e)/\log(8/7)}$ times. We thus have 
\[
\log d_{B_h} \leq \left(7/8\right)^{h} n \leq \e n.
\]
We obtain a set of $t \leq \left(\frac{n}{\e}\right)^{c\log(1/\e)}$ unitary transformations. Applying Proposition \ref{prop:metric-to-entropic}, we get
\begin{align*}
\frac{1}{t} \sum_{i=1}^{t} \entH(p_{U_k \ket{\psi}}) &\geq (1 - 8\e) (1-\e) n - 2\binent(2\e) \\
										&\geq (1-9\e) n - 2 \binent(2\e).
\end{align*}
\comment{
We added $\e \geq 10/\sqrt{n}$ to make sure that 
\[
\log d_{B_{h-1}} \geq \e n - 2 \log(8 \cdot 6^2 \cdot h^2/\e^2) \geq 10\sqrt{n} - 4\log(h/\e) - 18 \geq 2 \log(h/\e) + 10
\]
The reason this is an issue here is that the error $\e$ (in the MUR) and the size $\e n$ of the B system are taken to be the same.
}
\end{proof}

\chapter[Uncertainty relations: Applications]{Uncertainty relations for quantum measurements: Applications}
\label{chap:applications-ur}

\paragraph{Outline of the chapter} In this chapter, we give several applications of uncertainty relations. We start  in Section \ref{sec:locking} with applications related to information locking which all have a cryptographic flavour. In Section \ref{sec:qid}, we consider the communication problem called quantum identification.

\section{Locking classical information in quantum states}
\label{sec:locking}

\paragraph{Outline of the section}
We apply the results on metric uncertainty relations of the previous chapter to obtain locking schemes. After an introductory section on locking classical correlations (Section \ref{sec:locking-background}), we show how to obtain a locking scheme using a metric uncertainty relation in Section \ref{sec:locking-ur}. Using the constructions of the previous chapter, this leads to locking schemes presented in Corollaries \ref{cor:existence-locking} and   \ref{cor:explicit-locking}. Section \ref{sec:practical-locking} discusses the existence of error tolerant locking schemes. In Section \ref{sec:hiding-fingerprint}, we show how to construct quantum hiding fingerprints by locking a classical fingerprint. In Section \ref{sec:string-commitment}, we observe that these locking schemes can be used to construct efficient string commitment protocols. Section \ref{sec:ent-formation} discusses the link to locking entanglement of formation.

\subsection{Background}
\label{sec:locking-background}
Locking of classical correlations was first described by \cite{DHLST04} as a violation of the incremental proportionality of the maximal classical mutual information that can be obtained by local measurements on a bipartite state. More precisely, for a bipartite state $\omega^{AB}$, the maximum classical mutual information $\entI_c$ is defined by
\[
\entI_c(A;B)_{\omega} = \max_{\{M_i^A\}, \{M_i^B\} } \entI( I_{A}; I_{B}),
\]
where $\{M_i^A\}$ and $\{M_i^B\}$ are measurements on $A$ and $B$, and $I_{A}, I_B$ are the (random) outcomes of these measurements on the state $\omega^{AB}$.
Incremental proportionality is the intuitive property that $\ell$ bits of communication between two parties can increase their mutual information by at most $\ell$ bits. \citet{DHLST04} considered the states
\begin{equation}
\label{eq:lockingstate1}
\omega^{XKC} = \frac{1}{2d} \sum_{k=1}^2 \sum_{x=1}^{d} \proj{x}^X \ox \proj{k}^K \ox (U_k \proj{x} U_k^{\dagger})^C 
\end{equation}
where $U_1 = \1$ and $U_2$ is the Hadamard transform. 
It was shown by \citet{DHLST04} that the classical mutual information $\entI_c(XK; C)_{\omega} = \frac{1}{2} \log d$. However, if the holder of the $C$ system also knows the value of $k$, then we can represent the global state by the following density operator
\[
\omega^{XKCK'} = \frac{1}{2d} \sum_{k=1}^2 \sum_{x=1}^{d}  \proj{x}^X \ox \proj{k}^K \ox (U_k \proj{x} U_k^{\dagger})^C \ox \proj{k}^{K'}.
\]
It is easy to see that $\entI_c(XK;CK')_{\omega} = 1 + \log d$. This means that with only one bit of communication (represented by the register $K'$), the classical mutual information between systems $XK$ and $C$ jumped from $\frac{1}{2} \log d$ to $1+\log d$. In other words, it is possible to unlock $\frac{1}{2} \log d$ bits of information (about $X$) from the quantum system $C$ using a single bit.

\citet{HLSW04} proved an even stronger locking result. They generalize the state in equation \eqref{eq:lockingstate1} to
\begin{equation}
\label{eq:lockingstate2}
\omega^{XKCK'} = \frac{1}{t d} \sum_{x=1}^{d} \sum_{k=1}^{t} \proj{x}^X \ox \proj{k}^K \otimes (U_k \proj{x} U_k^{\dagger})^C \ox \proj{k}^{K'}
\end{equation}
where $U_k$ are chosen independently at random according to the Haar measure. They show that for any $\e > 0$, by taking $t = (\log d)^3$ and if $d$ is large enough, 
\[
\entI_c(X; C)_{\omega} \leq \e \log d \qquad \text{ and } \qquad \entI_c(XK; CK')_{\omega} = \log d+ \log t
\]
with high probability. Note that the size of the key measured in bits is only $\log t = O(\log \log d)$ and it should be compared to the $(1-\e) \log d$ bits of unlocked (classical) information. It should be noted that their argument is probabilistic, and it does not say how to construct the unitary transformations $U_k$. Standard derandomization techniques are not known to work in this setting. For example, unitary $t$-designs use far too many bits of randomness \citep{DCEL09}. Moreover, using a $\delta$-biased subset of the set of Pauli matrices fails to produce a locking scheme unless the subset has a size of the order of the dimension $d$ \citep{AS04,DD10} (see Section \ref{sec:app-pauli}).

Here, we view locking as a cryptographic task in which a message is encoded into a quantum state using a key whose size is much smaller than the message. Having access to the key, one can decode the message. However, an eavesdropper who does not have access to the key and has complete uncertainty about the message can extract almost no classical information about the message. 


\begin{definition}[$\e$-locking scheme]
\label{def:locking} Let $n$ be a positive integer, $\ell \in [0, n]$ and $\e \in [0,1]$.
An encoding $\cE : [2^n] \times [t] \to \cS(C)$ is said to be $(\ell, \e)$-\emph{locking} for the quantum system $C$ if:
\begin{itemize}
\item For all $x \neq x' \in [2^n]$ and all $k \in [t]$, $\tracedist{\cE(x,k),\cE(x', k)} = 1$.
\item Let $X$ (the message) be a random variable on $[2^n]$ with min-entropy $\entHmin(X) \geq \ell$, and $K$ (the key) be an independent uniform random variable on $[t]$. For any measurement $\{M_i\}$ on $C$ and any outcome $i$,
\begin{equation}
\label{eq:randomization}
\tracedist{p_{X|\event{I=i}}, p_{X}} \leq \e \ ,
\end{equation}
where $I$ is the outcome of measurement $\{M_i\}$ on the (random) quantum state $\cE(X, K)$. 

When the min-entropy bound $\ell$ is not specified, it should be understood that $\ell = n$ meaning that $X$ is uniformly distributed on $[2^n]$. The state $\cE(x, k)$ for $x \in [2^n]$ and $k \in [t]$ is referred to as the ciphertext.
\end{itemize}
\end{definition}
\begin{myremark}[]
The relevant parameters of a locking scheme are: the number of bits $n$ of the (classical) message, the dimension $d$ of the (quantum) ciphertext, the number $t$ of possible values of the key and the error $\e$. Strictly speaking, a classical one-time pad encryption, for which $t = 2^{n}$, is $(0,0)$-locking according to this definition. However, here we seek locking schemes for which $t$ is much smaller than $2^n$, say polynomial in $n$. This cannot be achieved using a classical encryption scheme.
\end{myremark}

In the remainder of this section, we comment on the definition.
We should stress first that this is not a composable cryptographic task, namely because an eavesdropper could choose to store quantum information about the message instead of measuring. In fact, as shown by \cite{KRBM07}, using the communicated message $X$ as a key for a one-time pad encryption might not be secure; see also \citep{FDHL10}.

Thus, a locking map destroys almost all classical correlations with the message, but it is impossible to erase all quantum correlations with a key significantly smaller than the message. For example, consider a map $\cE : \{0,1\}^n \times [t] \to \cS(C)$ such that the requirement \eqref{eq:randomization} is replaced by $\entI(X; C)_{\omega} \leq \delta$, where $\entI$ is the quantum mutual information computed for the state $\omega^{XKC} = \frac{1}{t2^n}\sum_{x, k} \proj{x}^X \otimes \proj{k}^K \otimes \cE(x,k)^C$. We have
\begin{align}
\entH(X) &= \entH(X) + \entH(C K) - \entH(C K) \notag \\
	&\leq \entH(X) + \entH(C) + \entH(K) - \entH(C K)  \notag.
\end{align}
Now we use the fact for all $k$, the states $\{\cE(x,k)\}_x$ are perfectly distinguishable. Thus, there exists an isometry that maps $\omega^{CK}$ to $\omega^{CKX}$.
Hence, $\entH(CK)_{\omega} = \entH(CKX)_{\omega}$. As a result,
\begin{align}
\entH(X)	&\leq \entH(X) + \entH(C) + \entH(K)  - \entH(C K X) \notag \\ 
	&\leq \entH(K) + \entH(X) + \entH(C)  - \entH(C X) \notag \\ \label{eq:shannon}
	&= \entH(K) + \entI(X;C).
\end{align}
This argument shows that if the key is much smaller than the message, then the quantum mutual information between the message and the ciphertext is large, it is in fact at least the size of the message minus the size of the key. It is basically the same argument that Shannon used to prove that any perfect encryption scheme has to use a key of size at least the message size \citep{Sha49}. The reason this argument fails for the classical mutual information $\entI_c$ is that the measurement to be made on the ciphertext to decode correctly depends on the value taken by the key. So replacing the system $C$ by the outcome $I$ of some fixed measurement on $C$, the inequality $\entH(IK) \geq \entH(IKX)$ does not hold.

One could compare a locking scheme to an entropically secure encryption scheme \citep{RW02,DS05}. These two schemes achieve the same task of encrypting a high entropy message using a small key. The security definition of a locking scheme is strictly stronger. In fact, for a classical eavesdropper (i.e., an eavesdropper that can only measure) an $\e$-locking scheme is secure in a strong sense. This additional security guarantee comes at the cost of upgrading classical communication to quantum communication. With respect to quantum entropically secure encryption \citep{Des09,DD10}, the security condition of a locking scheme is also more stringent (see Section \ref{sec:app-pauli} for an example of an entropically secure encryption scheme that is not $\e$-locking). However, a quantum entropically secure scheme allows the encryption of quantum states.

We mentioned that if the adversary has no quantum storage, then a message that is transmitted using a locking scheme can be used in subsequent protocols. In the following proposition, we show that it is still safe to re-use the transmitted message provided the adversary is only allowed to have a small quantum memory. We follow the same technique as in \cite[Corollary 2]{HMRRS10}.


\begin{proposition}
\label{prop:small-q-mem}
Let $\cE$ be an $\e$-locking scheme and $\cF : C \to YQ$ be a (eavesdropping) completely positive trace preserving map that sends all states $\cE(x,k)$ to states on $YQ$ that are classical on $Y$. Then, we have
\[
\tracedist{ \omega^{XYQ} , \omega^X \otimes \omega^{YQ}} \leq \e \cdot  c \sqrt{d_Q} \ ,
\]
where $\omega^{XKYQ} = \frac{1}{d_X d_K} \sum_{x,k} \proj{x} \ox \proj{k} \ox \cF( \cE(x,k) )$ and $c$ is a constant.
\end{proposition}
\begin{proof}
The idea is to use the fact that there exists a measurement that can be used to distinguish any pair of states reasonably well. More precisely, we use a result of \citet[Theorem 4]{AE07} that states that there exists a measurement map $\cN : \cL(Q) \to \cL(Z)$ such that for any $\omega_1, \omega_2$,
\begin{equation}
\label{eq:meas-random-basis}
\|\cN(\omega_1) - \cN(\omega_2)\|_1 \geq c d_Q^{-1/2} \| \omega_1 - \omega_2 \|_1 \ ,
\end{equation}
for some constant $c$; see also \cite{RRS09}.
We will need a slightly more general statement that applies to non-normalized states: for any $p_1 \geq p_2 \geq 0$,
\begin{equation}
\label{eq:meas-random-basis-p}
\|p_1\cN(\omega_1) - p_2\cN(\omega_2)\|_1 \geq c/4 \cdot d_Q^{-1/2} \| \omega_1 - \omega_2 \|_1.
\end{equation}
In order to prove this, we proceed as in \cite[Corollary 2]{HMRRS10}. We denote by $\{\mu_1(z)\}_z$ and $\{\mu_2(z)\}_z$ the outcome distributions for measurement $\cN$ on the states $\omega_1$ and $\omega_2$. We have
\begin{align*}
\| p_1 \cN(\omega_1) - p_2 \cN(\omega_2) \|_1 
&= \sum_z \left| p_1 \mu_1(z) - p_2 \mu_2 (z) \right| \\
&= \sum_z \left| p_1 (\mu_1(z) - \mu_2(z)) + (p_1 - p_2) \mu_2(z) \right|.
\end{align*} 
We now lower bound this expression in two different ways. First, we have
\begin{align*}
\sum_z \left| p_1 (\mu_1(z) - \mu_2(z)) + (p_1 - p_2) \mu_2(z) \right| &\geq \sum_{z : \mu_1(z) \geq \mu_2(z)} \left| p_1(\mu_1(z) - \mu_2(z)) \right| \\
&= p_1 \frac{\| \mu_1 - \mu_2 \|_1}{2} \ ,
\end{align*}
using the fact that $\mu_1$ and $\mu_2$ are probability distributions. Second, we have
\begin{align*}
\sum_z \left| p_1 (\mu_1(z) - \mu_2(z)) + (p_1 - p_2) \mu_2(z) \right| &\geq \left| \sum_{z} p_1(\mu_1(z) - \mu_2(z)) + (p_1 - p_2) \mu_2(z) \right| \\
&= | p_1- p_2 |.
\end{align*}
Thus, 
\begin{align*}
\| p_1 \cN(\omega_1) - p_2 \cN(\omega_2) \|_1 
&\geq \frac{|p_1 - p_2|}{2} + \frac{p_1 \| \cN(\omega_1) - \cN(\omega_2) \|_1}{4} \\
&\geq \frac{\|(p_1 - p_2) \omega_2 \|_1}{2} + \frac{p_1 c d_Q^{-1/2} \| \omega_1 - \omega_2 \|_1}{4} \\
&\geq cd_Q^{-1/2}/4 \cdot \| p_1 \omega_1 - p_2 \omega_2 \|_1.
\end{align*}
In the second inequality, we used the property \eqref{eq:meas-random-basis} of the measurement $\cN$ and in the third inequality, we used the triangle inequality. This proves property \eqref{eq:meas-random-basis-p}.

We are now in a position to prove the desired result. Let $\omega^{XYQ} = \frac{1}{d_X} \sum_{x,y} \proj{x} \ox p_{Y|X}(y|x) \proj{y} \ox \omega^Q_{x,y}$. We have
\begin{align*}
&\tracedist{\omega^{XYQ}, \omega^X \otimes \omega^{YQ}} \\
&=  \tracedist{ \sum_{x,y} \frac{ \proj{x} }{d_X} \otimes \proj{y} \ox p_{Y|X}(y|x)  \omega_{x,y},   \frac{ \id^X }{d_X} \otimes \sum_{x,y} \frac{p_{Y|X}(y|x)}{d_X} \proj{y} \otimes  \omega_{x,y}}.
\end{align*}
Letting $\omega_y  = \frac{1}{p_Y(y)} \sum_x \frac{p_{Y|X}(y|x)}{d_X} \omega_{x,y}$, we can write
\begin{align*}
&\tracedist{\omega^{XYQ}, \omega^X \otimes \omega^{YQ}} \\
&= \frac{1}{d_X} \sum_{x, y} \tracedist{  p_{Y|X}(y|x) \omega_{x,y},  p_Y(y) \omega_y} \\
&\leq \frac{1}{d_X} \sum_{x,y} \frac{4 d_Q^{1/2}}{c} \tracedist{  p_{Y|X}(y|x) \cN(\omega_{x,y}),  p_Y(y) \cN(\omega_y)} \\
&= \frac{4 d_Q^{1/2}}{c} \tracedist{\sum_{x,y} \frac{ \proj{x}}{ d_X }  \otimes p_{Y|X}(y|x) \proj{y} \otimes \cN( \omega_{x,y}), \frac{\id^X}{d_X} \otimes \sum_{y}  \proj{y}  \otimes p_Y(y) \cN(\omega_y) } \\
&= \frac{4 d_Q^{1/2}}{c} \tracedist{ p_{XYZ}, p_X \times p_{YZ} },
\end{align*}
where $Y, Z$ are obtained by performing the measurement defined by $(\id^Y \otimes \cN^{Q \to Z}) \circ \cF$ on the state $\cE(X,K)$. We conclude by using the fact that $\cE$ is $\e$-locking so that $\tracedist{ p_{XYZ}, p_X \times p_{YZ} } \leq \e$.
\end{proof}

Proposition \ref{prop:small-q-mem} is interesting for the scheme presented in Corollary \ref{cor:explicit-locking} below, for which the sender and the receiver do not use any quantum memory. One could then use such a scheme for key distribution in the bounded quantum storage model, where the adversary is only allowed to have a quantum memory of logarithmic size in $n$ but can have an arbitrarily large classical memory. Note that even though this is a strong assumption compared to the unconditional security of BB84 \citep{BB84}, one advantage of such a protocol for key distribution is that it only uses one-way communication between the two parties. In contrast, the BB84 quantum key distribution protocol needs interaction between the two parties.

Another remark about Definition \ref{def:locking} is that we used the statistical distance between $p_{X|\event{I=i}}$ and $p_X$ instead of the mutual information between $X$ and $I$ to measure the information gained about $X$ from a measurement. Using the trace distance is a stronger requirement as demonstrated by the following proposition.
\begin{proposition}
\label{prop:mutinfo}
Let $\e \in [0,1/2]$ and $\cE : [2^n] \times [t] \to \cS(C)$ be an $\e$-locking scheme. Define the state 
\[
\omega^{XKCK'} = \frac{1}{t d} \sum_{k=1}^t \sum_{x=1}^{2^n} \proj{x}^X \ox \proj{k}^K  \ox \cE(x, k)^C \ox \proj{k}^{K'}.
\] 
Then,
\[
\entI_c(X; C)_{\omega} \leq 8\e n + 2\binent(2\e) \qquad \text{ and } \qquad \entI_c(XK; CK')_{\omega} = n + \log t.
\]
\end{proposition}
\begin{proof}
First, we can suppose that the measurement performed on the system $X$ is in the basis $\{ \ket{x}\}_{x}$. In fact, the outcome distribution of any measurement on the $X$ system can be simulated classically  using the values of the random variables $X$.

Now let $I$ be the outcome of a measurement performed on the $C$ system. Using Fannes' inequality (a special case of Lemma \ref{lem:alicki-fannes}), we have for any $i$
\begin{align*}
\entH(X) - \entH(X|I=i) &\leq 8\tracedist{p_X, p_{X|\event{I=i}}} - 2\binent\left(2\tracedist{p_X, p_{X|\event{I=i}}}\right) \\
					&\leq 8\e n + 2\binent(2\e)
\end{align*}
using the fact that $\cE$ defines an $\e$-locking scheme. Thus,
\begin{align*}
\entI(X;I) &= \entH(X) - \sum_i \pr{I=i} \entH(X|I=i) \\
		&\leq 8\e n + 2\binent(2\e).
\end{align*}
As this holds for any measurement, we get $\entI_c(X; C)_{\omega} \leq 8\e n + 2\binent(2\e)$.
\end{proof}

The trace distance was also used in \cite{Dup09,FDHL10} to define a locking scheme. To measure the leakage of information about $X$ caused by a measurement, they used the probably more natural trace distance between the joint distribution of $p_{(X, I)}$ and the product distribution $p_{X} \times p_I$. Note that our definition is stronger, in that for all outcomes of the measurement $i$, $\tracedist{p_{X|\event{I=i}}, p_X} \leq \e$ whereas the definition of \cite{FDHL10} says that this only holds on average over $i$. The condition of \cite{FDHL10} is probably sufficient for most applications but our techniques naturally achieve the stronger form without degrading the parameters. We should finally note that the trace distance condition cannot be much stronger than the condition on the classical mutual information. In fact, using Pinsker's inequality, we can upper bound the trace distance using the mutual information: 
\[
\tracedist{p_{(X,I)}, p_X \times p_I} \leq \sqrt{\entI(X;I)/2}.
\]

For a survey on locking classical correlations, see~\cite{Leu09}.

\subsubsection{Other related work}
In a cryptographic setting, \citet{DPS04} used ideas related to locking to develop quantum ciphers that have the property that the key used for encryption can be recycled. In \cite{DPS05}, they construct a quantum key recycling scheme (see also \cite{OH05}) with near optimal parameters by encoding the message together with its authentication tag using a full set of mutually unbiased bases.

\subsection{Locking using a metric uncertainty relation}
\label{sec:locking-ur}

\begin{center}
\begin{figure}[h]
\begin{center}
\includegraphics[scale=0.9]{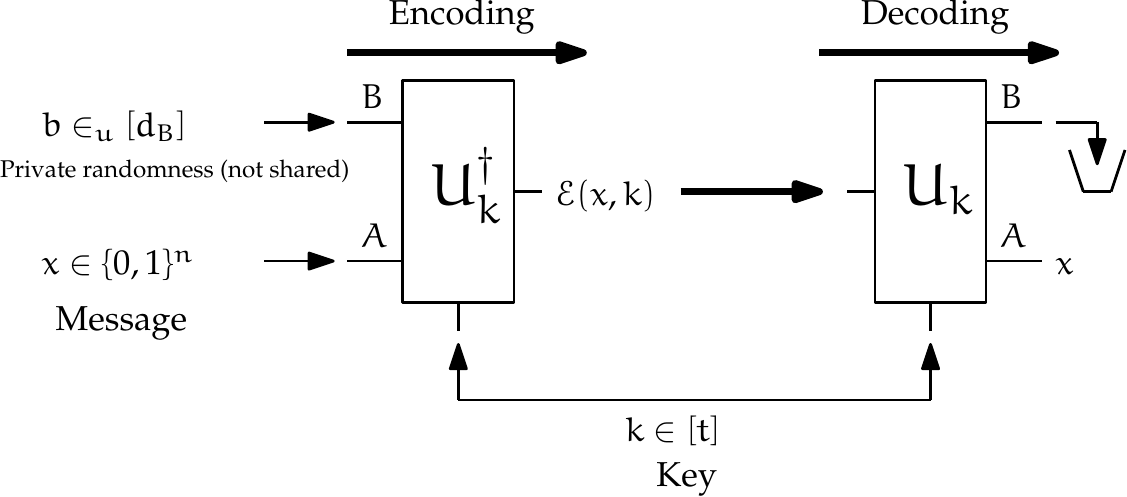}
\end{center}
\caption{Illustration of the locking scheme described in Theorem \ref{thm:ur-locking}.}
\label{fig:mur-locking}
\end{figure}
\end{center}

The following theorem shows that a locking scheme can easily be constructed using a metric uncertainty relation.
\begin{theorem}
\label{thm:ur-locking}
Let $\e \in (0,1)$ and $\{U_1, \dots, U_{t}\}$ be a set of unitary transformations of $A \otimes B$ that satisfies an $\e$-metric uncertainty relation on $A$, i.e., for all states $\ket{\psi} \in AB$,
\[
\frac{1}{t} \sum_{k=1}^t \tracedist{p^A_{U_k \ket{\psi}}, \unif([d_A])} \leq \e.
\]
Assume $d_A = 2^n$. Then, the mapping $\cE : [2^n] \times [t] \to \cS(A B)$ defined by
\[
\cE(x, k) = \frac{1}{d_B} \sum_{b=1}^{d_B} U_k^{\dagger} \left( \proj{x}^A \otimes \proj{b}^B \right) U_k. 
\]
is $\e$-locking. Moreover, for all $\ell \in [0, n]$ such that $2^{\ell - n} > \e$, it is $(\ell, \frac{2\e}{2^{\ell-n} - \e})$-locking.
\end{theorem}
\begin{myremark}[]
Figure \ref{fig:mur-locking} illustrates the locking scheme.
The state that the encoder inputs in the $B$ system is simply private randomness. The encoder chooses a uniformly random $b \in [d_B]$ and sends the quantum state $U_k^{\dagger} \ket{x}^A \ket{b}^B$. Note that $b$ does not need to be part of the key (i.e., shared with the receiver). This makes the dimension $d = d_A d_B$ of the ciphertext larger than the number of possible messages $2^n$. If one insists on having a ciphertext of the same size as the message, it suffices to consider $b$ as part of the message and apply a one-time pad encryption to $b$. The number of possible values taken by the key increases to $t \cdot d_B$.
\end{myremark}

\begin{proof}
First, it is clear that different messages are distinguishable. In fact, for $x \neq x'$ and any $k$,
\[
\tracedist{\cE(x,k), \cE(x', k)} = \frac{1}{2}\tr \left[ \sqrt{ \proj{x}^A \otimes \frac{\1^B}{d^2_B} - \proj{x'}^A \otimes \frac{\1^B}{d^2_B} } \right] = 1.
\]
We now prove the locking property. Let $X$ be the random variable representing the message. Assume that $X$ is uniformly distributed over some set $S \subseteq [d_A]$ of size $|S| \geq 2^{\ell}$. Let $K$ be a uniformly random key in $[t]$ that is independent of $X$. Consider a POVM $\{M_i\}$  on the system $AB$. Without loss of generality, we can suppose that the POVM elements $M_i$ have rank $1$. Otherwise, by writing $M_i$ in its eigenbasis, we could decompose outcome $i$ into more outcomes that can only reveal more information. So we can write the elements as weighted rank one projectors: $M_i = \xi_i\proj{e_i}$ where $\xi_i > 0$. Our objective is to show that the outcome $I$ of this measurement on the state $\cE(X,K)$ is almost independent of $X$. More precisely, for a fixed measurement outcome $I=i$, we want to compare the conditional distribution $p_{X|\event{I=i}}$ with $p_X$. The trace distance between these distributions can be written as
\begin{equation}
\label{eq:avgoutcome}
\frac{1}{2} \sum_{x=1}^{d_A} \big| \pr{X=x |  I=i} - \pr{X=x} \big|.
\end{equation}

Towards this objective, we start by computing the distribution of the measurement outcome $I$, given the value of the message $X=x$ (note that the receiver does not know the key):
\begin{align*}
\pr{ I = i | X = x} 
	&= \frac{\xi_i}{t d_B} \sum_{k=1}^t \sum_{b=1}^{d_B} \tr \big[U_k \proj{e_i} U_k^\dg \cdot \proj{x}^A \otimes \proj{b}^B \big] \\
	&= \frac{\xi_i}{t d_B} \sum_{k=1}^t \sum_{b=1}^{d_B} \bra{x}^A \bra{b}^B U_k \proj{e_i}  U^\dg_k \ket{x}^A \ket{b}^B \\
	&= \frac{\xi_i}{t d_B} \sum_{k=1}^t \sum_{b=1}^{d_B} \left|\bra{x}^A \bra{b}^B U_k \ket{e_i} \right|^2 \\
	&= \frac{\xi_i}{d_B} \frac{1}{t} \sum_{k=1}^t p^A_{U_k \ket{e_i}}(x).
\end{align*}
Since $X$ is uniformly distributed over $S$, we have that for all $x \in S$
\begin{align}
\pr{X=x|I=i} &= \frac{\pr{X=x} \pr{ I=i | X=x} }{ \sum_{x' \in S} \pr{X=x'} \pr{ I=i | X=x'} }   \notag \\ 
	&=  \frac{(1/t) \cdot\sum_k p^A_{U_k \ket{e_i}}(x) }{ (1/t) \cdot \sum_{x' \in S} \sum_k p^A_{U_k \ket{e_i}}(x') }.
\end{align}
Observe that in the case where $X$ is uniformly distributed over $[2^n]$ ($S = [2^n]$), it is simple to obtain directly that
\[
\tracedist{p_{X| \event{I=i}}, p_X} = \frac{1}{2} \sum_{x=1}^{d_A} \left|\frac{1}{t}\sum_{k=1}^t p^A_{U_k \ket{e_i}}(x) - \frac{1}{2^n} \right| \leq \e
\]
using the fact that $\{U_k\}$ satisfies a metric uncertainty relation on $A$.
Now let $S$ be any set of size at least $2^{\ell}$ and let $\alpha = \frac{1}{t} \sum_{x' \in S} \sum_k p^A_{U_k \ket{e_i}}(x')$. We then bound	
\begin{align*}
\frac{1}{2} \sum_{x=0}^{d_A-1} & \big| \pr{X=x |  I=i} - \pr{X=x} \big| \\
	 &= \frac{1}{2} \sum_{x \in S} \left| \frac{(1/t) \cdot \sum_k p^A_{U_k \ket{e_i}}(x) }{ \alpha } - \frac{1}{|S|}\right| \\
						&= \frac{1}{2 \alpha} \cdot  \sum_{x \in S} 
					\left| \frac{1}{t} \sum_{k=1}^t p^A_{U_k \ket{e_i}}(x) - \frac{\alpha}{|S|}\right| \\
						&\leq \frac{1}{2 \alpha} \cdot \frac{1}{t} \sum_k \left( \sum_{x \in S} 
					\left|  p^A_{U_k \ket{e_i}}(x) - \frac{1}{2^n} \right| + \left| \frac{1}{2^n} - \frac{\alpha}{|S|}\right| \right).
\end{align*}
We now use the fact that $\{U_k\}$ satisfies a metric uncertainty relation on $A$:  we get
\[
\frac{1}{t} \sum_k \frac{1}{2} \sum_{x \in S} \left|  p^A_{U_k \ket{e_i}}(x) - \frac{1}{2^n} \right| \leq \frac{1}{t} \sum_k \frac{1}{2} \sum_{x \in [d_A]} \left|  p^A_{U_k \ket{e_i}}(x) - \frac{1}{2^n} \right| \leq \e
\]
and
\begin{equation}
\label{eq:alpha}
\frac{1}{2} \left| \frac{|S|}{2^n} - \alpha\right| = \frac{1}{2} \left| \frac{|S|}{2^n} - \frac{1}{t} \sum_{x' \in S} \sum_{k=1}^t p^A_{U_k \ket{e_i}}(x') \right| \leq \e.
\end{equation}
As a result, we have
\[
\tracedist{p_{X| \event{I=i}}, p_X} \leq \frac{2\e}{\alpha}.
\]
Using \eqref{eq:alpha}, we have $\alpha \geq |S|2^{-n} - \e \geq 2^{\ell - n} - \e$. If $\e < 2^{\ell - n}$, we get
\[
\tracedist{p_{X| \event{I=i}}, p_X} \leq \frac{2\e}{2^{\ell - n} - \e}.
\]

In the general case when $X$ has min-entropy $\ell$, the distribution of $X$ can be seen as a mixture of uniform distributions over sets of size at least $2^{\ell}$. So there exist independent random variables $J \in \NN$ and $\{X_j\}$ uniformly distributed on sets of size at least $2^{\ell}$ such that $X = X_J$. One can then write
\begin{align*}
&\frac{1}{2} \sum_x \left| \pr{X = x| I = i} - \pr{X=x} \right| \\
&= \frac{1}{2} \sum_{x,j} \left| \pr{J = j} \left( \pr{X_j=x | I = i, J=j} - \pr{X_j = x | J=j} \right) \right| \\
								&\leq \frac{2\e}{2^{\ell-n} - \e}.
\end{align*}
\end{proof}

Using Theorem \ref{thm:ur-locking} together with the existence of metric uncertainty relations (Theorem \ref{thm:existence-ur}), we show the existence of $\e$-locking schemes whose key size depends only on $\e$ and not on the size of the encoded message. 
\begin{corollary}[Existence of locking schemes]
\label{cor:existence-locking}
Let $n$ be a large enough integer and $\e \in (0,1)$. Then there exists an $\e$-locking scheme encoding an $n$-bit message using a key of at most $2\log(1/\e) + O(\log \log(1/\e))$ bits into at most $n + 2 \log(18/\e)$ qubits.
\end{corollary}
\begin{myremark}[]
Observe that in terms of number of bits, the size of the key is only a factor of two larger (up to smaller order terms) than the lower bound of $\log(1/(\e + 2^{-n}))$ bits that can be obtained by guessing the key. In fact, consider the strategy of performing the decoding operation corresponding to the key value $0$. In this case, we have $\pr{X=i|I=i} \geq \pr{K=0} = 1/t$. Thus, $\tracedist{p_{X|I=i}, p_X} \geq 1/t - 2^{-n}$.

Recall that we can increase the size of the message to be equal to the number of qubits of the ciphertext at the cost of increasing the key size to at most $4\log(1/\e) + O(\log(\log(1/\e))$.
\end{myremark}
\begin{proof}
Use the construction of Theorem \ref{thm:existence-ur} with $d_A = 2^n$ and $d_B = 2^q$ such that $2^{q-1} < 9/\e^2 \leq 2^q$ and $d = d_A d_B$. Take $t = 2^p$ to be the power of two with $2^{p-1} \leq \frac{4 \cdot 18c\log (9/\e)}{\e^2} < 2^p$.
\end{proof}

To construct $(\ell, \e)$-locking schemes with $\ell < n$, it suffices to use Theorem \ref{thm:existence-ur} with say $\e' = 2^{\ell - n} \e/4$. In this case, we obtain a key of size $O((n - \ell) + \log(1/\e))$. We note that this increase in the key size is unavoidable because of the following proposition.


\begin{proposition}
\label{prop:lb-key-size}
Assume $\cE$ defines an $(\ell, \e)$-locking scheme with $\e \leq 1/4$ and a key of size $\log t$. Then $\log t \geq n - \ell - 2$.
\end{proposition}
\begin{proof}
We proceed as in the proofs of lower bounds on the key size in entropic security \citep{DS05}. 
The idea is to show that if $\cE$ is an $(\ell, \e)$-locking scheme, then it can be used to build an encryption scheme for messages of $n-\ell$ bits that has the following properties. Given the secret key, the encryption of $w$ and $w'$ are perfectly distinguishable, but without the key, the encryption of $w$ and $w'$ are almost indistinguishable. 
For such a scheme, we show in Proposition \ref{prop:lb-key-size-encryption} that the key size is at least the size of the message $\log t \geq n - \ell - 2$.


Define the random variables $X_{w}$ for $w \in \{0,1\}^{n-\ell}$ which are uniformly distributed on $\{w \} \times \{0,1\}^{\ell}$. Our encryption scheme encrypts $w$ using the key $k$ into $\cE(X_w, k)$. First, clearly a decoder having the key can determine $w$ using $\cE(X_w, k)$. Second, we show that the ciphertexts corresponding to $w$ and $w' \neq w$ are almost indistinguishable:
\begin{equation}
\label{eq:indinst-ciphers}
\tracedist{\rho_{w}, \rho_{w'}} \leq 2\e,
\end{equation}
where $\rho_w = \frac{1}{k 2^{\ell}} \sum_{k \in [t], y \in \{0,1\}^{\ell} }\cE( w \cdot y , k)$. To show this we let $\Lambda$ be a positive operator such that $\tracedist{\rho_{w}, \rho_{w'}} = \tr [ \Lambda (\rho_w - \rho_{w'}) ]$ (see equation \eqref{eq:tracedist-max}).

We then have
\begin{align*}
\tr [ \Lambda (\rho_w - \rho_{w'}) ] &\leq \left|\tr \left[ \Lambda \left(\frac{\rho_w + \rho_{w'}}{2} - \rho_{w'}\right) \right] \right| +\left |\tr \left[ \Lambda \left(\frac{\rho_w + \rho_{w'}}{2} - \rho_{w}\right) \right] \right| \\
&\leq \frac{1}{2^{\ell}} \sum_y \left|\tr \left[ \Lambda \left(\frac{\rho_w + \rho_{w'}}{2} - \frac{1}{k} \sum_{k \in [t]}\cE( w' \cdot y , k) \right) \right] \right| \\
&+ \frac{1}{2^{\ell}} \sum_{y} \left |\tr \left[ \Lambda \left(\frac{\rho_w + \rho_{w'}}{2} - \frac{1}{k} \sum_{k \in [t] }\cE( w \cdot y , k) \right) \right] \right| \\
&= 2\frac{1}{2^{\ell + 1}} \sum_{z \in \{w,w'\} \times \{0,1\}^{\ell}} \left |\tr \left[ \Lambda \left( \frac{\rho_w + \rho_{w'}}{2} - \frac{1}{k} \sum_{k \in [t] }\cE( z , k) \right) \right] \right| \\
&= 2 \sum_{z \in \{w,w'\} \times \{0,1\}^{\ell}} \pr{Z=z} \left|\pr{I=0} - \pr{I=0|Z=z} \right| \\
&= 2 \tracedist{p_{ZI}, p_{Z} \times p_I},
\end{align*}
where $Z$ is uniformly distributed on $\{w,w'\} \times \{0,1\}^{\ell}$ and $I$ is the outcome of the measurement $\{\Lambda, \1 - \Lambda\}$ performed on the state $\cE(Z, K)$. Inequality \eqref{eq:indinst-ciphers} follows from the fact that $\cE$ is an $(\ell, \e)$-locking scheme.

Using Proposition \ref{prop:lb-key-size-encryption}, we conclude that $\log t \geq n - \ell - 2$.
\end{proof}

The following corollary gives explicit locking schemes. We mention the constructions based on Theorems \ref{thm:explicit-ur1} and \ref{thm:explicit-ur2}. Of course, one could obtain a tradeoff between the key size and the dimension of the quantum system.
\begin{corollary}[Explicit locking schemes]
\label{cor:explicit-locking}
Let $\delta > 0$ be a constant, $n$ be a positive integer, $\e \in (2^{-c'n},1)$ ($c'$ is a constant independent of $n$). 
\begin{itemize}
\item Then, there exists an efficient $\e$-locking scheme encoding an $n$-bit message in a quantum state of $n' \leq (4+\delta) n + O(\log(1/\e))$ qubits using a key of size $O(\log(n/\e))$ bits. In fact, both the encoding and decoding operations are computable using a classical computation with polynomial running time and a quantum circuit with only Hadamard gates and preparations and measurements in the computational basis. 

\item There also exists an efficient $\e$-locking scheme $\cE'$ encoding an $n$-bit message in a quantum state of $n$ qubits using a key of size $O(\log(n/\e) \cdot \log n)$ bits. $\cE'$ is computable by a classical algorithm with polynomial runtime and a quantum circuit of size $O(n \polylog(n/\e))$.
\end{itemize}
\end{corollary}
\begin{proof}
For the first result, we observe that the construction of Theorem \ref{thm:ur-locking} encodes the message in the computational basis. Recall that the untaries $U_k$ of Theorem \ref{thm:explicit-ur1} are of the form $U_k = P_k V_k$ where $P_k$ is a permutation of the computational basis. Hence, it is possible to \emph{classically} compute the label of the computational basis element $P^{\dagger}_k \ket{x} \ket{b}$. One can then prepare the state $P^{\dagger}_k \ket{x} \ket{b}$ and apply the unitary $V^{\dagger}_k$ to obtain the ciphertext. The decoding is performed in a similar way. One first applies the unitary $V_k$, measures in the computational basis and then applies the permutation $P_k$ to the $n$-bit string corresponding to the outcome.


For the second construction, we apply Theorem \ref{thm:explicit-ur2} with $n' =  n + c'\ceil{\log\log n + \log(1/\e)}$ for some large enough constant $c'$.  We can then use a one-time pad encryption on the input to the $B$ system. This increases the size of the key by only $c'\ceil{\log\log n + \log(1/\e)}$ bits.
\end{proof}

As mentioned earlier (see equation \eqref{eq:lockingstate1}), explicit states that exhibit locking behaviour have been presented in \cite{DHLST04}. However, this is the first explicit construction of states $\omega$ that achieves the following strong locking behaviour: for any $\delta > 0$, for $n$ large enough, the state $\omega^{XCK}$ verifies $\entI_c(X; C)_{\omega} \leq \delta$ and $\entI_c(X; CK)_{\omega} = n + \log d_K$ where $K$ is a classical $O(\log(n/\delta))$-bit system. This is a direct consequence of Corollary \ref{cor:explicit-locking} taking $\e = \delta/(20n)$, and Proposition \ref{prop:mutinfo}. 
We should also mention that \citet{KRBM07} explicitly construct a state exhibiting some weak locking behaviour. We summarize the different locking schemes in Table \ref{tbl:locking-results}.

\begin{table}
\caption{Comparison of different locking schemes. $n$ is the number of bits of the message. The information leakage and the size of the key are measured in bits and the size of the ciphertext in qubits.  Efficient locking schemes have encoding and decoding quantum circuits of size polynomial in $n$. The locking schemes of the first and next to last actually have encoding circuits that are in principle implementable with current technology; they only use classical computations and simple single-qubit transformations. It should be noted that our locking definition is stronger than all the previous definitions.
Note that the variable $\e$ can depend on $n$. For example, one can take $\e = \eta/n$ to make the information leakage arbitrarily small.  The symbol $O( \cdot )$ refers to constants independent of $\e$ and $n$, but there is a dependence on $\delta$ for the next to last row. The symbol $\mathrm{ll}(\cdot)$ refers to $O(\log \log(\cdot))$.}
\begin{tabular}{|c||c|c|c|c|}
\hline
			& Inf. leak. & Key & Ciphertext & {\small Efficient?} \\ \hline \hline
\citeauthor{DHLST04}
	& $n/2$ & $1$ & $n$ & yes \\ \hline
\citeauthor{HLSW04} 	
& $3$ & $4 \log(n)$ & $n$ & no \\ \hline
\citeauthor{FDHL10} 	
& $\e n$ & $2\log(n/\e^2) + O(1)$ & $n$ & no \\ \hline \hline
Corollary \ref{cor:existence-locking}	& $\e n$  & $2 \log(1/\e) + \mathrm{ll}(1/\e)$ & $n + 2 \ceil{\log(9/\e)}$ & no \\ \hline
Corollary \ref{cor:existence-locking} 	& $\e n$  & $4 \log(1/\e) + \mathrm{ll}(1/\e)$ & $n$ & no \\ \hline
Corollary \ref{cor:explicit-locking}	& $\e n$ & $O_{\delta}(\log(n/\e))$ & $(4+\delta) \cdot n$ & yes \\ \hline
Corollary \ref{cor:explicit-locking}	& $\e n$ & $O(\log(n/\e) \log(n))$ & $n$ & yes \\ \hline
\end{tabular}
\label{tbl:locking-results}
\end{table}

\subsection{Impossibility of locking using Pauli operators}
\label{sec:app-pauli}
The objective of this section is to give an example of a construction that is not a locking scheme to illustrate what is needed to obtain a locking scheme. The $2 \times 2$ Pauli matrices are the four matrices $\{\1, \sigma_x, \sigma_z, \sigma_x\sigma_z\}$ where
\[
\sigma_x =
\left( \begin{array}{cc}
0 & 1  \\
1 & 0  \\
\end{array} \right) 
\qquad \textrm{ and } \qquad 
\sigma_z =
\left( \begin{array}{cc}
1 & 0  \\
0 & -1  \\
\end{array} \right).
\]
For bit strings $u,v \in \{0,1\}^n$, we define the unitary operation $\sigma_x^u \sigma_z^v$ on $\left(\CC^{2}\right)^{\otimes n}$ by
\[
\sigma_x^u \sigma_z^v = \sigma_x^{u_1} \sigma_z^{v_1} \otimes \dots \otimes \sigma_x^{u_n} \sigma_z^{v_n}.
\]
It was shown by \cite{AMTW00} that one can encrypt an $n$-qubit state $\ket{\psi}$ perfectly using a key $(U,V)$ of $2n$ bits. To encrypt $\ket{\psi}$, one simply applies $\sigma_x^{U} \sigma_z^{V}$ to $\ket{\psi}$, where $U$ and $V$ are uniformly distributed on $\{0,1\}^n$. This can be thought of as a quantum version of one-time pad encryption. Of course, this encryption scheme also defines a $(0,0)$-locking scheme, but the size of the key is $2n$ bits. Recall that we want to use the assumption that the message is random to reduce the key size to $O(\polylog(n))$ bits.

\citet*{AS04} showed that to achieve approximate encryption, it is sufficient to choose the key uniformly at random from a well-chosen subset $S \subseteq \{0,1\}^{2n}$ of size only $O(n^2 2^n)$. Such pseudorandom subsets are called $\delta$-biased sets and have also been used to construct entropically secure encryption schemes \citep{DS05,DD10}. For example, \citet{DD10} showed that it is possible to encrypt a uniformly random state by applying $\sigma_x^{U}\sigma_z^{V}$ where $(U,V)$ is chosen uniformly from a set $S \subset \{0,1\}^n$ of size $O(n^2)$ (see \citep{DS05,DD10} for a precise definition of entropic security). Such a set of transformations can seem like a good candidate for a locking scheme. The following proposition shows that this scheme is far from being $\e$-locking. Note that this also shows that the notion of entropic security defined in \citep{Des09,DD10} is weaker than the definition of locking.

\begin{proposition}
\label{prop:subset-pauli}
Consider an $\e$-locking scheme $\cE$ of the form $\cE(x,k = (u,v)) = \sigma_x^u \sigma_y^v \ket{x}$ where the message $x \in \{0,1\}^n$ and the key $u,v \in \{0,1\}^{n}$ (see Definition \ref{def:locking}). Suppose the secret key $K$ is chosen uniformly from a set $S \subseteq \{0,1\}^{2n}$. Then $|S| \geq (1-\e) 2^n$.
\end{proposition}
\begin{proof}
Let $X$ be the message ($X$ is uniformly distributed over $\{0,1\}^n$) and $(U,V)$ be the key. The key is uniformly distributed on $S$.
We show that a measurement in the computational basis gives a lot of information about $X$ . Let $I$ be the outcome of measuring $\cE(X,K)$ in the computational basis. We have for $x, i \in \{0,1\}^n$,
\begin{align*}
\pr{X = x| I = i} 	&= \pr{I = i | X=x} \\
			&= \frac{1}{|S|} \sum_{(u,v) \in S} \left| \bra{i} \sigma_x^u \sigma_z^v \ket{x} \right|^2.
\end{align*}
Observing that the term $\left| \bra{i} \sigma_x^u \sigma_z^v \ket{x} \right|^2 \in \{0,1\}$, we have that for any fixed $i$, there are at most $|S|$ different values of $x$ for which $\pr{X=x|I=i} > 0$. Thus, defining $T = \{x \in \{0,1\}^n : \pr{X=x|I=i} = 0\}$, we have
\[
\tracedist{p_{X|\event{I=i}}, p_X} \geq \pr{X \in T} - \pr{X \in T | I = i} = \frac{|T|}{2^n} \geq 1 - \frac{|S|}{2^n}.
\]
By the definition of a locking scheme, we should have 
\[
\tracedist{p_{X|\event{I=i}}, p_X} \leq \e
\]
which concludes the proof.
\end{proof}

\subsection{Error-tolerant information locking}
\label{sec:practical-locking}

The first protocol in Corollary \ref{cor:explicit-locking} can in principle be implemented using current technology. We say ``in principle'' because a locking scheme as defined here does not allow for any error in the transmission of the ciphertext. Can we construct a locking scheme that can tolerate a reasonable rate of errors?

One simple approach to build a protocol that tolerates errors is to use a \emph{quantum} error correcting code (QECC) to encode the ciphertext. Depending on the properties of the code, this would allow the receiver to correct some fraction of errors. Moreover, the security is preserved because an eavesdropper could perform an encoding into a QECC as part of his attack. Thus, it is possible to make any locking scheme error-tolerant provided we can perform encoding and decoding operations for a good QECC. Unfortunately, the encoding and decoding maps of interesting quantum error correcting codes are beyond the reach of current technology. But note that our objective is not necessarily to recover the quantum ciphertext correctly, we only want to be able to recover the classical message. Can we construct a locking scheme that can tolerate a reasonable rate of errors and that can be implemented with current technology?



In the remainder of this section, we show that some natural class of error tolerant protocols cannot be good locking schemes. Consider a locking scheme of the following form. The key is written as $k \in [t]$ and the message $x \in \{0,1\}^n$ is locked in the following way:
\begin{itemize}
	\item A classical (possibly randomized) function determined by the key $k$ is applied to $x$ . $x$ is mapped to $P_k(x, r) \in \{0,1\}^{n'}$, where $r$ is a random string private to Alice.
	\item The bitstring $P_k(x,r)$ is then encoded in a code $C_k$ possibly depending on the key $k$. The codes $C_k$ are assumed to have minimum distance $\alpha$ for all $k$. This bitstring is denoted $z=C_k(P_k(x,r)) \in \{0,1\}^{m}$. We denote by $C_k$ the set of bitstrings in this code.
	\item A quantum encoding of the form $H^{v_k}$ where $v_k \in \{0,1\}^m$ is performed on the computational basis element $\ket{C_k(P_k(x,z))}$. 
\end{itemize}

We start with a lemma that says that given a set of vectors that are almost orthogonal, they can be well approximated by orthogonal vectors. It was first proven by \cite{Sho66}; see also \cite[Claim 20]{KV11}.
\begin{lemma}
\label{lem:orth}
Let $s \leq d$ and $\ket{u_1}, \dots, \ket{u_s} \in \CC^d$ be unit vectors such that $\frac{1}{s} \sum_{i \neq j} |\braket{u_i}{u_j}|^2 \leq \e$. Then there exist orthogonal unit vectors $\ket{v_1}, \dots, \ket{v_s}$ such that $\frac{1}{s} \sum_i \| \ket{u_i} - \ket{v_i} \|_2^2 \leq \e$.
\end{lemma}
\begin{proof}
We start by fixing a set of orthonormal vectors $\ket{w_1}, \dots, \ket{w_s}$ such that the span of $\ket{u_1}, \dots, \ket{u_s}$ is included in the span of $\ket{w_1}, \dots, \ket{w_s}$. We then define the matrix $X$ whose columns represent the vectors $\ket{u_i}$ using the vectors $\ket{w_1}, \dots, \ket{w_s}$. Write the SVD decomposition of $X$ as $X = U \Sigma V^{\dagger}$ and let the singular values of $X$ be $\sigma_1, \dots, \sigma_s$. We have
\begin{align*}
\sum_{i \in [s]} (1-\sigma_i^2)^2 = \| \1 - \Sigma^{\dg} \Sigma \|^2_2 = \| \1 - X^{\dg} X \|_2^2 &= \sum_{i \neq j} | \braket{u_i}{u_j}|^2 + \sum_{i \in [s]} (1 - \braket{u_i}{u_i})^2 \\
& \leq s\e
\end{align*}
by assumption. We now define $Y = UV^{\dagger}$, look at the columns of this matrix and call these vectors $\ket{v_i}$ (of course the underlying basis used is still $\ket{w_i}$). We have by writing the desired expression in the basis $\ket{w_i}$ and then multiplying by $U^{\dg}$ on the left and $V$ on the right:
\begin{align*}
\sum_{i} \| \ket{u_i} - \ket{v_i} \|_2^2 =  \| X - Y \|_2^2 = \| \Sigma - \1 \|_2^2 &= \sum_i (\sigma_i - 1)^2 \\
 &\leq \sum_i (1-\sigma_i)^2 (1 + \sigma_i)^2 \leq s \e.
\end{align*}
\end{proof}

\begin{proposition}
For any encoding of the form above such that the random variable $(Z, V) = (P_K(X, R), v_K)$ is uniformly distributed on its support, there exists a measurement on the ciphertext that gives an outcome $I$ such that
\[
\entI(I; X) \geq (1-16 \cdot t2^{-\alpha+1}) n  - \log t - 2,
\]
provided $t 2^{-\alpha+1} \leq 1$.
\end{proposition}
\begin{remark}
Recall that we want $t$ to be sub-exponential (even polynomial) in $n$. Moreover, to be able to correct a constant fraction of errors, we want the minimum distance $\alpha$ to be linear in $n'$. In this case, $t2^{-\alpha+1} \ll 1$ and the measurement given by the proposition completely breaks the locking scheme. 
\end{remark}
\begin{proof}
We will apply Lemma \ref{lem:orth} to the set of vectors $\{H^v \ket{z} : (z,v) \in S\}$ with $S = \{ (z, v) : \exists k : z \in C_k, v = v_k\}$. 
Note that a pair $(z, v)$ that appears for different values of $k$ is counted only once. Fix any $z'$ and $v'$ (not necessarily in $S$). We have
\begin{align}
\sum_{(z,v_k) \in S, (z,v_k) \neq (z', v')} |\bra{z'} H^{v'} H^{v_{k}} \ket{z}|^2 &= 
\sum_{v_k=v', z \neq z'} |\braket{z'}{z}|^2
+
\sum_{v_k \neq v', z \in C_k} |\bra{z'} H^{v' + v_{k}} \ket{z}|^2 \notag\\
&= 0 + \sum_{v_k \neq v', z \in C_k \cap B_{v_k + v'}(z')} |\bra{z'} H^{v' + v_{k}} \ket{z}|^2 \label{eq:sum-dot-product}
\end{align}
where $B_{v_k + v'}(z') =  \{ y \in \{0,1\}^m : y_i = z'_i \text{ whenever } (v_k + v')_i = 0\}$. Observe that for all $z$ and $v$, $|B_v(z)| \leq 2^{\weight(v)}$ and in this case $\weight(v_k + v') = \hamdist(v_k, v')$. We now fix some $k$ such that $v_{k} \neq v'$ and bound the size of $C_k \cap B_{v_k + v'}$ assuming that $C_k$ is a code of minimum distance $\alpha$. The strings of $C_k \cap B_{v_k + v'}$ agree on $2^{n-\hamdist(v_k, v')}$ bits. This means it induces a code of minimum distance $\alpha$ on strings of length $\hamdist(v_k, v')$. Using the Singleton bound, we get
\[
|C_k \cap B_{v_k + v'}| \leq 2^{\hamdist(v_k, v') - \alpha+ 1}.
\]
To bound the expression in \eqref{eq:sum-dot-product}, we observe that $|\bra{z'} H^{v' + v_{k}} \ket{z}|^2 \leq 2^{-\hamdist(v_k, v')}$. Thus, for a fixed $v_k$,
\begin{align*}
\sum_{z \in C_k \cap B_{v_k+v'}(z')} |\bra{z'} H^{v' + v_k} \ket{z}|^2 &\leq 2^{-\hamdist(v_k, v')} \cdot 2^{\hamdist(v_k, v')-\alpha+1} \\
	&\leq 2^{-\alpha + 1}.
\end{align*}
As a result, we can bound the average over $(z', v') \in S$
\[
\frac{1}{|S|}\sum_{(z', v') \in S} \sum_{(z,v) \in S, (z,v) \neq (z', v')} |\bra{z'} H^{v' + v} \ket{z}|^2 \leq t \cdot 2^{-\alpha + 1}.
\]
Using Lemma \ref{lem:orth} for the set of vector $\{H^v \ket{z} : (z,v) \in S\}$, we obtain a set of orthonormal vectors $\ket{w_{z, v}}$ for $(z,v) \in S$ such that $\frac{1}{|S|} \sum_{(z,v) \in S} \| H^{v} \ket{z} - \ket{w_{z,v}} \|_2^2 \leq t 2^{-\alpha+1}$. We can rewrite this inequality as
\[
\frac{1}{|S|} \sum_{(z,v) \in S} \re \bra{w_{z,v}} H^v \ket{z} \geq 1 - t 2^{-\alpha+1}.
\]
Using the Cauchy-Schwarz inequality, we get $(\sum_{z,v} \textrm{Re}\bra{w_{z,v}} H^v \ket{z})^2 \leq |S| \cdot \sum_{z,v} (\re\bra{w_{z,v}} H^v \ket{z})^2$. 
It follows that
\begin{align}
\frac{1}{|S|} \sum_{(z,v) \in S} |\bra{w_{z,v}} H^{v} \ket{z}|^2 &\geq \frac{1}{|S|} \cdot \frac{1}{|S|} \cdot \left( \sum_{(z,v) \in S} \re \bra{w_{z,v}} H^v \ket{z} \right)^2 \notag \\
&\geq (1 - t 2^{-\alpha+1})^2. \label{eq:prob-recover-z}
\end{align}
The attack on the locking scheme is defined by the projective measurement of the orthonormal set $\{\ket{w_{z,v}}\}_{(z,v) \in S}$. Note that this is a valid attack because this set does not depend on the private randomness $r$ and only uses the description of the protocol (the codes $C_k$ and the bitstrings $v_k$). Let $I$ be the outcome of the measurement (only the $z$ part, so more precisely we perform the projective measurement whose elements are $\{\sum_{v} \proj{w_{z,v}}\}_z$). We have
\begin{align*}
\pr{I = Z} &= \exc{x,r,k}{ \sum_{v' : (C_k(P_k(x,r)),v') \in S} | \bra{w_{C_k(P_k(x,r)),v'}} H^{v_k} \ket{C_k(P_k(x,r))} |^2 } \\
		&= \sum_{z, v} \pr{Z=z, V=v} \sum_{v': (z,v') \in S} |\bra{w_{z,v'}} H^{v} \ket{z} |^2 \\
		&\geq \sum_{z, v} \pr{Z=z, V=v} |\bra{w_{z,v}} H^{v} \ket{z} |^2  \\
		&= \frac{1}{|S|} \sum_{(z, v) \in S} |\bra{w_{z,v}} H^{v} \ket{z} |^2 \\
		&\geq (1-t2^{-\alpha+1})^2.
\end{align*}
Here, we used the assumption that $Z, V$ is uniformly distributed on its support so that $\pr{Z=z, V=v} = 1/|S|$. This condition is satisfied for example by the scheme of Corollary \ref{cor:explicit-locking}. It follows that $\frac{1}{2} \| p_{XI} - p_{XZ} \|_1 \leq 2 t 2^{-\alpha+1}$. 
\begin{align*}
\entI(I;X) &= \entH(X) - \entH(X|I) \\
		&\geq \entH(X) - \entH(X|Z) - 4 \cdot 4 t 2^{-\alpha+1}n - 2\binent(4 t 2^{-\alpha+1})  \\
		&\geq n - 16t 2^{-\alpha+1}n - \left(\entH(X|ZK) + \entH(K) \right) - 2\binent(4 t 2^{-\alpha+1}) \\
		&= n - 16 t 2^{-\alpha+1} n - \log t - 2\binent(4 t 2^{-\alpha+1}).		
\end{align*}
In the first inequality, we used the Alicki-Fannes inequality (Lemma \ref{lem:alicki-fannes}).
For the second inequality, we used the fact that $\entH(X|Z) \leq \entH(X|ZK) + \entH(K)$ and for the last equality, we used the fact that given $Z$ and $K$, we can decode $X$ so that $\entH(X|ZK) = 0$.
\end{proof}

This result says that if we want to build a locking scheme with a small key that can be implemented with current technology and that tolerates some errors, we should look for schemes that do not lie in the class described above. However, if a key of size $c n$ for some constant $c < 1$ is acceptable (where $n$ is the size of the message), it is possible to construct a locking scheme that is tolerant to errors. In fact, we could for instance after applying the permutation to the $n$-bit message together with the private random string obtaining a bit string $z$ of length roughly $4n$, compute some parities of $z$. Let $y$ denote the string of parities obtained. As in the first protocol of Corollary \ref{cor:explicit-locking}, Alice then encodes $z$ by performing some Hadamard gates according to the key and sends this ciphertext to Bob. In addition, Alice uses another part of the key to encrypt $y$ (using a one-time pad) and sends it to Bob. Using the key, Bob can recover $z'$ a noisy version of $z$ and can recover $y$ perfectly (as it is sent through a classical channel). Using $z'$ and $y$, Bob can recover $z$ provided $z'$ didn't have too many errors.

To obtain a smaller key size, one idea is to use the method described above as a key expansion procedure and repeat it many times. Typically the number of times we would like to expand the key is $O(\log n)$, and it is thus possible to choose independently the permutation and Hadamard that are be applied at each step. But it seems difficult to analyse how these protocols compose. It is not clear what kind of information can be obtained from a measurement that acts on the big protocol.


\subsection{Quantum hiding fingerprints}
\label{sec:hiding-fingerprint}

In this section, we show that the locking scheme of Corollary \ref{cor:existence-locking} can be used to build mixed state quantum hiding fingerprints as defined by \citet{GI10}. A quantum fingerprint encodes an $n$-bit string into a quantum state $\rho_x$ of $n' \ll n$ qubits such that given $y \in \{0,1\}^n$ and the fingerprint $\rho_x$, it is possible to decide with small error probability whether $x=y$ \citep{BCWW01}. The additional hiding property ensures that measuring $\rho_x$ leaks very little information about $x$.
\citet*{GI10} used the accessible information\footnote{The accessible information about $X$ in a quantum system $C$ refers to the maximum over all measurements of the system $C$ of $\entI(X; I)$ where $I$ is the outcome of that measurement.} as a measure of the hiding property. Here, we strengthen this definition by imposing a bound on the total variation distance instead (see Proposition \ref{prop:metric-to-entropic}).
\begin{definition}[Quantum hiding fingerprint]
Let $n$ be a positive integer, $\delta, \e \in (0,1)$ and $C$ be a Hilbert space. An encoding $f : \{0,1\}^n \to \cS(C)$ together with a set of measurements $\{M^y, \1 - M^y\}$ for each $y \in \{0,1\}^n$ is a $(\delta, \e)$-hiding fingerprint if 
\begin{enumerate}
\item (Fingerprint property) For all $x \in \{0,1\}^n$, $\tr \left[ M^x f(x) \right] = 1$ and for $y \neq x$, $\tr \left[ M^y f(x) \right] \leq \delta$.
\item (Hiding property) Let $X$ be uniformly distributed on $\{0,1\}^n$. Then, for any POVM $\{N_i\}$ on the system $C$ whose outcome on $f(X)$ is denoted $I$, we have for all possible outcomes $i$,
\[
\tracedist{p_{X| \event{I=i}}, p_X} \leq \e.
\]
\end{enumerate}
\end{definition}
We usually want the Hilbert space $C$ to be composed of $O(\log n)$ qubits. \citet*{GI10} proved that for any constant $c$, there exist efficient quantum hiding fingerprinting schemes for which the number of the qubits in system $C$ is $O(\log n)$ and both the error probability $\delta$ and the accessible information are bounded by $1/n^c$. Here, we prove that the same result can be obtained by locking a classical fingerprint. The general structure of our quantum hiding fingerprint for parameters $n, \delta$ and $\e$ is as follows:
\begin{enumerate}
\item Choose a random prime $p \in \cP_{n,\e, \delta}$ uniformly from the set $\cP_{n, \e, \delta}$.
\item Set $t = \ceil{c \log(1/\e) \e^{-2}}$, $d_A = p$ and $d_B = \ceil{c'/\e^2}$ and generate $t$ random unitaries $U^p_1, \dots, U^p_{t}$ acting on $A \otimes B$.
\item The fingerprint consists of the random prime $p$ and the state $(U^p_k)^{\dagger} \ket{x\bmod{p}}^A \ket{b}^B$ where $k \in [t]$ and $b \in [d_B]$ are chosen uniformly and independently. The density operator representing this state is denoted $f(x) \eqdef \frac{1}{t d_B} \sum_{k, b} (U^p_k)^{\dagger} \proj{x\bmod{p}}^A \otimes \proj{b}^B U^p_k$.
\end{enumerate}
Observe that even though this protocol is randomized because the unitaries are chosen at random, it is possible to implement it with resources polynomial in $n$ as the size of the message to be locked is $O(\log n)$ bits. In fact, one can approximately sample a random unitary in dimension $2^{O(\log n)}$ using a polynomial number of public random bits. The mixed state protocol of \cite{GI10} achieves roughly the same parameters. Their construction is also randomized but it uses random codes instead of random unitaries. For this reason, the protocol of \cite{GI10} would probably be more efficient in practice.

\begin{theorem}
\label{thm:hiding-fingerprint}
There exist constants $c,c'$ and $c''$, such that for all positive integer $n$, $\delta, \e \in (0,1/4)$ if we define $\cP_{n,\delta, \e}$ to be the set of primes in the interval $[l, u]$ where 
\[
l = \left( \frac{c''}{\delta} \cdot \frac{\log^2(1/\e)}{\e^{8}} \right)^{1/0.9} + 10n \quad \text{ and } \quad u = l + (2n/\delta)^2
\]
and provided $u \leq 2^{n-2}$, the scheme described above is a $(\delta, \e)$-hiding fingerprint with probability $1-2^{-\Omega(n)}$ over the choice of random unitaries.
%
\end{theorem}
The proof of this result involves two parts. First, we need to show that the fingerprint of a uniformly distributed $X \in \{0,1\}^n$ does not give away much information about $X$. This follows easily from Theorem \ref{thm:existence-ur} and Theorem \ref{thm:ur-locking}. We also need to show that for every $y \in \{0,1\}^n$, there is a measurement that Bob can apply to the fingerprint to determine with high confidence whether it corresponds to a fingerprint of $y$ or not.  
In order to prove this we use Lemma \ref{lem:orth} that gives a way of approximating a set of almost orthogonal vectors by a set of orthogonal vectors.
\begin{lemma}
\label{lem:fingerprint-error}
Let $\{U_1, \dots, U_{t}\}$ be a set of unitary transformations on $AB$ that 
define $\gamma$-MUBs and $d \eqdef d_A d_B$. Define for $y \in [d_A]$ the subspace $F_y = \textrm{span} \{ U_k^{\dagger} \ket{y} \ket{b}, k \in [t], b \in [d_B] \}$. Then for any $x \in [d_A]$, $y \neq x$, $k_0 \in [t]$ and $b_0 \in [d_B]$,
\[
\tr \left[ \Pi_{F_y} U^{\dagger}_{k_0} \ket{x} \ket{b_0} \right] \leq 3 (t d_B)^2 d^{-\gamma}.
\]
where $\Pi_{F}$ is the projector on the subspace $F$.
\end{lemma}
\begin{proof}
Consider the set of vectors $\{U_k^{\dagger} \ket{y} \ket{b}\}_{k \in [t], b \in [d_B]}$. We have for all $(k,b) \neq (k',b')$, 
\[
|\bra{y} \bra{b'} U_{k'} U^{\dagger}_{k} \ket{y} \ket{b}| \leq d^{-\gamma/2},
\]
and as a result,
\[
\frac{1}{td_B} \sum_{(k,b) \neq (k', b')} |\bra{y} \bra{b'} U_{k'} U^{\dagger}_{k} \ket{y} \ket{b}|^2 \leq td_B d^{-\gamma}.
\]
Using Lemma \ref{lem:orth}, we obtain a set of orthonormal vectors $\{\ket{e_{k,b}(y)}\}_{k,b}$ such that
\[
\frac{1}{td_B} \sum_{k,b} \|\ket{e_{k,b}(y)} - U_k^{\dagger} \ket{y} \ket{b}\|_2^2 \leq td_B d^{-\gamma}.
\]
Note that $\{\ket{e_{k,b}(y)}\}_{k,b}$ is an orthonormal basis for $F_y$ so we can write $\Pi_{F_y}  = \sum_{k,b} \proj{e_{k,b}(y)}$.
Now observe that, using the Cauchy Schwarz inequality and the fact that the vectors have unit norm, we have $|\bra{e_{k,b}(y)} U^{\dagger}_{k_0} \ket{x} \ket{b_0}| \leq |\bra{y} \bra{b} U_{k} U^{\dagger}_{k_0} \ket{x} \ket{b_0}| + \norm{\ket{e_{k,b}(y)} - U_k^{\dagger} \ket{y} \ket{b}}$.
As a result, we have
\begin{align*}
&\tr \left[ \Pi_{F_y} U^{\dagger}_{k_0} \ket{x} \ket{b_0} \right] \\
&= \sum_{k,b} |\bra{e_{k,b}(y)} U^{\dagger}_{k_0} \ket{x} \ket{b_0}|^2 \\
											&\leq \sum_{k,b} \left| | \bra{y} \bra{b} U_{k} U^{\dagger}_{k_0} \ket{x} \ket{b_0}| + \norm{\ket{e_{k,b}(y)} - U_k^{\dagger} \ket{y} \ket{b}} \right|^2 \\
											&\leq t d_B d^{-\gamma} + (td_B)^2 d^{-\gamma} + \sum_{k,b} | \bra{y} \bra{b} U_{k} U^{\dagger}_{k_0} \ket{x} \ket{b_0}| \cdot \norm{\ket{e_{k,b}(y)} - U_k^{\dagger} \ket{y} \ket{b}} \\
											&\leq 2 (t d_B)^2 d^{-\gamma} + d^{-\gamma/2} \sqrt{td_B} \sqrt{ \sum_{k,b}  \| \ket{e_{k,b}(y)} - U_k^{\dagger} \ket{y} \ket{b} \|_2^2 }  \\
											&\leq 3 (t d_B)^2 d^{-\gamma}.
\end{align*}
\end{proof}

\begin{proof}[of Theorem \ref{thm:hiding-fingerprint}]
We start by proving the hiding property. For any fixed $p$, the random variable $Z \eqdef X \bmod{p}$ is almost uniformly distributed on $\{0, \dots, p-1\}$. In fact, we have for any $z \in \{0, \dots, p-1\}$, $\pr{Z = z} \leq \frac{2^n/p + 1}{2^n}$. In other words, $\entHmin(Z) \geq \log p - \log(1+p2^{-n})$. Thus, using Theorem \ref{thm:existence-ur} and Theorem \ref{thm:ur-locking}, we have that except with probability exponentially small in $n$ (on the choice of the random unitary), the fingerprinting scheme satisfies for any measurement outcome $i$
\[
\tracedist{p_{Z|\event{I=i}}, p_Z} \leq \frac{2\e}{\frac{1}{1+p2^{-n}} - \e} \leq 4\e
\]
where $I$ denotes the outcome of a measurement on the state $f(X)$.
Recall that we are interested in the information leakage about $X$ not $Z$. For this, we note that the random variables $X, Z, I$ form a Markov chain. Thus,
\begin{align*}
&\tracedist{p_{X|\event{I=i}}, p_X} \\
 &= \sum_{x \in \{0,1\}^n} \Big| \sum_{z} \pr{Z=z|I=i} \pr{X=x|I=i, Z=z} \\
 &\qquad - \pr{Z=z} \pr{X=x|Z=z}  \Big| \\
 &= \sum_{x \in \{0,1\}^n} \Big| \sum_{z} \pr{Z=z|I=i} \pr{X=x|Z=z} \\
 &\qquad - \pr{Z=z} \pr{X=x|Z=z}  \Big| \\
						&\leq \sum_{z} \left| \pr{Z=z|I=i} - \pr{Z=z} \right| \sum_{x \in \{0,1\}^n} \pr{X=x|Z=z}   \\
						&= \tracedist{p_{Z|\event{I=i}}, p_Z} \leq 4\e.
\end{align*}
This proves the hiding property.

We now analyse the fingerprint property. Let $x, y \in [2^n]$ and $p$ be the random prime of the fingerprint. We define the measurements by $M^y = \Pi_{F_y}$ for all $y \in \{0,1\}^n$ where $\Pi_{F_y}$ is the projector onto the subspace $F_y = \textrm{span} \{ {U^p_k}^{\dagger} \ket{y \bmod{p}} \ket{b}, k \in [t], b \in [d_B] \}$. If $x = y$, then $f(x)$ is a mixture of states in $\textrm{span} \{ {U^p_k}^{\dagger} \ket{y \bmod{p}} \ket{b}, k \in [t], b \in [d_B] \}$. Thus $\tr [ M^y f(x) ] = 1$.

We now suppose that $x \neq y$. First, we have for a random choice of prime $p \in \cP_{n, \e, \delta}$, $\pr{ x \bmod{p} = y \bmod{p} } = \pr{x-y \bmod{p} = 0} \leq \delta/2$ as the number of distinct prime divisors of $x-y$ is at most $n$ and the number of primes in $[l, u]$ is at least $2n/\delta$ for $n$ large enough. Then, whenever $x \bmod{p} \neq y \bmod{p}$, Lemma \ref{lem:fingerprint-error} with $\gamma = 0.9$ gives \begin{align*}
\tr \left[ \Pi_{F_y} f(x) \right] &\leq 3 (t d_B)^2 (d_A d_B)^{-0.9} \\
					&\leq 3 \cdot 4 c^2 c'^2 \frac{\log^2(1/\e)}{\e^8} \cdot  \frac{\delta \e^8}{c'' \log^2(1/\e)} \\
					&\leq \delta/2
\end{align*}
for $c''$ large enough with probability $1-2^{-\Omega(d_A d_B)} = 1-2^{-\Omega(n)}$ over the choice of the random unitaries (using Theorem \ref{thm:existence-ur}).
Finally, we get $\tr \left[ \Pi_{F_y} f(x) \right] \leq \delta$ with probability $1-2^{-\Omega(n)}$.
\end{proof}

\subsection{String commitment}
\label{sec:string-commitment}

In this section, we show how to use a locking scheme to obtain a weak form of bit commitment \citep{BCHLW06}. Bit commitment is an important two-party cryptographic primitive defined as follows. Consider two mutually distrustful parties Alice and Bob who are only allowed to communicate over some channel. The objective is to be able to achieve the following: Alice secretly chooses a bit $x$ and communicates with Bob to convince him that she fixed her choice, without revealing the actual bit $x$. This is the commit stage. At the reveal stage, Alice reveals the secret $x$ and enables Bob to open the commitment. Bob can then check whether Alice was honest. 

Using classical or quantum communication, unconditionally secure bit commitment is known to be impossible \citep{May97,LC97}. However, commitment protocols with weaker security guarantees do exist \citep{SR01,DFSS05,BCHLW06,BCHLW08}. Here, we consider the string commitment scenario studied in \cite[Section III]{BCHLW08}.
In a string commitment protocol, Alice commits to an $n$-bit string. Alice's ability to cheat is quantified by the number of strings she can reveal successfully. The ability of Bob to cheat is quantified by the information he can obtain about the string to be committed. One can formalize these notions in many ways. 
We use a security criterion that is similar to the one of \cite{BCHLW08} except that we use the statistical distance between the outcome distribution and the uniform distribution, instead of the accessible information. Our definition is slightly stronger by virtue of Proposition \ref{prop:mutinfo}. For a detailed study of string commitment in a more general setting, see \citep{BCHLW08}.

\begin{definition} 
An \emph{$(n,\alpha,\beta)$-quantum bit string commitment} is a quantum communication protocol between Alice (the committer) and Bob (the receiver) which has two phases. When both players are honest the protocol takes the following form.
\begin{itemize}
\item (Commit phase) Alice chooses a string $X \in \{0,1\}^n$ uniformly. Alice and Bob communicate, after which Bob holds a state $\rho_X$.
\item (Reveal phase) Alice and Bob communicate and Bob learns $X$.
\end{itemize}
The parameters $\alpha$ and $\beta$ are security parameters.
\begin{itemize}
\item If Alice is honest, then for any measurement performed by Bob on her state $\rho_X$, we have $\tracedist{p_X, p_{X|\event{I=i}}} \leq \frac{\beta}{n}$ where $I$ is the outcome of the measurement.
\item If Bob is honest, then for all commitments of Alice: $\sum_{x \in \{0,1\}^n} p_x \leq 2^{\alpha}$, where $p_x$ is the probability that Alice successfully reveals $x$.
\end{itemize}
\end{definition}

Following the strategy of \cite{BCHLW08}, the following protocol for string commitment can be defined using a locking scheme $\cE$.
\begin{itemize}
\item Commit phase: Alice has the string $X \in \{0,1\}^n$ and chooses a key $K \in [t]$ uniformly at random. She sends the state $\cE(X, K)$ to Bob.
\item Reveal phase: Alice announces both the string $X$ and the key $K$. Using the key, Bob decodes some value $X'$. He accepts if $X = X'$.
\end{itemize}
A protocol is said to be efficient if both the communication (in terms of the number of qubits exchanged) is polynomial in $n$ and the computations performed by Alice and Bob can be done in polynomial time on a quantum computer. The protocol presented in \cite{BCHLW08} is not efficient in terms of computation and is efficient in terms of communication only if the cost of communicating a (random) unitary in dimension $2^n$ is disregarded. Using the efficient locking scheme of Corollary \ref{cor:explicit-locking}, we get
\begin{corollary}
Let $n$ be a positive integer and $\beta \in (n2^{-cn}, n)$ ($c$ is a constant independent of $n$). There exists an efficient $(n, c \log(n^2/\beta) , \beta)$-quantum bit string commitment protocol for some constant $c$ independent of $n$ and $\beta$.
\end{corollary}
\begin{proof}
We use the first construction of Corollary \ref{cor:explicit-locking} with $\e = \beta/n$.
If Bob is honest, the security analysis is exactly the same as in \cite{BCHLW08}. If Alice is honest, the security follows directly from the definition of the locking scheme.
\end{proof}

\subsection{Locking entanglement of formation}
\label{sec:ent-formation}
The entanglement of formation is a measure of the entanglement in a bipartite quantum state that attempts to quantify the number of singlets required to produce the state in question using only local operations and classical communication
\citep{BDSW96}. For a bipartite state $\rho^{XY}$, the entanglement of formation is defined as
\begin{equation}
\label{eq:ent-formation}
\entF(X;Y)_{\rho} = \min_{ \{p_i, \ket{\psi_i} \}} \sum_i p_i \entH(X)_{\psi_i}.
\end{equation}
where the minimization is taken over all possible ways to write $\rho^{XY} = \sum_i p_i \proj{\psi_i}$ with $\sum_i p_i = 1$.
Entanglement of formation is related to the following quantity:
\[
\entIarrow(X;Y')_{\rho} = \max_{ \{M_i\} } \entI(X; I)
\]
where the maximization is taken over all measurements $\{M_i\}$ performed on the system $Y'$ and $I$ is the outcome of this measurement. This quantity is sometimes referred to as a classical correlations between $X$ and $Y$ \citep{HV01}. As mentioned previously, when the system $X$ is classical, this correlation measure is called accessible information. \citet*{KW04} showed that for a pure state $\ket{\rho}^{XYY'}$, a simple identity holds:
\begin{equation}
\label{eq:info-entf}
\entF(X;Y)_{\rho} + \entIarrow(X;Y')_{\rho} = \entH(X)_{\rho}.
\end{equation}
Let $\{U_1, \dots, U_{t}\}$ be a set of unitary transformations of $A \otimes B \isom C$ and define
\[
\ket{\rho}^{ABCA'K} = \frac{1}{\sqrt{t d_A d_B}} \sum_{k \in [t],a \in [d_A], b \in [d_B]} \ket{a}^A \ket{b}^B \left(U^{\dagger}_k \ket{a} \otimes \ket{b} \right)^C \ket{a}^{A'} \ket{k}^K.
\]
If $\{U_1, \dots, U_{t}\}$ satisfies an $\e$-metric uncertainty relation, then we get a locking effect using Theorem \ref{thm:ur-locking} and Proposition \ref{prop:mutinfo}. In fact, we have $\entIarrow(A; C)_{\rho} \leq 8\e \log d_A + 2\binent(2\e) $ and $\entIarrow(A; CK) = \log d_A$. Thus, using  \eqref{eq:info-entf}, we get
\[
\entF(A;A'BK)_{\rho} = \entH(A)_{\rho} - \entIarrow(A; C)_{\rho} \geq (1-8\e)\log d_A - 2\binent(2\e)
\]
and discarding the system $K$ of dimension $t$ we obtain a separable state 
\[
\entF(A;A'B)_{\rho} = 0.
\]
Explicit states exhibiting weak locking behaviour of the entanglement of formation have been presented in \cite{HHHO05}. Strong but non-explicit instances of locking the entanglement of formation were derived in \cite{HLW04}. Here, using Theorem \ref{thm:explicit-ur1}, we obtain explicit examples of strong locking behaviour.

One could also consider other quantities related to classical correlations, such as the popular quantum discord \citep{OZ01}, and they would exhibit a similar locking behaviour.


\section{Quantum identification codes}
\label{sec:qid}

Consider the following quantum analogue of the equality testing communication problem. Alice is given an $n$-qubit state $\ket{\psi} \in C$ and Bob is given $\ket{\ph} \in C$. Namely, Bob wants to output $1$ with probability in the interval $[|\braket{\psi}{\ph}|^2 - \e, |\braket{\psi}{\ph}|^2 + \e]$ and $0$ with probability in the interval $[1 - |\braket{\psi}{\ph}|^2 - \e, 1 - |\braket{\psi}{\ph}|^2 + \e]$. This task is referred to as quantum identification \citep{Win04}. Note that communication only goes from Alice to Bob. There are many possible variations to this problem. One of the interesting models is when Alice receives the quantum state $\ket{\psi}$ and Bob gets a classical description of $\ket{\ph}$. An $\e$-quantum-ID code is defined by an encoder, which is a quantum operation that maps Alice's quantum state $\ket{\psi}$ to another quantum state which is transmitted to Bob, and a family of decoding POVMs $\{D_{\ph}, \1 - D_{\ph}\}$ for all $\ket{\ph}$ that Bob performs on the state he receives from Alice.
\begin{definition}[Quantum identification \citep{Win04}]
\label{def:qid-code} Let $\cH_1, \cH_2, C$ be Hilbert spaces and $\e \in (0,1)$.
An \emph{$\e$-quantum-ID code} for the space $C$ using the channel $\cN : \cS(\cH_1) \to \cS(\cH_2)$ consists of an encoding map $\cE : \cS(C) \to \cS(\cH_1)$ and a set of POVMs $\{D_{\ph}, \1 - D_{\ph}\}$ acting on $\cS(\cH_2)$, one for each pure state $\ket{\ph}$ such that
\[
\forall \ket{\psi}, \ket{\ph} \in C, \qquad \Big| \tr \left[ D_{\ph} \cN(\cE(\psi)) \right]  - | \braket{\ph}{\psi} |^2 \Big| \leq \e.
\]
\end{definition}
Here we consider channels $\cN$ transmitting noiseless qubits and noiseless classical bits. We also say that $\e$-quantum identification of $n$-qubit states can be performed using $\ell$ bits and $m$ qubits when there exists an $\e$-quantum-ID code for the space $C = (\CC^2)^{\otimes n}$ using the channel $\cN = \bitchannel^{\otimes \ell} \otimes  \qubitchannel^{\otimes m}$, where $\bitchannel$ and $\qubitchannel$ are the noiseless bit and qubit channels.  \citet*{HW10} showed that classical communication alone cannot be used for quantum identification. However, a small amount of quantum communication makes classical communication useful. Using our metric uncertainty relations, we prove better bounds on the number of qubits of communication and give an efficient encoder for this problem.

Our protocol is based on a duality between quantum identification and approximate forgetfulness of a quantum channel demonstrated in \cite[Theorem 7]{HW10}. Specialized to our setting, the direction of the duality we use
states that if  $V : C \to A\otimes B$ defines a low-distortion embedding of $(C, \ell_2)$ into $(AB, \ell^A_1(\ell^B_2))$, then the maps $\Gamma_a: C \to B$ for $a \in [d_A]$ defined by $\ket{\psi} \mapsto \sum_{b \in d_B} (\bra{a} \bra{b} V \ket{\psi}) \ket{b}$ approximately preserve inner products on average. The following lemma gives a precise statement. We give an elementary proof in the interest of making the presentation self-contained.
\begin{lemma}
\label{lem:dvo-jl}
Let $V : C \to A \otimes B$ be an isometry, i.e., for all $\ket{\psi} \in C$, $\| V \ket{\psi} \|_2 = \| \ket{\psi} \|_2$. For any vector $\ket{\psi} \in C$, we define the vectors $\ket{\psi_a} \in B$ by $V \ket{\psi} = \sum_{a \in [d_A]} \ket{a} \ket{\psi_a}$. Assume that $V$ satisfies the following property:
\begin{equation}
\label{eq:dvo-like}
\forall \ket{\psi} \in C \qquad \sum_{a \in [d_A]} \left| \| \ket{\psi_a} \|^2_2 - \frac{\| \ket{\psi} \|^2_2}{d_A} \right| \leq \e \| \ket{\psi} \|_2^2.
\end{equation}
Then we have for all unit vectors $\ket{\psi}, \ket{\ph} \in C$ with $V \ket{\psi} = \sum_{a \in [d_A]} \ket{a} \ket{\psi_a}$ and $V \ket{\ph} = \sum_{a \in [d_A]} \ket{a} \ket{\ph_a}$
\begin{equation}
\label{eq:jl-like}
\frac{1}{d_A} \sum_{a \in [d_A]} \left| \frac{|\braket{\psi_a}{\ph_a} |^2}{\| \ket{\psi_a} \|_2 \|\ket{\ph_a}\|_2} - |\braket{\psi}{\ph}|^2 \right| \leq 12 \e + 2 \sqrt{\e}.
\end{equation}
\end{lemma}
\begin{proof}
Let $\ket{\psi}$ and $\ket{\ph}$ be unit vectors in $C$. 
We use the triangle inequality to get 
\begin{align}
&\frac{1}{d_A} \sum_{a \in [d_A]} \left| \frac{|\braket{\psi_a}{\ph_a}|^2}{\| \ket{\psi_a} \|_2 \| \ket{\ph_a} \|_2} - |\braket{\psi}{\ph}|^2 \right| \notag \\
&\qquad \leq  \sum_{a \in [d_A]} \left| \frac{|\braket{\psi}{\ph}|^2}{d_A} - |\braket{\psi_a}{\ph_a}|^2 \right|
+  \sum_{a \in [d_A]} \left| |\braket{\psi_a}{\ph_a}|^2 - \frac{|\braket{\psi_a}{\ph_a}|^2}{d_A \| \ket{\psi_a} \|_2  \| \ket{\ph_a} \|_2} \right|. \label{eq:jl-dvo-1}
\end{align}

We start by dealing with the first term in \eqref{eq:jl-dvo-1}. Observe that
\begin{align}
\left| |\braket{\psi_a}{\ph_a}|^2 - \frac{|\braket{\psi}{\ph}|^2}{d_A} \right|
& \leq \left| (\re \braket{\psi_a}{\ph_a})^2 - \frac{(\re \braket{\psi}{\ph})^2 }{d_A} \right| \notag \\
&\qquad +\left| (\im \braket{\psi_a}{\ph_a})^2 - \frac{(\im \braket{\psi}{\ph})^2 }{d_A} \right| \notag \\
& \leq 2 \left| \re \braket{\psi_a}{\ph_a} - \frac{\re \braket{\psi}{\ph} }{d_A} \right| +
2 \left| \im \braket{\psi_a}{\ph_a} - \frac{\im \braket{\psi}{\ph} }{d_A} \right|.
 \label{eq:jl-dvo-2}
\end{align}
In the last inequality, we used the fact that $|x^2 - y^2| \leq 2 |x-y|$ whenever $|x+y| \leq 2$. To bound these terms, we apply the assumption about $V$ (equation \eqref{eq:dvo-like}) to the vector $\ket{\psi} - \ket{\ph}$:
\[
\sum_{a \in [d_A]} \left| \| \ket{\psi_a} - \ket{\ph_a} \|^2_2 - \frac{\| \ket{\psi} - \ket{\ph} \|^2_2}{d_A} \right| \leq \e \| \ket{\psi} - \ket{\ph} \|^2_2 \leq 4\e.
\]
By expanding $\| \ket{\psi_a} - \ket{\ph_a} \|^2_2$ and $\| \ket{\psi} - \ket{\ph} \|^2_2$, we obtain using the triangle inequality
\begin{align*}
&\sum_{a \in [d_A]} \left| 2 \re \braket{\psi_a}{\ph_a} - \frac{2 \re \braket{\psi}{\ph}}{d_A} \right| \\
&\leq 4\e + \sum_{a \in [d_A]} \left| \| \ket{\psi_a} \|^2_2 - \frac{\| \ket{\psi} \|^2_2}{d_A} \right| + \left| \| \ket{\ph_a} \|^2_2 - \frac{\| \ket{\ph} \|^2_2}{d_A} \right| \\
					&\leq 6\e.
\end{align*}
In the last inequality, we used equation \eqref{eq:dvo-like} for $\ket{\psi}$ and $\ket{\ph}$.
The same argument can be applied to $i \ket{\psi}$ and $\ket{\ph}$ to get 
\[
2 \sum_{a \in [d_A]} \left| \im \braket{\psi_a}{\ph_a} - \frac{\im \braket{\psi}{\ph}}{d_A} \right| \leq 6\e
\]
Thus, substituting in equation \eqref{eq:jl-dvo-2} we obtain
\[
\left| |\braket{\psi_a}{\ph_a}|^2 - \frac{|\braket{\psi}{\ph}|^2}{d_A} \right| \leq 12\e.
\]

We now consider the second term in \eqref{eq:jl-dvo-1}. We have, using the Cauchy-Schwarz inequality,
\begin{align*}
& \sum_{a \in [d_A]} \left| |\braket{\psi_a}{\ph_a}|^2 - \frac{|\braket{\psi_a}{\ph_a}|^2}{d_A \| \ket{\psi_a} \|_2  \| \ket{\ph_a} \|_2} \right| \\
&\leq \sum_{a \in [d_A]} \left| \| \ket{\psi_a} \|_2  \| \ket{\ph_a} \|_2 - \frac{1}{d_A} \right| \\
&\leq \sum_{a \in [d_A]} \| \ket{\psi_a} \|_2 \left| \| \ket{\ph_a} \|_2 - \frac{1}{\sqrt{d_A}} \right|
+  \sum_{a \in [d_A]} \left| \frac{\| \ket{\psi_a} \|_2}{\sqrt{d_A}}  - \frac{1}{d_A} \right| \\
&\leq \sqrt{ \sum_{a \in [d_A]} \| \ket{\psi_a} \|^2_2} \sqrt{\sum_{a \in [d_A]} \left| \| \ket{\ph_a} \|_2 - \frac{1}{\sqrt{d_A}} \right|^2} + \sqrt{\sum_{a \in [d_A]} \left| \| \ket{\psi_a} \|_2 - \frac{1}{\sqrt{d_A}} \right|^2} \\
&\leq \sqrt{  \sum_{a \in [d_A]} \left| \| \ket{\ph_a} \|^2_2 - \frac{1}{d_A} \right| } + \sqrt{  \sum_{a \in [d_A]} \left| \| \ket{\psi_a} \|^2_2 - \frac{1}{d_A} \right| } \\
&\leq 2 \sqrt{\e}.
\end{align*}
For the third inequality, we used once again the Cauchy-Schwarz inequality and for the fourth inequality, we used the fact that $\sum_{a \in [d_A]} \| \ket{\psi_a} \|^2_2 = \| V \ket{\psi} \|^2_2 = 1$ and the inequality $|x - y|^2 \leq |x-y||x+y| = |x^2 - y^2|$ for all nonnegative $x,y$. Plugging this bound into equation \eqref{eq:jl-dvo-1}, we obtain the desired result.
\end{proof}

\begin{center}
\begin{figure}[h]
\begin{center}
\includegraphics[scale=0.9]{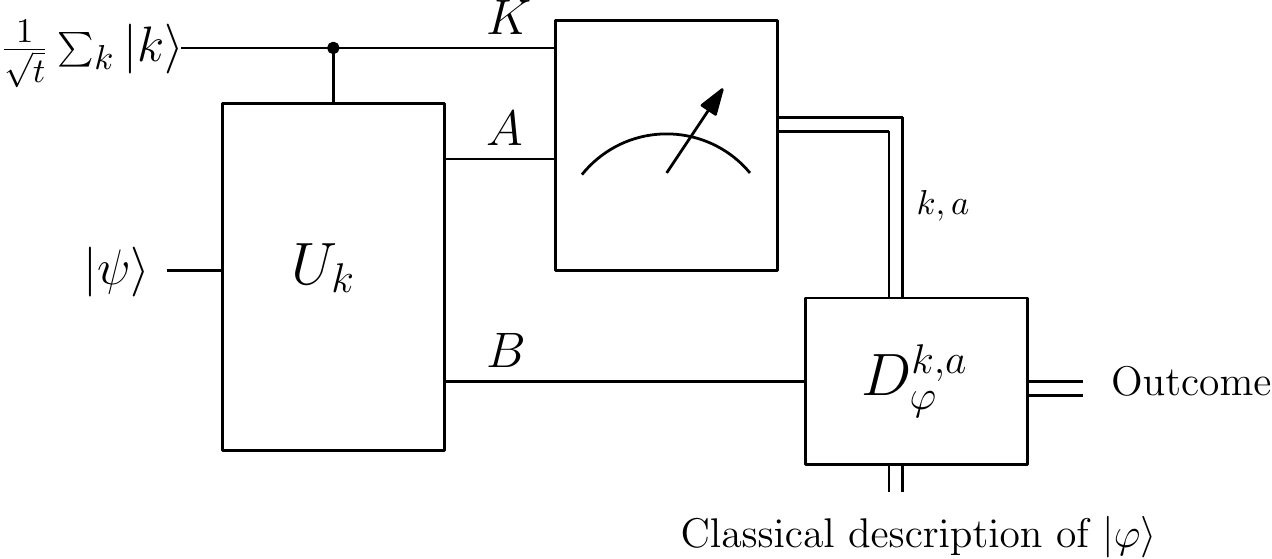}
\end{center}
\caption{Quantum identification based on a metric uncertainty relation. The system $K$ is prepared in a uniform superposition state $\frac{1}{\sqrt{t}} \sum_k \ket{k}$. Then, controlled by system $K$, the unitary $U_k$ is applied to $C = A \otimes B$, where the unitary transformations $\{U_k\}$ satisfy a metric uncertainty relation. The $KA$ system is then measured in its computational basis. The outcome $k,a$ of this measurement is sent through the classical channel. The system $B$ is sent using the noiseless quantum channel. The receiver constructs a POVM $D^{k,a}_{\ph}$ based on a classical description of the state $\ket{\ph}$ he wishes to test for and the classical communication $k,a$ he receives.}
\label{fig:qid-code}
\end{figure}
\end{center}
\begin{theorem}[Quantum identification using classical communication]
Let $n$ be a positive integer and $\e \in (2^{-c'n}, 1)$ where $c'$ is a constant independent of $n$. Then for some $m = O(\log(1/\e))$, $\e$-quantum identification of $n$-qubit states can be performed using a single message of $n$ bits and $m$ qubits.

Moreover, for some $m = O(\log(n/\e) \cdot \log(n))$, $\e$-quantum identification of $n$-qubit states can be performed using a single message of $n$ bits and $m$ qubits with an encoding quantum circuit of polynomial size.
\end{theorem}
\begin{proof}
Let $\{U_1, \dots, U_{t}\}$ be a set of unitaries on $n$ qubits verifying an $\e'$- metric uncertainty relation with $\e' = 1/2 \cdot (\e/28)^2$.
We start by preparing the uniform superposition $\frac{1}{\sqrt{t}} \sum_{k=1}^t \ket{k}^K$ and apply the unitary $U_k$ on system $C$ controlled by the register $K$. We get the state $\frac{1}{\sqrt{t}} \sum_k \ket{k}^K (U_k \ket{\psi})^{AB} = \sum_{k,a} \ket{k}^K \ket{a}^A \ket{\psi_{k,a}}^B$ for some non-normalized vectors $\ket{\psi_{k,a}} \in B$. Alice then measures the system $KA$ in the computational basis obtaining an outcome $k,a$ and sends $k,a$ and $\ket{\hat{\psi}_{k,a}}$ to Bob, where $\ket{\hat{\psi}_{k,a}} = \ket{\psi_{k,a}}/\| \ket{\psi_{k,a}} \|_2$. Observe that $\sum_{k,a} \| \ket{\psi_{k,a}} \|^2_2 = 1$ and $\| \ket{\psi_{k,a}} \|^2_2 = \frac{1}{t} \cdot p^A_{U_k \ket{\psi}}(a)$ so that the metric uncertainty relation property can be written as
\begin{equation}
\label{eq:identification-1}
\frac{1}{2} \sum_{k,a} \left| \| \ket{\psi_{k,a}} \|^2_2 - \frac{1}{t d_A} \right| \leq \e'.
\end{equation}
This shows that the isometry $\ket{\psi} \mapsto \frac{1}{\sqrt{t}} \sum_k \ket{k}^K (U_k \ket{\psi})^{AB}$ satisfies the condition \eqref{eq:dvo-like} of Lemma \ref{lem:dvo-jl}.

The decoding POVMs for received classical information $k,a$ and state $\ket{\ph}$ are defined by $D^{k,a}_{\ph} = \proj{\hat{\ph}_{k,a}}$ where $\frac{1}{\sqrt{t}} \sum_k \ket{k}^K (U_k \ket{\ph})^{AB} = \sum_{k,a} \ket{k}^K \ket{a}^A \ket{\ph_{k,a}}^B$ and $\ket{\hat{\ph}_{k,a}} = \ket{\ph_{k,a}}/\| \ph_{k,a} \|_2$. 
The protocol is illustrated in Figure \ref{fig:qid-code}.

We now analyse the probability that Bob outputs $1$. Recall that outcome $1$ corresponds to the projector $\proj{\ph}$. The probability that the protocol in Figure \ref{fig:qid-code} outputs $1$ is  
\[
\sum_{k,a} \| \ket{\psi_{k,a}} \|^2_2 \cdot \tr\left[ D^{k,a}_{\ph} \proj{\hat{\psi}_{k,a}}\right] = \sum_{k,a} \| \ket{\psi_{k,a}} \|^2_2 |\braket{\hat{\psi}_{k,a}}{\hat{\ph}_{k,a}} |^2.
\]
Applying Lemma \ref{lem:dvo-jl}, we have
\begin{equation}
\label{eq:identification-2}
\frac{1}{td_A} \sum_{k,a} \left| |\braket{\hat{\psi}_{k,a}}{\hat{\ph}_{k,a}} |^2 - |\braket{\psi}{\ph}|^2\right| \leq 14 \sqrt{2\e'} = \e/2.
\end{equation}
Using the triangle inequality, equations \eqref{eq:identification-2} and \eqref{eq:identification-1}, we obtain
\begin{align*}
&\sum_{k,a}  \| \ket{\psi_{k,a}} \|^2_2 \left| |\braket{\hat{\psi}_{k,a}}{\hat{\ph}_{k,a}} |^2
- |\braket{\psi}{\ph}|^2\right| \\
&\leq \sum_{k,a} \frac{1}{td_A} \left| |\braket{\hat{\psi}_{k,a}}{\hat{\ph}_{k,a}} |^2 - |\braket{\psi}{\ph}|^2\right| + \sum_{k,a} \left| \| \ket{\psi_{k,a}} \|^2_2 - \frac{1}{td_A} \right| \cdot 2 \\
&\leq \e/2 + 4 \e' \leq \e.
\end{align*}
Thus, the probability of obtaining outcome $1$ is in the interval $[|\braket{\psi}{\ph}|^2 - \e, |\braket{\psi}{\ph}|^2 + \e]$. 

We conclude by using the metric uncertainty relations of Theorems \ref{thm:existence-ur} and \ref{thm:explicit-ur2}. For the explicit construction, we still need to argue that the encoding can be computed by a quantum circuit of size $O(n^2 \polylog(n/\e))$ and depth $O(n\polylog(n/\e))$ using classical precomputations. To obtain this running time, we actually use the $1$-MUBs of Lemma \ref{lem:explicitmub} in the construction of Theorem \ref{thm:explicit-ur2}. The only thing we need to precompute is an irreducible polynomial of degree $n$ over $\FF_2[X]$. Then, using the same argument as in the proof of Lemma \ref{lem:explicitmub}, we can compute the unitary operation that takes as input the state $\ket{j} \otimes \ket{\psi}$ and outputs the state $\ket{j} \otimes V_j \ket{\psi}$ using a circuit of size $O(n^2 \polylog n)$ and depth $O(n \polylog n)$. Since the permutation extractor we use can be implemented by a quantum circuit of size $O(n \polylog(n/\e))$, the unitary transformation $\ket{k} \otimes \ket{\psi} \mapsto \ket{k} \otimes U_k \ket{\psi}$ can be computed by a quantum circuit of size $O(n^2 \polylog(n/\e))$ and depth $O(n \polylog(n/\e))$.
\end{proof}




This result can be thought of as an analogue of the well-known fact that the public-coin randomized communication complexity of equality is $O(\log(1/\e))$ for an error probability $\e$ \citep{KN97}. Quantum communication replaces classical communication and classical communication replaces public random bits. Classical communication can be thought of as an extra resource because on its own it is useless for quantum identification \cite[Theorem 11]{HW10}.

\chapter[URs in the presence of quantum side information]{Uncertainty relations in the presence of quantum side information}
\label{chap:qc-extractor}


\paragraph{Outline of the chapter.} 
In the previous chapters, it was assumed that the adversary trying to predict the outcome of the measurement is not entangled with the quantum system being measured. In Section \ref{sec:qc-ext-intro}, we explain what it means for an uncertainty relation to hold when the adversary has quantum side information. After that, in Section \ref{sec:qc-ext-main}, we introduce metric uncertainty relations with quantum side information that we call QC-extractors. We also give several efficient constructions of QC-extractors. We finally show how using such uncertainty relations, we can relate the security of two-party computations to the quantum capacity of the quantum storage of the adversary (Section \ref{sec:noisy}).

\section{Introduction}
\label{sec:qc-ext-intro}

Let us consider uncertainty relations in the form of a game, called the uncertainty game by \cite{BCCRR10}. Bob prepares a system called $A$ and sends it to Alice. Alice chooses a projective measurement $i$ at random from a set of possible measurements to perform on system $A$. She obtains an outcome that we denote $X$. She then sends $i$ to Bob whose goal is to predict $X$. In Chapter \ref{chap:uncertainty-relations}, we saw several constructions of measurements for which Bob has a lot of uncertainty about $X$. But in a fully quantum world, Bob might keep a quantum system $E$ that is \emph{entangled} with $A$ that could help him in predicting $X$. As an example, imagine that Bob prepares the maximally entangled state $\ket{\Phi} = \frac{1}{\sqrt{d_A}} \sum_{j} \ket{j}_A \ket{j}_E$. Assume the measurements that Alice performs on $A$ are obtained by first applying a unitary transformation $U_i$ on $A$ followed by a measurement in the computational basis. It is simple to see that if Bob, upon receiving the index $i$, applies $U^{*}_i$ on his system $E$ and performs a measurement in the computational basis will get the exact same outcome as Alice. Thus, if Alice and Bob share a maximally entangled state, then Bob can perfectly predict the outcome that Alice obtains: there is no uncertainty at all. This makes it clear that the amount of uncertainty depends on the information available for the adversary Bob. In the previous two chapters, we considered the case where Bob prepares a quantum state and sends it completely to Alice, i.e., the $E$ system is not present. In this chapter, we will construct uncertainty relations that hold even when Bob holds a quantum system.

\comment{Maybe figure here.}

As was discussed above, when the systems $A$ and $E$ are maximally entangled, all the measurement outcomes can be predicted perfectly. Thus, any uncertainty relation should take into account the amount of entanglement between $A$ and the adversary $E$. After being conjectured by \cite{RB09}, it was proven by \cite{BCCRR10} that for any state $\rho$ on $AB$, the following inequality holds:
\begin{equation}
\label{eq:ur-quantum-mem}
\frac{1}{2}\left(\entH(X|E)_{\rho^1} + \entH(X|E)_{\rho^2}\right) \geq \log(1/c) + \frac{1}{2} \entH(A|E)_{\rho}
\end{equation}
where $\rho^i_{XE} = \cM_{A \to X} (U_i \rho_{AE} U_i^{\dg})$ for $i \in \{1,2\}$ is the state obtained when measuring the system $A$ of the state $U_i \rho_{AE} U_i^{\dg}$ in the computational basis and $c$ is the maximum overlap between the vectors defined by $U_1$ and $U_2$, $c = \max_{a,a'} | \bra{a} U_1 U^{\dagger}_2 \ket{a'} |$. $\cM_{A \to X}$ refers to the measurement in the computational basis map: $\cM_{A \to X}(\sigma) = \sum_{a} \bra{a} \sigma \ket{a} \proj{a}$. Note that the reason we renamed the system $X$ after the measurement is simply to emphasize that it is a classical system. If the state $\rho_{AE}$ is a pure state on $A$, we have $\entH(A|E) = 0$ and recover the uncertainty relation of \cite{MU88} in \eqref{eq:entropic-ur-intro}. In the case where $\rho_{AE}$ is maximally entangled, then $\entH(A|E) = -\log d_A$, and $c$ cannot be smaller than $1/\sqrt{d_A}$.\footnote{To see this, just write one of the vector of basis $1$ in basis $2$: one of the squared coefficients has to be at least as large as the average of $1/d_A$.} This implies that the lower bound in \eqref{eq:ur-quantum-mem} is nonpositive, which as discussed earlier is unavoidable. For cryptographic applications, the most interesting case is usually when $\rho_{AE}$ is entangled but not maximally so, i.e., $-\log d_A < \entH(A|E) < 0$.

We should mention that quantum side information is usually much harder to handle than classical side information. This is due to the fact that it is not clear how to describe a conditional state. Consider the example of the study of randomness extractors. It is not hard to prove that an extractor can handle any classical adversary as long as it can handle a classical adversary holding a trivial system;\footnote{Provided the conditional entropy is the same of course.} see e.g., \cite[Proposition 1]{KT08}. The situation is quite different for quantum adversaries. In fact, \cite{GKKRW07} gave an example of an extractor that completely fails when quantum side information is available. This is not to say that quantum adversaries can break any extractor, but that quantum side information can behave in unexpected ways. We now know of many constructions of extractors that do work even when the adversary holds a quantum memory \citep{KT08, RK05, TSSR10, DPVR09, TS09}.

As was mentioned in Chapter \ref{chap:uncertainty-relations}, if we want a larger average measurement entropy, we need to consider more measurements. Unfortunately, up to this day, we only know of uncertainty relations that hold in the presence of quantum memory for \emph{two} measurements~\citep{BCCRR10,RB09,CYZ11,CCYZ12,CYGG10,CW05,TR11}. For two measurements, the incompatibility is directly related to a simple function of the pairwise inner products between vectors in the two bases. For more measurements, controlling the pairwise inner products between the different bases elements is not sufficient to guarantee a good lower bound on the uncertainty \citep{BW07}. In this chapter, we will give several constructions of strong uncertainty relations for many measurements.

Our strategy will be to follow the idea introduced in Chapter \ref{chap:uncertainty-relations} of quantifying the uncertainty in a set of measurement outcomes by the distance to the uniform distribution. In order to account for the possible side information that the adversary $E$ has, we also require the output to be independent of the adversary. More precisely, the condition for a set of unitaries $U_1, \dots, U_t$ will be of the form
\begin{align}
\label{eq:def-qc-extractor-intro}
\frac{1}{t} \sum_{i=1}^t \left\| \cT_{A\rightarrow A_{1}}(U_i \rho_{AE} U_i^{\dagger})- \frac{\id_{A_1}}{d_{A_1}} \ox \rho_E \right\|_1 \leq \eps\ ,
\end{align}
where the map $\cT$ performs a measurement in the computational basis and then discards the subsystem $A_2$ while keeping $A_1$:
\begin{align}
\label{eq:meas-map}
\cT(.)_{A\rightarrow A_1}= \sum_{a_{1}a_{2}} \bra{a_{1}a_{2}}(.)\ket{a_{1}a_{2}} \proj{a_1}\ ,
\end{align}
where $\{\ket{a_{1}}\},\{\ket{a_{2}}\}$ are the computational bases of $A_{1},A_{2}$ respectively. Here, $A_1$ plays the role of the hard to predict outcome called $X$ in the earlier discussion. Equation \eqref{eq:def-qc-extractor-intro} is analogous to Definition \ref{def:metric-ur}, except that we also require that the outcome $A_1$ be decoupled from the adversary $E$. Motivated by the similarity between equation \eqref{eq:def-qc-extractor-intro} and \emph{randomness extractors} (already introduced in Definition \ref{def:perm-extractor}), we call such a set of unitaries a QC-extractor. More details on randomness extractors and related constructions are given in Section \ref{sec:qc-ext-main}, where we also give constructions of QC-extractors. We will show in Section \ref{sec:URbounds} that if $U_1, \dots, U_t$ satisfy \eqref{eq:def-qc-extractor-intro}, they also satisfy an entropic uncertainty relation, as was done in Chapter \ref{chap:uncertainty-relations}. Section \ref{sec:noisy} is devoted to cryptographic applications of these uncertainty relations.

\section{Quantum to Classical randomness extractors (QC-extractors) }
\label{sec:qc-ext-main}
Randomness extractors\footnote{Throughout this thesis, we will only deal with what are known as seeded extractors.  For an overview of the different kinds of extractors for different kinds of sources, see \citep{Sha02}.} were introduced by \cite{NZ96} in the context of derandomization. An extractor is a function that transforms a weak source of randomness into almost uniform random bits. The initial motivating applications were related to complexity theory, e.g., derandomization of space-bounded computations \citep{NZ96}, simulating randomized algorithms with a weak random source \citep{Zuc96bpp} or also as a tool for proving hardness of approximation \citep{Zuc96approx}. The definition of randomness extractors was actually predated by the similar idea of privacy amplification introduced in a cryptographic context, more precisely for quantum key distribution \citep{BBR88, BBCM95}. There, the setting is as follows. Suppose Alice and Bob share a bitstring $X$ about which Eve might have some information $E$. They want to extract a secret key about which Eve has almost no information. Here $X$ viewed from the point of view of $E$ is a weak source of randomness from which we want to distill almost perfect random bits. It is particularly clear in this picture that an extractor should work subject only to the assumption that the source contains \emph{some} randomness, and not make any assumption on where this randomness is. The reason is that depending on her attack, Eve can obtain information about different parts of $X$. In Chapters \ref{chap:uncertainty-relations} and \ref{chap:applications-ur}, we saw yet other applications of randomness extractors to uncertainty relations and low-distortion norm embeddings. For more background on extractors, their constructions and applications, see the surveys \citep{Sha02, Vad07}.


Classical sources of randomness are described by probability distributions and the randomness extractors are families of (deterministic) functions taking each possible value of the source to a binary string. 
To understand the definition of quantum extractors, it is convenient to see a classical extractor as a family of \emph{permutations} acting on the possible values of the source. This family of permutations should satisfy the following property: for any probability distribution on input bit strings with high min-entropy, applying a typical permutation from the family to the input induces an almost uniform probability distribution on a prefix of the output; see \ref{def:perm-extractor} for a definition. We define a quantum to quantum extractor in a similar way by allowing the operations performed to be general unitary transformations and the input to the extractor to be quantum.

\begin{definition}[QQ-Extractors]\label{def:QQ-extractor}
Let $A = A_1 A_2$ with $n = \log d_A$. 

For $k \in [- n , n ]$ and $\eps \in [0,1]$, a $(k, \eps)$-QQ-extractor is a set $\{U_1, \dots, U_{t}\}$ of unitary transformations on $A$ such that for all states $\rho_{AE} \in \cS(AE)$ satisfying $\entHmin(A|E)_{\rho} \geq k$, we have
\begin{equation}\label{eq:QQ-extractor}
\frac{1}{t} \sum_{i=1}^t \left\| \tr_{A_2}\left[U_i \rho_{AE} U_i^{\dagger}\right] - \frac{\id_{A_1}}{d_{A_1}} \ox \rho_E \right\|_1 \leq \eps \ .
\end{equation}
$\log t$ is called the seed size of the $QQ$-extractor.
\end{definition}
Before making a few remarks on the definition, we recall the definition of CC-extractors which are simply randomness extractors that work in the presence of a quantum adversary.

\begin{definition}[CC-Extractors]\label{def:CC-extractor}
For $k \in [0 , n ]$ and $\eps \in [0,1]$, a $(k, \eps)$-CC-extractor is a set $\{f_1, \dots, f_{t}\}$ of functions from $\{0,1\}^n$ to $\{0,1\}^m$ such that for all states $\rho_{XE} \in \cS(XE)$ satisfying $\entHmin(X|E)_{\rho} \geq k$, we have
\begin{equation}\label{eq:CC-extractor}
\frac{1}{t} \sum_{i=1}^t \left\| \rho_{f_i(X)E} - \frac{\id_{Z}}{d_{Z}} \ox \rho_E \right\|_1 \leq \eps \ ,
\end{equation}
where the system $f_i(X)$ is obtained by applying the function $f_i$ to the system $X$.
\end{definition}

First, we should stress that the same set of unitaries should satisfy \eqref{eq:QQ-extractor} \emph{for all} states $\rho_{AE}$ that meet the conditional min-entropy criterion $\entHmin(A|E)_{\rho} \geq k$. In particular, the system $E$ can have arbitrarily large dimension. The quantity $\entHmin(A|E)_{\rho}$ measures the uncertainty that an adversary has about the system $A$. As it is usually impossible to model the knowledge of an adversary, a bound on the conditional min-entropy is often all one can get. A notable difference with the classical setting is that the conditional min-entropy $k$ can be negative when the systems $A$ and $E$ are entangled. 

A statement of the form of equation~\eqref{eq:QQ-extractor} is more commonly known as a decoupling result \citep{HOW05, HOW06, HHYW08, Dup09, ADHW09, DBWR10}. Such statements play an important role in quantum information theory and many coding theorems amount to proving a decoupling theorem. In fact, it was shown that a set of unitaries forming a unitary 2-design (see Definition \ref{def:two-design}) define a $(k,\eps)$-QQ-extractor as long as the output size $\log d_{A_1}\leq(n+k)/2-\log(1/\eps)$. The decoupling theorem of \cite{Dup09,DBWR10} is actually more general than this: it holds even if we replace $\tr_{A_2}$ by any completely positive trace preserving map $\cT$. Of course, then the value of $\e$ depends on an entropic quantity that is a function of the map $\cT$ in addition to the term $\entHmin(A|E)_{\rho}$, which was already present for QQ-extractors. 

A definition of quantum extractors was also proposed in~\cite[Definition 5.1]{BTS10}. A set of unitaries $\{U_1, \dots, U_t\}$ acting on $A$ is a $(k, \e)$-quantum extractor if for all $\rho \in \cS(A)$ with $\entHmin(A)_{\rho} \geq k$, we have
\begin{equation}
\label{eq:def-extractor-bts}
\left\| \frac{1}{t} \sum_{i=1}^t U_i \rho U_i^{\dg} - \frac{\id_A}{d_A} \right\|_1 \leq \e.
\end{equation}
First we note that in their definition, the extractor outputs the whole system (nothing is discarded). This is only possible  because they measure the distance between the randomized state $\frac{1}{t} \sum_i U_i \rho U_i^{\dagger}$ and the maximally mixed state (this condition refers to \emph{weak} extractors), whereas in our definition we ask for the average of the trace distances to be small (\emph{strong} extractors). It is easy to see that by the triangle inequality, a strong extractor is also a weak extractor. For strong extractors, the seed can be made public, i.e., even conditioned on the value of the random seed, the outcome is close to random. This is not the case for weak extractors. In cryptography, weak extractors are usually not good enough because the seed $i$ is made public during the protocol (see for example Section \ref{sec:noisy}). Our definition is also stronger in another respect, we require the extractor to decouple the $A$ system from any \emph{quantum} side information held in the system $E$.


\cite{BTS10} introduced their definition in the context of studying quantum expanders. In fact, they obtain  extractors for high min-entropy sources using their construction of quantum expanders\footnote{In general, an expander can always be used to construct an extractor for very high entropy sources.} as well as the construction of \cite{AS04}; see also \cite[Theorem 3]{DD10} where the construction based on \cite{AS04} is studied the language of approximately randomizing maps \citep{HLSW04}. \cite{BTS10} applied their extractor construction to prove that the quantum entropy difference problem is in the complexity class QSZK; see the paper for more details.


In the context of cryptography, a QQ-extractor is often more than one needs. In fact, it is usually sufficient to extract random \emph{classical} bits, which is in general easier to obtain than random qubits. This motivates the following definition, which differs from a QQ-extractor in that the output system $A_1$ is measured in the computational basis. In particular, any $(k, \eps)$-QQ-extractor is also a $(k, \eps)$-QC-extractor. 

\begin{definition}[QC-Extractors]\label{def:qc-extractor}
Let $A = A_1 A_2$ with $n = \log d_A$, and let $\cT_{A\rightarrow A_{1}}$ be the measurement map defined in equation~\eqref{eq:meas-map}.

For $k \in [-n, n]$ and $\eps \in [0,1]$, a $(k, \eps)$-QC-extractor is a set $\{U_1, \dots, U_{t}\}$ of unitary transformations on $A$ such that for all states $\rho_{AE} \in \cS(AE)$ satisfying $\entHmin(A|E) \geq k$, we have
\begin{align}\label{eq:qc-extractor}
\frac{1}{t} \sum_{i=1}^t \left\| \cT_{A\rightarrow A_{1}}(U_i \rho_{AE} U_i^{\dagger})- \frac{\id_{A_1}}{d_{A_1}} \ox \rho_E \right\|_1 \leq \eps\ .
\end{align}
$\log t$ is called the seed size of the $QC$-extractor.
\end{definition}

Observe that Definition~\ref{def:qc-extractor} only allows a specific form of measurements obtained by applying a unitary transformation followed by a measurement in the computational basis of $A_1$. The reason we restrict the measurements to be of this form is that we want the output of the extractor to be perfectly determined by the source and the choice of the seed. In the classical setting, an extractor is a family of deterministic functions of the source and the seed. In the quantum setting, a natural way of translating this requirement is by imposing that an adversary holding a system that is maximally entangled with the source can perfectly predict the output. This condition is satisfied by the form of measurements dictated by Definition~\ref{def:qc-extractor}. Allowing generalized measurements (POVMs) already (implicitly) allows the use of randomness for free. Note also, that in the case where the system $E$ is trivial, a $(0, \eps)$-QC-extractor is the same as an $\eps/2$-metric uncertainty relation (Definition \ref{def:metric-ur}).

Our definition of QC-extractors has some connections with some recent work on device independent randomness generation \citep{Col06, CK11, Pir10, AMP11, VV11, FGS11, PM11}. The objective of this line of work is to build protocols to generate bits that are \emph{certified} to be random. The setting is as follows. The system we consider has a special structure: it is composed of two parts $A$ and $B$ that are spatially separated. Using a small random seed, a pair of measurements is chosen to be performed on $A$ and $B$ obtaining outcomes $X$ and $Y$. Then, statistical tests are performed on $X$ and $Y$ to record a Bell inequality violation. Such a violation is then evidence that the systems $A$ and $B$ are entangled, which implies in particular that say $\entHmax(A|B)$ is significantly smaller than $\log d_A$. But this means that if the adversary holds a purification of the system $AB$, we have $\entHmin(A|E) = - \entHmax(A|B) \gg -\log d_A$. Thus, by applying a QC-extractor, one can generate almost perfect random bits. The challenge in that context is to detect Bell inequality violations with a small number of measurements.


\subsection{Examples and limitations of QC-extractors}

Universal (or two-independent) hashing is probably one of the most important extractor constructions, which even predates the general definition of extractors \citep{ILL89}. Unitary 2-designs can be seen as a quantum generalization of two-independent hash functions.

\begin{definition}\label{def:two-design}
A set of unitaries $\{U_1, \dots, U_t\}$ acting on $A$ is said to be a 2-design if for all $M \in \cL(A)$, we have
\begin{align}
\frac{1}{t} \sum_{i=1}^t U_i^{\otimes 2} M (U_i^{\dagger})^{\otimes 2} = \int U^{\otimes 2} M (U^{\dagger})^{\otimes 2} dU
\end{align}
where the integration is with respect to the Haar measure on the unitary group.
\end{definition}

Many efficient constructions of unitary 2-designs are known~\citep{DCEL09, GAE07}, and in an $n$-qubit space, such unitaries can typically be computed by circuits of size $O(n^2)$. However, observe that the number of unitaries of a 2-design is at least $t \geq d_A^4-2d_A^2+2$~\citep{GAE07}. The following is immediate using a general decoupling result of~\cite{Dup09,DBWR10} (see Lemma~\ref{lem:2design}).

\begin{corollary}\label{thm:qc-ext-two-design}
Let $A = A_1 A_2$ with $n = \log d_A$. For all $k \in [-n, n]$ and all $\eps > 0$, a unitary 2-design $\{U_1, \dots, U_t\}$ on $A$ is a $(k, \eps)$-QC-extractor with output size
\begin{align}
\log d_{A_1} = \min(n, n+k-2\log(1/\eps)).
\end{align}
\end{corollary}

Similar results also hold for almost unitary 2-designs; see \citep{SDTR11}. Using the results of \cite{HL09}, this shows for instance that random quantum circuits of size $O(n^2)$ are QC-extractors with basically the same parameters as in Corollary~\ref{thm:qc-ext-two-design}. 

Proposition \ref{lem:optimality} below shows that the output size of these QC-extractors is basically optimal. In fact, even if we are looking for a QC-extractor that works for a particular state $\rho_{AE}$, the output size is at most $n+H_{\min}^{\sqrt{\eps}}(A|E)_{\rho}$, where $n$ denotes the size of the input. In order to do that, we start by proving that when we measure a quantum system $A$, the min-entropy increases by at most the logarithm of the dimension of the system being measured.

\begin{lemma}\label{lem:qcvscc}
Let $\rho_{AB} \in \states(AB)$, $\eps\geq0$, and $\{P_{x}\}_{x=1}^{d_X}$ be a projective rank-one measurement on $A$. Then
\begin{align*}
\entHmin^{\eps}(X|B)_{\rho}\leq \entHmin^{\eps}(A|B)_{\rho}+\log d_X\ .
\end{align*}
\end{lemma}
\begin{proof}
Let $V_{A\rightarrow XX'}$ be an isometric purification of $\{P_{x}\}$ and $\rho_{XX'BB'}$ a purification of $\rho_{XX'B}=V\rho_{AB}V^{\dagger}$. By the invariance of the min-entropy under local isometries~\cite[Lemma 13]{TCR10} and the duality between the min- and max-entropy (Lemma~\ref{lem:duality}), the proposition becomes equivalent to
\begin{align*}
\entHmax^{\eps}(XX'|B')_{\rho}\leq \entHmax ^{\eps}(X|X'B')_{\rho}+\log d_X\ .
\end{align*}
By the definition of the smooth max-entropy (equations~\eqref{eq:maxentropy}-\eqref{eq:smoothmaxentropy}), there exists $\hat{\rho}_{XX'B'}\in\cB^{\eps}(\rho_{XX'B'})$ and $\hat{\sigma}_{X'B'}\in\cS(X'B')$ such that
\begin{align*}
\entHmax ^{\eps}(X|X'B')_{\rho}=\log F(\hat{\rho}_{XX'B},\id_{X}\otimes\hat{\sigma}_{X'B'})^{2}\ ,
\end{align*}
as well as $\bar{\rho}_{XX'B'}\in\cB^{\eps}(\rho_{XX'B'})$ and $\bar{\sigma}_{B}\in\cS(B)$ such that
\begin{align*}
\entHmax^{\eps}(XX'|B')_{\rho}=\log F(\bar{\rho}_{XX'B},\id_{XX'}\otimes\bar{\sigma}_{B'})^{2}\ .
\end{align*}
Now observe that
\begin{align*}
\entHmax^{\eps}(XX'|B')_{\rho} &\leq\log \left( d_X\cdot F(\hat{\rho}_{XX'B'},\id_{X}\otimes\frac{\id_{X'}}{d_X}\otimes\bar{\sigma}_{B'})^{2} \right) \\
&\leq \max_{\sigma_{X'B'} } \log F(\hat{\rho}_{XX'B'},\id_{X}\otimes \sigma_{X'B'})^{2}+\log d_X \\
&= \entHmax^{\e}(X|X'B')_{\rho}+\log d_X \ .
\end{align*}
\end{proof}

\begin{proposition}[Upper bound on the output size]\label{lem:optimality}
	Let $A=A_{1}A_{2}$, $\rho_{AE}\in\cS(AE)$, $\{U_1, \dots, U_t\}$ a set of unitaries on $A$, and $\cT_{A\rightarrow A_{1}}$ defined as in equation~\eqref{eq:meas-map}, such that
	\begin{align}
	\label{eq:opt-ext-cond}
		\frac{1}{t} \sum_{i=1}^{t} \left\| \cT_{A\rightarrow A_{1}}\left( U_i\rho_{AE} U_i^{\dagger} \right) - \frac{\id_{A_1}}{d_{A_1}} \otimes \rho_E\right\|_1 \leq \eps\ .
	\end{align}
	Then,
	\begin{align*}
		\log d_{A_1} \leq \log d_A + \entHmin^{\sqrt{\eps}}(A|E)_{\rho} \ . 
	\end{align*}
\end{proposition}
\begin{proof}
Consider the projective rank-one measurements $\{P^i_x\}$ obtained by performing $U_i$ followed by a measurement in the computational basis of $A$. 
As a result, we can apply Lemma \ref{lem:qcvscc} and obtain for all $i \in \{1, \dots, t\}$
\begin{align*}
\entHmin^{\sqrt{\eps}}(A|E)_{\rho} + \log d_A \geq \entHmin^{\sqrt{\eps}}(X_i|E)_{\rho}\ ,
\end{align*}
where $X_i$ denotes the outcome of the measurement $\{P^i_x\}$. But condition~\eqref{eq:opt-ext-cond} implies that there exists $i \in \{1, \dots, t\}$ such that 
\begin{align*}
\left\| \cT_{A\rightarrow A_{1}}\left( U_i\rho_{AE} U_i^{\dagger} \right) - \frac{\id_{A_1}}{d_{A_1}} \otimes \rho_E\right\|_1 \leq \eps.
\end{align*}
In other words, $\frac{\id_{A_1}}{d_{A_1}} \otimes \rho_E \in \cB^{\sqrt{\e}}\left(\cT_{A\rightarrow A_{1}}\left( U_i\rho_{AE} U_i^{\dagger} \right) \right)$.
By monotonicity of the min-entropy for classical registers~\cite[Lemma C.5]{BFS11}, we have that
\begin{align*}
\entHmin^{\sqrt{\eps}}(X_i|E)_{\rho} \geq \entHmin^{\sqrt{\eps}}(A_1|E)_{\cT_{A\rightarrow A_{1}}\left( U_i\rho_{AE} U_i^{\dagger} \right)} \geq \log d_{A_1}\ ,
\end{align*}
which proves the desired result.
\end{proof}

We now study the seed size property. We prove that choosing a reasonably small set of unitaries at random defines a QC-extractor with high probability. The seed size in this case is of the same order as the output size of the extractor. We expect that a much smaller seed size would be sufficient. However, as will be proved in Proposition \ref{prop:2norm-ext} below, different methods would have to be used in order to prove that.

\begin{theorem}\label{thm:small-decoupling-set}
Let $A=A_{1}A_{2}$ with $n = \log d_A$ and $\cT_{A\rightarrow A_{1}}$ be the measurement map defined in equation~\eqref{eq:meas-map}. Let $\eps > 0$, $c$ be a sufficiently large constant,
\begin{align*}
\log d_{A_1} \leq n + k - 4 \log (1/\eps) - c \quad \text{and} \quad \log t \geq \log d_{A_1} + \log n + 4 \log(1/\eps) + c \ .
\end{align*}
Then, choosing $t$ unitaries $\left\{U_1, \dots, U_t\right\}$ independently according to the Haar measure defines a $(k, \eps)$-QC-extractor with high probability. See \eqref{eq:prob-decoupling-set} for a probability bound.
\end{theorem}
\begin{proof}
The proof uses one-shot decoupling techniques~\citep{DBWR10,SDTR11,Dup09} combined with an operator Chernoff bound~\citep{AW02} (see Lemma~\ref{lem:operator-chernoff}).

Let $U$ be a unitary on $A$. We use the H\"{o}lder-type inequality (see e.g., ~\cite[Corollary IV.2.6]{Bha97})
\begin{align*}
\|\alpha\beta\gamma\|_{1}\leq\||\alpha|^{r}\|_{1}^{1/r}\||\beta|^{s}\|_{1}^{1/s}\||\gamma|^{r'}\|_{1}^{1/r'}
\end{align*}
where $1/r + 1/s + 1/r' = 1$. We use it with $r=r'=4$, $s=2$, and $\alpha=\gamma=(\id_{A_{1}}\otimes\rho_{E})^{1/4}$, $\beta=(\id_{A_{1}}\otimes\rho_{E})^{-1/4}\left(\cT(U \rho_{AE} U^{\dagger})-\frac{\id_{A_{1}}}{d_{A_1}} \ox \rho_E\right)(\id_{A_{1}}\otimes\rho_{E})^{-1/4}$ to get that\footnote{The inverses are generalized inverses.}
\begin{align*}
&\left\| \cT\left(U\rho_{AE}U^{\dagger}\right) - \frac{\id_{A_{1}}}{d_{A_1}} \ox \rho_E \right\|_1 \\
&\leq
d_{A_1}^{1/4}\sqrt{\tr\left[\left(\id_{A_{1}}\otimes\rho_{E}\right)^{-1/4}\left(\cT\left(U\rho_{AE}U^{\dagger}\right) - \frac{\id_{A_{1}}}{d_{A_1}} \ox \rho_E\right)\left(\id_{A_{1}}\otimes\rho_{E}\right)^{-1/4}\right]^{2}}d_{A_1}^{1/4} \\
&=d_{A_1}^{1/2}\left\|\left(\id_{A_{1}}\otimes\rho_{E}\right)^{-1/4}\left(\cT\left(U\rho_{AE}U^{\dagger}\right) - \frac{\id_{A_{1}}}{d_{A_1}} \ox \rho_E\right)\left(\id_{A_{1}}\otimes\rho_{E}\right)^{-1/4}\right\|_{2} \\
&=d_{A_1}^{1/2}\left\|\cT\left(U\tilde{\rho}_{AE}U^{\dagger}\right)-\frac{\id_{A_{1}}}{d_{A_1}}\otimes\tilde{\rho}_{E}\right\|_{2}\ ,
\end{align*}
where $\tilde{\rho}_{AE}=(\id_{A}\otimes\rho_{E})^{-1/4}\rho_{AE}(\id_{A}\otimes\rho_{E})^{-1/4}$. Together with the concavity of the square root function, this implies
\begin{align}
\frac{1}{t} \sum_{i=1}^{t} & \left\| \cT\left(U_i \rho_{AE}U_i^{\dagger}\right) - \frac{\id_{A_{1}}}{d_{A_1}} \ox \rho_E \right\|_1 \notag \\
&\leq\sqrt{\frac{1}{t} \sum_{i=1}^{t}\left\| \cT\left(U_i \rho_{AE}U_i^{\dagger}\right) - \frac{\id_{A_{1}}}{d_{A_1}} \ox \rho_E \right\|_1^{2}} \notag \\
&\leq\sqrt{d_{A_1}\frac{1}{t} \sum_{i=1}^{t}\left\| \cT\left(U_i \tilde{\rho}_{AE}U_i^{\dagger}\right) - \frac{\id_{A_{1}}}{d_{A_1}} \ox \tilde{\rho}_E \right\|_2^{2}} \notag \\
&=\sqrt{d_{A_1}\frac{1}{t} \sum_{i=1}^{t}\tr\left[\cT\left(U_i \tilde{\rho}_{AE}U_i^{\dagger}\right) - \frac{\id_{A_{1}}}{d_{A_1}} \ox \tilde{\rho}_E\right]^{2}}\ .\label{eq:concave}
\end{align}
We continue with
\begin{align}
&\frac{1}{t} \sum_{i=1}^{t}\tr\left[\cT\left(U_i \tilde{\rho}_{AE}U_i^{\dagger}\right) - \frac{\id_{A_{1}}}{d_{A_1}} \ox \tilde{\rho}_E\right]^{2} \notag \\
&=\frac{1}{t} \sum_{i=1}^{t}\tr\left[\cT\left(U_i \tilde{\rho}_{AE} U_i^{\dagger}\right)\right]^{2}-2\tr\left[\cT\left(U_i \tilde{\rho}_{AE}U_i^{\dagger}\right)\left(\frac{\id_{A_{1}}}{d_{A_1}} \ox \tilde{\rho}_E\right)\right]+\tr\left[\frac{\id_{A_{1}}}{d_{A_1}} \ox \tilde{\rho}_E\right]^{2} \label{eq:expandsquare}
\end{align}
and first compute the cross term
\begin{align*}
\tr\left[\cT\left(U_i \tilde{\rho}_{AE}U_i^{\dagger}\right)\left(\frac{\id_{A_{1}}}{d_{A_1}} \ox \tilde{\rho}_E\right)\right]&
=\frac{1}{d_{A_1}}\tr\left[\tr_{A_{1}}\left[\cT\left(U_i \tilde{\rho}_{AE}U_i^{\dagger}\right)\left(\id_{A_{1}} \ox \tilde{\rho}_E\right)\right]\right]\\
&=\frac{1}{d_{A_1}}\tr\left[\tilde{\rho}_{E}^{2}\right]\ .
\end{align*}
Going back to equation~\eqref{eq:expandsquare}, we obtain
\begin{align}
\frac{1}{t} \sum_{i=1}^{t}\tr\left[\cT\left(U_i \tilde{\rho}_{AE}U_i^{\dagger}\right) - \frac{\id_{A_{1}}}{d_{A_1}} \ox \tilde{\rho}_E\right]^{2}=\frac{1}{t} \sum_{i=1}^{t}\tr\left[\cT\left(U_i \tilde{\rho}_{AE}U_i^{\dagger}\right)\right]^{2}-\frac{1}{d_{A_1}}\tr\left[\tilde{\rho}_{E}^{2}\right]\ .\label{eq:middle}
\end{align}
Let $F_{AA'}$ denote the swap operator $F_{AA'} = \sum_{aa'} \ket{a a'} \bra{a' a}$. We now compute the first term using the ``swap trick'' (Lemma~\ref{lem:swap-trick})
\begin{align*}
&\tr\left[\cT(U\tilde{\rho}_{AE}U^{\dagger}) \right]^2 \\
&= \tr\left[ \sum_{a_{1}a_{2}} \bra{a_{1}a_{2}} U\tilde{\rho}_{AE}U^{\dagger} \ket{a_{1}a_{2}} \proj{a_{1}} \right]^2 \\
&= \tr\left[ \sum_{a_{1}a_{2}a_{1}'a_{2}'}\bra{a_{1}a_{2}a_{1}'a_{2}'}U^{\ox 2} \tilde{\rho}^{\ox 2}_{AE} (U^{\ox 2})^{\dagger} \ket{a_{1}a_{2}a_{1}'a_{2}'} \proj{a_{1}a_{1}'}\left(F_{A_{1}A_{1}'} \otimes F_{EE'}\right)\right] \\
&= \sum_{a_{1}a_{2}a_{1}'a_{2}'} \tr\left[ \tilde{\rho}^{\ox 2}_{AE} (U^{\ox 2})^{\dagger} \ket{a_{1}a_{2}a_{1}'a_{2}'} \bra{a_{1}a_{1}'}\left(F_{A_{1}A_{1}'} \otimes F_{EE'}\right) \ket{a_{1}a_{1}'} \bra{a_{1}a_{2}a_{1}'a_{2}'} U^{\ox 2} \right]\ .
\end{align*}
In the last equality, we used the fact that $\ket{a_1 a'_1}$ commutes with the scalar $\bra{a_{1}a_{2}a_{1}'a_{2}'}U^{\ox 2} \tilde{\rho}^{\ox 2}_{AE} (U^{\ox 2})^{\dagger} \ket{a_{1}a_{2}a_{1}'a_{2}'}$ and the cyclicity of the trace.
Taking the average over the set $\left\{U_1, \dots, U_t\right\}$, we get
\begin{align}
&\frac{1}{t} \sum_{i=1}^{t}\tr\left[\cT\left(U_i \tilde{\rho}_{AE}U_i^{\dagger}\right)\right]^{2} \notag \\
&= \sum_{a_{1}a_{2}a_{1}'a_{2}'} \tr\left[ \tilde{\rho}^{\ox 2}_{AE} \frac{1}{t} \sum_{i=1}^{t}\left\{\left(U_i^{\ox 2}\right)^{\dagger} \ket{a_{1}a_{2}a_{1}'a_{2}'} \bra{a_{1}a_{1}'}F_{A_{1}A_{1}'}\ket{a_{1}a_{1}'} \bra{a_{1}a_{2}a_{1}'a_{2}'}U_i^{\ox 2}\right\}  \otimes F_{EE'}\right] \notag \\
&=  \tr\left[ \tilde{\rho}^{\ox 2}_{AE} \frac{1}{t} \sum_{i=1}^{t}\left\{\left(U_i^{\dagger}\right)^{\ox 2} \sum_{a_{1}a_{2}a_{2}', a_1' = a_1} \ket{a_{1}a_{2}a_{1}'a_{2}'} \bra{a_{1}a_{2}a_{1}'a_{2}'}U_i^{\ox 2}\right\}  \otimes F_{EE'}\right]\ .\label{eq:average}
\end{align}
Using for example \cite[Lemma 3.4]{DBWR10}, if $U$ is distributed according to the Haar measure on the group of unitaries acting on $A$, then
\begin{align*}
&\exc{U}{\left(U^{\dagger}\right)^{\ox 2} \sum_{a_{1}a_{2}a_{2}'} \ket{a_{1}a_{2}a_{1}a_{2}'} \bra{a_{1}a_{2}a_{1}a_{2}'}U^{\ox 2}} \\
&= \left( \frac{d_Ad_{A_2} - 1}{d_A^{2} -1} \right) \id_{AA'} + \frac{d_A-d_{A_2}}{d_A^{2}-1}F_{AA'} \ .
\end{align*}
We use the shorthand $\Gamma_{AA'}$ for the expression above. Now we note that $\frac{d_Ad_{A_2} - 1}{d_A^{2} -1}\geq\frac{1}{2d_{A_1}}$, and apply the operator Chernoff bound (Lemma~\ref{lem:operator-chernoff}) to get
\begin{align}
&\prob{\frac{1}{t} \sum_{i=1}^{t}  (U_i^{\dagger})^{\ox 2} \sum_{a_{1}a_{2}a_{2}'} \ket{a_{1}a_{2}a_{1}a_{2}'} \bra{a_{1}a_{2}a_{1}a_{2}'} U_i^{\ox 2} \leq (1+\eta) \Gamma } \notag \\
&\geq1-d_A\exp{-\frac{t\eta^2}{d_{A_1}4 \ln 2} }\ . \label{eq:prob-decoupling-set}
\end{align}
This shows that if $t \geq 2 \cdot 4\ln2\cdot d_{A_1}\log d_A/\eta^2$, the unitaries $U_1, \dots, U_t$ satisfy the above operator inequality with high probability. In the rest of the proof, we show that such unitaries define QC-extractors. Putting these unitaries in equation~\eqref{eq:average}, we get
\begin{align*}
\frac{1}{t} \sum_{i=1}^{t}\tr\left[\cT\left(U_i \tilde{\rho}_{AE} \left(U_i\right)^{\dagger}\right) \right]^2
\leq (1+\eta)\left( \frac{d_Ad_{A_2}-1}{d_A^{2}-1}\tr\left[ \tilde{\rho}_E^2 \right]+\frac{d_A-d_{A_2}}{d_A^{2}-1}\tr\left[ \tilde{\rho}_{AE}^2 \right] \right)\ .
\end{align*}
Plugging this expression in equation \eqref{eq:middle} and then in equation~\eqref{eq:concave}, we get
\begin{align*}
&\frac{1}{t} \sum_{i=1}^{t}\left\| \cT\left(U_i \rho_{AE} \left(U_i\right)^{\dagger}\right) - \frac{\id_{A_{1}}}{d_{A_1}} \ox \rho_E \right\|_1 \\
&\leq \sqrt{(1+\eta)\left(\frac{d_A^2 - d_{A_1}}{d_A^2 - 1}\right) \tr\left[\tilde{\rho}_E^2\right]+(1+\eta)\left(\frac{d_{A_1}d_A-d_A}{d_A^2 - 1} \right)\tr\left[ \tilde{\rho}_{AE}^2 \right]-\tr\left[\tilde{\rho}_{E}^{2}\right]}\\
&\leq \sqrt{\eta+(1+\eta)\frac{d_{A_1}}{d_A+1} \tr\left[\tilde{\rho}_{AE}^2\right]}\ ,
\end{align*}
since $\tr \left[ \tilde{\rho}_{E}^2 \right] = \tr \left[\tr_A \left[\left(\id_A \ox {\rho}_E^{-1/4}\right) \rho_{AE} \left(\id_A \ox {\rho}_E^{-1/4}\right)\right]^{2}\right] = \tr \left[ \rho_E \right]=1$. By the definition of the conditional collision entropy (equation~\eqref{eq:coll}) and Lemma~\ref{lem:hmin-h2-rhorho}, it follows that,
\begin{align}
\frac{1}{t} \sum_{i=1}^{t}\left\| \cT\left(U_i \rho_{AE} \left(U_i\right)^{\dagger}\right) - \frac{\id_{A_{1}}}{d_{A_1}} \ox \rho_E \right\|_1
&\leq\sqrt{\eta+(1+\eta)\frac{d_{A_1}}{d_A+1}2^{-\entHtwo(A|E)_{\rho|\rho}}} \notag \\
&\leq\sqrt{\eta+(1+\eta)\frac{d_{A_1}}{d_A+1}2^{-\entHmin(A|E)_{\rho|\rho}}}\ .
\label{eq:c25}
\end{align}
Now let $\rho'_{AE}\in\cB^{\delta+\delta'}(\rho_{AE})$ be such that $\entHmin^{\delta+\delta'}(A|E)_{\rho|\rho}=\entHmin(A|E)_{\rho'|\rho'}$. Since we have $\|\rho_{AE}'-\rho_{AE}\|_{1}\leq2(\delta+\delta')$ (by equation~\eqref{eq:purifiedVStrace}), we know that by the triangle inequality and the monotonicity of the trace distance,
\begin{align*}
&\left| \| \cT(U \rho_{AE} U^{\dagger}) - \frac{\id_{A_{1}}}{d_{A_1}} \ox \rho_E \|_1 - \| \cT(U \rho'_{AE} U^{\dagger}) - \frac{\id_{A_{1}}}{d_{A_1}} \ox \rho_E \|_1 \right| \\
&\leq \| \cT(U \rho_{AE} U^{\dagger}) - \cT(U \rho'_{AE} U^{\dagger})  \|_1 \\
&\leq \| \rho'_{AE} - \rho_{AE} \|_1 \leq2(\delta+\delta')\ ,
\end{align*}
and hence applying \eqref{eq:c25} to $\rho'_{AB}$, we get
\begin{align*}
\frac{1}{t} \sum_{i=1}^{t} & \left\| \cT\left(U_i \rho_{AE} \left(U_i\right)^{\dagger}\right) - \frac{\id_{A_{1}}}{d_{A_1}} \ox \rho_E \right\|_1 \\
&\leq\sqrt{\eta+(1+\eta)\frac{d_{A_1}}{d_A+1}2^{-\entHmin^{\delta+\delta'}(A|E)_{\rho|\rho}}}+2(\delta+\delta')\ .
\end{align*}
We then use Lemma~\ref{lem:hmin-rhorho} about the equivalence of the different conditional min-entropies to get
\begin{align}
\frac{1}{t} \sum_{i=1}^{t} & \left\| \cT\left(U_i \rho_{AE} \left(U_i\right)^{\dagger}\right) - \frac{\id_{A_{1}}}{d_{A_1}} \ox \rho_E \right\|_1 \notag \\
&\leq\sqrt{\eta+(1+\eta)\frac{d_{A_1}}{d_A+1}2^{-\entHmin^{\delta}(A|E)_{\rho} + z}}+2(\delta+\delta')\ ,
\label{eq:small-decoupling-set-exp}
\end{align}
with $z = \log(2/\delta'^2 + 1/(1-\delta))$. Setting $\eta = \eps^2/4$, $\delta = 0$, $\delta' = \eps/4$, and assuming $\log d_{A_1} \leq n+ k - 4 \log(1/\eps) - c$ with $k = \entHmin(A|E)_{\rho}$, we can upper bound equation \eqref{eq:small-decoupling-set-exp} for large enough $c$ by
\begin{align*}
&\eps/2+\sqrt{\eps^2/4 + 2 \cdot 2^{k - 4 \log(1/\eps) - c -k + \log(8/\eps^2+1)}}\\
&\leq\eps/2+\sqrt{\eps^2/4 +\eps^2\cdot2^{1-c+4}}\leq\eps\ .
\end{align*}
\end{proof}


The following simple argument shows that the number of unitaries of a QC-extractor has to be at least $1/\eps$.

\begin{proposition}[Lower bound on seed size]\label{prop:coneps}
Let $A=A_{1}A_{2}$. Any $(k,\eps)$-QC-extractor with $k \leq \log d_A - 1$ is composed of a set of unitaries on $A$ of size at least $t\geq1/\eps$.
\end{proposition}
\begin{proof}
Let $S \subseteq [d_{A_1}]$ be an arbitrary subset of $d_{A_1}/2$ basis elements of $A_1$. Then consider the state
\begin{align*}
\rho_{A} = \frac{2}{d_A}\cdot\sum_{a_1 \in S, a_2 \in [d_{A_2}]} U_1^{\dagger} \proj{a_1a_2} U_1\ .
\end{align*}
Note that $\cT(U_1 \rho_A U_1^{\dagger}) = \frac{2}{d_{A_1}}\sum_{a_1 \in S} \proj{a_1}$ and thus $\| \cT(U_1 \rho_A U_1^{\dagger}) - \frac{\id_{A_1}}{d_{A_1}} \|_1 = 1$. This implies the claim.
\end{proof}

Observe that in the case where the system $E$ is trivial (or classical), we showed in Theorem \ref{thm:existence-ur} that there exists QC-extractors composed of $t = O(\log(1/\eps)\eps^{-2})$ unitaries. This is a difference with classical extractors, for which the number of possible values of the seed has to be at least $\Omega((n - k) \eps^{-2})$ \citep{RTS00} and a probabilistic construction shows that this is tight. It is not clear whether this is an important difference or whether it simply comes from the fact that the analogue for $(0,\e)$-QC-extractors should be $(n, \e)$-CC-extractors. An interesting question in this regard is to see whether one can prove an analogous lower bound on the seed size for $(k,\e)$-QC-extractors with negative $k$.

Observe that in the analysis of Theorem \ref{thm:small-decoupling-set}, we actually proved something stronger than condition \eqref{eq:qc-extractor}. There and actually in all the constructions given in this chapter, we prove that the stronger condition \eqref{eq:l2-qc-extractor} below holds. The following proposition shows that with such a strong definition, the seed has to be quite large. In particular, to show the existence of QC-extractors (or even QQ or CC-extractors) with a short seed, one should use different techniques to bound the trace distance directly.
\begin{proposition}[Extractors for the $2$-norm]\label{prop:2norm-ext}
Let $A=A_{1}A_{2}$. Let $\{U_1, \dots, U_t\}$ be unitaries such that 
\begin{equation}
\label{eq:l2-qc-extractor}
\frac{1}{t} \sum_{i=1}^{t} \left\| \cT_{A\rightarrow A_{1}}(U_i \rho_{AE} U_i^{\dagger})- \frac{\id_{A_1}}{d_{A_1}} \ox \rho_E \right\|^2_2 \leq \frac{\eps^2}{d_{A_1}} \ .
\end{equation}
Then, $t \geq 1/\eps^2 \cdot \min\left( \frac{d_A}{2^{k+1}}, d_{A_1}/4 \right)$. 
\end{proposition}
\begin{proof}
Let $S \subseteq [d_{A_1}]$ be an arbitrary subset of $\max(1, \ceil{2^k / d_{A_2}})$ basis elements of $A_1$. Then consider the state
\begin{align*}
\rho_{A} = \frac{1}{|S| \cdot d_{A_2}}\cdot\sum_{a_1 \in S, a_2 \in [d_{A_2}]} U_1^{\dagger} \proj{a_1a_2} U_1\ .
\end{align*}
We have $\entHmin(A)_{\rho} \geq k$ and $\cT(U_1 \rho_A U_1^{\dagger}) = \frac{1}{|S|} \sum_{a_1 \in S} \proj{a_1}$. We can then compute
\begin{align*}
\left\| \cT(U_1 \rho_A U_1^{\dagger}) - \frac{\id_{A_1}}{d_{A_1}} \right\|^2_2 &\geq 
\sum_{a_1 \in S} \left( \frac{1}{|S|} - \frac{1}{d_{A_1}} \right)^2 \\
&= |S|\left(\frac{1}{|S|} - \frac{1}{d_{A_1}}\right)^2.
\end{align*}
As a result, we have $\frac{1}{t} \cdot |S|\left(\frac{1}{|S|} - \frac{1}{d_{A_1}}\right)^2 \leq \frac{\e^2}{d_{A_1}}$. In the case $2^k/d_{A_2} \leq 1$, we obtain a lower bound of $t \geq 1/\eps^2 \cdot d_{A_1} (1 - 1/d_{A_1})^2$. In the other case, we get
\begin{align*}
t &\geq 1/\eps^2 \cdot d_{A_1}/ |S| \\
	&\geq 1/\e^2 \cdot \frac{d_{A_1}}{2 \cdot 2^k/d_{A_2}} \\
	&= 1/\e^2 \cdot \frac{d_A}{2 \cdot 2^k}.
\end{align*}
\end{proof}

Our results about QC-extractors are summarized in Table~\ref{tab:ext-summary}.
\begin{table}[ht]
\centering
\begin{tabular}{|l|l|c|c|}
	\hline
	\multicolumn{2}{|c|}{} & CC-extractors & QC-extractors \\
	\hline
\multirow{3}{*}{Seed} & LB &  $\log(n-k) + 2 \log(1/\eps)$ \ \citeyearpar{RTS00}  & $\log(1/\eps)$ \\[3mm]
  & \multirow{2}{*}{UBs} & $\log(n-k) + 2 \log(1/\eps)$ & $m + \log n + 4 \log(1/\eps)$ \ [\ref{thm:small-decoupling-set}] \   \\
  &  & $c \cdot \log(n/\eps)$ \ \citeyearpar{GUV09}   & $3n$ \ [\ref{thm:full-set-mub}]  \\ \hline
\multirow{2}{*}{Output} & UB & $k-2\log(1/\eps)$ \ \citeyearpar{RTS00} & $n+\entHmin^{\sqrt{\eps}}(A|E)$ \ [\ref{lem:optimality}] \\
  & LB & $k - 2 \log(1/\eps)$ \ \citeyearpar{ILL89,RK05}  & $n + k - 2\log(1/\eps)$ \ [\ref{thm:full-set-mub}]    \\ \hline
\end{tabular}
\caption{Known lower bounds (LB) and upper bounds (UB) on the seed size and output size in terms of (qu)bits for different kinds of $(k, \eps)$-randomness extractors. $n$ refers to the number of input (qu)bits, $m$ the number of output (qu)bits and $k$ the min-entropy of the input $\entHmin(A|E)$. Note that for QC-extractors, $k$ can be as small as $-n$. Additive absolute constants are omitted. We note that the constructions corresponding to the second line are non-explicit.}
\label{tab:ext-summary}
\end{table}


\subsection{Full set of mutually unbiased bases (MUBs)}\label{sec:QCFromAllMUBS}

We saw that unitary 2-designs define QC-extractors. As unitary 2-designs also define QQ-extractors, it is natural to expect that we can build smaller and simpler sets of unitaries if we are only interested in extracting random classical bits. To that end, in this section, we construct simpler sets of unitaries that define QC-extractors. Two ingredients are used: a full set of mutually unbiased bases and a family of pairwise independent permutations.

A set of unitaries $\{U_1, \dots, U_t\}$ acting on $A$ is said to define \emph{mutually unbiased bases} if for all elements $\ket{a}, \ket{a'}$ of the computational basis of $A$, we have $| \bra{a'} U_j U_i^{\dagger} \ket{a} |^2 \leq d_A^{-1}$ for all $i \neq j$. In other words, a state described by a vector $U_i^{\dagger} \ket{a}$ of the basis $i$ gives a uniformly distributed outcome when measured in basis $j$ for $i \neq j$. For example the two bases, sometimes called computational and Hadamard bases (used in most quantum cryptographic protocols), are mutually unbiased. There can be at most $d_A+1$ mutually unbiased bases for $A$. Constructions of full sets of $d_A+1$ MUBs are known in prime power dimensions~\citep{WF89, BBRV02}. Such unitaries can be implemented by quantum circuits of almost linear size; see Lemma \ref{lem:explicitmub}. Mutually unbiased bases also have applications in quantum state determination~\citep{Ivo81, WF89}. 

To state our result, we will need one more notion.
A family $\cP$ of permutations of a set $X$ is \emph{pairwise independent} if for all $x_1 \neq x_2$ and $y_1 \neq y_2$, and if $\pi$ is uniformly distributed over $\cP$, $\prob{\pi(x_1) = y_1, \pi(x_2) = y_2} = \frac{1}{d_X(d_X-1)}$. If $X$ has a field structure, i.e., if $d_X$ is a prime power, it is simple to see that the family $\cP = \{ x \mapsto a \cdot x + b : a \in X^*, b \in X\}$ is pairwise independent. In the following, a permutation of basis elements of a Hilbert space $A$ should be seen as a unitary transformation on $A$.

\begin{theorem}\label{thm:full-set-mub}
Let $A=A_{1}A_{2}$ with $n = \log d_A$, $d_A$ a prime power, and consider the map $\cT_{A\rightarrow A_{1}}$ as defined in equation~\eqref{eq:meas-map}. Then, if $\left\{U_1, \dots, U_{d_A+1}\right\}$ defines a full set of mutually unbiased bases, we have for $\delta\geq0$,
\begin{align}\label{eq:full-set-mub}
 \frac{1}{|\cP|}  \frac{1}{d_A+1}  \sum_{P \in \cP} \sum_{i=1}^{d_A+1} & \left\| \cT_{A\rightarrow A_{1}}\left(PU_i \rho_{AE} \left(PU_i\right)^{\dagger}\right) - \frac{\id_{A_{1}}}{d_{A_1}} \ox \rho_E \right\|_1 \notag \\
&\leq \sqrt{\frac{d_{A_1}}{d_A+1} 2^{-\entHmin^{\delta}(A|E)_{\rho}}}+2\delta\ ,
\end{align}
where $\cP$ is a set of pairwise independent permutation matrices. In particular, the set $\{PU_i : P \in \cP, i \in [d_A+1]\}$ defines a $(k, \eps)$-QC-extractor provided
\begin{align*}
\log d_{A_1} \leq n + k - 2 \log(1/\eps)\ ,
\end{align*}
and the number of unitaries is
\begin{align*}
t = (d_A+1) |\cP|\ ,
\end{align*}
which for the pairwise independent permutations described above gives $t = (d_A+1)d_A (d_A-1)$.
\end{theorem}
\begin{proof}
The idea is to bound the trace norm in equation~\eqref{eq:full-set-mub} by the Hilbert-Schmidt norm of some well-chosen operator. This term is then computed exactly using the fact that the set of all the MUB vectors form a \emph{complex projective} 2-design (Lemma~\ref{lem:mub-design}), and the fact that the set of permutations is pairwise independent.

Similar to the proof of Theorem~\ref{thm:small-decoupling-set}, but with the difference that now $\tilde{\rho}_{AE}=\left(\id_{A}\otimes\sigma_{E}\right)^{-1/4}\rho_{AB}\left(\id_{A}\otimes\sigma_{E}\right)^{-1/4}$ for some $\sigma_E \in \cS(E)$ to be chosen later, we get
\begin{align}
&\frac{1}{|\cP|}  \frac{1}{d_A+1} \sum_{P \in \cP} \sum_{i=1}^{d_A+1} \left\| \cT\left(PU_i \rho_{AE} \left(PU_i\right)^{\dagger}\right) - \frac{\id_{A_{1}}}{A_{1}} \ox \rho_E \right\|_1
\notag \\
&\leq\sqrt{d_{A_1}\sum_{a_{1}a_{2}a_{2}'}\tr\left[\tilde{\rho}_{AE}^{\otimes2} \Gamma_{a_1, a_2, a'_2} \otimes F_{EE'}\right]-\tr\left[\tilde{\rho}_{E}^{2}\right]}\ ,\label{eq:start}
\end{align}
where $\Gamma_{a_1, a_2, a'_2} = \exc{P,i}{\left(U_i^{\dagger}P^{\dagger}\right)^{\otimes2}\proj{a_{1}a_{2}a_{1}a_{2}'}\left(PU_i\right)^{\otimes2}}$ and $F_{EE'}$ is the swap operator. We now compute $\Gamma_{a_1, a_2, a'_2}$ handling the case $a_{2}=a_{2}'$ and the case $a_{2}\neq a_{2}'$ differently. When $a_{2}=a_{2}'$, we have $(U_i^{\dagger})^{\ox 2} \ket{aa} \bra{aa} U_i^{\ox 2} = (U_i^{\dagger}\ket{a} \bra{a} U_i)^{\ox 2}$, where $a = P^{-1}(a_{1}a_{2})$. As $\{U_1,\ldots,U_{d_A+1}\}$ form a full set of mutually unbiased bases, the vectors $\{U_i \ket{a}\}_{i,a}$ define a complex projective 2-design (Lemma~\ref{lem:mub-design}), and we get
\begin{align}
\sum_{a_{1}a_{2}, a_{2}'=a_{2}} & \exc{P,i}{ \left(U_i^{\dagger}P^{\dagger}\right)^{\ox 2} \ket{a_{1}a_{2}a_{1}a_{2}'} \bra{a_{1}a_{2}a_{1}a_{2}'}\left(PU\right)^{\ox 2}} \notag \\
&=\sum_{a} \exc{i}{\left(U_i^{\dagger}\right)^{\ox 2} \ket{aa} \bra{aa} U_i^{\ox 2}} \notag \\
&=d_A\frac{2\Pi^{\sym}_{AA'}}{(d_A+1)d_A} \notag \\
&= \frac{\id_{AA'} + F_{AA'}}{d_A+1}\ ,\label{eq:next}
\end{align}
where $\Pi^{\sym}_{AA'}$ is the projector onto the symmetric subspace of $AA'$, i.e., the subspace spanned by vectors $\ket{a'a} + \ket{aa'}$.
We now consider $a_{2}\neq a_{2}'$. We have
\begin{align}
&\exc{P}{\left(P^{\dagger}\right)^{\ox 2} \ket{a_{1}a_{2}a_{1}a_{2}'} \bra{a_{1}a_{2}a_{1}a_{2}'} P^{\ox 2} } \notag \\
&= \exc{P}{\proj{P^{-1}(a_{1}a_{2})} \ox \proj{P^{-1}(a_{1}a_{2}')}} \notag \\
&= \sum_{a \neq a'} \probc{P}{P^{-1}(a_{1}a_{2}) = a, P^{-1}(a_{1}a_{2}')=a'} \proj{a} \ox \proj{a'} \notag \\
&= \frac{1}{d_A(d_A-1)}\sum_{a \neq a'}\proj{a} \ox \proj{a'} \notag \\
&= \frac{\id_{AA'}}{d_A(d_A-1)} - \frac{1}{d_A(d_A-1)} \sum_a \proj{aa}\ .\label{eq:after}
\end{align}
Going back to equation~\eqref{eq:next}, we get together with equation~\eqref{eq:after} that for any $a_2 \neq a_2'$,
\begin{align*}
&\exc{P,i}{\left(U_i^{\dagger}\right)^{\otimes 2} \left(P^{\dagger}\right)^{\ox 2} \ket{a_{1}a_{2}a_{1}a_{2}'} \bra{a_{1}a_{2}a_{1}a_{2}'} P^{\ox 2} U_i^{\ox 2} } \\
&=
\frac{\id_{AA'}}{d_A(d_A-1)}-\frac{1}{d_A(d_A-1)}\sum_a \exc{i}{\left(U_i^{\dagger}\right)^{\otimes 2} \proj{aa} U_i^{\otimes 2}} \\
&= \frac{\id_{AA'}}{d_A\left(d_A-1\right)}-\frac{\id_{AA'} + F_{AA'}}{d_A(d_A-1)(d_A+1)} \\
&= \frac{d_A\id_{AA'} - F_{AA'}}{d_A(d_A^{2}-1)}\ .
\end{align*}
This being true for all $a_{1},a_{2},a_{2}'$, it follows with equation~\eqref{eq:start} that,
\begin{align}
&\exc{P,i}{\left\| \cT\left(PU_i \rho_{AE} \left(PU_i\right)^{\dagger}\right) - \frac{\id_{A_{1}}}{d_{A_{1}}} \ox \rho_E \right\|_1}
\notag \\
&=\sqrt{d_{A_1}\tr\left[ \tilde{\rho}_{AE}^{\ox 2} \left( \frac{\id_{AA'} + F_{AA'}}{d_A+1} + d_A(d_{A_2}-1)\frac{d_A\id_{AA'}-F_{AA'}}{d_A(d_A^{2}-1)}\right)\ox F_{EE'}\right]-\tr\left[\tilde{\rho}_{E}^{2}\right]}. \label{eq:sq-root-cont}
\end{align}
Expanding the expression inside the square root, we obtain
\begin{align*}
&d_{A_1}\left( \frac{1}{d_A + 1} + \frac{d_A(d_{A_2}-1)}{d_A^{2}-1} \right)\tr\left[\tilde{\rho}_{AE}^{\ox 2} \left(\id_{AA'} \ox F_{EE'}\right) \right] \\
&+d_{A_1}\left(\frac{1}{d_A + 1} - \frac{d_{A_2}-1}{d_A^2 - 1} \right) \tr \left[ \tilde{\rho}_{AE}^{\ox 2}\left(F_{AA'} \ox F_{EE'}\right)\right]-\tr\left[\tilde{\rho}_{E}^{2}\right] \\
&= d_{A_1}\frac{d_Ad_{A_2} - 1}{d_A^2 - 1}\tr\left[ \tr_{AA'} \left[\tilde{\rho}^{\ox 2}_{AE}\left(\id_{AA'} \ox F_{EE'}\right)\right]\right] \\
&+d_{A_1}\left(\frac{d_A-d_{A_2}}{d_A^2 - 1} \right)\tr[ \tilde{\rho}_{AE}^2 ]-\tr\left[\tilde{\rho}_{E}^{2}\right].
\end{align*}
Continuing from \eqref{eq:sq-root-cont}, we get
\begin{align*}
&=\sqrt{\left(\frac{d_A^{2}-d_{A_1}}{d_A^2 - 1}-1\right)\tr[ \tilde{\rho}_E^2 ] + \left(\frac{d_{A_1}d_A-d_A}{d_A^2 - 1} \right)\tr[ \tilde{\rho}_{AE}^2 ]} \\
&\leq\sqrt{\frac{d_{A_1}}{d_A+1}\tr\left[\tilde{\rho}_{AE}^2\right]}=\sqrt{\frac{d_{A_1}}{d_A+1} 2^{-\entHtwo(A|E)_{\rho|\sigma}}}\ ,
\end{align*}
where we used the definition of the conditional collision entropy (equation~\eqref{eq:coll}) in the last step. Now, by choosing $\sigma_{E}$ appropriately, and an argument analogous to the very end of the proof of Theorem~\ref{thm:small-decoupling-set}, we conclude that,
\begin{align*}
\exc{P,i}{\left\| \cT\left(PU_i \rho_{AE} \left(PU_i\right)^{\dagger}\right) - \frac{\id_{A_{1}}}{d_{A_{1}}} \ox \rho_E \right\|_1}
\leq \sqrt{\frac{d_{A_1}}{d_A+1} 2^{-\entHmin^{\delta}(A|E)_{\rho}}}+2\delta\ .
\end{align*}
\end{proof}

In terms of output size, this construction is almost optimal, but the number of unitaries is again much larger than we expect should be possible.


\subsection{Bitwise QC-extractor}\label{sec:singlequdit}

The unitaries we construct in this section are even simpler. They are composed of unitaries $V$ acting on single qudits followed by permutations $P$ of the computational basis elements. Note that this means that the measurements defined by these unitaries can be implemented with current technology. As the measurement $\cT$ commutes with the permutations $P$, we can first apply $V$, then measure in the computational basis and finally apply the permutation to the (classical) outcome of the measurement. In addition to the computational efficiency, the fact that the unitaries act on single qudits, is often a desirable property for the design of cryptographic protocols. In particular, the application to the noisy storage model that we present in Section \ref{sec:noisy} does make use of this fact. But the price we pay is that the parameters (both output and seed size) are worse than the previous construction.

Let $d \geq 2$ be a prime power so that there exists a complete set of mutually unbiased bases in dimension $d$. We represent such a set of bases by a set of unitary transformations $\left\{V_0,V_1,\dots,V_d\right\}$ mapping these bases to the standard basis. For example, for the qubit space ($d=2$), we can choose 
\begin{equation*}
V_0 = \left( \begin{array}{cc}
1 & 0 \\
0 & 1
\end{array} \right)
\qquad V_1 = \frac{1}{\sqrt{2}} \left( \begin{array}{cc}
1 & 1 \\
1 & -1
\end{array} \right)
\qquad V_2 = \frac{1}{\sqrt{2}} \left( \begin{array}{cc}
1 & i \\
i & -1
\end{array} \right)\ .
\end{equation*}
We define the set $\cV_{d,n}$ of unitary transformations on $n$ qudits by $\cV_{d,n} \eqdef \left\{V=V_{u_1}\ox\cdots\ox V_{u_n}|u_i\in\left\{0,\dots,d\right\}\right\}$. As in the previous section, $\cP$ denotes a family of pairwise independent permutations.

\begin{theorem}\label{thm:singleQuditExtract}
Let $A=A_{1}A_{2}$ with $d_A=d^{n}$, $d_{A_1}=d^{\xi n}$, $d_{A_2}=d^{(1-\xi)n}$, and $d$ a prime power. Consider the map $\cT_{A\rightarrow A_{1}}$ as defined in equation~\eqref{eq:meas-map}. Then for $\delta\geq0$ and $\delta'>0$,
\begin{align}
\notag
\frac{1}{|\cP|} \frac{1}{(d+1)^n} \sum_{P \in \cP}\sum_{V \in \cV_{d,n}} &\left\| \cT_{A\rightarrow A_{1}}\left(PV \rho_{AE} \left(PV\right)^{\dagger}\right)- \frac{\id_{A_{1}}}{d_{A_1}} \ox \rho_E \right\|_1 \\
&\leq \sqrt{2^{\left(1-\log (d+1)+\xi\log d\right)n}(1+2^{-\entHmin^{\delta}(A|E)_{\rho}+z})}+2(\delta+\delta')\ ,
\label{eq:singleQuditExtract}
\end{align}
where $\cV_{d,n}$ is defined as above, $\cP$ is a set of pairwise independent permutation matrices, and $z=\log\left(\frac{2}{\delta'^2}+\frac{1}{1-\delta}\right)$. In particular, the set $\{ PV : P \in \cP, V \in \cV_{d,n}\}$ is a $(k, \eps)$-extractor provided
\begin{align*}
\log d_{A_1} \leq (\log(d+1) - 1) n + \min \left\{0, k\right\}  - 4 \log(1/\eps) - 7\, 
\end{align*}
and if we choose the pairwise independent permutations described in Theorem \ref{thm:full-set-mub}, the number of unitaries is
\begin{align*}
t = (d+1)^n d^n (d^n - 1)\ .
\end{align*}
\end{theorem}

The analysis uses the same technique as in the proof of Theorem~\ref{thm:full-set-mub}. The main difference is that we were not able to express the Hilbert-Schmidt norm exactly in terms of the conditional min-entropy $\entHmin(A|E)_{\rho}$. Instead, we use some additional inequalities, which account for the slightly more complicated expression 
we obtain.

\begin{proof}
We use the same strategy as in the proofs of Theorem~\ref{thm:small-decoupling-set} and Theorem~\ref{thm:full-set-mub}; here again with $\tilde{\rho}_{AE}=\left(\id_{A}\otimes\rho_{E}\right)^{-1/4}\rho_{AE}\left(\id_{A}\otimes\rho_{E}\right)^{-1/4}$. As in \eqref{eq:start} and \eqref{eq:after}, we get
\begin{align}
&\exc{P,V}{ \left\| \cT\left(PV \rho_{AE} \left(PV\right)^{\dagger}\right) - \frac{\id_{A_{1}}}{d_{A_1}} \ox \rho_E \right\|_1} \notag \\
&\leq\sqrt{d_{A_1}\tr\left[\tilde{\rho}_{AE}^{\otimes 2}\left(\sum_{a}\Gamma_a+d_A\left(d_{A_2}-1\right)\frac{\id_{AA'}}{d_A\left(d_A-1\right)}\right)\otimes F_{EE'}\right]-\tr\left[\tilde{\rho}_{E}^{2}\right]}\ , \label{eq:singleq1}
\end{align}
where $\Gamma_a = \exc{V}{\left(V^{\dagger}\proj{a}V\right)^{\otimes2}}$. We calculate
\begin{align}
\sum_{a} \exc{V}{\left(V^{\dagger}\proj{a}V\right)^{\otimes2}} &= \frac{1}{(d+1)^n} \sum_{a_1, a_2, \dots, a_n} \sum_{V_1, \dots, V_n} \bigotimes_i \left( ( V^{\dagger}_i \proj{a_i} V_i)^{\otimes 2} \right) \notag \\
&= \frac{1}{(d+1)^n} \bigotimes_i \left(\sum_{a_i, V_i} \left(V_i^{\dagger} \proj{a_i} V_i \right)^{\otimes 2} \right)\ . \label{eq:singleq2}
\end{align}
As $\left\{V_0,\dots,V_d\right\}$ form a maximal set of mutually unbiased bases in dimension $d$, and with this form a complex projective 2-design (Lemma~\ref{lem:mub-design}), we have
\begin{align*}
\sum_{a \in \{0,\ldots,d\}, V \in \cV_{d,1} } \left(V^{\dagger} \proj{a} V \right)^{\otimes 2} = 2 \Pi^{\sym}\ ,
\end{align*}
where $\Pi^{\sym}$ is the projector onto the symmetric subspace, i.e., the subspace spanned by vectors $\ket{a'a} + \ket{aa'}$. Furthermore $(\Pi^{\sym}_{B})^{\otimes n} \leq \Pi^{\sym}_{B^{\ox n}}$ for any inner product space $B$, and hence we obtain
\begin{align}
\frac{1}{(d+1)^n} \bigotimes_i \left(\sum_{a_i, V_i} \left(V_i^{\dagger} \proj{a_i} V_i \right)^{\otimes 2} \right)
&\leq \left(\frac{2}{d+1}\right)^n \Pi^{\sym}_{AA'} \notag \\
&= \left(\frac{2}{d+1}\right)^n \frac{\id_{AA'} + F_{AA'} }{2}\ .\label{eq:Mdef}
\end{align}
Plugging equation \eqref{eq:Mdef} into the expression inside the square root in equation~\eqref{eq:singleq1}, we can bound it by
\begin{align*}
&\leq d_{A_1}\tr\left[\tilde{\rho}_{AE}^{\ox 2}\left(\left(\frac{2}{d+1}\right)^n\frac{\id_{AA'}+F_{AA'}}{2}+d_A(d_{A_2}-1)\frac{\id_{AA'} }{d_A(d_A-1)}\right)\ox F_{EE'} \right] \\
&\qquad -\tr\left[\tilde{\rho}_{E}^{2}\right] \\
&= \left(\frac{d_A-d_{A_1}}{d_A-1}+\frac{d_{A_1}}{2}\left(\frac{2}{d+1}\right)^n\right)\tr\left[\tilde{\rho}_{AE}^{\ox 2}\left(\id_{AA'} \ox F_{EE'}\right)\right] \\
& \qquad +\frac{d_{A_1}}{2}\left(\frac{2}{d+1}\right)^n\tr\left[\tilde{\rho}_{AE}^{\ox 2}\left(F_{AA'}\ox F_{EE'}\right)\right]-\tr\left[\tilde{\rho}_{E}^{2}\right] \\
&\leq \left(1+2^{(1-\log(d+1)+\xi\log d)n}\right)\tr\left[\tr_{AA'}\left[\tilde{\rho}^{\ox 2}_{AE}\left(\id_{AA'}\ox F_E\right)\right]\right] \\
& \qquad + 2^{(1-\log(d+1)+\xi\log d)n}\tr[\tilde{\rho}_{AE}^2]-\tr\left[\tilde{\rho}_{E}^{2}\right].
\end{align*}
Continuing from \eqref{eq:singleq1}, we get
\begin{align*}
&\leq \sqrt{2^{(1-\log(d+1)+\xi\log d)n}\tr\left[\tilde{\rho}_E^2\right]+2^{(1-\log(d+1)+\xi\log d)n}\tr\left[\tilde{\rho}_{AE}^2\right]} \notag \\
&=\sqrt{2^{(1-\log(d+1)+\xi\log d)n}\left(1+2^{-\entHtwo(A|E)_{\rho|\rho}}\right)}\ , \notag
\end{align*}
where we used the definition of the conditional collision entropy (equation~\eqref{eq:coll}) in the last step. Again by an argument analogous to the very end of the proof of Theorem~\ref{thm:small-decoupling-set}, we conclude that,
\begin{align*}
&\exc{P,V}{ \left\| \cT\left(PV \rho_{AE} \left(PV\right)^{\dagger}\right) - \frac{\id_{A_{1}}}{d_{A_1}} \ox \rho_E \right\|_1} \\
&\leq\sqrt{2^{\left(1-\log (d+1)+\xi\log d\right)n}(1+2^{-\entHmin^{\delta}(A|E)_{\rho} + z})}+2(\delta+\delta')\ ,
\end{align*}
where $z=\log\left(\frac{2}{\delta'^2}+\frac{1}{1-\delta}\right)$.
Setting $\delta = 0$ and $\delta' = \eps/4$, we conclude that the set $\{PV : P \in \cP, V \in \cV_{d,n}\}$ is a $(k, \eps)$-QC-extractor provided 
\begin{align*}
\log d_{A_1} &= n \cdot \xi \log d \\
&\leq (\log(d+1) - 1) n - \log(1 + 2^{-k + \log(8/\eps^2 + 1)}) + \log((\eps/2)^2)  \\
&\leq (\log(d+1) - 1) n + \min \left\{0, k - \log(8/\eps^2 + 1)\right\} - 2\log(1/\eps) - 3 \\
&\leq (\log(d+1) - 1) n + \min \left\{0, k\right\}  - 4 \log(1/\eps) - 7.
\end{align*}
\end{proof}
It seems that the parameters proved for this QC-extractor construction are not optimal. In fact, equation \eqref{eq:singleQuditExtract} does not give anything non-trivial when $\entHmin(A|E)_{\rho} < - (\log(d+1) - 1)n$. We believe however that it should be possible to obtain a non-trivial statement for any min-entropy as long as it is larger than $-c n \log d$ for some $c < 1$. Such an improvement would be quite interesting for the application we provide to two-party secure computation in Section \ref{sec:noisy} (see the discussion following Theorem \ref{thm:para}). We think that the place where the analysis should be improved is the inequality \eqref{eq:Mdef}. If we do not use this inequality, we end up with having to handle an expression of the form
\begin{equation}
\label{eq:sum-subsets}
\sum_{S \subseteq [n]} \tr\left[\tilde{\rho}^{2}_{A_SE}  \right],
\end{equation}
where $A_S$ refers to the qudits of $A$ indexed by elements of $S$. Because we have $\tr [\tilde{\rho}^2_{A_SE}] \leq 2^{|S| \log d}$ for any $S$, the sum in equation \eqref{eq:sum-subsets} is always bounded by $(d+1)^n$. It would be interesting for example to show that whenever $\entHmin(A|E) \geq -c n \log d$ for some $c < 1$, then there exists some $\beta < (d+1)$ such that the sum in \eqref{eq:sum-subsets} is bounded by $\beta^n$. This kind of statement is related to min-entropy sampling \citep{KR07}. The problem there is to prove that for most subsets $S$ of $[n]$ of size $r$, we have $\entHmin(A_S|E) \gtrsim \frac{r}{n} \entHmin(A|E)$. Such a statement was proved by \cite{KR07} in the case where $A$ is classical. It would be interesting to see if such a result holds when $A$ is a general quantum system.

\section{Entropic uncertainty relations with quantum side information}\label{sec:URbounds}

In this section, we show how to obtain entropic uncertainty relations from general QC-extractors. 

\subsection{Min-entropy uncertainty relations}
We start by proving uncertainty relations for the smooth min-entropy, which is usually the relevant measure in the context of cryptography. Consider the state $\rho_{XEJ}=\frac{1}{t}\sum_{j=1}^{t}\cM_{A\rightarrow X}( U_j \rho_{AE} U_j^{\dg}) \otimes\proj{j}_{J}$. We note that unlike for the von Neumann entropy, the conditional entropy $\entHmin(X|EJ)_{\rho}$ is not the same as the average $\frac{1}{t} \sum_{j=1}^{t} \entHmin(X|E)_{\rho^j}$, where $\rho^j = \cM_{A\rightarrow X}( U_j \rho_{AE} U_j^{\dg})$. However, by concavity of the logarithm, we have
\begin{align*}
	\frac{1}{t}\sum_{j=1}^{t} \hmin(X|E)_{\rho^j} &\geq
	- \log\left[\frac{1}{t} \sum_{j=1}^{t} 2^{- \hmin(X|E)_{\rho^{j}}}\right] = \hmin(X|EJ)_{\rho}\ .
\end{align*}
Here, we used the expression for the conditional min-entropy in \eqref{eq:cond-min-entropy-classical}. It follows that proving lower bounds for $\entHmin(X|EJ)$ is stronger and directly gives lower bounds for the average measurement entropy. For this reason, we use the conditional min-entropy $\entHmin(X|EJ)$ in place of the average entropy.

The following lemma shows that a QC-extractor directly satisfies an uncertainty relation for the smooth min-entropy. The idea is simple: if the outcome of the QC-extractor is $\e$-close to $\frac{\id_{A_1}}{d_{A_1}} \otimes \rho_E$, then by the definition of the smoothed entropies, the smooth min-entropy of the outcome has to be at least $\log d_{A_1}$.

\begin{lemma}\label{lem:minentropy}
Let $\rho_{AE}\in\states(AE)$, and $\left\{U_1, \dots, U_{t}\right\}$ be a set of unitaries on $A$  such that,
\begin{align}	
\label{eq:qc-extractor-cond}
\frac{1}{t}\sum_{j=1}^{t}\left\|\cT_{A\rightarrow A_{1}}(U_j \rho_{AE} U_j^{\dagger})-\frac{\id_{A_{1}}}{d_{A_1}}\otimes\rho_{E}\right\|_{1}\leq\eps(\rho)\ ,
\end{align}
for some $\eps(\rho)$ depending on the input state $\rho_{AE}$. Then
\begin{align*}
\hmin^{\sqrt{2\eps(\rho)}}(X|EJ)_{\rho} \geq \log d_{A_1}\ ,
\end{align*}
where $\rho_{XEJ}=\frac{1}{t}\sum_{j=1}^{t}\cM_{A\rightarrow X}( U_j \rho_{AE} U_j^{\dg}) \otimes\proj{j}_{J}$ and $\cM_{A \to X}$ is the measurement in the computational basis.
\end{lemma}
\begin{proof}
	By the definition of the smooth min-entropy and the inequality \eqref{eq:purifiedVStrace} between the purified and trace distance, condition \eqref{eq:qc-extractor-cond} directly translates into
	\[
	\entHmin^{\sqrt{2\e(\rho)} }(A_1|EJ)_{\rho} \geq \entHmin(A_1|EJ)_{\frac{\id_{A_{1}}}{d_{A_1}}\otimes\rho_{E}\otimes \frac{\id_J}{t} } = \log d_{A_1} \ ,
	\]
	where $\rho = \frac{1}{t} \sum_{j=1}^{t} \proj{j}_J \ox \cT_{A\rightarrow A_{1}}(U_j \rho_{AE} U_j^{\dagger})$. 	
	Recall that $\cT$ performs a measurement in the computational basis and then discards a (classical) system called $A_2$. Because we are only discarding a classical system, the min-entropy of the whole measurement outcome is at least the min-entropy in the register $A_1$~(see Lemma \ref{lem:monotonicity-minentropy-classical}). As a result,
	\[
	\entHmin^{\sqrt{2\e(\rho)} }(X|EJ)_{\rho} \geq \entHmin(A_1|EJ)_{\frac{\id_{A_{1}}}{d_{A_1}}\otimes\rho_{E}\otimes \frac{\id_J}{t} } = \log d_{A_1}.
	\]
\end{proof}

This allows us to translate all our constructions from Section \ref{sec:qc-ext-main} into a min-entropy uncertainty relation form. Note that conversely, we can convert a min-entropy uncertainty relation into a QC-extractor simply by applying a CC-extractor. We state below uncertainty relations for mutually unbiased bases and for ``single-qudit'' bases. 

\begin{corollary}
\label{cor:min-ent-ur}
Let $A$ be a Hilbert space such that $d_A$ is a prime power. Let $\{U_1, \dots, U_{d_A+1}\}$ be a full set of mutually unbiased bases. For any state $\rho_{AE}$, we have for all $\e > 0$ and $0 \leq \delta < \e^2/4$, 
\[
\hmin^{\eps}(X|EJ)_{\rho} \geq \log\left(d_A + 1\right)+\hmin^{\delta}(A|E)_{\rho}-\log\left(\frac{1}{\left(\eps^{2}/2-2\delta\right)^{2}}\right) - 1,
\]
where $\rho_{XEJ} = \frac{1}{d_A+1} \sum_{j} \cM_{A \to X}(U_j \rho U_j^{\dagger}) \otimes \proj{j}_J$.
\end{corollary}
\begin{proof}
Recall that the unitaries of the QC-extractor of Theorem \ref{thm:full-set-mub} are composed of mutually unbiased bases $U_1, \dots, U_{d_A+1}$ but also some permutations $P \in \cP$. Letting $\rho_{XEJ} = \frac{1}{d_A+1} \sum_{j} \cM_{A \to X}(U_j \rho U_j^{\dagger}) \otimes \proj{j}_J$, Lemma \ref{lem:minentropy} gives $\entHmin^{\e}(P(X)|EJP)_{\rho} \geq \log d_{A_1}$. But $P$ is a permutation that simply relabels the measurement outcomes, and thus does not change the entropy. It follows that
\[
\entHmin^{\e}(X|EJ)_{\rho} \geq \log d_{A_1}.
\]
Now it only remains to choose the dimension $d_{A_1}$. We pick
\[
d_{A_1} = \floor{ (\e' - 2 \delta)^2 \frac{d_A+1}{2^{-\entHmin^{\delta}(A|E)}} }.
\]
Plugging this value of $d_{A_1}$ in \eqref{eq:full-set-mub}, we get
\begin{align*}
 \frac{1}{|\cP|}  \frac{1}{d_A+1} \sum_{P \in \cP} \sum_{i=1}^{d_A+1} \left\| \cT_{A\rightarrow A_{1}}\left(PU_i \rho_{AE} \left(PU_i\right)^{\dagger}\right) - \frac{\id_{A_{1}}}{d_{A_1}} \ox \rho_E \right\|_1
\leq \e'.
\end{align*}
As a result, condition \eqref{eq:qc-extractor-cond} is satisfied with $\e(\rho) = \e'$. The desired result follows from the fact that $\log d_{A_1} \geq \log(1/2) + \log\left( (\e' - 2 \delta)^2 \frac{d_A+1}{2^{-\entHmin^{\delta}(A|E)}} \right)$.
\end{proof}

From the point of view of applications, the following entropic uncertainty relation for single-qudit measurements is probably the most interesting. It can be seen as a generalization to allow for quantum side information of uncertainty relations obtained by \cite{DFRSS07}. The proof is very similar to the proof of the previous corollary.

\begin{corollary}
\label{cor:min-ent-ur-bitwise}
Let $d \geq 2$ be a prime power. For any state $\rho_{AE}$, we have
\begin{align*}
\hmin^{\eps}(X|EJ)_{\rho} &\geq n\cdot\left(\log(d+1)-1\right) + \min\left\{0,\hmin^{\delta}(A|E)_{\rho}-\log\left(\frac{2}{\delta'^2} +\frac{1}{1-2\delta}\right)\right\} \\
&-\log\left(\frac{1}{\left(\eps^{2}/2-2(\delta+\delta')\right)^2}\right)-2,
\end{align*}
where $\rho_{XEJ} = \frac{1}{(d+1)^n} \sum_{j} \cM_{A \to X}( V_j \rho V_j^{\dagger} ) \otimes \proj{j}_J$ and $\{V_j\}_j = \cV_{d,n}$ as defined in Theorem \ref{thm:singleQuditExtract}.
\end{corollary}
\begin{proof}
We choose the dimension $d^{\xi n}$ of the $A_1$ system of Theorem \ref{thm:singleQuditExtract} to be
\[
d_{A_1} = \floor{\left(\e' - 2(\delta+\delta') \right)^2 \frac{ 2^{(\log(d+1) - 1)n} }{1 + 2^{-\entHmin^{\delta}(A|E)_{\rho} + z} }  }.
\]
We also compute
\begin{align*}
&\log d_{A_1} \\
&\geq -1 -\log\left(\frac{1}{\left(\eps'-2(\delta+\delta')\right)^2}\right) + n \cdot (\log (d+1) - 1) - \log \left( 1 + 2^{-\entHmin^{\delta}(A|E)_{\rho} + z} \right)  \\
		&= n\cdot\left(\log(d+1)-1\right) + \min\left\{0,\hmin^{\delta}(A|E)_{\rho}-z\right\} -\log\left(\frac{1}{\left(\eps'-2(\delta+\delta')\right)^2}\right)-2.
\end{align*}
Setting $\sqrt{2\e'} = \e$, we achieve the desired result.
\end{proof}

\subsection{Uncertainty relations for the von Neumann entropy}

Uncertainty relations for the conditional von Neumann entropy can also be obtained as in Proposition \ref{prop:metric-to-entropic}.

\begin{lemma}\label{lem:shannon}
Let $\rho_{AE}\in\states(AE)$, and $\left\{U_1, \dots, U_{t}\right\}$ be a set of unitaries on $A$  such that,
\begin{align*}	
\frac{1}{t}\sum_{j=1}^{t}\left\|\cT_{A\rightarrow A_{1}}(U_j \rho_{AE} U_j^{\dagger})-\frac{\id_{A_{1}}}{d_{A_1}}\otimes\rho_{E}\right\|_{1}\leq\eps(\rho)\ ,
\end{align*}
for some $\eps(\rho)$ depending on the input state $\rho_{AE}$. Then
\begin{align*}
		\frac{1}{t} \sum_{j=1}^{t} \entH(X|E)_{\rho^{j}}= \entH(X|EJ)_{\rho} \geq (1-4\eps(\rho))\log d_{A_1}-2\binent(\eps(\rho))\ ,
	\end{align*}
	where $\rho^{j}=\meas_{A\rightarrow X}(U_j \rho_{AE} U_j^{\dagger})$ and $\rho_{XEJ}=\frac{1}{t}\sum_{j=1}^{t}\rho^{j}\otimes\proj{j}$.
\end{lemma}

\begin{proof}
The argument is the same as the proof of Lemma \ref{lem:minentropy}, except that instead of just obtaining a bound on the smooth entropy, we use the Alicki-Fannes inequality (Lemma~\ref{lem:alicki-fannes}).
\end{proof}

This lemma can naturally be applied directly to all the constructions of QC-extractors. For example, for a full set of mutually unbiased bases, by choosing $d_{A_1} = \floor{ (\e - 2 \delta)^2 (d_A+1) 2^{\entHmin^{\delta}(A|E)} }$, we can get
\begin{align*}
	\frac{1}{t} & \sum_j  \entH(X|E)_{\rho^{j}} \\
	&\geq (1-4\eps)\left(\log\left(d_A + 1\right) + \hmin^{\delta}(A|E)_{\rho} - \log\left(\frac{1}{(\eps-2\delta)^2}\right)-1\right) - 2\binent(\eps).
\end{align*}
Using the asymptotic equipartition property for the smooth min-entropy, we can obtain an uncertainty relation only in terms of von Neumann entropies.

\begin{proposition}\label{thm:neumann}
Let $d \geq 2$ be a prime power, and $\left\{V_0,V_1,\dots,V_d\right\}$ define a complete set of MUBs of $\CC^d$. Consider the set of measurements on the $n$ qudit space $A$ defined by the unitary transformations $\left\{V=V_{u_1}\ox\cdots\ox V_{u_n}|u_i\in\left\{0,\dots,d\right\}\right\}$ that we index by numbers from $1$ to $(d+1)^n$ as 
$\{V_j\}_{j \in \{1, \dots, (d+1)^n\}}$. Then for all $\rho_{AE} \in \cS(AE)$, we have
\begin{align*}
\frac{1}{(d+1)^n}\sum_{j=1}^{(d+1)^n} \entH(X|E)_{\rho^{j}}\geq n\cdot\left(\log(d+1)-1\right)+\min\left\{0,\entH(A|E)_{\rho}\right\}\ ,
\end{align*}
where $\rho^j = \cM_{A \to X}(V_j \rho V_j^{\dagger})$.
\end{proposition}
\begin{proof}
Using the QC-extractor for the single-qudit MUB of Theorem \ref{thm:singleQuditExtract} with
\[
d_{A_1} = \floor{\left(\e - 2\delta \right)^2 \frac{ 2^{(\log(d+1) - 1)n} }{1 + 2^{-\entHmin^{\delta}(A|E)_{\rho|\rho}} }  } \ ,
\]
we get 
\begin{align}
	&\frac{1}{(d+1)^n}\sum_{j=1}^{(d+1)^n} \entH(X|E)_{\rho^j} \notag \\
	&\geq(1-4\eps)\left(n\left(\log(d+1)-1\right)-\log\left(1+2^{-\hmin^{\delta}(A|E)_{\rho|\rho}}\right)-\log\left(\frac{1}{(\eps-2\delta)^{2}}\right) - 1\right) \notag \\
	&\qquad -2\binent(\eps) \notag \\
	&\geq(1-4\eps)\left(n\left(\log(d+1)-1\right)+\min\left\{0,\hmin^{\delta}(A|E)_{\rho|\rho}\right\}-2-\log\left(\frac{1}{(\eps-2\delta)^{2}}\right)\right) \notag \\
	&\qquad -2\binent(\eps) \label{eq:aepqudit}\ .
\end{align}
Here, we use a version with $\hmin^{\delta}(A|E)_{\rho|\rho}$ instead of $\hmin^{\delta}(A|E)_{\rho}$ that is in the statement of Theorem~\ref{thm:singleQuditExtract}. The expression with $\hmin^{\delta}(A|E)_{\rho|\rho}$ is however easily obtained by looking at the proof. Evaluating equation~\eqref{eq:aepqudit} on the $m$-fold tensor product of the original input system $d^{n}$, and multiplying both sides with $1/m$, we obtain
\begin{align*}
&\frac{1}{(d+1)^n}\sum_{j=1}^{(d+1)^n} \entH(X|E)_{\rho^j} \\
&\geq(1-4\eps)\left(n\left(\log(d+1)-1\right)+\min\left\{0,\frac{1}{m}\hmin^{\delta}(A|E)_{\rho^{\otimes m}|\rho^{\otimes m}}\right\}\right)\\
&-\frac{1-4\eps}{m}\left(2+\log\left(\frac{1}{(\eps-2\delta)^{2}}\right)\right)-\frac{2\binent(\eps)}{m}\\
&\geq(1-4\eps)\left(n\left(\log(d+1)-1\right)+\min\left\{0,\entH(A|E)_{\rho}-\frac{4\sqrt{1-2\log\delta}\left(2+\frac{n}{2}\right)}{\sqrt{m}}\right\}\right)\\
&-\frac{1-4\eps}{m}\left(2+\log\left(\frac{1}{(\eps-2\delta)^{2}}\right)\right)-\frac{2\binent(\eps)}{m}\ .
\end{align*}
Here we used the fully quantum asymptotic equipartition property for the smooth conditional min-entropy (Lemma~\ref{lem:aep}). By first letting $m\rightarrow\infty$ and then $\eps\rightarrow0$, we obtain the desired result.
\end{proof}

Note that for $n=1$, this again gives an uncertainty relation for the full set of MUBs only in terms of von Neumann entropies
\begin{align}\label{eq:finalvonneumman2}
\frac{1}{d+1}\sum_{j=1}^{d+1}\entH(X|E)_{\rho^{j}}\geq \log(d+1)-1+\min\left\{0,\entH(A|E)_{\rho}\right\}\ .
\end{align}
In the special case when $E$ is trivial, we arrive at
\begin{align}\label{eq:sanchez_H2}
\frac{1}{d+1}\sum_{j=1}^{d+1}\entH(X)_{\rho^{j}}\geq \log(d+1)-1\ ,
\end{align}
which is the best known bound for a full set of MUBs and general $d$~\citep{Lar90, Iva92,San93}. But without side information and when $d$ is even, this was improved by~\cite{Sanchez95} to
\begin{align}
\frac{1}{d+1}\sum_{j=1}^{d+1}\entH(X)_{\rho^{j}}\geq \frac{1}{d+1} \left(\frac{d}{2}\log\left(\frac{d}{2}\right)+\left(\frac{d}{2}+1\right)\log\left(\frac{d}{2}+1\right)\right)\ .
\end{align}
For one qubit ($d=2$) the latter gives $2/3$ (which is best possible for three measurements), whereas our bound gives $\log3-1\approx 0.585$. 


%


\section{Applications to security in the noisy-storage model}\label{sec:noisy}

We use the min-entropy uncertainty relation of Corollary \ref{cor:min-ent-ur-bitwise} to prove the security of secure function evaluation in the noisy storage model.


\subsection{Introduction}
Consider two mutually distrustful parties Alice and Bob who want to collaborate to perform a distributed computation in a secure fashion. Typically, Alice holds $x$ and Bob holds $y$, and they both want to figure out $f(x,y)$ in such a way that each party does not learn too much about the other party's input. Unfortunately, if we are looking for information theoretic security, it turns out that even quantum communication does not allow us to solve general two-party secure function evaluation~\citep{lo:insecurity}. For example, it is known that only weak variants of (information theoretically secure) bit commitment are possible; see Section~\ref{sec:string-commitment} and~\citep{LC97, May97,DKSW07}. 


The natural question then is under which assumptions can we obtain secure protocols for two-party computations.
Classically, these assumptions typically limit the computational power of a party. One then assumes that a particular problem requires a lot of computational resources to solve in some precise complexity theoretic sense, and then one  proves using this assumption that a cheating strategy needs more computational resources than what is available. It goes without saying that the computational assumptions are almost always not proven. As computation is such a complicated notion to understand, a natural question is then whether one can make simpler assumptions on the devices of the parties.
 

Classically, it is possible to obtain security when we are willing to assume that the adversary's memory is limited in size~\citep{Maurer92b,cachin:bounded}. But unfortunately, \cite{maurer:imposs} showed that \emph{any} classical protocol in which the honest players need to store $n$ classical bits to execute can be broken by an adversary who can store $O(n^2)$ bits.


Motivated by this unsatisfactory gap, it was thus suggested to assume that the attacker's \emph{quantum} storage was bounded~\citep{DFSS05, DFRSS07}. The central assumption in this model is that during waiting times $\Delta t$ introduced in the protocol, the adversary can only store a limited number of qubits $N$. 
This is the only assumption on the adversary, who is otherwise all powerful. In particular, he can store an unlimited amount of classical information, and perform any operation instantaneously. The latter implies that he is able to perform any encoding and decoding operation before and after using his memory device. \cite{KWW09} based on \cite{DFSS05,DFRSS07} constructed a protocol for bit commitment using BB84 encoded qubits that is secure whenever Bob is only allowed to store $N$ qubits while Alice and Bob exchange more than (roughly) $2N$ qubits during the protocol.

A natural question then is to characterize the property of Bob's storage device that allows him and Alice to implement secure two-party function evaluation. The \emph{noisy-storage model} introduced by \cite{WST08, STW08} is a generalization of the bounded storage model. As in the bounded storage model, during waiting times $\Delta t$, the adversary can only keep quantum information in his quantum storage device $\cF$. Mathematically, such a quantum storage device is simply a quantum channel $\cF:\states(\hil_{\rm in}) \rightarrow \states(\hil_{\rm out})$ mapping input states on the space $\hil_{\rm in}$ to some noisy output states on the space $\hil_{\rm out}$. An example of a storage device would be $N$ $d$-dimensional identical memory cells, so that $\cF$ takes the form $\cF = \cN^{\otimes N}$.
In particular, in the bounded quantum storage model, the adversary is only allowed to store $N$ qubits, which means $\cF = \qubitchannel^{\otimes N}$~\citep{DFSS05,DFRSS07}. The kind of statement one proves in this framework is of the following form: Provided $\cF$ cannot be used to reliably transmit $n$ bits or qubits of information, the protocol $\cP_n$ is secure. We describe precise versions of this statement in the following sections.



\subsection{The noisy storage model}

\subsubsection{Weak string erasure}

\cite{KWW09} showed that bit commitment and oblivious transfer,\footnote{Oblivious transfer is an important primitive that was shown to be complete for two-party computation by \cite{Kil88}. The exact definition is not important here.} and hence any two-party secure 
computation, can be implemented securely against an \emph{all-powerful} quantum adversary given access to a simple primitive called \emph{weak string erasure (WSE)}. Hence, it suffices to construct a protocol for WSE that is secure under the assumption that the storage devices of the parties are noisy, and we will follow that approach here. 

The motivation behind the weak string erasure primitive is to create a basic \emph{quantum} protocol that builds up \emph{classical} correlations between Alice and Bob which are 
later used to implement more interesting cryptographic primitives. Informally, weak string erasure achieves the following task. WSE takes no inputs from Alice and Bob. Alice receives as output a randomly chosen string $X^n = X_1,\ldots,X_n \in \{0,1\}^n$. Bob
receives a randomly chosen subset $\cI \subseteq [n]$ and the substring $X_{\cI}$ of $X^n$ corresponding to the bits  in positions indexed by $\cI$. For each $i \in [n]$, we decide independently to put $i$ in the set $\cI$ with probability $p$. Originally, $p=1/2$~\citep{KWW09}, but any probability $0 < p <1$ allows for the implementation of oblivious 
transfer~\citep{prabha:limits}. The security requirements of weak string erasure are that Alice does not learn $\cI$, and Bob's min-entropy given all of his information $B$ is 
bounded as $\hmin(X|B) \geq \lambda n$ for some parameter $\lambda > 0$. To summarize all relevant parameters, we thereby speak of an $(n,\lambda,\eps,p)$-WSE scheme. 



The precise requirement of security is stated in terms of (approximate) indistinguishability between the states obtained in an execution of the real protocol and some ideal states. We should highlight that the notion of distance we use here is the trace distance, which is more relevant than the purified distance in the context of cryptography because of its interpretation it terms of distinguishing probability~\citep{Hel67}.
It will be convenient to express the distribution of the random subset $\cI$ by a density operator:
\begin{align}
\label{eq:probDist}
	\Psi(p)= \sum_{\cI \subseteq 2^{[n]}} p^{|\cI|} (1-p)^{n-|\cI|} \proj{\cI}\ .
\end{align}

\begin{definition}[\textbf{Non-uniform WSE}]\label{def:wse}
An $(n,\lambda, \varepsilon,p)$-weak string erasure scheme is a protocol between A and B satisfying the following properties:

\textbf{Correctness:} If both parties are honest, then there exists an ideal state $\sigma_{X^{n}\mathcal{I}X_{\mathcal{I}}}$ such that
\begin{enumerate}
\item The joint distribution of the $n$-bit string $X^{n}$ and subset $\mathcal{I}$ is given by
\begin{equation}\label{eq:wsecorrect}
\sigma_{X^{n}\cI} = \frac{\id_{X^{n}}}{2^n} \otimes \Psi(p)\ ,
\end{equation}
\item The joint state $\rho_{AB}$ created by the real protocol is equal to the ideal state: $\rho_{AB} = \sigma_{X^{n}\cI X_{\cI}}$ where we identify $(A,B)$ with $(X^{n},\cI X_{\cI})$.
\end{enumerate}

\textbf{Security for Alice:} If A is honest, then there exists an ideal state $\sigma_{X^{n}B'}$ such that 
\begin{enumerate}
\item The amount of information $B'$ gives Bob about $X^{n}$ is limited:
\begin{equation}\label{eq:honestA}
\frac{1}{n}\hmin(X^{n}|B')_{\sigma} \geq \lambda
\end{equation}
\item The joint state $\rho_{AB'}$ created by the real protocol is $\eps$-close to the ideal state in trace distance, where we identify $(X^{n},B')$ with $(A,B')$.
\end{enumerate}

\textbf{Security for Bob:} If B is honest, then there exists an ideal state $\sigma_{A'\hat{X}^{n}\cI}$ where $\hat{X}^{n} \in \{0,1\}^{n}$ and $\cI \subseteq [n]$ such that
\begin{enumerate}
	\item The random variable $\cI$ is independent of $A'\hat{X}^{n}$ and distributed over $2^{[n]}$ according to the probability distribution given by~\eqref{eq:probDist}:
\begin{equation}\label{eq:honestB}
\sigma_{A'\hat{X}^{n}\cI} = \sigma_{A'\hat{X}^{n}} \otimes \Psi(p)\ .
\end{equation}
\item The joint state $\rho_{A'B}$ created by the real protocol is equal to the ideal state: $\rho_{A'B} = \sigma_{A'(\cI\hat{X}_{\cI})}$, where we identify $(A',B)$ with $(A',\cI\hat{X}_{\cI})$.
\end{enumerate}
\end{definition}
Note that any positive $\lambda$ allows one to build a protocol for bit commitment and oblivious transfer but of course, larger values of $\lambda$ naturally lead to better parameters. To give an example, \cite{prabha:limits} prove that using a $(n, \lambda, \eps, 1/3)$-WSE, one can obtain an 1-2 oblivious transfer of strings of length about $\lambda/24 \cdot n$.

\subsubsection{Protocol for weak string erasure}

\begin{center}
	\begin{figure}[h]
	\begin{center}
		\includegraphics[scale=0.7]{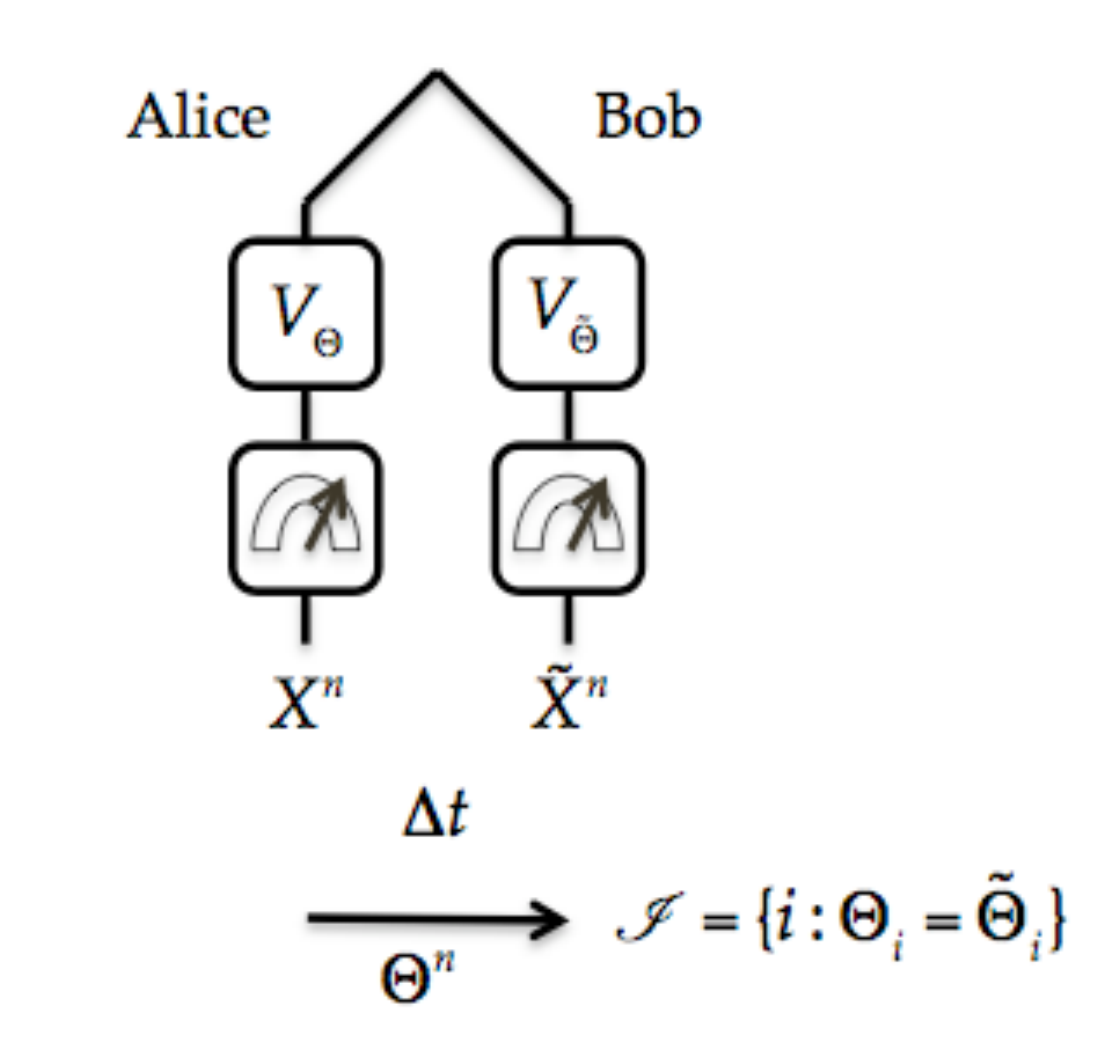}
		\caption{Illustration of the protocol for weak string erasure}
		\label{fig:wse-protocol}
	\end{center}
	\end{figure}
\end{center}

We now construct a very simple protocol for weak string erasure, and prove its security using our bitwise QC-randomness extractor. 
The only difference to the protocol proposed in~\cite{KWW09} is that we will use three MUBs per qubit instead of only two.
For sake of argument, we state the protocol in a purified form where Alice generates the EPR-pairs and later measures them. Note, however, that the protocol
is entirely equivalent to Alice creating single qubits and sending them directly to Bob. That is, honest Alice and Bob do not need any quantum memory to implement the protocol below. In the purified protocol, the choice of bit she encodes is determined randomly 
by her measurement outcome in the chosen basis on the EPR-pair. 
The protocol is illustrated in Figure \ref{fig:wse-protocol}.

\begin{protocol}{Weak string erasure (WSE)}{Outputs: $x^n \in \01^n$
	to Alice, $(\cI,z^{|\cI|}) \in 2^{[n]} \times \01^{|\cI|}$ to Bob.}{}
\item[1.] {\bf Alice:} Creates $n$ EPR-pairs $\Phi$, and sends half of each pair to Bob.
\item[2.] {\bf Alice:} 
	Chooses a bases-specifying string $\theta^n \in_R \{0,1,2\}^n$ uniformly at random. 
	For all $i$, she measures the $i$-th qubit in the basis $\theta_i$ to obtain outcome $x_i$.

\item[3.] {\bf Bob: } Chooses a basis string $\tilde{\theta}^n \in_R \{0,1,2\}^n$ uniformly at random. When receiving the $i$-th qubit, Bob measures it in the basis
given by $\tilde{\theta}_i$ to obtain outcome $\tilde{x}_i$.

\item[Both parties wait time $\Delta t$.]

\item[4.] {\bf Alice: } Sends the basis information $\theta^n$ to Bob, and outputs $x^n$.
\item[5.] {\bf Bob: } Computes $\cI = \{i \in [n] \mid \theta_i = \tilde{\theta}_i\}$, and outputs $(\cI,z^{|\cI|}):=(\cI,\tilde{x}_{\cI})$.
\end{protocol}

\begin{center}
	\begin{figure}
	\begin{center}
		\includegraphics[scale=0.7]{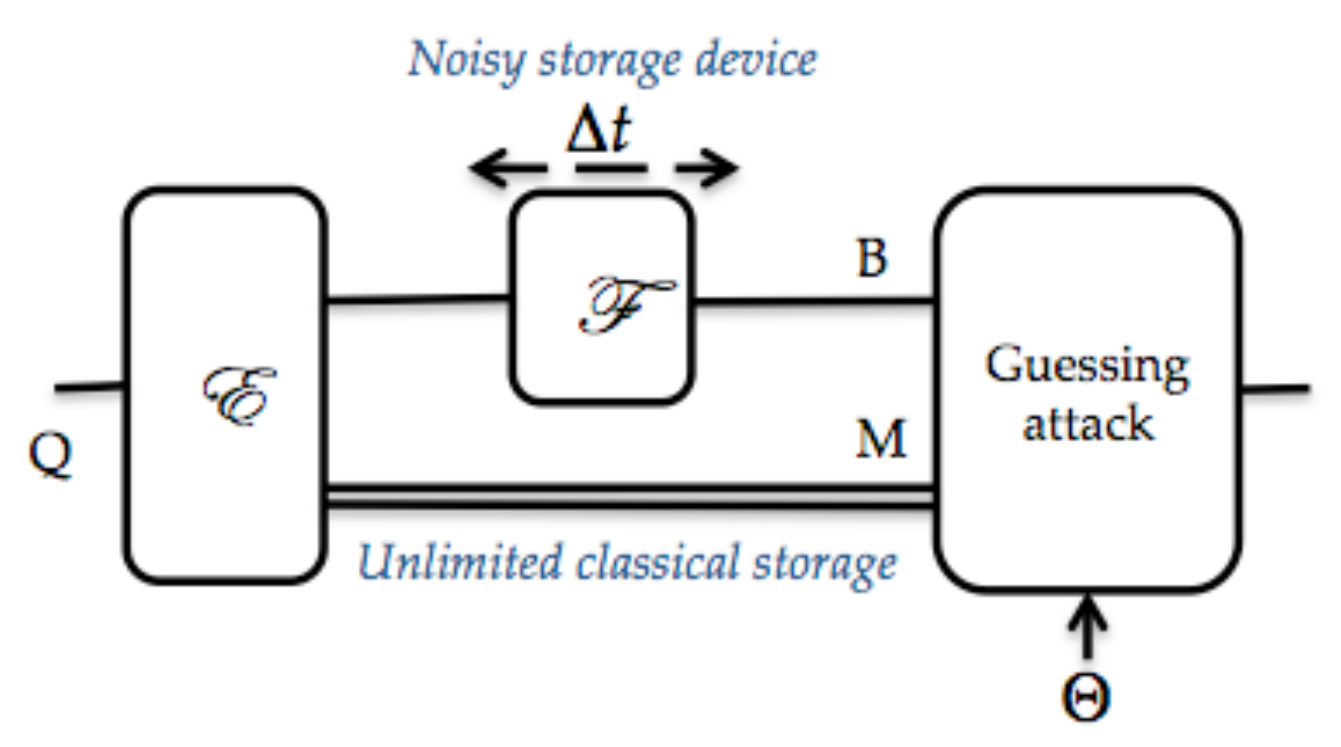}
		\caption{Any attack of dishonest Bob is described by an encoding attack $\cE$ and a guessing attack, since for classical $X$ the min-entropy $\hmin(X|BM\Theta)$ is
		directly related to the probability that Bob guesses $X$. As we will see below, it is however sufficient to consider how well a decoding attack $\cD$ can preserve entanglement
		between Alice and Bob, where $\cD$ acts on $BM$ on the state $\rho_{ABM}$ from equation~\eqref{eq:rho}.}
		\label{fig:BobAttack}
	\end{center}
	\end{figure}
\end{center}

The proof of correctness of the protocol, and security against dishonest Alice is identical to~\cite{KWW09,prabha:limits}. It essentially follows from the fact that
Bob never sends any information to Alice. The main difficulty lies in 
proving security against dishonest Bob. Before embarking on a formal proof, let us first consider the general form that any attack of Bob takes (see Figure~\ref{fig:BobAttack}).
First of all, note that the noisy-storage model only assumes that Bob has to use his storage device during waiting times $\Delta t$. 
Let $Q$ denote Bob's quantum register containing all $n$ qubits that he receives. 
Note that since there is no communication between Alice and Bob during the transmission of these $n$ qubits, we can without loss of generality 
assume that Bob first waits for all $n$ qubits to arrive before mounting any form of attack.

As any operation in quantum theory is a quantum channel, Bob's attack can be described by a quantum channel $\cE: \states(Q) \rightarrow \states(\hin \otimes \msg)$. This map takes $Q$, to some quantum state on the input of Bob's storage device ($\hin$), and some arbitrarily large amount of classical information ($\msg$). For example, $\cE$ could be an encoding into an error-correcting code. By assumption of 
the noisy-storage model, Bob's quantum memory is then affected by noise $\cF: \states(\hin) \rightarrow \states(\hout)$.
After the waiting time, the joint state held by Alice and Bob in the purified version of the protocol, i.e., before Alice measures, is thus of the form
\begin{align}\label{eq:rho}
	\rho_{ABM} = \id_A \otimes \left[\left(\cF \otimes \clchannel_{\msg}\right) \circ \cE\right](\Phi^{\otimes n})\ ,
\end{align}
where $\Phi$ is an EPR-pair.
After the waiting time, Bob can perform any form of quantum operation to try and recover information about $X$ from the storage device. 


\subsection{Security and the quantum capacity}\label{sec:securityCap}

Recall from the definition above that our goal is to show that $\hmin^\eps(X|BM\Theta)_{{\rho}} \geq \lambda \cdot n$ for some parameter $\lambda > 0$. Although it was always clear that security should be related to the channel's ability to store quantum information, i.e., the \emph{quantum capacity} of $\cF$, proving this fact has been a challenge for several years. Partial progress to answering this question was made in~\cite{KWW09} and~\cite{entCost}, where security was linked to the \emph{classical capacity} and \emph{entanglement cost}
of $\cF$, respectively. We informally state these results. Both of them prove security of the protocol described above except that two mutually unbiased bases are chosen instead of three. \cite{KWW09} prove the protocol is secure whenever for some $R < 1/2$, the channel $\cF$ is such that any attempt to transmit $nR$ \emph{classical bits} across $\cF$ is bound to fail with probability exponentially close to $1$. \cite{entCost} consider channels of the form $\cF = \cN^{\otimes N}$ and they show that the protocol is secure provided $\frac{N \cdot E_C(\cN)}{n}$ is bounded away from $1/2$, where $E_C(\cN)$ is the entanglement cost of the channel $\cN$. The entanglement cost is the amount of entanglement needed to simulate the channel $\cN$ when classical communication is given for free. $E_C(\cN)$ is a measure of how good the channel is for sending quantum information but it is in general larger than the quantum capacity.

Note that our objective is to make a statement about some \emph{classical} information $X$ obtained by measuring $A$ in a randomly selected basis $\Theta$.
That is, we effectively ask for an uncertainty relation for these measurements. Previously, however, suitable uncertainty relations were only known for \emph{classical} side information. 
The missing ingredient was an uncertainty relation with \emph{quantum} side information, linked to the channel's ability to preserve quantum information. Here is where our uncertainty relation of Corollary \ref{cor:min-ent-ur-bitwise} comes in.

To state the result, we first define the notion of \emph{channel fidelity} introduced by \cite{Barnum00} which is perhaps the most widely used quantity to measure how good a channel is at sending quantum information. For a channel $\cN : \cS(Q) \to \cS(Q')$, the channel fidelity $F_c$ quantifies how well $\cN$ preserves entanglement with a reference:
\begin{equation}
\label{eq:channel-fidelity}
F_c(\cN) = F( \Phi_{QA}, \left[\cN \otimes \id_{A} \right] (\Phi_{Q'A}) ),
\end{equation}
where $\Phi_{QA}$ is a maximally entangled state.
For example, one way of defining the (one-shot) quantum capacity with free classical forward communication of a channel $\cF$ is by the maximum of $\log d_A$ over all encodings $\cE : \cS(Q) \to \cS(\hin \times M)$ and decodings $\cD : \cS( B \otimes M) \to \cS(Q')$ such that $F_c(\cD \circ (\cF \otimes \clchannel_M) \circ \cE) \geq 1-\e$ for small enough $\e$. Here $\clchannel_M$ refers to a noiseless classical channel.

\begin{theorem}\label{thm:para}
	Let Bob's storage device be given by $\cF : \cS(\hin) \to \cS(B)$. Let $\e \in (0,1)$, $\kappa = 8 \log(4/\e)$, $\lambda \leq \log 3 - 1$. Assume that we have
	\begin{equation}
	\label{eq:fid-condition}
	\max_{\cD,\cE} F_c(\cD \circ (\cF \otimes \clchannel_{\msg}) \circ \cE) \leq 2^{-(2 - \log 3 + \lambda ) n - \kappa}
	\end{equation}
	where the maximum is over all quantum channels $\cE : \cS\left((\CC^2)^{\otimes n}\right) \to \cS(\hin \otimes M)$ and $\cD:\cS(B \otimes M) \to \cS((\CC^2)^{\otimes n})$.
	
	Then, Protocol 1 implements a $(n,\lambda,\eps,1/3)$-WSE.
\end{theorem}
\begin{proof}
	The proof of correctness of the protocol, and security against dishonest Alice is identical to~\cite{KWW09,prabha:limits} and does not lead to
	any error terms. 

Using the uncertainty relation of Corollary \ref{cor:min-ent-ur-bitwise}, with $E=BM \Theta$ on ${\rho}_{ABM\Theta}$ we get
\begin{align}\label{eq:minBound}
	\entHmin^{\e}(X|BM\Theta)_{{\rho}} \gtrsim (\log(3)-1) n + \min\{0,\hmin(A|BM\Theta)_{{\rho}}\}\ .
\end{align}
Note that because $\Theta$ is independent of $ABM$, we have $\hmin(A|BM\Theta)_{\rho} = \entHmin(A|BM)$. To place a bound on~\eqref{eq:minBound}, we would like to obtain a lower bound on
\begin{align*}
	\min_{\cE} \hmin(A|BM)_{{\rho}}\ ,
\end{align*}
where the minimization is taken over all encoding attacks as described above. 
We will use condition \eqref{eq:fid-condition} to obtain such a lower bound.
We now use an operational interpretation of the conditional min-entropy due to~\cite{krs:operational}:
\begin{align}\label{eq:minOp}
	\hmin(A|BM)_{{\rho}} = - \log d_A \max_{\Lambda_{BM\rightarrow A'}} F(\Phi_{AA'},\id_A \otimes \Lambda(\rho_{ABM}))\ ,
\end{align}
where $\Phi_{AA'}$ is the maximally entangled state accross $AA'$. That is, the min-entropy is directly related to the ``amount'' of entanglement between $A$ and $E=BM$. 
The map $\Lambda$ in~\eqref{eq:minOp} can be understood as a decoding attack $\cD$ aiming to restore entanglement with Alice.
Further, note that $|A'| = |Q|$ and we can equivalently upper bound
\begin{align}\label{eq:chanFid}
	\max_{\cD,\cE} F\left(\Phi_{AB},\id_A \otimes \left[\cD \circ (\cF \otimes \clchannel_{\msg}) \circ \cE\right](\Phi_{AQ})\right) = 
	\max_{\cD,\cE} F_c(\cD \circ (\cF \otimes \clchannel_{\msg}) \circ \cE)\ .
\end{align}

	By the assumption on the storage device $\cF$, we obtain that for any encoding $\cE$ and decoding $\cD$ attack of Bob
	\begin{align*}
		\hmin(A|BM)_{\rho}  &\geq - \log 2^n F_c(\cD \circ (\cF \otimes \clchannel_{\msg}) \circ \cE)\\
						&\geq - \left(n - (2 - \log 3) n - \lambda n - \kappa \right) \\
						&= - (\log 3 - 1) n + \lambda n + \kappa. 
	\end{align*}
	Then, using the uncertainty relation for 3 MUBs per qubit of Corollary~\ref{cor:min-ent-ur-bitwise} (with $\delta=0$ and $\delta' = \e^2/8$), we get
	\begin{align*}
		\entHmin^{\eps}(X|BM\Theta)_{{\rho}}
		\geq \lambda n - \log\left(2 \cdot 64/\e^4 + 1 \right) - \log (16/\e^4) - 2 + 8 \log(4/\e) \geq \lambda n .
	\end{align*}
\end{proof}
Note that ideally, we would want a statement of the form: if
\begin{equation}
\label{eq:better-fid-condition}
\max_{\cD,\cE} F_c(\cD \circ (\cF \otimes \clchannel_{\msg}) \circ \cE) \lesssim 2^{-\lambda n} \ ,
\end{equation}
then the Protocol 1 implements $(n, \lambda, \e, 1/3)$-WSE. Unfortunately, we have a stronger constraint in equation \eqref{eq:fid-condition} with an additional positive factor of $2 - \log 3$. If we wanted to prove security with the condition \eqref{eq:better-fid-condition}, we would need to prove a stronger uncertainty relation than in Corollary \ref{cor:min-ent-ur-bitwise}. In particular, observe that if $\entHmin(A|E) \leq - (\log(3) - 1) n$, our uncertainty relation does not give any useful bound. It would be very interesting to improve it so that we can get a non-trivial lower bound for any $\entHmin(A|E) \geq -cn$ with $c < 1$. Note that, as in \citep{prabha:limits}, we can get arbitrarily close to proving security under a condition of the form \eqref{eq:better-fid-condition} by using higher dimensional encodings (i.e., using the uncertainty relation of Corollary \ref{cor:min-ent-ur-bitwise} with larger values of $d$), but it becomes hard to implement the protocol with current technology.

\subsubsection{Example: bounded storage}\label{sec:why}
We look at the simple case where $\cF = \id^{\otimes N}$, known as the bounded storage model. 
In this case, it is simple to prove that if you try to send more than $N$ qubits of information using $\cF$, the channel fidelity will decrease exponentially in $n$~\citep{entCost}. 

\begin{lemma}
	For $n \geq N$, we have
	\[
	\max_{\cD,\cE} F_c(\cD \circ (\id_2^{\otimes N} \otimes \clchannel_{\msg}) \circ \cE) \leq 2^{-n + N},
	\]
	where the maximum is over all quantum channels $\cE : \cS\left((\CC^2)^{\otimes n}\right) \to \cS((\CC^2)^{\otimes N} \otimes M)$ and $\cD:\cS((\CC^2)^{\otimes N}  \otimes M) \to \cS((\CC^2)^{\otimes n})$
\end{lemma}

\begin{proof}
Consider a decomposition of the encoding and decoding map in terms of their Kraus operators as $\cE(\rho) = \sum_j E_j \rho E_j^\dagger$ and
$\cD(\rho) = \sum_{k,m} \hat{D}_{k,m} \rho \hat{D}_{k,m}^\dagger$ where $\hat{D}_{k,m} = D_{k,m} \otimes \proj{m}$. Note 
that without loss of generality, the latter has this form since it is processing classical forward communication on $\msg$.
Let $\Pi_{k,m}$ denote the projector onto the subspace that $\hat{D}_{k,m}$ maps to.
We can now bound
\begin{align*}
	&F_c(\cD \circ (\qubitchannel^{\otimes N} \otimes \clchannel_{\msg}) \circ \cE) \\
	&= \frac{1}{2^n} \sum_{i} \bra{i i} \left( \sum_{\ell, \ell'} \sum_{jkm} \hat{D}_{k,m} E_j \ket{ \ell } \bra{ \ell' } E_j^{\dagger} \hat{D}^{\dagger}_{k,m} \otimes \ket{ \ell } \bra{ \ell'} \right) \sum_{i} \ket{ i i} \\ 
	&= \frac{1}{2^n} \left( \sum_{\ell, \ell'} \sum_{jkm} \bra{\ell} \hat{D}_{k,m} E_j \ket{ \ell } \bra{ \ell' } E_j^{\dagger} \hat{D}^{\dagger}_{k,m} \ket{\ell'} \right) \\	
	&= \sum_{jkm} \left|\tr\left[\hat{D}_{k,m}E_j\left(\frac{\id}{2^{n/2}}\right) \cdot \left(\frac{\id}{2^{n/2}}\right) \Pi_{m,k}\right]\right|^2\\
	&\leq \sum_{jkm} \tr\left[\hat{D}_{k,m}E_j \left(\frac{\id}{2^{n}}\right) E_j^\dagger \hat{D}_{k,m}^\dagger\right] \tr\left[\Pi_{k,m}\left(\frac{\id}{2^{n}}\right)\right]\\
	&\leq 2^{-n + N} \tr\left[\cD\circ\cE\left(\frac{\id}{2^{n}}\right)\right]\\
	&= 2^{-n + N}\ ,
\end{align*}
where in the third equality, we used the cyclicity of the trace and fact that $\Pi_{m,k} \hat{D}_{k,m} = \hat{D}_{k,m}$. We used the Cauchy-Schwarz inequality for the first inequality, and the fact that $\tr[\Pi_{k,m}] = \rank[\hat{D}_{k,m}] = \rank[D_{k,m} \otimes \proj{m}] \leq 2^N$. For the last equality, we used the fact that $\cD$ and $\cE$ are trace preserving.
\end{proof}

%

It then follows from Theorem \ref{thm:para} that if $N/n$ is bounded away from $\log 3 - 1$, then the described protocol is secure. We note that the parameters obtained here are slightly worse than what was obtained in~\cite{prabha:limits}, where security was shown to be possible for $N/n$ bounded away from $2/3$ instead of $0.585$. This is due to the fact that the lower bound $0.585$ in our uncertainty relation stems from an expression involving the collision entropy rather than the Shannon entropy. We emphasize however, that due to finite size effects our bound is still better in the practically relevant regime of $n \lesssim10^{6}$ (for the same security parameters).

\section{Concluding remarks}
In this chapter, we considered uncertainty relations that take into account an adversary that is potentially entangled with the system being measured. As in Chapter \ref{chap:uncertainty-relations}, the measure of uncertainty we used is the distance to the uniform distribution. But in addition, we also asked for the joint state of the outcome together with the system of the adversary to be close to a product state. The advantage of this measure is that we were able to apply techniques similar to the decoupling theorem. We first use a H\"{o}lder type inequality to work with the $\ell_2$-norm, which is much easier to handle. Then, we use symmetry properties of the unitaries to obtain bounds on these norms. This allowed us to analyse several constructions of basis, but as we saw in Proposition \ref{prop:2norm-ext}, this technique cannot be used to show uncertainty relations for small sets of bases. Handling quantum side information using the $\ell_1$ norm directly seems like a difficult technical challenge. In the context of CC-extractors, there are constructions that have a small seed \citep{KT08, TS09, DPVR09}. It would be interesting to use these ideas to construct QQ or QC-extractors with small seed.

We then used one of our uncertainty relations for single-qubit measurements to relate the security of two-party secure function evaluation to the capacity of the storage device to store quantum information reliably. We showed that provided the storage device is ``very bad'' at storing $n$ qubits, there is a protocol for performing secure function evaluation in which Alice and Bob communicate $n$ qubits. This is the first time the security is related to the capacity of the channel to send quantum information. As explained in the discussion following Theorem \ref{thm:para}, this is not totally satisfying, but is hopefully a step towards proving the ideal result which would be that we get security as soon as the storage device is just ``bad'' at storing quantum information.


%


\chapter{Discussion}
\label{chap:discussion}


\section{Summary}

In this thesis, we considered uncertainty relations for several observables and their applications to quantum information theory. We have first seen how the problem of finding uncertainty relations is closely related to the problem of finding large almost Euclidean subspaces of $\ell_1(\ell_2)$. Even though we did not use any norm embedding result directly, many of the ideas presented here come from the proofs and constructions in the study of the geometry of normed spaces. In particular, we obtained an explicit family of bases that satisfy a strong metric uncertainty relation by adapting a construction of \cite{Ind07}. Moreover, using standard techniques from asymptotic geometric analysis, we were able to prove a strong uncertainty relations for random bases.

We used these uncertainty relations to exhibit strong locking effects. In particular, we obtained the first explicit construction of a method for encrypting a random $n$-bit string in an $n$-qubit state using a classical key of size polylogarithmic in $n$. Moreover, our non-explicit results give better key sizes than previous constructions while simultaneously meeting a stronger locking definition. In particular, we showed that an arbitrarily long message can be locked with a constant-sized key. Our results on locking are summarized in Table \ref{tbl:locking-results}. We should emphasize that, even though we presented information locking from a cryptographic point of view, it is not a composable primitive because an eavesdropper could choose to store quantum information about the message instead of measuring. For this reason, a locking scheme has to be used with great care when composed with other cryptographic primitives.

We also used uncertainty relations to construct quantum identification codes. We proved that it is possible to identify a quantum state of $n$ qubits by communicating $n$ classical bits and $O(\log(1/\e))$ quantum bits. We also presented an efficient encoder for this problem that uses $O(\log^2 (n/\e))$ qubits of communication instead. The main weakness of this result is that the decoder uses a classical description of the state $\ket{\ph}$ that is in general exponential in the number of qubits of $\ket{\ph}$. One cannot hope to avoid this difficulty because, as shown by \cite{Win04}, if Bob was to receive a copy of the quantum state $\ket{\ph}$, the task of quantum identification becomes the same as the task of transmission of quantum information.

%
%




We then considered uncertainty relations that hold even in the presence of quantum side information. For this, we defined QC-extractors which are sets of unitary transformations that have the following property: for any state $\rho_{AE}$ for which $\entHmin(A|E)_{\rho}$ is sufficiently larger than $- \log d_A$, applying a typical unitary on $A$ followed by a measurement of some prefix of the output gives an outcome that is almost uniformly distributed and independent of $E$. Such a definition fits in the general framework of the decoupling theorem of \cite{Dup09, DBWR10} and we use similar techniques to analyse the different constructions we propose; see Table~\ref{tab:ext-summary} for a summary. All these constructions lead to strong min-entropy uncertainty relations. We used them to prove the security of two-party function evaluation under a condition on the capacity of the parties' storage device to maintain quantum information. We also proved von Neumann entropy uncertainty relations with quantum side information for a full set of mutually unbiased bases, thus generalizing the results of \cite{Iva92, San93}.

\section{Open questions}

We expect to see more applications to quantum information theory of the tools used in the theory of pseudorandomness. An interesting open question is whether these techniques can be helpful in constructing explicit subspaces of highly entangled states. Such subspaces are related to one of the central problems in quantum information theory:  the classical capacity of a quantum channel. An explicit construction of such spaces would lead to explicit channels that violate additivity of the minimum output entropy \citep{HW08,Has09}, but also explicit protocols for superdense coding of quantum states \citep{HHL04}. As shown by \cite{ASW10,ASW10b}, this problem  amounts to finding explicit almost Euclidean sections for matrix spaces endowed with Schatten $p$-norms, which corresponds to the $\ell_p$ norm of the singular values. In addition to the applications in quantum information theory, such almost Euclidean sections are closely related to rank minimization problems for which the nuclear norm heuristic allows exact recovery \citep{DF10}.

Addressing this question is related to finding explicit constructions of $(0,\e)$-QQ-extractors with output size close to $n/2$ (which is optimal) and with small (say sublinear in $n$) seed size. In fact, by applying the unitaries of the QQ-extractor in superposition, all input pure states get mapped to highly entangled output states. More generally, it would be very interesting to understand what kinds of sets of unitaries other than unitary two-designs satisfy the decoupling theorem. Is it possible to use a number of unitaries that is smaller than the output dimension? Even non-explicit constructions would be interesting. In the special case of QC-extractors, do the metric uncertainty relations of Chapter \ref{chap:uncertainty-relations} remain valid in the presence of quantum side information?

From a computational complexity point of view, I think it would be also interesting to study the hardness of some natural problems related to uncertainty relations. For example, given a set of unitaries as an input, can one compute efficiently how good uncertainty relations they define? Does quantum side information make things significantly harder?

We might also wonder whether the decoupling theorem, or the different notions of quantum extractors defined here have applications to complexity theory, just as classical extractors have applications in derandomization for example.

There is also an intriguing general question on the power of the second moment. We know that pairwise independent permutations are good (classical) extractors. We also know that a full set of mutually unbiased bases --- which defines a state 2-design --- satisfies good uncertainty relations. In addition, the decoupling theorem says that unitary 2-designs satisfy a strong decoupling statement. All these results are based on a second moment argument. Is there a precise way of unifying these results?

On the cryptography side, are there cryptographic applications of locking schemes? For example, suppose that we authenticate the message before encoding it. Then the receiver can check whether an eavesdropper has altered the encoded message. Conditioned on passing the authentication test, is it true that the state held by the eavesdropper is independent of the message? If this is the case, then the security guarantee would be composable and we could use a locking scheme as a key distribution protocol that only uses communication from Alice to Bob.

\appendix

\appendixpage

\chapter{Deferred proofs}


%
%

\section{Existence of metric uncertainty relations}

In this section, we prove the lemmas used for proving Theorem \ref{thm:existence-ur}.

\label{sec:app-existence-ur}

\newtheorem*{lemma-expl1l2}{Lemma \ref{lem:expl1l2} }
\begin{lemma-expl1l2}[Average value of $\ell^A_1(\ell^B_2)$ on the sphere] 
Let $\ket{\ph}^{AB}$ be a random pure state on $AB$. Then,
\[
\ex{ \| \ket{\ph}^{AB} \|_{\ell^A_1(\ell^B_2)} } = \frac{\Gamma(d_B+\frac{1}{2})}{\Gamma(d_B)} \frac{\Gamma(d_A d_B)}{\Gamma(d_A d_B + \frac{1}{2})} \geq \sqrt{1 - \frac{1}{d_B}} \sqrt{d_A}.
\]
where $\Gamma$ is the Gamma function $\Gamma(z) = \int_{0}^{\infty} u^{z-1} e^{-u} du$ for $z \geq 0$.
\end{lemma-expl1l2}
\begin{proof}
The presentation uses methods described in \cite{Bal97}.

Observe that the random variable $\| \ket{\ph}^{AB} \|_{12}$ is distributed as the $\ell_1^{d_A}(\ell_2^{2d_B})$ norm of a \emph{real} random vector chosen according to the rotation invariant measure on the sphere $\Sp^{2d_Ad_B-1}$. We define for integers $n$ and $m$ the norm $\ell_1^{n}(\ell_2^{m})$ of a real $n+m$-dimensional vector $\{v_{i,j}\}_{i \in [n], j \in [m]}$ as for the complex case (Definition \ref{def:l1l2})
\[
\| v \|_{\ell_1^{n}(\ell_2^{m})} = \sum_i \sqrt{ \sum_j |v_{i,j}|^2 }.
\]
Note that we only specify the dimension of the systems as the systems themselves are not relevant here. In the rest of the proof, we use $\| \cdot \|_{12}$ as a shorthand for $\| \cdot \|_{\ell^{d_A}_1(\ell^{2d_B})}$. 
Our objective is to evaluate the expected value $\ex{\| \Theta \|_{12}}$ where $\Theta$ has rotation invariant distribution on the real sphere $\Sp^{s-1}$ and $s = 2d$  with $d = d_Ad_B$.  For this, we start by relating the $\ex{\|Z\|_{12}}$ and $\ex{\| \Theta \|_{12}}$ where $Z$ has a standard Gaussian distribution on $\RR^{s}$ . By changing to polar coordinates, we get
\begin{align*}
\ex{ \| Z \|_{12} }	 &= \int_{\RR^s} \| x \|_{12} \frac{e^{-\frac{1}{2}\sum_{i=1}^s x^2_i}}{(2\pi)^{s/2}} dx \\
			 &= \int_{0}^{\infty} \int_{\mathbb{S}^{s-1}} \| r \theta \|_{12} \frac{e^{-r^2/2}}{(2\pi)^{s/2}} \cdot \frac{s \pi^{s/2} d\sigma(\theta) }{\Gamma(\frac{s}{2}+1)} r^{s-1} dr
\end{align*}
where $\sigma$ is the normalized Haar measure on $\mathbb{S}^{s-1}$. The term $\frac{s \pi^{s/2}}{\Gamma(\frac{s}{2}+1)}$ is the surface area of the sphere in dimension $s-1$. Using the equality $\Gamma(z+1) = z \Gamma(z)$, we have $\frac{s \pi^{s/2}}{\Gamma(\frac{s}{2}+1)} = \frac{2 \pi^{s/2}}{\Gamma(\frac{s}{2})}$. Thus,
\begin{align*}
\ex{ \| Z \|_{12} }	 &= \frac{2 \pi^{s/2} }{(2\pi)^{s/2}\Gamma(\frac{s}{2})} \int_{0}^{\infty} r^{s} e^{-r^2/2} dr \cdot \int_{\mathbb{S}^{s-1}} \| \theta \|_{12} d\sigma(\theta) \\
				&= \frac{1}{2^{s/2-1}\Gamma(\frac{s}{2})} \int_{0}^{\infty} r^{s} e^{-r^2/2} dr \cdot \int_{\mathbb{S}^{s-1}} \| \theta \|_{12} d\sigma(\theta) \\
\end{align*}
We then perform a change of variable $u = r^2/2$:
\begin{align}
\ex{ \| Z \|_{12} }	 &=  \frac{1}{2^{s/2-1}\Gamma(\frac{s}{2})} \int_{0}^{\infty} (2u)^{(s-1)/2} e^{-u} du \cdot \int_{\mathbb{S}^{s-1}} \| \theta \|_{12} d\sigma(\theta) \notag \\
			&= \frac{2^{(s-1)/2} \Gamma(\frac{s-1}{2}+1)}{2^{s/2-1} \Gamma(\frac{s}{2})} \cdot \int_{\mathbb{S}^{s-1}} \| \theta \|_{12} d\sigma(\theta) \notag \\
			&= \frac{ \sqrt{2} \Gamma(\frac{s+1}{2})}{\Gamma(\frac{s}{2})} \cdot \ex{\| \Theta \|_{12}}. \label{eq:gaussian-haar}
\end{align}
Now, we compute
\begin{align*}
\ex{ \| Z \|_{12} }	 &= \int_{\RR^s} \| x \|_{12} \frac{e^{-\frac{1}{2}\| x \|_2^2}}{(2\pi)^{s/2}} dx \\
			&= \sum_{i=1}^{d_A} \int_{\RR^s}  \| x_i \|_2  \frac{e^{-\frac{1}{2} \| x \|_2^2 }}{(2\pi)^{s/2}} dx
\end{align*}
where we decomposed $x = (x_1, \dots, x_{d_A})$ where $x_i \in \RR^{2d_B}$. As all the terms of the sum are equal
\begin{align*}
\ex{ \| Z \|_{12} }&= d_A \int_{\RR^{2d_B}}  \| x_0 \|_2  \frac{e^{-\frac{1}{2} \| x_0 \|_2^2 }}{(2\pi)^{d_B}} dx_0 \left(\int_{\RR^{2d_B}}  \frac{e^{-\frac{1}{2} \| x_1 \|_2^2 }}{(2\pi)^{d_B}} dx_1 \right)^{d_A-1}  \\
			&= d_A \frac{\sqrt{2} \Gamma(\frac{2d_B+1}{2})}{\Gamma(d_B)} \int_{\mathbb{S}^{2d_B-1}} \|\theta \|_2 d\sigma(\theta) \\
			&= d_A \frac{ \sqrt{2} \Gamma(\frac{2d_B + 1}{2})}{\Gamma(d_B)}.
\end{align*}
To get the second equality, we use the same argument as for equation \eqref{eq:gaussian-haar}.
We conclude using equation \eqref{eq:gaussian-haar}
\begin{align*}
\ex{\| \ket{\ph} \|_{\ell_1^A(\ell_2^B)}} &= \ex{\| \Theta \|_{12}} \\
	&= d_A \frac{\Gamma(d_B+\frac{1}{2})}{\Gamma(d_B)} \cdot \frac{\Gamma(d_A d_B)}{\Gamma(d_A d_B + \frac{1}{2})}.
\end{align*}

We now prove the inequality in the statement of the lemma. We use the following two facts about the $\Gamma$ function: $\log \Gamma$ is convex and for all $z > 0$, $\Gamma(z+1) = z \Gamma(z)$. The first property can be seen by using H\"older's inequality for example and the second using integration by parts. Using these properties, we have
\begin{align*}
\log \Gamma\left(x+\frac{1}{2}\right) &\leq \frac{1}{2} \log \Gamma(x) + \frac{1}{2} \log \Gamma(x+1) \\
						&= \frac{1}{2} \log \left( x \Gamma(x)^2 \right) \\
						&= \log \left( \sqrt{x} \Gamma(x) \right).
\end{align*}
Thus, $\frac{\Gamma(x+\frac{1}{2})}{\Gamma(x)} \leq \sqrt{x}$. Similarly, we have $\frac{\Gamma(x)}{\Gamma(x-\frac{1}{2})} \leq \sqrt{x-\frac{1}{2}}$ which implies that $\frac{\Gamma(x+\frac{1}{2})}{\Gamma(x)}  \geq \sqrt{x-\frac{1}{2}}$ when writing $\Gamma(x+1/2) = (x-1/2) \Gamma(x-1/2)$.

We conclude that 
\begin{align*}
\ex{\| \ket{\ph} \|_{\ell_1^A(\ell_2^B)}} &\geq d_A \cdot \sqrt{d_B - \frac{1}{2}} \frac{1}{\sqrt{d_A d_B}} \\
		&= \sqrt{d_A} \cdot \sqrt{1 - \frac{1}{2d_B}}.
\end{align*}
\end{proof}

\newtheorem*{lemma-levy}{Lemma \ref{lem:levy} }
\begin{lemma-levy}[Levy's lemma]
Let $f : \CC^d \to \RR$ and $\eta > 0$ be such that for all pure states $\ket{\ph_1}, \ket{\ph_2}$ in $\CC^d$, 
\[
| f(\ket{\ph_1}) - f(\ket{\ph_2}) | \leq \eta \| \ket{\ph_1} - \ket{\ph_2} \|_2.
\]
Let $\ket{\ph}$ be a random pure state in dimension $d$. Then for all $0 \leq \delta \leq \eta$,
\[
\pr{| f(\ket{\ph}) - \ex{f(\ph)} | \geq \delta } \leq 4 \exp{- \frac{\delta^2 d}{c \eta^2} }
\]
where $c$ is a constant. We can take $c = 9 \pi^2$.
\end{lemma-levy}
\begin{proof}
We can instead study the concentration of a Lipschitz function on the real sphere $\Sp^{2d-1}$. Note that the induced function (that we also call $f$) is still $\alpha$-Lipschitz. Concentration on $\Sp^{2d-1}$ can be proved in a simple way using concentration of the standard Gaussian distribution. This proof is due to Maurey and Pisier and can be found in \cite[Appendix V]{MS86}. Specifically, using \cite[Corollary V.2]{MS86}, we get 
\begin{align*}
\pr{| f(Z) - \ex{f(Z)} | \geq t} &\leq 2 \exp{-\frac{\delta^2 (2d)}{18 \pi^2 \eta^2}} + 2 \exp{-\frac{2d}{2 \pi^2}} \\
						&\leq 4\exp{-\frac{\delta^2 d}{9 \pi^2 \eta^2}}. 
\end{align*}
In the notation of the proof of \cite[Corollary V.2]{MS86}, we have set $\delta = 1/2$. This can be done because using the same arguments as in the proof of Lemma \ref{lem:expl1l2}, we can show that the expected $\ell_2$ norm of the standard Gaussian distribution in dimension $n$ at least $\sqrt{2}\sqrt{n-\frac{1}{2}} > \sqrt{n}$ for $n\geq 2$.

We used this version of Levy's lemma because it has an elementary proof and it gives directly the concentration about the expected value. Different versions involving the median of $f$ and giving better constants can be found in  \cite[Corollary 2.3]{MS86} or \cite[Proposition 1.3]{Led01} for example.
\end{proof}

\newtheorem*{lemma-net}{Lemma \ref{lem:eps-net} }
\begin{lemma-net}[$\delta$-net]
Let $\delta \in (0,1)$. There exists a set $\cN$ of pure states in $\CC^d$ with $|\cN| \leq (3/\delta)^{2 d}$ such that for every pure state $\ket{\psi} \in \CC^d$ (i.e., $\| \ket{\psi} \|_2 = 1$), there exists $\ket{\tilde{\psi}} \in \cN$ such that
\[
\| \ket{\psi} - \ket{\tilde{\psi}} \|_{2} \leq \delta.
\] 
\end{lemma-net}
\begin{proof}
A proof can be found in \cite[Lemma II.4]{HLSW04}. We repeat it here for completeness.
Let $\cN$ be a maximal set of pure states satisfying $\| \ket{\psi_1} - \ket{\psi_2} \|_2 \geq \delta$ for all pure states $\ket{\psi_1}, \ket{\psi_2} \in \cN$. This set can be constructed iteratively by adding at each step a state that is at distance at least $\delta$ from all states already in the set. First, we show that this procedure terminates by bounding the size of such a set. We do this using a volume argument. For this it is simpler to look at vectors $\ket{\psi} \in \cN$ as real vectors in dimension $2d$. The open balls of radius $\delta/2$ centered at each $\ket{\psi} \in \cN$ are disjoint and are contained in the open ball of radius $1+\delta/2$ centered at the origin. Therefore,
\[
|\cN| \left(\frac{\delta}{2}\right)^{2d} \leq \left(1 + \frac{\delta}{2}\right)^{2d} \leq \left(\frac{3}{2}\right)^{2d}.
\]
We conclude by observing that such a set has the desired property. In fact, if there exists a state $\ket{\psi} \in \cH$ such that for all $\ket{\tilde{\psi}} \in \cN$, $\| \ket{\psi} - \ket{\tilde{\psi}} \|_{2} > \delta$, then $\ket{\psi}$ can be added to $\cN$ and contradict the fact that $\cN$ is maximal.
\end{proof}

\comment{Note that the constants are not optimized in the following Lemma. It is probably possible to do much better.}

\newtheorem*{lemma-avgconc}{Lemma \ref{lem:avgconc} }
\begin{lemma-avgconc}[Concentration of the average]
Let $a, b \geq 1$, $\delta \in (0,1)$ and $t$ a positive integer. Suppose $X$ is a random variable with $0$ mean satisfying the tail bounds
\[
\pr{X \geq \eta} \leq a e^{-b\eta^2} \quad \text{ and } \quad \pr{X \leq -\eta} \leq ae^{-b \eta^2}.
\]
Let $X_1, \dots X_t$ be independent copies of $X$. Then if $\delta^2 b \geq 16 a^2 \pi $,
\[
\pr{\left| \frac{1}{t} \sum_{k=1}^t X_k \right| \geq \delta} \leq \exp{-\frac{\delta^2 bt }{2}}.
\]
\end{lemma-avgconc}
\begin{proof}
For any $\lambda > 0$, using Markov's inequality
\begin{align*}
\pr{\sum_{k=1}^t X_k \geq t \delta} &= \pr{\exp{\lambda \sum_{k=1}^t X_k} \geq \exp{\lambda t \delta} } \\
							&\leq \ex{\exp{\lambda \sum_{k=1}^t X_k}} e^{-\lambda t \delta} \\
							&= \ex{e^{\lambda X}}^t e^{-\lambda t \delta}.
\end{align*}
We now bound the moment generating function $\ex{e^{\lambda X}}$ of $X$ using the tail bounds.
\begin{align*}
\ex{e^{\lambda X}} &= \int_{0}^{\infty} \pr{e^{\lambda X} \geq u} du \\
				&= \int_{0}^{\infty} \pr{X \geq \frac{\ln u}{\lambda}} du \\
				&= \int_{0}^1 \pr{X \geq \frac{\ln u}{\lambda}} du +  \int_{1}^{\infty} \pr{X \geq \frac{\ln u}{\lambda}} du \\
				&\leq 1 + \int_{1}^{\infty} a \exp{-\frac{b \ln^2 u}{\lambda^2}} du \\
				&= 1 + a \int_{0}^{\infty} \exp{-\frac{b z^2}{\lambda^2}} e^{z} dz
\end{align*}
by making the change of variable $z = \log u$.
\begin{align*}				
\ex{e^{\lambda X}}			&\leq 1 + a \int_{0}^{\infty} \exp{-\frac{b}{\lambda^2}\left(z - \frac{\lambda^2}{2 b} \right)^2 + \frac{\lambda^2}{4b}} dz \\
				&\leq 1 + a \exp{\frac{\lambda^2}{4 b}} \int_{-\infty}^{\infty} \exp{-\frac{b}{\lambda^2}\left(z - \frac{\lambda^2}{2 b} \right)^2 } dz \\
				&= 1 + a \exp{\frac{\lambda^2}{4 b}} \frac{\lambda}{\sqrt{2 b}} \int_{-\infty}^{\infty} \exp{-\frac{u^2}{2} } du \\
				&= 1 + a\frac{\sqrt{2\pi}\lambda}{\sqrt{2b}} \cdot \exp{\frac{\lambda^2}{4 b}} \\
				&\leq 2 \max \left(1, a \frac{\sqrt{\pi}\lambda}{\sqrt{b}} \cdot \exp{\frac{\lambda^2}{4 b}} \right).
\end{align*}
We choose $\lambda = 2 \delta b$ (this is not the optimal choice but it makes expressions simpler),
\begin{align*}
\pr{\sum_{k=1}^t X_k \geq t \delta} &\leq \max\left( 2^t, \left(2 a \frac{\sqrt{\pi}\lambda}{\sqrt{b}}\right)^t \cdot \exp{\frac{\lambda^2 t}{4 b}} \right) \exp{-\lambda t \delta} \\
							&= \max \left( \exp{-2 \delta^2 bt + t \ln 2}, \exp{ \delta^2 bt - 2 \delta^2 bt + t \ln ( 4 a \sqrt{\pi} \delta \sqrt{b} ) } \right) \\
							&= \max \left\{ \exp{ \left(-2 \delta^2 b + \ln 2\right) t}, \exp{ \left( -\delta^2 b + \ln ( 4 a \sqrt{\pi} \delta \sqrt{b} ) \right) t } \right\}.
\end{align*}
\begin{claim}[]
For all $c \geq 1$ and $x \geq c$
\[
\frac{1}{2} \ln(cx) - x \leq -\frac{x}{2}.
\]
\end{claim}
The function $x \mapsto \frac{x}{2} - \frac{1}{2}\ln(cx)$ is increasing for $x \geq 1$. It suffices to show that it is nonnegative for $x = c$. To see that, we differentiate the function $y \mapsto y - \ln(y^2)$ to prove that for all $y \geq 1$, 
we have $y - \ln(y^2) \geq 0$. This proves the claim.

Using this inequality, we have for $\delta^2 b \geq 16 a^2 \pi$,
\[
-\delta^2 b + \ln ( 4 a \sqrt{\pi} \delta \sqrt{b} ) \leq -\frac{\delta^2b}{2} \quad \text{ and } \quad -2 \delta^2 b + \ln 2 \leq -\frac{\delta^2b}{2}.
\]
Finally,
\[
\pr{\sum_{k=1}^t X_k \geq t \delta} \leq \exp{-\frac{\delta^2bt}{2}}.
\]
\end{proof}



\section{Permutation extractors}
\label{sec:app-perm-extractor}

In order to prove the existence of strong permutation extractors with good parameters, we use the construction of \citet*{GUV09} which is inspired by list decoding. Their main construction is a lossless condenser based on Parvaresh-Vardy codes. Using this condenser, they build an explicit extractor with good parameters. However, this lossless condenser based on Parvaresh-Vardy codes does not seem to be easily extended into a permutation condenser. The same paper also presents a lossy condenser based on Reed-Solomon codes, which can indeed be transformed into a permutation condenser. This permutation condenser can then be used in the extractor construction instead of the lossless condenser giving a strong permutation extractor. In this section, we describe this construction. For completeness, we reproduce most of the proof here, except the results that are used exactly as stated in \cite{GUV09}. 

It is also worth mentioning that to obtain metric uncertainty relations, we want strong extractors. Even though the extractors in \cite{GUV09} are not directly described as strong, they are essentially strong. In this section, we describe all the condensers and extractors as strong.
  
\begin{definition}[Condenser]
\label{def:condenser}
A function $C : \{0,1\}^n \times S \to \{0,1\}^{n'}$ is an $(n,k) \to_{\e} (n', k')$ \emph{condenser} if for every $X$ with min-entropy at least $k$, $C(X, U_S)$ is $\e$-close to a distribution with min-entropy $k'$ when $U_S$ is uniformly distributed on $S$. A condenser $C$ is \emph{strong} if $(U_S, C(X, U_S))$ is $\e$-close to $(U_S, Z)$ for some random variable $Z$ such that for all $y \in S$, $Z|_{U_S = y}$ has min-entropy at least $k$.

A condenser is \emph{explicit} if it is computable in polynomial time in $n$.
\end{definition}
\begin{myremark}[]
The set $S$ is usually of the form $\{0,1\}^d$ for some integer $d$. Here, it is convenient to take sets $S$ not of this form to obtain permutation extractors. Note also that an extractor is an $(n,k) \to_{\e} (m,m)$ condenser.
\end{myremark}

\begin{definition}[Permutation condenser]
A family  $\{P_y\}_{y \in S}$ of permutations of $\{0,1\}^n$  is an  $(n,k) \to_{\e} (n', k')$ \emph{strong permutation condenser} if the function $P^C: (x,y) \mapsto P^C_y(x)$ where $P_y^C(x)$ refers to the first $n'$ bits of $P_y(x)$ is an
$(n,k) \to_{\e} (n',k')$ strong condenser. 

A strong permutation condenser is \emph{explicit} if for all $y \in S$, both $P_y$ and $P_y^{-1}$ are computable in polynomial time.
\end{definition}

The following theorem describes the condenser that will be used as a building block in the extractor construction. It is an analogue of Theorem 7.2 in \cite{GUV09}.
\begin{theorem}
\label{thm:perm-condenser}
For all positive integers $n$ and $\ell \leq n$, as well as $\alpha, \e \in (0,1/2)$, there exists an explicit family of permutations $\{RS_y\}_{y \in S}$ of $\FF_{2^t}^n$ that is an
\[
(nt, (\ell+1)t) \to_{\e} (\ell t, (1-\alpha) \ell t - 4)
\]
strong permutation condenser with $t = \ceil{1/\alpha \cdot \log(24n^2/\e)}$ and $\log |S| \leq t$. Moreover, the functions $(x, y) \mapsto RS_y(x)$ and $(x, y) \mapsto RS^{-1}_y(x)$ can be computed by a circuit of size $O(n \polylog(n/\e))$.
\end{theorem}
\begin{proof}
Set $q = 2^t$ and $\e_0 = \e/6$. Consider the function $C' : \FF_q^n \times \FF_q \to \FF_q^{\ell+1}$ defined by
\[
C'(f, y) = [y, f(y), f(\zeta y), \dots, f(\zeta^{\ell-1} y)]
\]
where $\FF_q^n$ is interpreted as the set of polynomials over $\FF_q$ of degree at most $n-1$ and $\zeta$ is a generator of the multiplicative group $\FF_q^{*}$. First, we compute the input and output sizes in terms of bits. The inputs can be described using $\log |\FF_q^n| = n \log q = nt$ bits, the seed using $\log |\FF_q| = t$ bits and the output using $\log |\FF_q^{\ell+1}| = (\ell + 1) t$. Using \cite[Theorem 7.1]{GUV09}, for any integer $h$, $C'$ is a 
\begin{equation}
\label{eq:rs-condenser}
\left(nt, \log \left( \frac{q^{\ell}-1}{\e_0} \right) \right) \to_{2\e_0} \left(\ell t + t, \log \left( \frac{Ah^{\ell}-1}{2\e_0} \right) \right)
\end{equation}
condenser where $A \eqdef \e_0 q - (n-1)(h-1)\ell$. We now choose $h = \ceil{q^{1-\alpha}}$. 
As $q \geq (4n^2/\e_0)^{1/\alpha}$, we have $A \geq \e_0 q - n^2 h \geq \e_0 q - \e_0 q^{\alpha}/4 \cdot (q^{1-\alpha} + 1) \geq \e_0 q/2$. Thus, we can compute the bounds we obtain on the condenser $C'$:
\[
\log \left( \frac{q^{\ell}-1}{\e_0} \right) = \ell t + \log(1/\e_0) \leq (\ell+1) t
\]
and 
\begin{align*}
\log \left( \frac{Ah^{\ell}-1}{2\e_0} \right) &= \log \left( \frac{Ah^{\ell}}{2\e_0} \right) + \log \left(1 - \frac{1}{Ah^\ell} \right) \\
							&\geq \log (q/4) + \ell \log h - 1 \\
							&\geq t + (1-\alpha) \ell t - 3.
\end{align*}
Plugging these values in equation \eqref{eq:rs-condenser}, we get that $C'$ is a 
\begin{equation}
\label{eq:paramscond}
\left(nt, (\ell+1) t \right) \to_{2\e_0} \left( \ell t + t, (1-\alpha)\ell t + t - 3) \right)
\end{equation}
condenser.

Observe that the seed $y$ is part of the output of the condenser. As we want to construct a strong condenser, we do not consider the seed as part of the output of the condenser. For this, we define $C : \FF_q^n \times \FF_q \to \FF_q^{\ell}$ by $C(f,y) = [f(y), \dots, f(\zeta^{\ell-1} y)]$. Moreover, as will be clear later when we try to build a permutation condenser, we take the seed to be uniform on $S \eqdef \FF_q^{*} = \FF_q - \{0\}$ instead of being uniform on the whole field $\FF_q$. Note that this increases the error of the condenser by at most $2^{-t} \leq \e_0$ (because one can choose $U_{\FF_q^{*}} = U_{\FF_q}$ with probability $1-2^{-t}$). Here and in the rest of this proof, we will be using Doeblin's coupling lemma (see Chapter \ref{chap:prelim}).


Equation \eqref{eq:paramscond} then implies that if $X$ has min-entropy at least $(\ell + 1)t$ and $U_{S}$ is uniform on $S$, then the distribution of $(U_S, C(X, U_S))$ is $3\e_0$-close to a distribution with min-entropy at least $(1-\alpha) \ell t + t - 3$. Let $Y \in S$ and $Z \in \{0,1\}^{(\ell+1)t}$ be random variables such that $\entHmin(Y,Z) \geq (1-\alpha) \ell t + t - 3$ and $(U_S, C(X, U_S)) = (Y,Z)$ with probability at least $1-3\e_0$. If $Y$ was uniformly distributed on $S$, then it would follow directly that for all $y \in S$, $\entHmin(Z|Y=y) \geq (1-\alpha) \ell t - 3$. However, $Y$ is not necessarily uniformly distributed. We define a new random variable $Z'$ by
\[
Z' = \left\{ \begin{array}{ll}
Z & \textrm{if $Y=U_S$}\\
U' & \textrm{if $Y \neq U_S$}\\
\end{array} \right.
\]
where $U'$ is uniformly distributed on $\{0,1\}^{(\ell+1)t}$ and independent of all the other random variables. We have for any $z \in \{0,1\}^{(\ell+1) t}$ and $y \in S$,
\begin{align*}
\pr{Z'=z|U_S=y} 	&= \frac{1}{\pr{U_S = y}} \big( \pr{Z' = z, Y=y, Y = U_S}   \\
				& \quad	+ \pr{Z'=z, U_S = y, Y \neq U_s} \big) \\
				&\leq \frac{1}{\pr{U_S = y}} \left(  2^{-(1-\alpha) \ell t - t + 3} + 2^{-(\ell+1)t} \cdot \frac{1}{|S|} \right) \\
				&\leq 2 \cdot 2^{-(1-\alpha) \ell t + 3}.
\end{align*}
Moreover, we have $(U_S, C(X, U_S)) = (U_S,Z')$ with probability at least $1 - 6\e_0$.

We conclude that $C$ is a
\begin{equation}
\label{eq:strongcond}
\left(nt, (\ell+1) t \right) \to_{\e} \left( \ell t, (1-\alpha)\ell t - 4) \right)
\end{equation}
strong condenser. 


To define our permutation condenser, we set the first $n' = \ell t$ bits $RS^C_y(x)$ of $RS_y(x)$ to be $RS^C_y(x) = C(x,y)$. 
We then define the remaining bits by $RS^R_y(f) = [f(\zeta^{\ell} y), \dots, f(\zeta^{n-1} y)]$. As $q \geq n-1$ and $\zeta$ is a generator of $\FF^*_q$, the elements $y, \zeta y, \dots, \zeta^{n-1} y$ are distinct provided $y \neq 0$. So for $y \neq 0$, $(RS^C,RS^R)_y(f)$ is the evaluation of the polynomial $f$ of degree at most $n-1$ in $n$ distinct points. Thus, $f \mapsto RS_y(f)$ is a bijection in $\FF_q^n$ for all $y \neq 0$. This is why the value $0$ for the seed was excluded earlier.

Concerning the computation of the functions $RS^C_y$ and $RS^R_y$, they only require the evaluation of a polynomial on elements of the finite field $\FF_q$. Computations in the finite field $\FF_q$ can be performed efficiently by finding an irreducible polynomial of degree $\log q$ over $\FF_2$ and doing computations modulo this polynomial. In fact, finding an irreducible polynomial of degree $\log q$ over $\FF_2$ can be done in time polynomial in $\log q$ (see for example \cite{Sho90} for a deterministic algorithm and Corollary 14.43 in the book \cite{GG99} for a simpler randomized algorithm). Since addition, multiplication and finding the greatest common divisor of polynomials in $\FF_2[X]$ can be done using a number of operations in $\FF_2$ that is polynomial in the degrees, we conclude that computations in $\FF_q$ can be implemented in time $O(\polylog(n/\e))$. Moreover, one can efficiently find a generator $\zeta$ of the group $\FF^*_q$. For example, Theorem 1.1 in \cite{Sho92} shows the existence of a deterministic algorithm having a runtime $O(\poly(\log(q))) = O(\polylog(n/\e))$.

To evaluate $RS_y$ at a polynomial $f$, we compute the field elements $y, \zeta y, \dots, \zeta^{n-1} y$, and then evaluate the polynomial $f$ on these points. Using a fast multipoint evaluation, this step can be done in $O(n \polylog n)$ number of operations in $\FF_q$ (see Corollary 10.8 in \cite{GG99}). Moreover, given a list $[f(y), \dots, f(\zeta^{n-1} y)]$ for $y \neq 0$, we can find $f$ by fast interpolation in $\FF_q[X]$ (see Corollary 10.12 in \cite{GG99}). As a result $RS_y^{-1}$ can also be computed in $O(n \polylog n)$ operations in $\FF_q$.
\end{proof}

This condenser will be composed with other extractors, the following lemma shows how to compose condensers.

\begin{lemma}[Composition of strong permutation condensers]
\label{lem:compositioncond}
Let $(P_{1,y_1})_{y_1 \in S_1}$ be an $(n, k) \to_{\e} (n', k')$ strong permutation condenser and $(P_{2,y_2})_{y_2 \in S_2}$ be an $(n', k') \to_{\e} (n'', k'')$ strong permutation condenser. Then $(P_{y})_{y=(y_1, y_2) \in S_1 \times S_2} = (P^C_y, P^R_y)$ where $P^C_{y_1y_2} = P_{2,y_2}^C \comp P^C_{1,y_1} $ and $P^R_{y_1y_2} = (P^R_{2,y_2} \comp P^C_{1, y_1}) \concat P^R_{1, y_1}$ is an
$(n, k) \to_{2\e} (n'', k'')$ strong permutation extractor.
\end{lemma}
\begin{proof}
$P_y$ is clearly a permutation of $\{0,1\}^n$. We only need to check that $P^C$ is a strong condenser. By definition, if $\entHmin(X) \geq k$, $(U_{S_1}, P^C_{1, U_{S_1}}(X))$ is $\e$-close to $(U_{S_1}, Z)$ where $Z|_{U_{S_1} = y_1}$ has min-entropy at least $k'$. Now putting $Z$ into the condenser $P^C_{2}$, we get that for any $y_1$, $(U_{S_2}, P^C_{2, U_{S_2}}(Z_{U_{S_1}}))$ is $\e$-close to $(U_{S_2}, Z_2)$ where $Z_2|_{U_{S_2} = y_2}$ has min-entropy at least $k''$ for any $y_2 \in S_2$. Thus, $Z_2|_{U_{S_1}U_{S_2} = y_1y_2}$ has min-entropy at least $k''$. Moreover, by the triangle inequality, we have $\tracedist{ (U_{S_1},U_{S_2}, P^C_{U_{S_1}U_{S_2}}(X)), (U_{S_1},U_{S_2}, Z_2) } \leq 2\e$.
\end{proof}

Next, we present one of the standard extractors that are used as a building block in many constructions. 

\begin{lemma}[``Leftover Hash Lemma'' extractor \citep{ILL89}]
\label{lem:leftoverext}
For all positive integers $n$ and $k \leq n$, and $\e > 0$, there exists an explicit family $(P_y)_{y \in S}$ of permutations of $\{0,1\}^n$ that is an $(n,k) \to_{\e} m$ strong permutation extractor with $\log |S| = \log (2^n - 1)$ and $m \geq k - 2 \log(2/\e)$.
\end{lemma}
\begin{proof}
We view $\{0,1\}^n$ as the finite field $\FF_{2^n}$ and the set $S = \FF_{2^n}^{*}$. We then define the permutation $P_y(x) = x \cdot y$ where the product $x \cdot y$ is taken in the field $\FF_{2^n}$. The family of functions $P_y$ is pairwise independent. Applying the Leftover Hash Lemma \citep{ILL89}, we get that if $Y$ uniform on $\FF_{2^n}$, the distribution of the first $\ceil{k - 2\log(1/\e)}$ bits of $P_Y(X)$ together with $Y$ is $\e$-close to uniform. Now if $U_S$ is only uniform in $\FF_{2^n}^*$, $(U_S, P_{U_S}(X))$ is $\e + 2^{-n}$-close to the uniform distribution. The result follows from the fact that we can suppose $\e \geq 2^{-n}$ (otherwise, $k-2\log(1/\e) \leq 0$ and the theorem is true).
\end{proof}

The problem with this extractor is that it uses a seed that is as long as the input. Next, we introduce the notion of a block source.

\begin{definition}[Block source]
$X = (X_1, X_2, \dots, X_s)$ is a $(k_1, k_2, \dots, k_s)$ block source if for every $i \in \{1, \dots, s\}$ and $x_1, \dots, x_{i-1}$, $X|_{X_1=x_1, \dots, X_{i-1}=x_{i-1}}$ is a $k_i$-source. When $k_1 = \dots = k_s = k$, we call $X$ a $s \times k$ source.
\end{definition}

A block source has more structure than a general source. However, for a source of large min-entropy $k$ (or equivalently with small entropy deficiency $\Delta = n - k$), one does not lose too much entropy by viewing a general source as a block source where each block has entropy deficiency roughly $\Delta$. See \cite[Corollary 5.9]{GUV09} for a precise statement.

\begin{lemma}[{\cite[Lemma 5.4]{GUV09}}]
\label{lem:baseext}
Let $s$ be a (constant) positive integer. For all positive integers $n$ and $\ell \leq n$ and all $\e > 0$, setting $t = \ceil{8s \log (24n^2 \cdot (4s+1)/\e)}$, there is an explicit family $\{L_y\}_{y \in S}$ of permutations of $\{0,1\}^n$ that is an
\[
(n, 2 \ell t) \to_{\e} \ell t
\]
strong permutation extractor with $\log |S| \leq 2 \ell t / s + t$.
\end{lemma}
\begin{proof}
As the extractor is composed of many building blocks, each generating some error, we define $\e_0 = \e/(4s+1)$ where $\e$ is the target error of the final extractor. The idea is to first apply the condenser $RS$ of Theorem \ref{thm:perm-condenser} with $\alpha = \frac{1}{8s}$  to obtain a string $X' = RS^{C}(X,U_{\FF^*_{2^t}})$ of length $n' = (2\ell-1) t$ which is $\e_0$-close to a $k'$-source where  
\[
k' = \left(1-\frac{1}{8s}\right)(2\ell-1) t - 4
\]
The entropy deficiency $\Delta$ of this $k'$-source can be bounded by $\Delta = n' - k' \leq \frac{(2\ell -1) t}{8s} + 4$.  
Then, we partition $X' = (X'_1, \dots, X'_{2s})$ (arbitrarily) into $2s$ blocks of size $n'' = \floor{n'/2s}$ or $n'' + 1$ . Using \cite[Corollary 5.9]{GUV09}, $(X'_1, \dots, X'_{2s})$ is $2s \e_0$-close to some $2s \times k''$-source where $k'' = (n'' - \Delta-\log(1/\e_0))$. 

We have $\Delta \leq \ell t/(4s) + 3 \leq \ell t/(3s)$ for $n$ large enough. Thus,
\[
k'' \geq \frac{2 \ell t}{2s} - \frac{\ell t}{3s} - \log(1/\e_0) = \frac{2}{3s} \ell t - \log(1/\e_0).
\]
We can then apply the extractor of Lemma \ref{lem:leftoverext} to all the $2s$ blocks using the same seed of size $n''+1$. Note that we can reuse the same seed because we have a strong extractor and the seed is independent of all the blocks. This extractor extracts almost all the min-entropy of the sources. More precisely, if we input to this extractor a $2s \times k''$-source, the output distribution is $2s \e_0$-close to $m$ uniform bits where 
\[
m \geq 2s \cdot (k'' - 2 \log(2/\e_0)) \geq \frac{4}{3} \ell t - 6s \log(2/\e_0) \geq \ell t.
\]

Overall, the output of this extractor is $\e_0 + 2s \e_0 + 2s \e_0 = \e$-close to the uniform distribution on $m$ bits.

It only remains to show that the extractor we just described is strong and can be extended to a permutation. This follows from Lemma \ref{lem:compositioncond} and the fact the condensers (coming from Theorem \ref{thm:perm-condenser} and Lemma \ref{lem:leftoverext}) are strong permutation condensers.
\end{proof}
\begin{myremark}[]
As pointed out in \cite{GUV09}, a stronger version of this lemma (i.e., with larger output) can be proved by using the condenser of Theorem \ref{thm:perm-condenser} and the high min-entropy extractor in \cite{GW97} with a Ramanujan expander (for example, the expander of \cite{LPS88}). This construction can also give a strong permutation extractor. However, using this extractor would slightly complicate the exposition and does not really influence the final extractor construction presented in Theorem \ref{thm:perm-extractor}.
\end{myremark}

The following lemma basically says that the entropy is conserved by a permutation extractor. It is an adapted version of  \cite[Lemma 26]{RRV99}.
\begin{lemma}
\label{lem:manyextractions}
Let $\{P_y\}_{y \in S}$ be a $(n, k) \to_{\e} m$ strong permutation extractor. Let $X$ be a $k$-source, then $(U_S, P^E_{U_S}(X), P^R_{U_S}(X))$ is $2\e$-close to $(U'_{S}, U'_{\{0,1\}^m}, W)$ where $U'_{S}$ and $U'_{\{0,1\}^m}$ are independent and uniformly distributed over $S$ and $\{0,1\}^m$ respectively, and for all $y \in S, z \in \{0,1\}^m$
\[
\entHmin(W| (U'_{S}, U'_{\{0,1\}^m}) = (y,z)) \geq k - m - 1.
\]
\end{lemma}
\begin{proof}
As $\{P_y^E\}$ is a strong extractor, there exist random variables $U'_{S}$ and $U'_{\{0,1\}^m}$ uniformly distributed on $S$ and $\{0,1\}^m$ such that $\pr{(U_S, P^E_{U_S}(X)) \neq (U'_{S}, U'_{\{0,1\}^m})} \leq \e$.
Define $\Gamma = \{ (y, z) \in S \times \{0,1\}^m : \pr{P^E_y(X) = z} < \frac{1}{2} \cdot 2^{-m} \}$. We have for every $(y,z) \notin \Gamma $ and $x \in \{0,1\}^{n-m}$,
\begin{align*}
\pr{P^R_y(X) = x | P^E_y(X) = z} 	&\leq \frac{\pr{P^R_y(X) = x, P^E_y(X) = z}}{2^{-m-1}} \\
							&\leq 2^{m+1} \pr{X = P^{-1}_y(x,z)} \\
							&\leq 2^{-(k - m - 1)}.
\end{align*}
We then show that $\pr{(U_S, P^E_{U_S}) \in \Gamma} \leq \e$. Using the fact that $\{P^E_y\}$ is a strong extractor, we have
\[
\left|\pr{U'_S, U'_{\{0,1\}^m} \in \Gamma} - \pr{(U_S, P^E_{U_S}) \in \Gamma} \right| \leq \e.
\]
But recall that, by definition of $\Gamma$, $\pr{(U_S, P^E_{U_S}) \in \Gamma} < \frac{1}{2} \pr{(U'_S, U'_{\{0,1\}^m}) \in \Gamma}$, so we get
\[
\pr{(U_S, P^E_{U_S}) \in \Gamma} \leq \e.
\]

Finally we define
\[
W = \left\{ \begin{array}{ll}
P^R_{U_S}(X) & \textrm{if $(U_S, P^E_{U_S}(X)) \notin \Gamma $}\\
U^* & \textrm{if $(U_S, P^E_{U_S}(X)) \in \Gamma $}\\
\end{array} \right.
\]
where $U^*$ is uniform on $\{0,1\}^{n-m}$ and independent of all other random variables. We conclude by observing that with probability at least $1-2\e$, we have $(U_S, P^E_{U_S}(X)) = (U'_{S}, U'_{\{0,1\}^m})$ and $P^R_{U_S}(X) = W$.
\end{proof}

 
We then combine these results to obtain the desired extractor. The proof of the following theorem closely follows  \cite[Theorem 5.10]{GUV09} but using the lossy condenser presented in Theorem \ref{thm:perm-condenser} and making small modifications to obtain a permutation extractor.
\begin{theorem}
\label{thm:perm-ext1}
For all integers $n \geq 1$, all $\e \in (0,1/2)$, and all $k \in \left[ 200 \ceil{200 \log(24 n^2/\e)}, n \right]$ there is an explicit $(n, k) \to_{\e} \floor{k/4}$ strong permutation extractor $\{P_y\}_{y \in S}$ with $\log |S| \leq 200 \ceil{200 \log(24 n^2/\e)}$. Moreover, the function $(x,y) \mapsto P_y(x)$ can be computed by a circuit of size $O(n \polylog(n/\e))$.
\end{theorem}
\begin{proof}
If $n \leq 2 \cdot 10^6$, we can use the extractor of Lemma \ref{lem:baseext} with $s = 200$ and $\ell \geq 1$ such that $2 \ell t \leq k \leq 2 (\ell+1) t$. This gives an extractor whose seed has size $\frac{k}{200} \leq 10^4 \leq 200 \ceil{200 \log(24 n^2/\e)}$ and that extracts $\ell t \geq \frac{1}{4} \cdot 2 (\ell+1) t \geq \frac{k}{4}$ bits, so the statement still holds true. In the rest of the proof, we assume $n > 2 \cdot 10^6$.

\comment{With this condition on $k$ we have $k \geq 2t = 8 \cdot 200 \cdot \log(24n^2 \cdot 801/\e)$.}

The idea of the construction is to build for an integer $i \geq 0$ an explicit $(n, 2^i \cdot 8d) \to_{\e} 2^{i-1} \cdot 8d$ extractor using $d$ bits of seed by induction on $i$. Fix $t(\e) = \ceil{200\log( 24n^2/\e)}$ and $d(\e) = 200 t(\e)$. The induction hypothesis for an integer $i \geq 0$ is as follows: For all integers $i' \leq i$ and $n$ and $\e > 0$, there is an explicit 
\[
(n, 2^{i'} \cdot 8d(\e)) \to_{\e} 2^{i'-1} \cdot 8d(\e)
\]
strong permutation extractor with seed size $d(\e)$. This extractor is called $\{P^{(i)}_y\}_{y \in S_i}$. 

For both $i=0$ and $i=1$, we can use the extractor of Lemma \ref{lem:baseext} with $s = 20$. 
For $i \in \{0,1\}$, this gives an extractor with seed $\frac{2^i \cdot 8d(\e/81)}{20} + t \leq \frac{16}{20} d(\e) + \frac{16}{20} 200 \ceil{200 \log(81)} \leq d(\e)$.

We now show for $i \geq 2$ how to build the extractor $\{P^{(i)}_y\}$ using the extractors $\{P^{(i')}_y\}$ for $i' < i$. Using the induction hypothesis, we construct the following extractor, which will be applied four times to extract the necessary random bits to prove the induction step. The choice of the form of the min-entropy values will become clear later. Set $\e_0 = \e/20$.
\comment{Probably the most natural thing would be to apply the extractor of $i-1$ just two times. We can do the first step by putting $2d$ entropy to generate the seed and $4d$ entropy to to apply the extractor $i-1$ to it, but the problem is that in the following step, we would only have only $6d$ bits of entropy which is exactly what is need so we cannot do it because there will be some losses. One solution is to use the expander based construction of GW}
\begin{center}
\begin{figure}[h]
\begin{center}
\includegraphics[width=.7\textwidth]{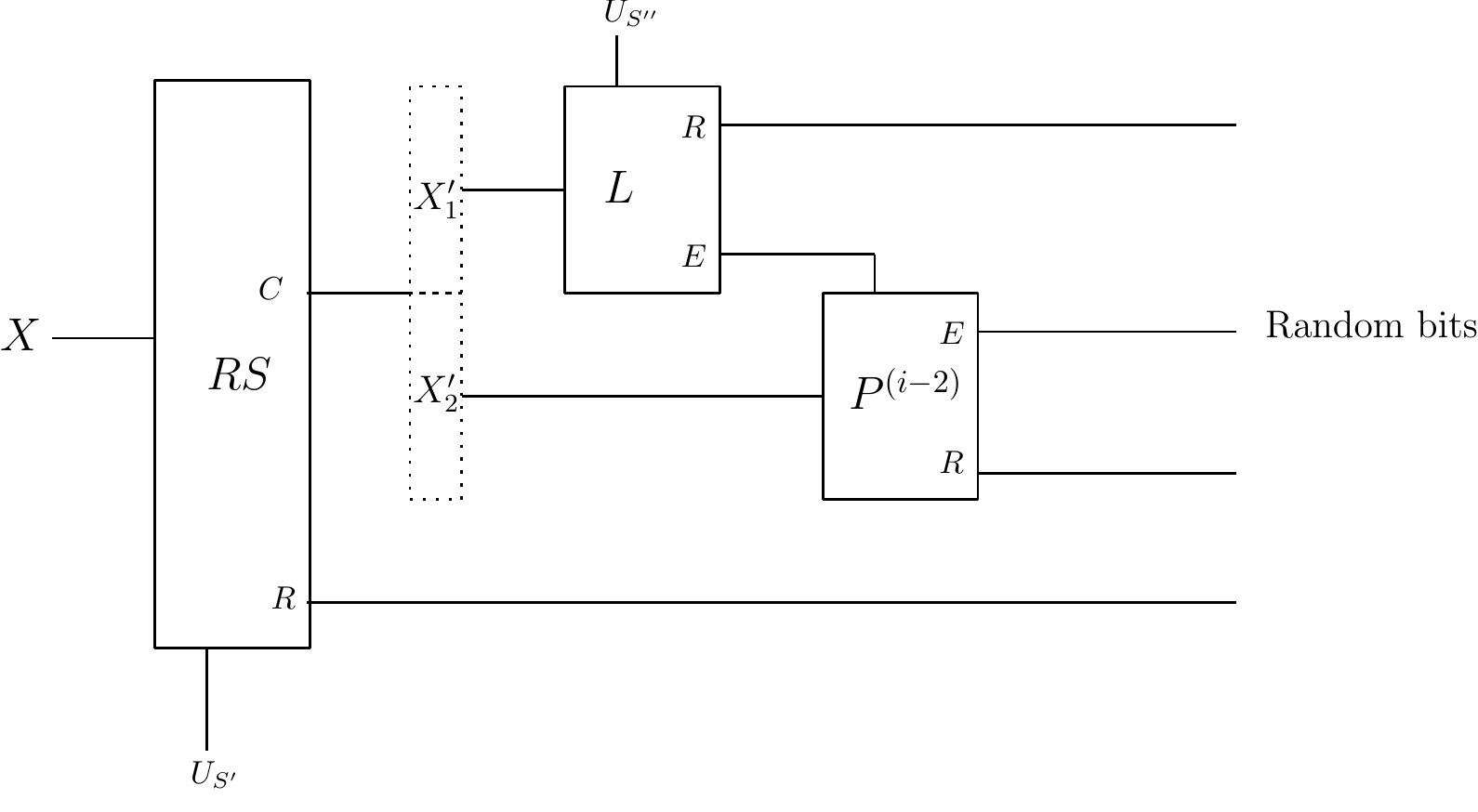}
\end{center}
\caption{The extractor $Q$ is obtained by first applying the condenser of Theorem \ref{thm:perm-condenser} and decomposing the output into two parts. The Leftover Hash Lemma extractor (Lemma \ref{lem:baseext}) is applied to the first half and its output is used as a seed for the extractor $\{P^{(i-2)}_y\}$ coming from the induction hypothesis.}
\label{fig:extractorQ}
\end{figure}
\end{center}

\begin{claim}[]
There exists an
\[
(n, 2^i \cdot 4.5d(\e_0)) \to_{5\e_0} 2^{i} \cdot d(\e_0)
\]
strong permutation extractor $\{Q_y\}_{y \in T}$ with seed size $\log |T| \leq \frac{d(\e_0)}{8}$. 
\end{claim}
To prove the claim, we start by applying the condenser of Theorem \ref{thm:perm-condenser} with $\alpha = 1/200$ and $\e = \e_0$ (so we use a seed of size $t(\e_0)$). The output $X'$ of size at most $2^i \cdot 4.5d(\e_0)$ is then $\e_0$-close to having min-entropy at least $(1-\alpha)2^i \cdot 4.5 d(\e_0) - t(\e_0)$. The entropy deficiency of this distribution is $\alpha 2^i \cdot 4.5 d(\e_0) + \frac{d(\e_0)}{200} \leq \frac{2^i \cdot 4.5d(\e_0)}{100}$. We then divide $X'$ into two equal blocks $X' = (X'_1, X'_2)$, and we know that it is $2 \e_0$ close to being a $2 \times k'$-source for
\[
k' = \frac{2^{i} \cdot 4.5 d(\e_0)}{2} - \frac{2^{i} \cdot 4.5 d(\e_0)}{100} - \log(1/\e_0) \geq \left(\frac{49}{100} \cdot 2^{i} \cdot 4.5 - \frac{1}{200}\right) d(\e_0)
\]
as $\log(1/\e_0) \leq t(\e_0) = \frac{d(\e_0)}{200}$.
For the extractors we will apply next to this source, we should note that $k' \geq 2d(\e_0)$ and that $2^{i} \cdot 4 d(\e_0) \leq k' < 2^{i} \cdot 8d(\e_0)$.

We now apply the extractor of Lemma \ref{lem:baseext} to $X'_1$ (viewed as a $2d(\e_0)$-source) using a seed of size $\frac{2d(\e_0)}{20}$ and obtaining $X''$ that is $\e_0$ close to uniform on $d(\e_0)$ bits. We then use the extractor $\{P^{(i-2)}_y\}$ obtained by induction for $i-2$ to the $X'_2$ (of size $2^i \cdot 4.5d(\e_0) \leq n$) with seed $X''$ (of size $d(\e_0)$): it is an $(n,2^{i-2} \cdot 8d(\e_0)) \to_{\e_0} 2^{i} \cdot d(\e_0)$ permutation extractor.

The construction is illustrated in Figure \ref{fig:extractorQ}. Note that the number of bits of the seed is $\log |T| \leq t(\e_0) + \frac{2d(\e_0)}{20} \leq \frac{d(\e_0)}{8}$. This concludes the proof of the claim.

\vfill

\begin{center}
\begin{figure}[h]
\begin{center}
\includegraphics[width=.7\textwidth]{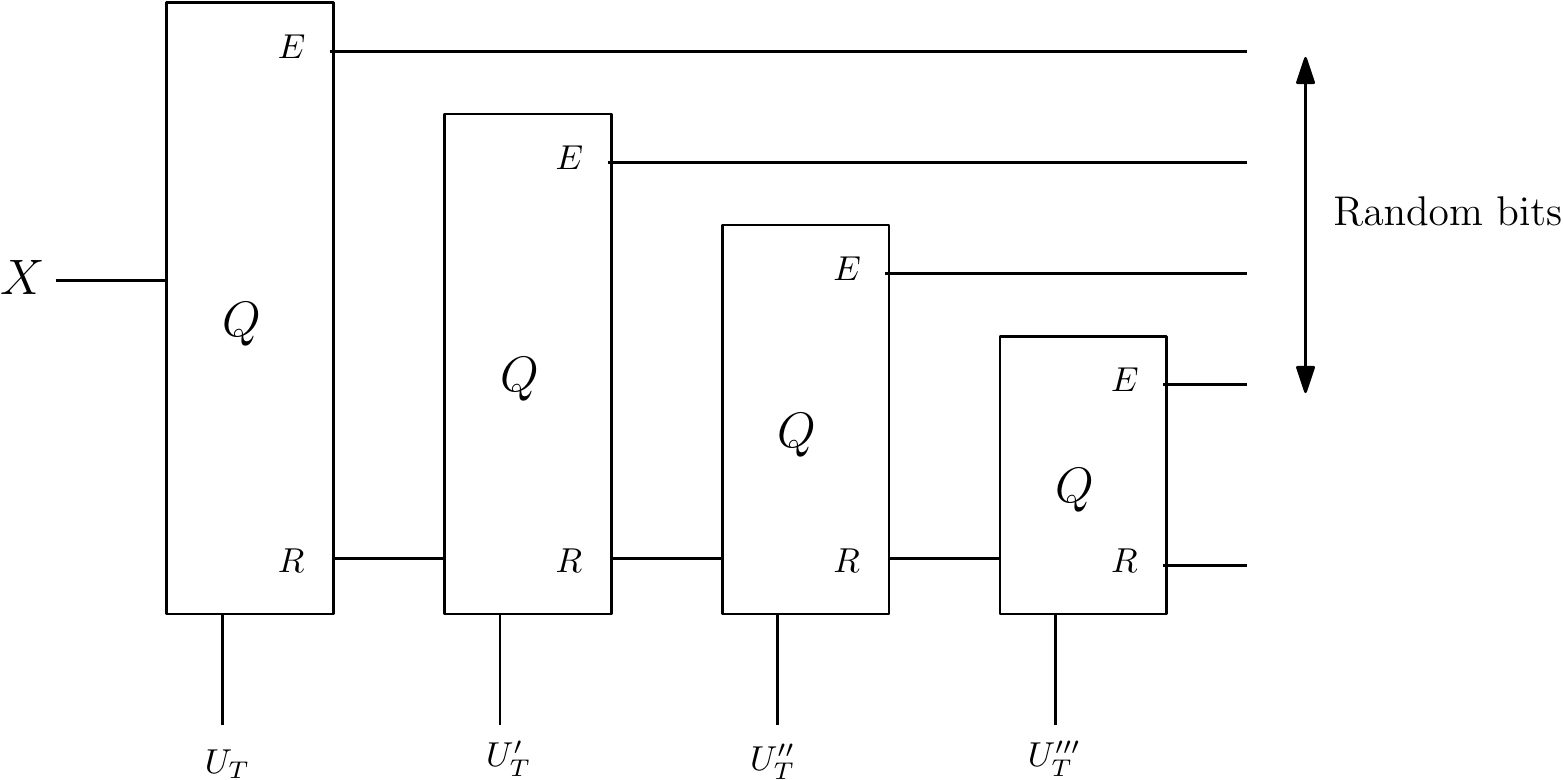}
\end{center}
\caption{The permutation extractor $\{Q_y\}$ described in the claim is applied four times with independent seeds in order to extract $2^{i-1} \cdot 8 d(\e)$ random bits.}
\label{fig:repeatextractor}
\end{figure}
\end{center}

The source $X$ we begin with is a $2^i \cdot 8 d(\e)$-source. But we have $2^i \cdot 8d(\e) \geq 2^{i} \cdot 8d(\e_0) - 2^i \cdot 8 \cdot 200^2 \log 20 \geq 2^{i} \cdot 4.5 d(\e_0)$ so that we can apply the permutation extractor $(Q_y)_{y \in T}$ of the claim. We obtain $Q^{E}_{U_T}(X)$ which is $\e_0$-close to $2^{i} \cdot d(\e_0)$ random bits. As $Q^E$ is part of a permutation extractor, the remaining entropy is not lost: it is in $Q^R_{U_T}(X)$. More precisely, applying Lemma \ref{lem:manyextractions}, we get $Q^R_{U_T}(X)$ is $\e_0$-close to a source of min-entropy at least $2^{i} \cdot 8d(\e) - 2^{i} \cdot d(\e_0) - 1$. As $2^{i} \cdot 8d(\e) - 2^{i} \cdot d(\e_0) - 1 \geq 2^i \cdot 4.5d(\e_0)$, we can apply the extractor $\{Q_y\}_{y \in T}$ of the claim to this source. Note that the input size has decreased but this only makes it easier to extract random bits as one can always encode in part of the input space. To apply $Q$, we use a fresh new seed that outputs a bit string that is close to uniform on $2^{i-3} \cdot 8d(\e_0)$ bits and the remaining entropy can be found in the $R$ register. We apply this procedure four times in total as shown in Figure \ref{fig:repeatextractor}. Note that the reason we can apply it four times is that at the last application $2^i \cdot 8d(\e) - 3 \cdot 2^{i-3} \cdot 8 d(\e_0) - 3 \geq 2^i \cdot 4.5d(\e_0)$. As the extractor $\{Q_y\}_{y \in T}$ has error at most $5\e_0$, the total error is bounded by $20\e_0 = \e$.

\comment{The reason all these inequalities hold is that $2^i \cdot 8d(\e) - 3 \cdot 2^{i-3} \cdot 8 d(\e_0) - 3 \geq 2^i \cdot 5d(\e_0) - 2^i \cdot 8 \cdot 200^2 \log 20$. In fact we want $\frac{24 n^2}{\e_0} \geq 20^{16}$, which is garenteed if $n \geq \sqrt{\frac{1}{20 \cdot 24} 20^{16}}$. So we basically want $n \geq 2 \cdot 10^6$.}

We thus obtain an
\[
(n, 2^{i} \cdot 8d(\e)) \to_{\e} 4 \cdot 2^{i-3} \cdot 8d(\e_0)
\]
strong permutation extractor with seed set $S = T^4$ so that $\log |S| \leq 4 \cdot \frac{d(\e_0)}{8} \leq d(\e)$.
This proves the induction step. To obtain the theorem, we simply choose the smallest $i$ such that $2^i \cdot 8 d(\e) \geq k$.
\comment{Why don't we use the same seed? Because it is not independent of the source}
\end{proof}

By a repeated application of the previous theorem, we can extract a larger fraction of the min-entropy.
\newtheorem*{thm-perm-extractor}{Theorem \ref{thm:perm-extractor}}
\begin{thm-perm-extractor}[]
For all (constant) $\delta \in (0,1)$, there exists $c > 0$, such that for all positive integers $n$, all $k \in [c \log(n/\e), n]$, and all $\e \in (0,1/2)$, there is an explicit $(n, k) \to_{\e} (1-\delta) k$ strong permutation extractor $\{P_y\}_{y \in S}$ with $\log |S| = O(\log(n/\e))$. Moreover, the functions $(x,y) \mapsto P_y(x)$ and $(x,y) \mapsto P^{-1}_y(x)$ can be computed by circuits of size $O(n \polylog(n/\e))$.
\end{thm-perm-extractor}
\begin{proof}
We start by applying the extractor of Theorem \ref{thm:perm-ext1}. We extract part of the min-entropy of the source and the remaining min-entropy is in the $R$ system (Lemma \ref{lem:manyextractions}). This min-entropy can be extracted using once again the extractor of Theorem \ref{thm:perm-ext1}. After $O(\log(1/\delta))$ applications of the extractor, we obtain the desired result.
\end{proof}

\section{Various technical results}\label{app:technical}

This section contains various technical results. We start by a lower bound on the key size for an encryption scheme.
\begin{proposition}
 \label{prop:lb-key-size-encryption}
Let $\cE : \{0,1\}^n \times [t] \to \cS(A)$ be an encryption scheme with the following properties: there exists a decoding map $\cD_k$ for every $k \in \{0,1\}^s$ such that $\cD_k(\cE(x,k)) = x$ and for all $x \neq x'$, we have
\begin{equation}
\label{eq:def-enc-scheme}
\tracedist{\frac{1}{t} \sum_k \cE(x,k), \frac{1}{t} \sum_k \cE(x',k) } \leq \e.
\end{equation}
Then, $\log t \geq n-2$ provided $\e \leq 1/2$.
\end{proposition}
\comment{The reason we don't get $n-1$ is because we only asked for a pair-wise small distance instead of having a small distance to some fixed guy. Actually, the bound we get is $2^s \geq (1-\e)2^n - 1$.}
\begin{proof}
The argument we use is quite similar to \citep[Theorem 6]{DD10}. First by averaging \eqref{eq:def-enc-scheme} over all $x'$, we obtain
\[
\tracedist{\rho^{XA}, \rho^X \otimes \rho^A } \leq \e + 2^{-n},
\]
where $\rho^{XKA} = \frac{1}{t \cdot 2^n} \sum_{x,k} \proj{x}^X \ox \proj{k}^K \ox \cE(x,k)^A$. Using the relation between the trace distance and fidelity (equation \eqref{eq:purifiedVStrace}), we get
\begin{equation}
\label{eq:lb-fidelity}
\fid{\rho^{XA}, \rho^X \otimes \rho^A} \geq 1-\e - 2^{-n}.
\end{equation}
Now, using the key $K$, one should be able to recover $X$ from $A$: this will allow us to get an upper bound on $\fid{\rho^{XA}, \rho^X \otimes \rho^A}$. Using Uhlmann's theorem (Theorem \ref{thm:uhlmann}), we can find a purification $\ket{\rho}^{XKAR}$ of $\rho^{XKA}$ and a purification $\ket{\sigma}^{XKAR}$ of $\rho^X \otimes \rho^A$ such that 
\begin{align*}
\fid{\rho^{XA}, \rho^X \ox \rho^A} &= \fid{\rho^{XKAR}, \sigma^{XKAR}} \\
						&\leq \fid{\rho^{XKA}, \sigma^{XKA}} \\
						&\leq \fid{\cD(\rho^{XKA}), \cD(\sigma^{XKA}) } \\
						&= \fid{\bar{\Phi}^{XX'}, \cD(\sigma^{XKA})}.
\end{align*}
Here, $\cD = \sum_k \proj{k} \ox \cD_k$ acts on $KA$ and $\bar{\Phi}^{XX'} = \frac{1}{2^n} \sum_x \proj{x} \ox \proj{x}$. The two inequalities follow from the monotonicity of the fidelity (equation \eqref{eq:monotonicity-pd}). The last equality comes from the fact that $\cD$ decodes $X$ correctly given $K$ and $A$. Note that we can assume that $\sigma^{XKA}$ is classical on the $XK$ system (otherwise, you can simply measure $XK$ in the computational basis and use the monotonicity of the fidelity). We can then write $\sigma^{XKA} = \frac{1}{2^n} \sum_x \proj{x} \ox \sigma_x^{KA}$. In this case $\fid{\bar{\Phi}^{XX'}, \cD(\sigma^{XKA})}$ is simply the probability of successfully guessing $X$ given the system $KA$ by applying $\cD$, the underlying state being $\sigma^{XKA}$. In fact, expanding the fidelity, we have
\begin{align*}
 \fid{\bar{\Phi}^{XX'}, \cD(\sigma^{XKA})} &=  \left\| \frac{1}{2^{n/2}} \sum_{x} \proj{x} \ox \proj{x} \cdot \frac{1}{2^{n/2}} \sum_x \proj{x} \ox \sqrt{\cD(\sigma^{KA}_x)} \right\|_1 \\
 			&= \frac{1}{2^n} \sum_x \left\| \proj{x} \cdot \sqrt{\cD(\sigma_x^{KA})} \right\|_1 \\
			&= \frac{1}{2^n} \sum_x \tr \left[ \sqrt{\proj{x} \cD(\sigma_x^{KA}) \proj{x} } \right] \\
			&= \frac{1}{2^n} \sum_x \bra{x} \cD(\sigma_x^{KA}) \ket{x} \\
			&\leq P_{\textrm{guess}}(X|KA)_{\sigma} = 2^{-\entHmin(X|KA)_{\sigma}},
\end{align*}
where we used the operational interpretation of the min-entropy \eqref{eq:pguess-hmin} in the last line.
Now using a chain rule for the min-entropy in \citep[Lemma 7]{DD10}, we have $\entHmin(X|KA)_{\sigma} \geq \entHmin(X|A)_{\sigma} - \log t = n- \log t$ (note that it is important here that $K$ is classical). Combining with \eqref{eq:lb-fidelity}, we get $1 - \e - 2^{-n} \leq 2^{n}/t$, which leads to the desired result.
\comment{
In the case where $\cE$ is classical, then we can consider the following simpler argument. Let $X$ be uniformly distributed on $\{0,1\}^n$ and $K$ uniformly distributed on $[t]$. Using condition \eqref{eq:def-enc-scheme}, we obtain an event $\textsf{E}$ such that $X, \cE(X,K) = X', \cE(X'', K')$ on event $\textsf{E}$, where $X'$ and $X''$ are independent and distributed as $X$. This means that $\entI(X; \cE(X,K)|\textsf{E}) = 0$. Thus, using the chain rule
\begin{align*}
\entI(X; \cE(X,K) K |\textsf{E}) &= \entI(X; \cE(X,K) |\textsf{E}) + \entI(X; K|\textsf{E} \cE(X,K)) \\
&\leq \entH(K) = \log t.
\end{align*}
At the same time, $\entI(X; \cE(X, K) K | \textsf{E}) \geq \entI(X; \cD_K(\cE(X,K))|\textsf{E}) = \entH(X|\textsf{E}) \geq n-1$.
}
\end{proof}

Next, we state the general decoupling result of \cite{Dup09, DBWR10} for exact unitary 2-designs.
\begin{lemma}[{\cite[Theorem 3.7]{Dup09}}]
\label{lem:2design}
Let $A=A_{1}A_{2}$, and consider the map $\cT_{A\rightarrow A_{1}}$ as defined in Equation~\eqref{eq:meas-map}. Then, if $\left\{U_1, \dots, U_{t}\right\}$ defines an exact unitary 2-design (Definition~\ref{def:two-design}), we have for $\delta\geq0$,
\begin{align}\label{eq:2design}
\frac{1}{t}\sum_{i=1}^{t}\left\|\cT_{A\rightarrow A_{1}}(U_{i}\rho_{AE}U_{i}^{\dagger})-\frac{\id_{A_{1}}}{|A_{1}|}\otimes\rho_{E}\right\|_{1}\leq\sqrt{\frac{d_{A_1}}{d_A}2^{-\hmin^{\delta}(A|E)_{\rho}}}+2\delta\ .
\end{align}
\end{lemma}



We also use the fact that a full set of MUBs defines a complex projective 2-design.
\begin{lemma}[\cite{KR05}]\label{lem:mub-design}
Let $\left\{U_{1},\dots,U_{d_A+1}\right\}$ define a full set of mutually unbiased bases of $A$. Then
\begin{align*}
\frac{1}{d_A(d_A+1)} \sum_{i=1}^{d_A+1} \sum_{a \in [d]} (U_i \proj{a} U_i^{\dagger})^{\ox 2} = \frac{2 \Pi^{\sym}}{d_A(d_A+1)},
\end{align*}
where $\Pi^{\sym}$ is the projector onto the symmetric subspace of $A \otimes A'$ (with $A' \isom A$) spanned by the vectors $\ket{aa'} + \ket{a'a}$ for $a,a' \in [A]$. Note that $\Pi^{\sym} = \frac{\id_{AA'} + F_{AA'}}{2}$.
\end{lemma}

The following well known \lq swap trick\rq~is used to prove decoupling statements.

\begin{lemma}\label{lem:swap-trick}
Let $M,N\in\cL(A)$. Then,
\begin{align*}
\tr [ MN ] = \tr [(M_A \ox N_{A'})F_{AA'}],
\end{align*}
where $A' \isom A$ and $F_{AA'} = \sum_{aa'} \ket{aa'}\bra{a'a}$ is the swap operator.
\end{lemma}

The following is called operator Chernoff bound.

\begin{lemma}[{\cite[Theorem 19]{AW02}}]
\label{lem:operator-chernoff}
Let $X_1, \dots, X_t$ be independent and identically distributed operator valued random variables and $0 \leq X_i \leq \id$, $\ex{X_i} = \Gamma \geq \alpha \id$. Then
\begin{align*}
\prob{\frac{1}{t} \sum_{i=1}^t X_i \leq (1+\eta) \Gamma} \geq 1 - d \exp{- \frac{t \eta^2 \alpha}{4 \ln 2}}.
\end{align*}
\end{lemma}

\bibliographystyle{abbrvnat}
\bibliography{ur}

\end{document}